\documentstyle[aps,eqsecnum,preprint,tighten,floats,graphicx,amsmath]{revtex}
\newcommand{\sfrac}[2]{\mbox{$\frac{#1}{#2}$}}
\newcommand{\half}{\mbox{$\frac{1}{2}$}}
\newcommand{\trace}{\mbox{Tr}}

\newcommand{\deriv}[2]{\frac{d{#1}}{d{#2}}}

\newcommand{\ket}[1]{|#1\rangle}
\newcommand{\bra}[1]{\langle#1|}

\newcommand{\braket}[2]{\langle#1|#2\rangle}
\newcommand{\Ham}{\mbox{$\cal H$}}

\newenvironment{aside}{\vspace*{1mm}\par\hrule width\textwidth\vspace*{1mm}\par \footnotesize}{\vspace*{1mm}\par\hrule width\textwidth\vspace*{1mm}\par}

\renewcommand{\vec}[1]{{\mathbf{#1}}}
\begin{document}

\title{Self-consistent theory of large amplitude collective motion:
Applications to approximate quantization of non-separable systems and
to nuclear physics}
\author{Giu~Do~Dang\footnote{
Email: { Giu.Dodang@th.u-psud.fr}}}
\address{Laboratoire de Physique Th\'eorique,     
B\^at 211, Universit\'e de Paris-Sud, 91405 Orsay, France}

\author{Abraham Klein\footnote{
Email: {aklein@walet.physics.upenn.edu}}}
\address{Department of Physics, University of Pennsylvania,
Philaldelpia, PA 19104-6396, USA}

\author{Niels~R.~Walet\footnote{Email: {Niels.Walet@umist.ac.uk}} }
\address{Department of Physics, UMIST, P.O. Box 88,
Manchester M60 1QD, UK}

\maketitle

\bigskip
\begin{abstract}

The goal of the present account is to review our efforts to obtain and
apply a ``collective'' Hamiltonian for a few, approximately decoupled,
adiabatic degrees of freedom, starting from a Hamiltonian system with
more or many more degrees of freedom.  The approach is based on an
analysis of the classical limit of quantum-mechanical problems.
Initially, we study the classical problem within the framework of
Hamiltonian dynamics and derive a fully self-consistent theory of
large amplitude collective motion with small velocities. We derive a
measure for the quality of decoupling of the collective degree of
freedom.  We show for several simple examples, where the classical
limit is obvious, that when decoupling is good, a quantization of the
collective Hamiltonian leads to accurate descriptions of the low
energy properties of the systems studied.  In nuclear physics problems
we construct the classical Hamiltonian by means of time-dependent
mean-field theory, and we transcribe our formalism to this case.  We
report studies of a model for monopole vibrations, of $^{28}$Si with a
realistic interaction, several qualitative models of heavier nuclei,
and preliminary results for a more realistic approach to heavy nuclei.
Other topics included are a nuclear Born-Oppenheimer approximation for
an {\em ab initio} quantum theory and a theory of the transfer of
energy between collective and non-collective degrees of freedom when
the decoupling is not exact.  The explicit account is based on the
work of the authors, but a thorough survey of other work is included.
\end{abstract}

{\it PACS number(s)}: 21.60.-n, 21.60.Jz, 21.60.Ev


\newpage

\tableofcontents
\newpage

\section{Introduction}\label{sec:1}

\subsection{Why was this review written?\label{sec:1.1}}

The account that follows is a review of work that was stimulated by a
search for a solution to the following problem: We are given a
non-relativistic nuclear many-body Hamiltonian.  We assume that the
chosen Hamiltonian is capable of describing the low-energy properties
of medium to heavy nuclei.  We then set ourselves the task of deriving
from this Hamiltonian a description of large amplitude collective
motion (defined below) without the intervention of any additional {\it
ad hoc} elements.  It turns out that the solution developed for this
problem has additional applications to molecular structure and
reactions, to the general problem of the approximate quantization of
non-separable systems, and to field theories with soliton solutions.
We shall also include examples of these additional applications in the
following review, except for the soliton problem, which would take us
too far afield.

For nuclear physics, the ``simplest'' applications of such a theory
are to the rotational-vibrational
motion of deformed nuclei and to fission, though applications to 
reaction and transport processes are also of great interest.  
Historically, it was the first of the problems cited
that called attention to the need for a deeper theory than
the one that was first applied (successfully) to its solution.  We refer here
to the work of Kumar and Baranger \cite{1a,1b,1c,1d,1e}, that we judge to be the first
successful calculations, starting from a many-body Hamiltonian,
of the low-energy rotations
and quadrupole vibrations of deformed nuclei.
The main features of this work are:
\begin{enumerate}
\item  The stable equilibrium shape and the energy of a deformed nucleus
are given by a Hartree-Bogoliubov mean-field calculation.
\item  The potential energy for non-equilibrium shapes is obtained by
solving a constrained Hartree-Bogoliubov problem, the constraining operator
chosen plausibly as the (mass) quadrupole moment operator of the model.
This calculation is not restricted to small values of the quadrupole
distortion.  This plus the fact that the procedure ``decouples'' the
quadrupole degree of freedom, a collective operator depending on all the 
coordinates of the nucleus, from the many-body Hamiltonian accounts for
the name large amplitude collective motion.
\item  To complete the derivation of a model Hamiltonian, one also
needs a kinetic energy.  In the adiabatic
(small velocity) approximation, to which this approach is limited, this means 
an expression quadratic in the velocities.  The mass coefficients, which
can also be rather general functions of the collective coordinates, are 
computed by the cranking method \cite{RS}.
\item  The classical Hamiltonian thus obtained is quantized (a not
ambiguity-free procedure) and the associated Schr\"odinger equation solved
for various medium to heavy nuclei.  Results are compared to
the spectroscopic data.
\end{enumerate}

The method of Kumar and Baranger, with improved Hamiltonians and improved
numerical algorithms, has continued to be used up to the
present, for instance in the analysis of the fission process \cite{2}.
What then is ``wrong'' with this method?  The glaring fault is that the
potential energy of the nucleus is determined from a constrained search
for a minimum of the mean-field energy,
with a constraining operator that is an {\em ad hoc} choice.  
Baranger and others (Ref.~\cite{no3} and the literature cited in Sec.~\ref{sec:7})
realized that
it would be of interest to construct a theory of large amplitude collective
motion free of this {\em ad hoc} feature.  In short, the nucleus should
decide internally, on the basis of the given microscopic Hamiltonian,
how to deform itself, rather than have this property imposed
from the outside.  We refer to a formulation with this property as a 
self-consistent theory of large amplitude collective motion.

We restrict the account that follows largely to the solution of this
problem developed by the writers and their collaborators
\cite{3,4,5,6,7,8,9,10,11,12,13,14,15,16,17,18,19,20,21,22,23,24,25,26,27,%
28,29,30,30a,31,31_2,31a,31b}.  Since we entered the arena only after
a full decade of effort by others, this decision requires
justification.  We shall, however, postpone a detailed account of this
early work, as well as other contemporaneous work until
Sec.~\ref{sec:7}, i.e., until after we finish the exposition of our
work.  Such a review may then be more meaningful to the reader.  We
certainly benefited from some of this earlier research, as we shall
try to make clear at appropriate junctures.  Nevertheless, it is our
judgment that compared to the existing literature, for the problem as
we have defined it, our method is simpler in concept and more complete
in execution than can be found elsewhere.  This makes it suitable for
a coherent account, designed to explain the basic ideas of the
subject.

In addition to developing a complete theory, we have also carried
out a number of applications, of which the most important are included
in this account.
Though we hope that the reader will be impressed by the theoretical
structure and by some of the new applications, it is also true, unfortunately,
that so far we have not gotten back
to the applications that stimulated this research in the first place.
However, a new effort in this direction has been undertaken \cite{31a,31b}.

\subsection{Elementary physical considerations concerning decoupled motion\label{sec:1.2}}

When we enter the main text, it will
be easy at times to lose sight of the simple physical ideas that underlie 
the whole
development.  We shall therefore present those ideas here in an irreducibly
simple setting.  In the following we shall, using elementary
examples, distinguish three kinds of motion, exactly separable motion
(which is {\em a fortiori} decoupled), exactly decoupled motion that does not
correspond to a separable Hamiltonian, and approximately decoupled motion.

We first consider the separable Hamiltonian
\begin{eqnarray}
H &=& H_1(q_1, p_1) +H_2(q_2, p_2), \label{eq:1:Intro_1} \\
H_i &=& \frac{1}{2}(p_i^2 +\omega_i^2 q_i^2) +\frac{1}{4}\lambda_i q_i^4.
\label{eq:1:Intro_2}
\end{eqnarray}
Classically this separable Hamiltonian allows decoupled
motion in the following sense.  Consider the equations of motion
\begin{equation}
\ddot{q}_i = -\omega_i^2q_i - \lambda_i q_i^3.   \label{eq:1:Intro_3}
\end{equation}
These admit, for example, the class of solutions $q_2(t)=0$ for
the initial conditions
$q_2(0)=\dot{q}_2(0)=0$.  The motions $q_1=q_1(t)$ for these initial 
conditions then define a
one-dimensional decoupled manifold.  This motion is along the valley $q_2=0$
of the two-dimensional potential energy surface. (The precise definition
of a valley will be given below.)

There is also a
decoupled solution with indices 1 and 2 interchanged.  However,
we restrict our
attention to low-energy motion and assume (adiabatic approximation)
$\omega_2^2\gg \omega_1^2$, a condition whose implications we assume not to be
invalidated by the quartic coupling.  As a consequence of these assumptions,
the ``high-energy'' decoupled manifold will be of no primary interest in any
of the following discussion.

We now consider the associated
quantum theory.  The wave functions have the form
\begin{equation}
\psi_{n_1 n_2} =\phi_{n_1}(q_1)\chi_{n_2}(q_2).  \label{eq:1:Intro_4}
\end{equation}
In the adiabatic approximation, the lowest-lying states will be a set
$(n_1,0)$, with the high frequency oscillator in its ground state ($n_2=0$)
.  If we
further recall that the length $1/\sqrt{\omega_2}$ (in suitable units)
determines the very narrow width of the function $\chi_0(q_2)$, we
see that for this low-energy spectrum, the wave-functions are almost
one-dimensional.  This corresponds to the classical notion of a
one-dimensional decoupled manifold.  As far as energy {\em differences} are
concerned, it suffices to quantize the Hamiltonian $H_1$ of the
decoupled manifold, as long as we do not excite the second
degree of freedom.

A further remark about this trivial example is that classically, small
perturbations of the motion away from the decoupled manifold are stable
in the sense that the fluctuations remain small in the course of time.
This also reflects itself in the quantum theory in that the corresponding
low-lying excited states remain
well-localized in the $q_2$ direction.

Next we consider a slightly more complicated case where the Hamiltonian
is non-separable, but the system still admits exactly decoupled motion.
To the previous Hamiltonian, we now add the term
\begin{equation}
\delta_1 H = \frac{1}{2}\lambda_{12}q_1^2 q_2^2,   \label{eq:1:Intro_5}
\end{equation}
($\lambda_{12}$ positive) leading to the equations of motion
\begin{eqnarray}
\ddot{q}_1 &=& -\omega_1^2 q_1 -\lambda_1 q_1^3 -\lambda_{12}q_1 q_2^2,
\label{eq:1:Intro_6} \\
\ddot{q}_2 &=& -\omega_2^2 q_2 -\lambda_2 q_2^3 -\lambda_{12}q_2 q_1^2.
\label{eq:1:Intro_7}
\end{eqnarray}
Note that these equations still admit the solution $q_2(t)=0$ for the
initial conditions $q_2(0)=\dot{q}_2(0)=0$.  Just as in the separable
case, the motion takes place along a valley of the potential energy
surface, as discussed further below.

For small $q_2\neq 0$, the equation of motion in the $q_2$ direction may
be written (ignoring the cubic term)
\begin{eqnarray}
\ddot{q}_2 &=& -\Omega_2^2(q_1) q_2,  \;\; \Omega_2^2(q_1) = \omega_2^2
+ \lambda_{12} q_1^2.     \label{eq:1:Intro_8}
\end{eqnarray}
This may be viewed as the equation of motion for an harmonic oscillator
with a slowly varying (local) frequency, since the $q_1$ motion is
supposed to
be slow compared to the $q_2$ motion.  We still expect stability with respect
to small excursions away from the collective manifold.

Intuitively we still expect the eigenfunctions for the low-energy states
of the quantized system to be approximately one dimensional.  What shall
we take as the approximate quantum Hamiltonian of such a state?  The first
answer is that we quantize the Hamiltonian that describes the motion
on the decoupled manifold, namely $H_1(p_1, q_1)$, just as in the separable
case.  An improvement, that will be necessary for some of
the applications to be presented in the main text, is to include the
quantum corrections due to the zero point oscillations associated with the  
$q_2$ motion.  For the separable case we obtain just an
additive constant. For the approximation described by Eq.~(\ref{eq:1:Intro_8}), on the
other hand, the zero-point energy is $(1/2)\Omega_2(q_1)$, which thus
modifies the potential energy of the effective one-dimensional Hamiltonian.

Before going on to the third example, where one has neither a
separable Hamiltonian nor an exactly decoupled manifold, but where one
would like to explore the concept of an approximately decoupled
motion, let us consider more precisely what is meant by a valley.  We
have used this term twice to characterize the decoupled manifolds
previously identified.  We wish to present a precise definition of
this concept for the following reason: For the first two cases the
decoupled manifold was a valley, a result that will later prove to be
general.  Even where there is no exactly decoupled manifold, for all
examples of interest to us there will nevertheless be a valley, or its
multi-dimensional generalization, that we shall put forward as the
domain on which there may be approximately decoupled motion.  It is
therefore of interest to learn how to recognize and extract
analytically such a domain, i.e., formalize our intuitive
understanding of the concept.

Consider a two-dimensional potential energy surface $V(q_1,q_2)$ and
an equipotential curve $V=\text{constant}$.  We move along this curve, at
each point take a step of fixed length orthogonal to the curve, and then
measure the magnitude of the gradient of the potential. A local
minimum for this gradient is called an element of a valley, and a
local maximum an element of a ridge.  For example, for the separable
potential considered above, the line $q_2=0$, the decoupled manifold
chosen, is a valley, and the line $q_1=0$ is a ridge.  This will be
shown analytically below.

The definition of the valley (or ridge) given above can be
formalized as the solution of the constrained variational problem,
\begin{equation}
\delta[(\nabla V)^2 -\Lambda V] = 0, \label{eq:1:Intro_10}
\end{equation}
namely, we seek stationary values of the square of the gradient of the
potential (square chosen for analytical convenience) for fixed values of the
potential ($\Lambda$ is a Lagrange multiplier).

We study Eq.~(\ref{eq:1:Intro_10}) for the potential
\begin{equation}
V(q_1,q_2) =\frac{1}{2}(\omega_1^2 q_1^2 +\omega_2^2 q_2^2 +
\lambda_{12}q_1^2 q_2^2).   \label{eq:1:Intro_11}
\end{equation}
and derive the equations
\begin{eqnarray}
[F_1(q_1,q_2) - \Lambda\Omega_1^2(q_2)]q_1 &=& 0,  \label{eq:1:Intro_12} \\
{}[F_2(q_1,q_2) - \Lambda\Omega_2^2(q_1)]q_2 &=& 0, \label{eq:1:Intro_13}
\end{eqnarray}
where it is of no particular interest to record the detailed forms of
$F_i$ and $\Omega_i^2$.  It suffices to observe that these equations have two
solutions, the valley $q_2=0$, which satisfies Eq.~(\ref{eq:1:Intro_13}), and the
ridge $q_1=0$, which satisfies Eq.~(\ref{eq:1:Intro_12}).  The associated Lagrange
multiplier is evaluated from the other of the pair of equations.  This
calculation substantiates assertions about the relationship of the
chosen decoupled manifold to a valley.

As our third and last example, we consider the problem of two harmonic
oscillators  with the coupling term
\begin{equation}
\delta_2 H = \beta q_1^2 q_2.   \label{eq:1:Intro_14}
\end{equation}
The equations of motion
\begin{eqnarray}
\ddot{q}_1 &=& -\omega_1^2 q_1 -2\beta q_1 q_2  ,  \label{eq:1:Intro_15} \\
\ddot{q}_2 &=& -\omega_2^2 q_2 -\beta q_1^2,    \label{eq:1:Intro_16}
\end{eqnarray}
still admit one decoupled manifold, $q_1=0$, but this corresponds to 
high energy motion and is not the physics we are after.
There is no decoupled motion corresponding to our previous valley $q_2=0$.
We see this from Eq.~(\ref{eq:1:Intro_16}); by setting $q_2(0)=\dot{q}_2(0)=0$,
we obtain $\ddot{q}_2(0) = -\beta q_1^2(0)$.

This is an example of the general physical situation.  Whereas there is no
exactly decoupled manifold, what one finds in cases of interest is that
there continues to be a valley, satisfying Eq.~(\ref{eq:1:Intro_10}),
where the simple valley $q_2=0$ is replaced by the curve
\begin{equation}
q_2=f(q_1).                 \label{eq:1:Intro_17}
\end{equation}
We shall not present the details of the calculation of the result
(\ref{eq:1:Intro_17}) for the example under study, since a number of such
calculations will
be presented in the main text.  The important point is that the general
theory to be developed suggests that the optimum choice of an approximately
decoupled manifold is the valley (and suitable multi-dimensional
generalizations) and that the appropriate classical Hamiltonian to
quantize is the value of the full Hamiltonian on the valley.  It follows
from Eq.~(\ref{eq:1:Intro_17}) that by applying formulas for the mass to be derived
in the main text this has the form
\begin{eqnarray}
2H &=& M(q_1)p_1^2 + \omega_1^2 q_1^2 +2\beta q_1^2 f(q_1),  \label{eq:1:Intro_18}\\
M(q_1) &=& \frac{1+ (f')^2}{(f')^2}, \;\; f'=\frac{df}{dq_1}.  \label{eq:1:Intro_19}
\end{eqnarray}
One can quantize Eq.~(\ref{eq:1:Intro_18}) and as will be seen in the examples worked
out in the main text, include corrections for quantum fluctuations
orthogonal to the valley. (We shall not discuss here the ambiguity in the
quantization of the kinetic energy.)

\begin{aside}
We should remark that for the specific example (\ref{eq:1:Intro_15})
and (\ref{eq:1:Intro_16}),
another definition of the collective manifold suggests itself, namely,
by setting the left hand side of Eq.~(\ref{eq:1:Intro_16}) to zero, we obtain
the equation
\begin{equation}
-\omega_2^2 q_2 = \beta q_1^2  \label{eq:1:Intro_20}
\end{equation}
as a definition of the approximately decoupled manifold.  This is admittedly
simpler than the exercise of computing the valley. The reason we have chosen
not to pursue this suggestion is that though it may be simple
to apply to systems
with a few degrees of freedom, it is more difficult to systematize it for the
nuclear many-body problem.
\end{aside}

The review that follows consists first of the development of a
theoretical framework that formalizes the (relatively) simple concepts
that we have discussed above.  Secondly, it reports a series of
successively more elaborate applications. Following that, we turn to
more fundamental questions regarding the foundations of the quantum
theory of large amplitude collective motion.  Finally we consider the
problem of exchange of energy between the collective and the
non-collective degrees of freedom in non-equilibrium processes.

\newpage
\section{Theory of decoupling}\label{sec:2}

\subsection{Formal theory of decoupled motion within a classical
Hamiltonian framework\label{sec:2.1}}

\subsubsection{Introduction\label{sec:2.1.1}}

In this section, we develop both a formal and several versions of a
practical theory of decoupled large amplitude collective motion for
Hamiltonian systems.  Though our ultimate interest was in formulating
such a theory for many-particle fermion systems, especially nuclei, we
believe that it was an important step to divorce the considerations
initially from the statistics of the particles involved and to
concentrate on the Hamiltonian features of the problem.  This emphasis
can be found in several brief accounts in the earlier literature
\cite{32,33}.  The dominant physics of large amplitude collective
motion corresponds to a well-defined classical limit of the quantum
Hamiltonian.  The methods and results developed in this section emerge
from a Hamiltonian framework utilizing the theory of canonical
transformations.  In Sec.~\ref{sec:4.1} we shall show how to adapt the
resulting theory to the nuclear many-body problem.

Given a classical many-particle Hamiltonian, $H$, that we wish to
investigate for large amplitude collective motion, the first point to
recognize is that the existence of such motions may not be manifest
from the given form of $H$.  (This is almost universally the case for
nuclear physics.) Thus the first task is to carry out a canonical
transformation to a coordinate system in which the division into
collective (small velocity, adiabatic) and non-collective (high
velocity) coordinates can be made.  The adiabatic approximation limits
the generality of the transformations required, but does not restrict
them to point transformations, as assumed in all work prior to ours
and in our earliest work.  In studying the canonical transformations
and the conditions for decoupling in the general case, where the mass
coefficients are functions of the coordinates, care must be taken with
the tensor structure of the space.  The considerations of this section
is a revised (and improved) version of some of the contents of 
Ref.~\cite{22}, which reference in turn draws
from Refs.~\cite{6,12,16}.

\subsubsection{Decoupling conditions under a point transformation (adiabatic
approximation) \label{sec:2.1.2}}
 
We study a classical system with $N$ canonical pairs $\xi^{\alpha}$ and
$\pi_{\alpha}$ (the ``single-particle coordinates'')
described by the Hamiltonian,
\begin{equation}
   H = \frac{1}{2} \pi_{\alpha} B^{\alpha \beta}
       (\xi)\pi_{\beta} + V(\xi). \label{eq:2:2.1}
\end{equation}
The dynamics is thus characterized by a point function $V$ and by the
reciprocal mass tensor $B^{\alpha \beta}$ that also plays the role of
metric tensor for the considerations that follow.  (We shall sometimes
omit the adjective reciprocal in future allusions to this tensor.)
It is assumed that the system described by (\ref{eq:2:2.1}) admits motions that
are exactly decoupled, as defined precisely below, motions that 
are thus fully characterized
by fewer than $N$ canonical coordinates and momenta.  Included in such
motions are those we call collective.  It is further assumed that the
original coordinates are not natural ones for the description of these
motions, but that a suitable set can be introduced by means of a
canonical transformation.  In this initial discussion, we limit
ourselves to the special case of point transformations.
We thus study mappings of the form
\begin{subequations}
\begin{equation}
\xi^{\alpha} = g^{\alpha} (q^{1},...,q^{N}),    \label{eq:2:2.2}
\end{equation}
with inverse
\begin{equation}
q^{\mu} = f^\mu(\xi^1,...,\xi^N) .   \label{eq:2:2.3}
\end{equation}
\end{subequations}
The corresponding momenta are given by the formulas
\begin{eqnarray}
&&\pi_{\alpha} = f^{\mu}_{,\alpha}p_{\mu}, 
\;\;   p_{\mu} = g^{\alpha}_{, \mu} \pi_{\alpha},    \label{eq:2:2.5}
\end{eqnarray}
where the comma indicates a partial derivative.  The indices $\alpha,\;
\beta,...$ will enumerate the initial coordinates, and $\lambda,\;\mu,
...$ the final coordinates, although each set has the same range
$1,...,N$.
 
The point transformation (\ref{eq:2:2.2})
is the only canonical transformation that
exactly preserves the quadratic dependence
on the momenta.  It will be evident from the next section, however,
that for small collective momenta there exists a generalization of
(\ref{eq:2:2.2}) that approximately conserves the quadratic momentum
dependence.  We have chosen to treat the case of point transformations
separately, at least initially, because of the relative simplicity of
this case, and because most (but not all) serious applications 
carried out up to the present
are done within this framework. The
restriction of the original Hamiltonian to the form (2.1) and the
limitation to point transformations can justifiably be
designated as an adiabatic approximation.  However, this is not the most
general definition of adiabatic approximation to be utilized in this review.
For the latter, we start with a Hamiltonian with arbitrary dependence
on its momenta, but after carrying out the canonical transformation
designed to effect
the decoupling, we expand the new Hamiltonian to second order
in the collective momenta.
 
We assume that in the new coordinates we can identify a
decoupled surface, defined as follows: We divide the set $q^\mu$
into two subsets, $q^{i}, i=1...K,$ and $q^{a}, a=K+1...N$, and suppose this
division to be such that if at time $t=0$ both $q^{a}=0$ (by convention) and
$\dot{q}^{a}=0$, then $q^{a}(t)= 0$.  Such motions evolve on a
K-dimensional submanifold
\begin{equation}
     \xi^{\alpha} = g^{\alpha} (q^{1}...q^{K},0,...,0)
     \equiv g^{\alpha}(q),         \label{eq:2:2.6}
\end{equation}
designated as the surface $\Sigma $.  In geometrical terms, if
the system point is initially on $\Sigma $, and the initial velocity is on
$T \Sigma$, the tangent plane to $\Sigma $ at the given point, then
provided the subsequent motion of the system is confined to this surface,
$\Sigma$ is said to be decoupled.  It may be useful to imagine that
the system has a self-imposed set of holonomic constraints.
 
The main practical
aim of the research that follows is to develop methods of
discovering such manifolds or, more realistically, of finding manifolds
that are {\em approximately} decoupled.  For this
enterprise, it is essential to distinguish between the
totality of conditions that must be satisfied for exact decoupling
to occur and
the formulation of an algorithm for determining manifolds that are
candidates for approximate decoupling (based on some convenient
subset of this totality).  Such an algorithm should then include some
test of the extent of violation of the remaining conditions. The exact
conditions derived below are meant to determine the functions (\ref{eq:2:2.6}),
which are the defining equations of the collective submanifold and,
at the same time, can be viewed as the restriction of a point canonical
transformation to the surface $\Sigma$.
As remarked below, they determine a dynamics on this manifold.  In
some cases where the decoupling is almost exact, one may consequently
obtain a
rather accurate description of the collective motion without having
to be concerned with the non-collective coordinates.  In other cases
coupling to the non-collective coordinates becomes essential,
and one must thus seek a full point transformation (\ref{eq:2:2.2}),
or at least extend
(\ref{eq:2:2.6}) to some immediate neighborhood of the collective manifold.
We shall nevertheless,
in what follows immediately, restrict considerations to the collective
manifold itself.  The necessary extensions will be considered 
in Sec.~\ref{sec:2.2.4}.
 
Before deriving the conditions that characterize decoupled motion,
let us note that under the point transformation (\ref{eq:2:2.2}) and
(\ref{eq:2:2.3}), the Hamiltonian becomes
\begin{equation}
H(\xi,\pi) = \bar{H}(q,p) = \frac{1}{2} p_{\mu} \bar{B} ^{\mu \nu}(q)
                           p_{\nu}+ \bar{V}(q),          \label{eq:2:2.7}
\end{equation}
where
\begin{equation}
\bar{B} ^{\mu \nu} = f^{\mu}_{,\alpha}B^{\alpha \beta} f^{\nu}_{,\beta}
                                                         \label{eq:2:2.8}
\end{equation}
transforms like a tensor of second rank.
Also, note that of the chain rule relations
\begin{eqnarray}
(\partial\xi^\alpha/\partial\xi^\beta)= g^{\alpha}_{,\mu}f^{\mu}_{,\beta}
= \delta^{\alpha}_{\beta},        \nonumber\\
(\partial q^\mu/\partial q^\nu) = f^{\mu}_{,\alpha} g^{\alpha}_{,\nu}
= \delta^{\mu}_{\nu},             \label{eq:2:2.9}
\end{eqnarray}
(which also represent orthonormalization conditions for a complete set of
basis vectors), the first
permits the  re-expression of (\ref{eq:2:2.8}) as
\begin{equation}
\bar{B}^{\mu\nu} g^{\alpha} _{,\nu} = B^{\alpha\beta} f^{\mu}_{,\beta}.
                                                        \label{eq:2:2.10}
\end{equation}
Equations (\ref{eq:2:2.9}) are the residue for point transformations
of the canonicity conditions,
that require the Poisson or Lagrange brackets of the
new coordinates with respect to the old to have values appropriate to
canonical pairs, e.g., with braces referring to Poisson brackets,
\begin{eqnarray}
\{q^\mu,p_\nu\} & \equiv& \frac{\partial q^\mu}{\partial\xi^\alpha}
                         \frac{\partial p_\nu}{\partial\pi_\alpha}-
                         \frac{\partial q^\mu}{\partial\pi_\alpha}
                         \frac{\partial p_\nu}{\partial\xi^\alpha}=
                         \delta^\mu_\nu, \nonumber\\
\{q^\mu,q^\nu\}&=&\{p_\mu,p_\nu\}=0. \label{eq:2:2.38}
\end{eqnarray}
 
The conditions that characterize $\Sigma$ are derived  most readily from
the equations of motion for the set $q^{a}, p_{a}$, the canonical pairs
whose values are frozen on this surface.  These equations are
\begin{subequations}
\begin{eqnarray}
         \dot{q}^{a}& =& \partial \bar{H}/ \partial p_{a}
                     = \bar{B}^{ai} p_{i} + \bar{B}^{ab}p_{b} ,
\label{eq:2:2.11}\\
         -\dot{p}_{a} & = & \partial \bar{H}/\partial q^{a}
= \bar{V}_{,a} + \frac{1}{2}\bar{B}^{ij}_{,a} p_{i} p_{j}
+ \bar{B}^{bi}_{,a}p_{i} p_{b} +
\frac{1}{2}\bar{B}^{bc}_{,a}p_{b} p_{c} . \label{eq:2:2.12}
\end{eqnarray}
\end{subequations}
 
The requirements $\dot{q}^{a} = \dot{p}_{a} = 0$ can be compatible with
the requirements $q^{a} = p_{a} = 0$ only if the equations
\begin{subequations}
\begin{eqnarray}
\bar{B} ^{ai} p_{i}&=&0 ,  \label{eq:2:2.13}  \\
\bar{V}_{,a} + \frac{1}{2} p_{i} p_{j} \bar{B} ^{ij} _{,a}&=&0,
\label{eq:2:2.14}
\end{eqnarray}
\end{subequations}
are satisfied, as one sees from (\ref{eq:2:2.11}) and (\ref{eq:2:2.12}).
Equations (\ref{eq:2:2.13}) and (\ref{eq:2:2.14}) are equivalent to three
sets of conditions provided none of the $p_{i}$ are constants of the
motion, for in that case (\ref{eq:2:2.14}) yields two independent
conditions, and altogether we have
\begin{subequations}
\begin{eqnarray}
\bar{B} ^{ai}&=&0 ,   \label{eq:2:2.15}    \\
\bar{V} _{,a}&=&0,   \label{eq:2:2.16}     \\
\bar{B} ^{ij} _{,a}&=&0 .   \label{eq:2:2.17}
\end{eqnarray}
\end{subequations}
The modifications necessary when one or more of the $p_{i}$ is a
constant of the motion will be considered in Sec.~\ref{sec:2.2.5}.
For many-body applications, such cases are of fundamental interest.
 
The physical significance of the decoupling conditions is apparent.
The first tells us that the mass tensor must be block diagonal.
Since, in general, we deal with real, symmetric, positive-definite
mass tensors, this is no real restriction.
The remaining two equations then demand the absence of both ``real'' and
``geometrical'' (centripetal)
 forces orthogonal to the decoupled surface.  These conditions also
imply that an exactly decoupled surface is geodesic, as we shall prove
in Sec.~\ref{sec:2.1.5}.
 
It follows readily from the decoupling conditions that the Hamiltonian
that governs the motion on $\Sigma$, the ``collective'' Hamiltonian, is
the value of $\bar{H}$, Eq.~(\ref{eq:2:2.7}), on the surface. A proof of this
assertion can be found in Sec.~\ref{sec:2.1.4}.
 
Equations (\ref{eq:2:2.15})-(\ref{eq:2:2.17}) are the most transparent form of the
decoupling conditions, and in cases of exact decoupling can
be used to check that exact solutions have been found.
Though it may be feasible to develop an algorithm for obtaining
approximate solutions directly from these conditions, the methods that
have actually been developed depend on the
transformation of (\ref{eq:2:2.15})-(\ref{eq:2:2.17}) into several equivalent sets
described in Secs.~\ref{sec:2.2} and \ref{sec:2.3}.
 
A first stage of transformation is to replace
(\ref{eq:2:2.15}--\ref{eq:2:2.17}) by the equivalent set,
\begin{subequations}
\begin{eqnarray}
B^{\alpha \beta} f^{i} _{,\beta}&=&\bar{B} ^{ij}g^{\alpha} _{,j},
\label{eq:2:2.18}\\
{V} _{,\alpha}&=&\bar{V} _{,i} f^{i} _{,\alpha},
\label{eq:2:2.19}\\
\bar{B} ^{ij} _{,\alpha}&=&\bar{B} ^{ij} _{,k} f^{k} _{,\alpha}.
\label{eq:2:2.20}
\end{eqnarray}
\end{subequations}
Of these relations, (\ref{eq:2:2.19}) and (\ref{eq:2:2.20}) are chain rule
relations that have been simplified by the imposition of the decoupling
conditions (\ref{eq:2:2.16}) and (\ref{eq:2:2.17}),
respectively, whereas Eq.~(\ref{eq:2:2.18}) is a simplified version of
Eq.~(\ref{eq:2:2.10}),
obtained by remembering the block-diagonality of the mass tensor.
Geometrically, (\ref{eq:2:2.18}) states that the
quantities $g^{\alpha}_{,i}$ and $f^{i}_{,\alpha}$ are equivalent sets
of basis vectors for $T\Sigma$, and (\ref{eq:2:2.19}), e.g.,
affirms that the gradient of $V$ lies in $T\Sigma$.\\
 
\subsubsection{Extended adiabatic approximation \label{sec:2.1.3}}
 
In practical applications one often approximates the initial Hamiltonian
to obtain
a kinetic energy quadratic in the momenta.  (In conjunction with the
point canonical transformation, this has been defined, tentatively,
as the adiabatic approximation.)  In such a case, a point transformation
on the approximate Hamiltonian gives a different result than what is
found by first performing a general canonical transformation
on the exact Hamiltonian and then taking the adiabatic limit after
this transformation.  To understand this remark, we generalize the point
transformation (\ref{eq:2:2.2}) and (\ref{eq:2:2.3}) to the forms
\begin{subequations}
\begin{eqnarray}
 \xi^\alpha&=&g^{\alpha}(q) +
    \frac{1}{2} g^{(1)\alpha\mu\nu}(q)p_\mu p_\nu
              + {\cal O}(p^4),\label{eq:2:2.21}\\
\pi_\alpha&=&f^{\mu}_{,\alpha}p_\mu + \frac{1}{3!}F^{(1)\mu \nu
\lambda}_{\alpha}p_{\mu}p_{\nu}p_{\lambda} + {\cal
O}(p^5).\label{eq:2:2.22}
\end{eqnarray}
\end{subequations}
The inverse equations are
\begin{subequations}
\begin{eqnarray}
     q^{\mu}&=&f^{\mu}(\xi) +\frac{1}{2}f^{(1)\mu\alpha\beta}\pi_{\alpha}
    \pi_{\beta} + {\cal O}(\pi^4),\label{eq:2:2.23} \\
     p_{\mu}&=& g^{\alpha}_{,\mu}\pi_\alpha + \frac{1}{3!}G^{(1)\alpha
     \beta\gamma}_{\mu}\pi_\alpha\pi_\beta\pi_\gamma + {\cal O}(\pi^5).
     \label{eq:2:2.24}
\end{eqnarray}
\end{subequations}
In fact the terms cubic in the momenta do not play a role in the
modification of the results of the previous subsection and are included
only to point out, below, that they are determined
by the quadratic terms.

\begin{aside} 
 Of the four new sets of functions introduced in the above 
equations, only one
is independent.  What follows is a concise demonstration of this
assertion that depends on studying the canonicity conditions as
power series in the momenta.  In presentation of the results that follow,
it is understood that each condition studied is taken to yield a relation
at the first order in which it is not trivial, and that we never go beyond
terms of second order in momentum.
We rely in part on a form of the canonicity equations,
\begin{eqnarray}
&&\frac{\partial\xi^\alpha}{\partial q^\mu}=\frac{\partial p_\mu}
          {\partial\pi_\alpha},  \;\;\;
\frac{\partial\pi_\alpha}{\partial q^\mu} =-\frac{\partial p_\mu}
         {\partial\xi^\alpha}, \label{eq:2:2.25}
\end{eqnarray}
that can be deduced by comparison of the Poisson bracket conditions
(\ref{eq:2:2.38}) 
with the chain rule for partial differentiation in {\em phase space}.
We first note that from the latter it follows, according to the specified
conditions of reasoning,
that Eqs.~(\ref{eq:2:2.9}), the orthonormalization conditions
for the basis vectors in configuration space, continue to hold.
Next, from the first of Eqs.~(\ref{eq:2:2.25}), with the aid of
(\ref{eq:2:2.9}), we can derive
\begin{subequations}
\begin{equation}
G^{(1)\alpha\beta\gamma}_\mu=g^{(1)\alpha\nu\lambda}_{,\mu}
g^\beta_{,\nu}g^\gamma_{,\lambda}, \label{eq:2:2.27}
\end{equation}
and by interchanging old and new coordinates in (\ref{eq:2:2.25}), we find
equally
\begin{equation}
F^{(1)\mu\nu\lambda}_\alpha=f^{(1)\mu\beta\gamma}_{,\alpha}
f^\nu_{,\beta} f^\lambda_{,\gamma}. \label{eq:2:2.28}
\end{equation}
\end{subequations}
By requiring, finally,
that the application of a transformation followed by its
inverse should be equivalent to the identity transformation,
we deduce the relation
\begin{equation}
g^{(1)\alpha\nu\lambda}=-f^{(1)\mu\beta\gamma}g^\alpha_{,\mu}
f^\nu_{,\beta}f^\lambda_{,\gamma}. \label{eq:2:2.29}
\end{equation}
{}From (\ref{eq:2:2.27})-(\ref{eq:2:2.29}) we thus see
that only one of the four new functions
appearing in the generalized transformation
(\ref{eq:2:2.21})-(\ref{eq:2:2.24}) is independent.
\end{aside}
 
Returning now to the primary object of this subsection,
insertion of the transformation (\ref{eq:2:2.21}) and (\ref{eq:2:2.22}) into
the Hamiltonian (\ref{eq:2:2.1}) and neglect of
terms of higher than second order in the momenta cause
the transformed Hamiltonian to take the form
\begin{eqnarray}
\bar{H}(q,p)&=&\frac{1}{2} p_\mu \left(
      f^{(0)\mu}_{,\alpha} B^{\alpha\beta}(g(q))\,f^{(0)\nu}_{,\beta} +
 V_{,\gamma}(g(q))\,g^{(1)\gamma\mu\nu}\right) p_\nu  + V(g(q))
\nonumber \\
&\equiv&\frac{1}{2} p_\mu \bar{B}^{\mu\nu} p_\nu +\bar{V}(q).  \label{eq:2:2.31}
\end{eqnarray}
This establishes the main point concerning the
admission of extended transformations.  The result is that one encounters
an altered definition
of $\bar{B}^{\mu\nu}$, as discussed more fully below.  
 
It follows from the transformed Hamiltonian
(\ref{eq:2:2.31}) that the decoupling conditions (\ref{eq:2:2.15})-(\ref{eq:2:2.17})
are formally unmodified.  The same is not the case for the alternative
versions (\ref{eq:2:2.18})-(\ref{eq:2:2.20}). Though (\ref{eq:2:2.19}) and
(\ref{eq:2:2.20}) are formally
unchanged, the derivation of the mass condition
(\ref{eq:2:2.18}) has to be reconsidered.
The transformation of the metric tensor embodied
in (\ref{eq:2:2.31}) can be written in the form
\begin{subequations}
\begin{eqnarray}
\bar{B}^{\mu\nu}& =&f^\mu_{,\alpha}\tilde{B}^{\alpha\beta}f^\nu_{,\beta},
\label{eq:2:2.32}  \\
\tilde{B}^{\alpha\beta} &=& B^{\alpha\beta}+\tau^{\alpha\beta\gamma}
V_{,\gamma}, \label{eq:2:2.33} \\
\tau^{\alpha\beta\gamma} &=& g^{(1)\gamma\mu\nu}
g^{(0)\beta}_{,\mu}g^{(0)\alpha}_{,\nu} = -f^{(1)\mu\beta\gamma}
g^\alpha_{,\mu}.  \label{eq:2:2.30}
\end{eqnarray}
\end{subequations}
It follows that Eq.~(\ref{eq:2:2.18}) is supplanted by the relation
\begin{equation}
\tilde{B}^{\alpha\beta}f^i_{,\beta}=\bar{B}^{ij}g^\alpha_{,j}.
\label{eq:2:2.34}
\end{equation}
The replacement of the mass tensor $B$ by $\tilde{B}$ pinpoints the
major difference between the extended theory and that based purely
on point transformations.

The theory just presented will, in general, be more difficult to
implement than the theory of point transformations.  We shall find, however,
that for a class of non-trivial exactly solvable models, presented in
Sec.~\ref{sec:4.2}, the use of the extended theory is essential.
 
\subsubsection{Alternative equivalent formulations  \label{sec:2.1.4}}
 
We describe briefly two other arguments that lead to the
decoupling conditions in one or the other of the forms given above.
First suppose we expand the transformed Hamiltonian $\bar{H}(q,p)$,
Eq.~(2.7), about
its value on the (decoupled) surface $\Sigma$,
\begin{eqnarray}
\bar{H}&=&\left.\bar{H}\right|_\Sigma + 
\left.\frac{\partial\bar{H}}{\partial  q^a}\right|_\Sigma\, q^a +
\left.\frac{\partial\bar{H}}{\partial p_a}\right|_\Sigma\, p_a +
\text{quadratic terms} \nonumber\\
& =& \bar{H}|_\Sigma +
\Bigl(\bar{V}_{,a}+\frac{1}{2}p_i p_j
\bar{B}^{ij}_{,a}\Bigr)q^a + \bar{B}^{ia}p_i p_a + ...\;\;.\label{eq:2:2.35}
\end{eqnarray}
By comparison of this expression with the decoupling conditions
(\ref{eq:2:2.13}) and (\ref{eq:2:2.14}), we see that these follow from
the requirement that the coefficients of $q^a$ and $p_a$ vanish.
Thus, the decoupling conditions are equivalent to the requirement that
$\bar{H}$ be stationary with respect to small variations of the coordinates
and momenta perpendicular to $\Sigma$.  This approach also
calls one's attention to the importance of the quadratic terms
that determine whether and to what extent a decoupled motion
is locally stable.  We shall return to this point in Sec.~\ref{sec:2.2.4}.
 
A second approach, that yields the decoupling conditions directly in the
alternative
form (\ref{eq:2:2.18})-(\ref{eq:2:2.20}), is to study the expression of
$\dot{\xi}^\alpha$
and of $\dot{\pi}_\alpha$ in the equivalent forms
\begin{subequations}
\begin{eqnarray}
\left.\frac{\partial H}{\partial\pi_\alpha}\right|_\Sigma&=&\{\xi^\alpha ,
\bar{H}(q^i ,p_i)\}, \label{eq:2:2.36}\\
-\left.\frac{\partial H}{\partial\xi^\alpha}\right|_\Sigma& =&\{\pi_\alpha ,
\bar{H}(q^i ,p_i)\}, \label{eq:2:2.37}
\end{eqnarray}
\end{subequations}
where the braces define a Poisson bracket with respect to
the new coordinates.
In these expressions one evaluates the left hand sides in general
and subsequently specializes to values on $\Sigma$.  This evaluation
thus contains no information about decoupling. Upon expansion in the
collective momenta, what thus emerges
are the left hand sides of (\ref{eq:2:2.18})-(\ref{eq:2:2.20}).
On the right hand
sides, by contrast, one is instructed to restrict $\bar{H}$ to the
surface before evaluating the Poisson bracket.
This incorporates the
assumption that the Hamiltonian thus restricted is the time-development
operator on the surface, and as previously implied, is an
equivalent expression of the decoupling conditions.  This is verified
when the corresponding evaluation
yields the right hand sides of (\ref{eq:2:2.18})-(\ref{eq:2:2.20}).
 
\subsubsection{Decoupled motion as motion on a geodesic surface\label{sec:2.1.5}}
 
In this section, we present a proof that a decoupled surface is a
geodesic.  We recall the definition of a geodesic surface.
Suppose that (locally) we have coordinates parameterizing
the surface as well as a set of orthogonal coordinates completing the
specification of the full space. 
The formula for the surface area, $S_K$, of the $K$-dimensional surface $S_K$
with boundary $\partial S_K$ is \cite{Eis}
\begin{eqnarray}
S_K &=& \int \sqrt{D}\, dq^1.. dq^K , \label{eq:2:surfaceA}  \\
D &=& {\rm det}(\bar{B}^{ij}), \label{eq:2:D}
\end{eqnarray}
where $\bar{B}$ is the metric on the surface. (Note that in order
to write this equation we require block-diagonality of the
metric, $B_{ia} = 0$.)
 
A geodesic surface is defined as having minimal surface for fixed
boundary. We prove below that $\delta S_K  = 0$ implies
\begin{equation}
 \bar{B}^{ij}(g^\beta_{,ij}+\Gamma^\beta
_{\gamma\alpha}g^\gamma_{,i}g^\alpha_{,j}-\bar{\Gamma}^l_{ij}g^\beta_{,l})=0.
\label{eq:2:5.12}
\end{equation}
These equations are linear combinations of the equations
\begin{equation}
g^\beta_{,ij}+\Gamma^\beta
_{\gamma\alpha}g^\gamma_{,i}g^\alpha_{,j}-\bar{\Gamma}^l_{ij}g^\beta_{,l}=0,
\label{eq:2:5.13}
\end{equation}
whose significance will now be established.
In the above equations, the affine connections (Christoffel symbols)
$\Gamma$ and $\bar{\Gamma}$ are defined as
\begin{subequations}
\begin{eqnarray}
\Gamma^\alpha_{\beta \gamma}&=&\frac{1}{2}{B}^{\alpha\delta}
({B}_{\delta\beta,\gamma}+{B}_{\delta\gamma,\beta}-
{B}_{\beta\gamma,\delta}),
\label{eq:2:5.afc1} \\
\bar{\Gamma}^\lambda_{\mu \nu}&=&\frac{1}{2}\bar{B}^{\lambda \kappa}
(\bar{B}_{\kappa\mu,\nu}+\bar{B}_{\kappa\nu,\mu}-\bar{B}_{\mu\nu,\kappa})
.\label{eq:2:5.afc2}
\end{eqnarray}
\end{subequations}

We turn to the derivation of (\ref{eq:2:5.13}).
{}From the definitions of the affine connections,
(\ref{eq:2:5.afc1}) and (\ref{eq:2:5.afc2}) one can show that
their standard transformation\cite{Eis}
properties are expressed by the equations
\begin{subequations}
\begin{equation}
g^\beta_{,\mu\nu}+\Gamma^\beta
_{\gamma\alpha}g^\gamma_{,\mu}g^\alpha_{,\nu}-
\bar{\Gamma}^\kappa_{\mu\nu}g^\beta_{,\kappa}=0.       \label{eq:2:5.13a}
\end{equation}
{}From this equation one can derive the inverse transformation
\begin{equation}
f^\lambda_{,\beta\gamma}-\Gamma^\alpha_{\beta\gamma}f^\lambda_{,\alpha}
+\bar{\Gamma}^\lambda_{\mu\nu}f^\mu_{,\beta}f^\nu_{,\gamma}=0.
\label{eq:2:5.13b}     \end{equation}
\end{subequations}
{}From Eq.~(\ref{eq:2:5.13a}) we can verify Eq.~(\ref{eq:2:5.13}) if the condition
$\bar{\Gamma}^a_{ij} = 0$ holds.
If we take the block-diagonality of the metric for granted, the condition
$\bar{\Gamma}^a_{ij}=0$ implies and is conversely implied by the
third decoupling condition, $\bar{B}^{ij}_{,a} = 0$, as follows from 
(\ref{eq:2:5.afc2}).
Equation (\ref{eq:2:5.13a}) is thereby reduced
to the required form (\ref{eq:2:5.13}).
It then follows that any surface on which the geometrical forces
normal to the surface vanish is a geodesic.                                           
If the space is Euclidean, exactly decoupled surfaces are planes.

\begin{aside}
 We give here a proof of Eq.~(\ref{eq:2:5.12}).
The Euler-Lagrange equations for the minimization of the area
(\ref{eq:2:surfaceA}) are
\begin{equation}
\frac{\partial}{\partial q^i}\frac{\partial\sqrt{D}}{\partial g^\alpha_{,i}}
 -\frac{\partial\sqrt{D}}{\partial g^\alpha} =0.  \label{eq:2:g5}
\end{equation}
We calculate
\begin{eqnarray}
\frac{\partial\sqrt{D}}{\partial g^\alpha_{,i}} &=&
\frac{1}{2\sqrt{D}}\frac{\partial D}{\partial\bar{B}_{kl}}
\frac{\partial\bar{B}_{kl}}{\partial g^\alpha_{,i}},  \label{eq:2:g6}\\
\frac{1}{\sqrt{D}}\frac{\partial D}{\partial\bar{B}_{kl}} &=&
\sqrt{D}\frac{\text{cofactor}(\bar{B})_{kl}}{D} =\sqrt{D}\bar{B}^{kl},
\label{eq:2:g7} \\
\frac{\partial\bar{B}_{kl}}{\partial g^\alpha_{,i}} &=&
B_{\alpha\beta}(g^\alpha_{,k}\delta_{li} + g^\beta_{,l}\delta_{ki}).
\label{eq:2:g8}
\end{eqnarray}
Similarly
\begin{eqnarray}
\frac{\partial\sqrt{D}}{\partial g^\alpha} &=& \frac{1}{2}\sqrt{D}
B^{kj} g^\beta_{,k} g^\gamma_{,j}B_{\beta\gamma,\alpha}.  \label{eq:2:g9}
\end{eqnarray}

Combining (\ref{eq:2:g6}) and (\ref{eq:2:g9}), the Euler-Lagrange equation takes
the form
\begin{equation}
\frac{\partial}{\partial q^i}(\sqrt{D}\bar{B}^{ij}B_{\alpha\beta}
g^\beta_{,j}) -\frac{\sqrt{D}}{2}\bar{B}^{ij} B_{\beta\gamma,\alpha}
g^\beta_{,i} g^\gamma_{,j} =0.   \label{eq:2:g10}
\end{equation}
Using the condition $\bar{B}^{ik}\bar{B}_{kj}=\delta_{ij}$ and its
first derivatives, as well as the symmetry property of the metric tensor,
the first term of (\ref{eq:2:g10}) may be put into the form
\begin{eqnarray}
&&\sqrt{D}\bar{B}^{ij}(B_{\alpha\beta} g^\beta_{,ij} -B_{\alpha\beta}
g^\beta_{,k}\bar{\Gamma}^k_{ij} + B_{\alpha\beta,\gamma}
g^\gamma_{,i}g^\beta_{,j}).   \label{eq:2:g11} 
\end{eqnarray}
Finally, we can combine the second term of (\ref{eq:2:g10}) and the last term
of (\ref{eq:2:g11}) to bring in the other Christoffel symbol $\Gamma$.  Altogether
we verify Eq.~(\ref{eq:2:5.12}). 
\end{aside}

\subsection{Generalized valley formulation of the decoupling
problem \label{sec:2.2}}

\subsubsection{Fundamental theorem characterizing a decoupled motion
            \label{sec:2.2.1}}
 
In this section we study the transformation of the decoupling conditions
into one of the forms found useful in practice.
We first restrict the considerations to point transformations and
afterwards add the 
remarks necessary to include the extended adiabatic case.
The basic idea underlying the following considerations is that
we should be able to reconstruct the surface $\xi ^{\alpha} = g^{\alpha}
(q)$ provided we can specify the tangent plane at
each point.  We shall do this by discovering a complete set of basis
vectors for the tangent plane that can be computed from the elements of
the given Hamiltonian, namely, the potential energy and the (reciprocal)
mass tensor.
In what follows it is natural to consider the latter as metric tensor of
a Riemannian space.  However, all that is required is the recognition
that as a contravariant tensor of rank two under point
transformations, it can be contracted with any two covariant vector
fields to form a point function.
 
To carry out this program, we define a
sequence of single index point functions
according to the definitions
\begin{subequations}
\begin{eqnarray}
X^{(0)}&\equiv&V(\xi) = \bar{V}(q), \label{eq:2:3.1}   \\
X^{(1)} \equiv V_{,\alpha} B^{\alpha \beta} V_{,\beta}&=&
\bar{V} _{,\mu} \bar{B} ^{\mu \nu} \bar{V} _{,\nu},\label{eq:2:3.2} \\
&             \vdots  & \nonumber\\
X^{(\sigma + 1)}\equiv X^{(\sigma)}_{,\alpha} B^{\alpha \beta}
X^{(\sigma)}_{,\beta}&=&\bar{X} ^{(\sigma)} _{,\mu} \bar{B} ^{\mu\nu}
\bar{X} ^{(\sigma)} _{,\nu} .   \label{eq:2:3.3}
\end{eqnarray}
\end{subequations}
For $\sigma \neq \tau $, we can next
define a sequence of double index point functions,
\begin{equation}
X^{(\sigma \tau)}\equiv X^{(\sigma)} _{,\alpha} B^{\alpha\beta}
X^{(\tau)} _{,\beta} = \bar{X} ^{(\sigma)} _{,\mu} \bar{B} ^{\mu \nu}
\bar{X} ^{(\tau)} _{,\nu}  ,   \label{eq:2:3.4}
\end{equation}
etc.
Thus the single index sequence is constructed with the help of the
reciprocal
mass tensor by forming the gradient
of the previous point  function and then calculating the length of the
new vector.  The
double index scalars are scalar products of different gradients.  By
finding the gradients of these we can form still additional sequences
of point functions all of which are subsumed under the considerations
that follow.
 
We now prove that for a decoupled surface, $\Sigma$, the gradient
of every scalar belonging to the set defined above is a vector field
that lies in the tangent plane to $\Sigma$.
The proof follows by induction.
We first note that according to the fundamental decoupling condition
(\ref{eq:2:2.16}), the gradient of $X^{(0)}$   lies in the tangent plane, i.e.,
$\bar{X}^{(0)}_{,a} = 0$.  Now let us assume that $\bar{X}^{(\sigma)}
_{,a} = 0$ and show that in consequence of this assumption and all the
remaining decoupling conditions, $\bar{X} ^{(\sigma + 1)} _{,a} = 0$.
We simply compute
\begin{eqnarray}
\bar{X}^{(\sigma + 1)}_{,a} & = &
   2\bar{X}^{(\sigma)} _{,\mu a} \bar{X}^{(\sigma)}
_{,\nu} \bar{B}^{\mu \nu} + \bar{X}^{(\sigma)} _{,\mu}\bar{X}^{(\sigma)}
_{,\nu} \bar{B}^{\mu \nu} _{,a}\nonumber\\
&=& 2\bar{X}^{(\sigma)} _{,ba}\bar{X}^{(\sigma)} _{,i}\bar{B}^{bi}
     + 2 \bar{X} ^{(\sigma)} _{,ia} \bar{X}^{(\sigma)}_{,j}
\bar{B}^{ij} + \bar{X}^{(\sigma)} _{,i} \bar{X}^ {(\sigma)} _{,j}
\bar{B}^{ij} _{,a}
= 0 .   \label{eq:2:3.5}
\end{eqnarray}
In passing to the second line of (\ref{eq:2:3.5}),
we have used only the statement $\bar{X}
^{(\sigma)} _{,a} = 0$; in order to obtain zero overall, we have then
used (\ref{eq:2:2.15}) and (\ref{eq:2:2.17}) in the first and third terms,
respectively, to make these vanish,
whereas the second term vanishes because $\bar{X}^{(\sigma)} _{,ai}
= 0$ (as follows from $\bar{X}^{(\sigma)}_{,a}=0$).
The vanishing of the gradients of the multiply-indexed scalars
follows from the same mode of proof.
 
Before continuing the development, it is appropriate to ask if one can
provide a simple geometrical interpretation of the results just
established.  This is easily done if we remember that a decoupled
surface is geodesic.  In the case where the metric is flat,
the geodesic is a hyper-plane.  The condition that $\nabla V$ lies in
this hyper-plane at every point implies that the difference between
gradients at different points must also lie in the plane, as well as
the difference of differences, etc.  Our theorem is clearly the
generalization of this trivial observation to curved spaces where the
generalized notion of parallel transport becomes relevant. 

When we first considered the concepts under discussion, we already knew
from the results in Ref.~\cite{34} that
 (\ref{eq:2:3.1}) and (\ref{eq:2:3.2}) satisfied the theorem and from
the proof for (\ref{eq:2:3.2}), we were able to surmise the structure needed
for additional vectors.  For some applications, the series
chosen has the drawback that the inclusion of additional members requires the
computation of higher and higher derivatives of the potential.  
In Sec.~\ref{sec:2.3}
we shall learn that there is an alternative to the formulation of this
section which involves at most the second derivative of the potential.
This, in turn, suggests that we replace the set of scalars 
$X^{(\sigma)}$ by the following
set of vectors $X^{(\sigma)}_\alpha$, which, in general, do not
involve higher derivatives,
\begin{subequations}
\begin{eqnarray}
X^{(0)}_\alpha &\equiv& V_{,\alpha}   \label{eq:2:31.1} \\
X^{(1)}_\alpha &\equiv& V_{,\beta}B^{\beta\beta_1}V_{;\beta_1\alpha}
\label{eq:2:31.2} \\
&\vdots& \nonumber \\
X^{(\sigma +1)}_\alpha &\equiv& V_{,\beta}B^{\beta\beta_1}V_{;\beta_1\beta_2}
B^{\beta_2\beta_3} ... B^{\beta_\sigma\beta_{\sigma +1}}
V_{;\beta_{\sigma +1}\,\alpha},             \label{eq:2:31.3} 
\end{eqnarray}
\end{subequations}
where $V_{;\alpha\beta}$ is the covariant derivative
\begin{eqnarray}
V_{;\alpha\beta} &=& V_{,\alpha\beta} -\Gamma^\gamma_{\alpha\beta}V_{,\gamma},
\label{eq:2:31.4}  
\end{eqnarray}
with $\Gamma^\gamma_{\alpha\beta}$ defined in (\ref{eq:2:5.afc1}).

Using the decoupling conditions, the proof that these vectors lie in the
tangent plane is straightforward.  As before we work in the barred
(transformed) coordinate system, where we wish to prove that
$\bar{X}^{(\sigma)}_a=0$.  The essential element of the proof is already clear
from the case $\sigma=1$, where, utilizing two of the decoupling conditions,
we have
\begin{equation}
\bar{X}^{(1)}_a = \bar{V}_{,i}\bar{B}^{ij}\bar{V}_{;ja},  \label{eq:2:3aa.1}
\end{equation}
which vanishes because not only is $\bar{V}_{,ja}=0$, but so also is
$\bar{\Gamma}^i_{ja}$, in consequence of the third decoupling condition,
leading to the vanishing of the covariant derivative in (\ref{eq:2:3aa.1}).

The information contained in the fundamental theorem
may be summarized in two other equivalent forms,
alternative to the statement, $\bar{X}^{(\sigma)}_{,a}=0.$   (We continue
the discussion using the first formulation in terms of scalars, but
there are corresponding statements for the second formulation.)
Let us suppose for the moment that all the point functions of interest
have been arranged into a linear array designated $X^{(\sigma)}$,
in the notation used previously only for the single index scalars.
In the same way as the decoupling condition (\ref{eq:2:2.16}) implied its
equivalent, (\ref{eq:2:2.19}), namely, by the combination
of (\ref{eq:2:3.5}) with the chain rule for differentiation, we have more
generally
\begin{equation}
X^{(\sigma)}_{,\alpha} = \bar{X}^{(\sigma)}_{,i}
        f^{i}_{,\alpha}   .    \label{eq:2:3.6}
        \end{equation}
By using $B^{\alpha \beta}$ in the entire space and
$\bar{B}^{ij}$ on $\Sigma$ to raise indices, and remembering the mass
condition (\ref{eq:2:2.18}), (\ref{eq:2:3.6}) may be converted to 
the equivalent form
\begin{equation}
X^{(\sigma),\alpha} = \bar{X}^{(\sigma),i}
    g^{\alpha}_{,i}   .  \label{eq:2:3.7}
\end{equation}
Equations (\ref{eq:2:3.6}) and (\ref{eq:2:3.7}) both state, one in covariant,
the other in contravariant form, that the vector fields in question lie
in the tangent plane to $\Sigma$.
 
The theorem given above holds equally well for the extended
point transformation requiring only the replacement of $B$ by
$\tilde{B}$, Eq.~(\ref{eq:2:2.33}), in
(\ref{eq:2:3.2})--(\ref{eq:2:3.4}).
This obviously also introduces $\tilde{\Gamma}$, defined from
$\tilde{B}$ as in Eq.~(\ref{eq:2:5.afc1}).
For the practical implementation of the theorem
we must expect some complication in this more general case.
In Sec.~\ref{sec:2.2.2}, that follows, the algorithm described applies only to the
case of a point transformation.  Extended point transformations will be
dealt with best as a limiting case of general
canonical transformations or by special considerations applicable to the
problem at hand.
 
\subsubsection{Generalized valley algorithm for implementation of
fundamental theorem  \label{sec:2.2.2}}
 
We describe next how the theorem just established may be made the basis
for a construction of manifolds that are exactly decoupled if the given
Hamiltonian admits such solutions and are candidates for approximately
decoupled manifolds in all other cases. The algorithm to which one is
naturally led carries with it one or even several criteria for testing
goodness of decoupling.
 
Let us suppose that a system with $N$ coordinates admits a
$K$-dimensional decoupled manifold.  Then any $K+1$ of the vector
fields $X^{(\sigma)}_{,\alpha}$ must be linearly dependent at any
point of the manifold, since any $K$ of them constitutes a basis
for the tangent space.  This may be expressed by the condition,
choosing the ``simplest" $K+1$ vector fields, in the first form of the
fundamental theorem,
\begin{equation}
\Bigl(X^{(K)} - \sum_{i=0}^{K-1}\Lambda_iX^{(i)}\Bigr)_{\;,\alpha}=0,
                                                       \label{eq:2:3.8}
\end{equation}
where $\Lambda_i$ are a set of point functions that may
be identified as Lagrange multipliers.
This is because (\ref{eq:2:3.8}) is clearly
the differential expression of the following
constrained variational problem:  Find extremal values of $X^{(K)}$
subject to fixed values of $X^{(i)},\;i=0,...,K-1$.
By elimination of the Lagrange multipliers, one obtains just
enough equations
to determine $\xi^a$, $a=K+1,...,N$ in terms of the $\xi^i$, $i=1,...,K$,
a $K$-dimensional manifold,
or, by suitable parameterization, to express the manifold in the form
(\ref{eq:2:2.6}).  Since, in general, the equations encountered will be
non-linear, there is no reason not to expect multiple solutions (generalized
valleys and generalized ridges).

Assuming that we have found a solution of (\ref{eq:2:3.8}), there
exist various ways to determine the collective mass. We can, simply by
differentiation of  Eq.~(\ref{eq:2:2.6}), calculate the contravariant basis vectors
$g^\alpha_{,i}$.  These quantities allow us to compute a covariant
metric tensor for the collective manifold, 
\begin{equation}
\bar{B}_{ij} = (\bar{B}^{-1}) ^{ij} 
=g_{,i}^\alpha B_{\alpha\beta}g_{,j}^\beta.  \label{eq:2:32.1}
\end{equation}
Finally from (\ref{eq:2:2.18})
we find
\begin{equation}
f^i_{,\alpha} =B_{\alpha\beta}\bar{B}^{ij}g^\beta_{,j}. \label{eq:2:32.2}
\end{equation}
Thus we have described a calculation which determines a $K$-dimensional
surface, $\Sigma$, and the basis vectors for the tangent plane to any
point on $\Sigma$.  This calculation requires that we calculate the mass
tensor, $B_{\alpha\beta}$, the inverse of the metric tensor
$B^{\alpha\beta}$.  For the nuclear problem, this is often technically
difficult.  For most applications, we have therefore used an alternative definition of the mass that
we next describe.

Equation (\ref{eq:2:3.8}) has been presented as a necessary condition that
there be a decoupled manifold. It is also a consistency condition that a chosen
subset of $K+1$ equations of the set (\ref{eq:2:3.6}) or (\ref{eq:2:3.7})
determine a plane of dimension $K$ at each point of a $K$-dimensional
manifold.  Thus when the consistency condition is satisfied,
these equations will determine a set of basis vectors
$\breve{f}^i_{,\alpha}$ or a set $\breve{g}^\alpha_{,i}$.  The breve
has been added to emphasize that in the general case where the decoupling
is not exact, the basis vectors so determined are not the same as those
computed in the previous paragraph.  These vectors will then determine a 
breve form of the mass tensor.  The details of how such a computation
can be carried out is the essential subject matter of Sec.~\ref{sec:2.3} and therefore will not be
discussed here. 
Thus the calculation based directly on (\ref{eq:2:3.8})
determines a manifold $\Sigma$ together with the tangent plane at every
point.  At the same time an auxiliary calculation 
determines a second plane at each point, with
the breve basis.  In general, this second plane does not coincide with the 
tangent plane; if it does, we have exact decoupling. 
 
To understand the designation ``generalized valley'' for the contents of
this section, we consider the special case $K=1$. If we choose the
vector fields associated with
Eqs.~(\ref{eq:2:3.1}) and (\ref{eq:2:3.2}), then Eqs.~(\ref{eq:2:3.8}) are the
equations for a valley,
\begin{equation}
(|\nabla V|^2) - \lambda V)_{,\alpha}=0,  \label{eq:2:valley}
\end{equation}
a concept that we have already introduced in Sec.~\ref{sec:1.2}.
As explained briefly there, a simple mechanical picture can be associated
with the variational expression
of which Eq.~(\ref{eq:2:valley}) is the consequence.
We imagined that our direction is generally upward in motion along an
uneven terrain.  Having arrived at a certain value of the potential
energy, we were instructed to traverse the equipotential associated with
this value
and at each point check the amount of work we have to do for a
further ascent of {\em fixed} step length.
If we can locate a minimum value for this work,
i.e., a minimum value for the magnitude of the gradient of the potential,
then we have indeed found a segment of a valley.
This accords with our intuitive understanding of the meaning of valley,
except that the latter usually carries with it the extra, but
unnecessary, concept of continuity.  In the differential characterization
(\ref{eq:2:valley}), the valley is determined segment by segment and need not
be a continuous curve.  It can also fork at a given point into two or
more prongs.  Of course, Eq.~(\ref{eq:2:valley}),
as a first order variational condition, is only the condition for an
extremal rather than a minimum, and this fact will reflect itself in
its solutions.
 
Because of the identification
just made for the one-dimensional case, we refer to the multi-dimensional
case as the generalized valley formulation of
the problem of decoupled manifolds.
In this formulation, the
decoupling conditions Eqs.~(\ref{eq:2:2.19}) and
(\ref{eq:2:2.20}) are replaced by the ensemble of statements
(\ref{eq:2:3.6}) or (\ref{eq:2:3.7}), that normally includes (\ref{eq:2:2.19}),
and to these we continue to adjoin the mass-tensor condition
(\ref{eq:2:2.18}).
It is essential to distinguish, however, between
the exact formulation and the practical algorithm, called the generalized
valley approximation (GVA) that is summarized by
(\ref{eq:2:3.8}).
For exact decoupling, {\em all} of Eqs.~(\ref{eq:2:3.6}) must be satisfied.
As described above, a
solution of (\ref{eq:2:3.8}) can be exploited in two ways: On the one hand
it determines a surface $\Sigma$ and its tangent plane at every point.  
On the other hand, through the offices of a subset of Eqs.~(\ref{eq:2:3.6}) 
and (\ref{eq:2:3.7}), it determines a second plane at each point of $\Sigma$.
It remains to be tested to what extent this plane coincides
with the tangent plane, which is the condition for exact decoupling.

\subsubsection{Quality of decoupling; calculation of collective Hamiltonian
            \label{sec:2.2.3}}
 
Consideration of the quality of decoupling is closely tied to the problem
of determining the collective Hamiltonian.  The solution of Eq.~(\ref{eq:2:3.8})
provides the surface $\Sigma$ in the form of Eq.~(\ref{eq:2:2.6}), that in turn
fixes the collective potential energy,
\begin{equation}
\bar{V}(q)=V(\xi(q)). \label{eq:2:3.9}
\end{equation} 
The same solution also determines a value for the collective mass tensor
by means of the formula already recorded in (\ref{eq:2:32.1})
On the other hand, as previously remarked,
the GVA is also a consistency condition for the
associated sets of equations, (\ref{eq:2:3.6}) and (\ref{eq:2:3.7}) to
be solvable, respectively, for sets of basis vectors
$\breve{f}^i_{,\alpha}$ and $\breve{g}^\alpha_{,i}$. As remarked previously,
the notation is meant to emphasize that in general, i.e.
when the decoupling is approximate, these basis vectors do not
lie in the tangent space defined, for example by Eq.~(\ref{eq:2:32.2}),
but rather
in the tangent space determined by the physical vectors $X^{(\sigma)}
_{,\alpha}$ themselves, and thus the quantities $\breve{g}^\alpha_{,i}$
differ from the quantities ${g}^\alpha_{,i}$.
It follows that the
calculation of the mass tensor is ambiguous.  As an alternative to
(\ref{eq:2:32.1})
one may calculate the quantities (in the contravariant version)
\begin{equation}
\breve{B}_{ij}=\breve{g}^{\alpha}_{,i}B_{\alpha\beta}\breve{g}^{\beta}_{,j}.
\label{eq:2:3.12}
\end{equation}
The difference between the two mass tensors can be measured, for example,
by the point function
\begin{equation}
D\equiv K^{-1}|\trace[(\bar{B}^{-1}-\breve{B}^{-1})\bar{B}]|,
\label{eq:2:3.13}
\end{equation}
that for $K=1$ is simply the absolute value of the fractional differences of the masses.
Insofar as the quantity in Eq.~(\ref{eq:2:3.13}) is small compared to
unity, we may assert that the generalized valley algorithm has produced
an essentially unique result for the collective Hamiltonian.
 
In some applications, especially to nuclear physics, the inversion of
the mass tensor is a time-consuming task.  In such cases, one
may prefer to settle in
practice for a calculation of $\breve{B}$. In this
case it becomes desirable to have a criterion for the quality of
decoupling that does not require inversion of the mass tensor.  A
possible choice is
\begin{equation}
 E \equiv
K^{-1}\sum_i[\breve{f}^i_{,\alpha}(\breve{g}^\alpha_{,i}
-g^\alpha_{,i})]^2, \label{eq:2:3.14}
\end{equation}
since it is easily seen that
this quantity can be determined from the results of any of the
algorithms proposed without having to invert the
mass tensor.
 
\subsubsection{Conditions for local stability of collective motion
            \label{sec:2.2.4}}
 
It remains for us to discuss the problem of local
stability.  Given an exactly decoupled surface, suppose that there
is a small perturbation in the initial conditions that pushes the system
off the collective surface.  Will the system then remain in the
neighborhood of the surface?
 
To study this question, we consider a point $\xi^{\alpha}$ in the
neighborhood of the decoupled surface,
\begin{equation}
\xi^{\alpha} = \xi^{\alpha}_{0} + \delta\xi^{\alpha},
\label{eq:2:3.15}
\end{equation}
where $\xi^{\alpha}_{0}$ is a point on the surface, and the variation
is a vector orthogonal to the surface at every point.  To first order
it has the form                
\begin{equation}
\delta\xi^{\alpha} = g^{\alpha}_{,a}\delta q^{a}.    \label{eq:2:3.16}
\end{equation}
Thus, the specification of (\ref{eq:2:3.16}) requires calculation of a
set of basis vectors orthogonal to the decoupled surface (see below).
When we now expand the potential energy about the point $\xi^{\alpha}
_{0}$, the first order term vanishes because grad V is assumed to lie in
the collective surface (though in practice this is an approximation).
We thus obtain
\begin{eqnarray}
V({\vec{\xi}})&=&V({\vec{\xi}}_{0}) + \frac{1}{2}\bar{V}_{,ab}
\delta q^{a}\delta q^{b}  \nonumber \\
& \equiv  &V_{\text{C}} + V_{\text{NC}},
\label{eq:2:3.17}
\end{eqnarray}
i.e., the sum of the collective contribution and of a non-collective part
that is quadratic in the deviations of the coordinates away from the
starting surface.  For a prescribed deviation we have
\begin{equation}
\bar{V}_{,ab} = V_{,\alpha\beta}g^{\alpha}_{,a}g^{\beta}
_{,b}.  \label{eq:2:3.18}
\end{equation}
This result is not strictly correct to second order in $\delta q$, as
evidenced by the fact that the ordinary rather than the covariant
second derivative appears on the right hand side.  We shall not discuss
the point here, since it will be handled correctly by the methods
developed in Sec.~\ref{sec:2.3}.

Because of the decoupling condition on the mass matrix, the kinetic
energy, $T$, also decomposes in the immediate neighborhood
of the decoupled
surface into the sum of a collective and of a non-collective part,
\begin{eqnarray}
T&=&T_{\text{C}} + T_{\text{NC}}  \nonumber \\
&\equiv&\frac{1}{2}p_{i}\bar{B}^{ij}p_{j}+
\frac{1}{2}p_{a}\bar{B}^{ab}p_{b}, \label{eq:2:3.19}
\end{eqnarray}
where
\begin{equation}
\bar{B}^{ab} = f^{a}_{,\alpha}B^{\alpha\beta}f^{b}_{,\beta}.
\label{eq:2:3.20}
\end{equation}
 
We wish to study the non-collective energy,
\begin{equation}
H_{\text{NC}} = V_{\text{NC}} + T_{\text{NC}}, \label{eq:2:3.21}
\end{equation}
since wherever it is positive, we have local stability.  We see
from Eqs.~(\ref{eq:2:3.18}) and (\ref{eq:2:3.20}) that this requires the
specification at
each point of the surface of a coordinate system spanning the space
orthogonal to the collective space.  This is, in principle,
an elementary problem (see below for details)
which can be solved in such a way that at the same time the mass
matrix in the non-collective space can be chosen to be the unit matrix
at every point.  With such a choice, we then have only to diagonalize
the matrix $\bar{V}_{,ab}$ to check whether the resulting eigenvalues
are positive.  Since these values depend on the collective coordinates,
the associated zero-point energies may be considered as part of the
collective potential energy in an improved treatment of this quantity.
Situations may arise in which anharmonic corrections to the
non-collective motion will also be of interest.

\begin{aside}
We describe a procedure for the construction of a basis,
$f^a_{,\alpha}$, in the non-collective subspace.  In the following, we shall
suppose that we are working in the bar representation of the basis rather
than in the breve representation.  Formal aspects are the same for both,
but unless there is exact decoupling, detailed results will differ.
As an example suppose that
there are three coordinates labeled $1,2,3$, where $1$ is a collective
coordinate.  Assume that we have determined the collective basis vector
$f^1_{,\alpha}$ from the GVA procedure.  The requirement that
$f^2_{,\alpha}$ be orthogonal to
$f^1_{,\alpha}$ with respect to the metric $B^{\alpha\beta}$,
\begin{equation}
\bar{B}^{12}= f^1_{,\alpha}B^{\alpha\beta}f^2_{,\beta} =0, \label{eq:2:fnc1}
\end{equation}
fixes a two dimensional subspace for $f^2_{,\alpha}$.  Make a specific choice
and normalize by the mass condition
\begin{equation}
1 = \bar{B}^{22} =f^2_{,\alpha}B^{\alpha\beta}f^2_{,\beta}. \label{eq:2:fnc2}
\end{equation}
With these choices for $f^2_{,\alpha}$, we obtain a unique value of
$f^3_{\alpha}$ from the conditions
\begin{eqnarray}
\bar{B}^{13} &=& \bar{B}^{23} = 0, \label{eq:2:fnc3} \\
\bar{B}^{33} &=& 1.   \label{eq:2:fnc4}
\end{eqnarray}
In the bar basis, the contragradient basis vector $g^\alpha_{,1}$ is usually
the natural output of the GVA procedure.  Under these circumstances, we
would carry out a similar  construction as above but with the basis
vectors $g$ replacing the $f$, the former used in conjunction with the
reciprocal tensor $B_{\alpha\beta}$.  Such a calculation will be illustrated
in Sec.~\ref{sec:3.2.2}. 
\end{aside}

\subsubsection{Modification of the theory for conserved quantities
            \label{sec:2.2.5}}
 
We next reexamine how the theory developed thus far must be
modified when there are additional constants of the motion on the
decoupled surface.  What fails in the reasoning based on the decoupling
condition (\ref{eq:2:2.14}) is
that if some of the $p_{i}$  are constant, we can no longer equate
to zero separately the coefficients of unity and of terms quadratic
in the $p_{i}$.
Physically this means that the real  and geometric forces are no
longer separately zero.  Instead, for some degrees of freedom there
is a balance between dynamical and centripetal forces orthogonal
to the decoupled surface, a not unfamiliar circumstance.
 
In order to discuss this case, it is necessary to divide the coordinate
indices into three sets.  Indices $i,j,\,...$ will refer to the
non-cyclic collective coordinates, $l,m,\,...$
to the cyclic or ignorable collective coordinates, and $a,b,\,...$,
as before, to
the non-collective coordinates.  The decoupling conditions
(\ref{eq:2:2.15})--(\ref{eq:2:2.17}) are now
replaced by the equations
\begin{subequations}
\begin{eqnarray}
&&\bar{B}^{ai}=\bar{B}^{al}=0, \label{eq:2:34.6} \\
&&\bar{V}_{,l}=0,\; \;\; \bar{V}_{,a} +\frac{1}{2}p_l p_m\bar{B}^{lm}_{,a}=0,
\label{eq:2:34.7} \\
&&\bar{B}^{\mu\nu}_{,l}=\bar{B}^{ij}_{,a}=\bar{B}^{il}_{,a}=0.
\label{eq:2:34.8}
\end{eqnarray}
\end{subequations}
These equations follow from (\ref{eq:2:2.13}) and (\ref{eq:2:2.14}) and the
definition  of a cyclic coordinate.
 
We describe next how the generalized valley formulation is modified in
the present circumstances.  In place of
(\ref{eq:2:3.1})--(\ref{eq:2:3.3}), we define a sequence of point
functions,
\begin{subequations}
\begin{eqnarray}
\bar{Y}^{(0)}&\equiv &\bar{V}+\frac{1}{2}p_l p_m\bar{B}^{lm},
\label{eq:2:34.9} \\
\bar{Y}^{(\sigma+1)}&\equiv &\bar{Y}^{(\sigma)}_{,\mu}
\bar{B}^{\mu\nu}\bar{Y}^{(\sigma)}_{,\nu}.
\label{eq:2:34.10}
\end{eqnarray}
\end{subequations}
We may also define the multiple-index point functions
of the type represented by (\ref{eq:2:3.4}), and the following conclusions apply
as well to these.
All such quantities are independent of the cyclic coordinates, and
therefore we have, for example,
\begin{equation}
\bar{Y}^{(\sigma)}_{,l} = 0. \label{eq:2:34.11}
\end{equation}
In addition it is  straightforward to show, using the
decoupling conditions (\ref{eq:2:34.6})--(\ref{eq:2:34.8}), that
\begin{equation}
\bar{Y}^{(\sigma)}_{,a}=0. \label{eq:2:34.12}
\end{equation}
 
The combination of (\ref{eq:2:34.11}) and (\ref{eq:2:34.12}) allows
us to write the
analogue of (\ref{eq:2:3.6}) ($Y^{(\sigma)}\equiv \bar{Y}^{(\sigma)}$),
\begin{equation}
Y^{(\sigma)}_{,\alpha}=\bar{Y}^{(\sigma)}_{,i}f^i_{,\alpha},
\label{eq:2:34.13}
\end{equation}
as well as the corresponding analogue of the contravariant
form (\ref{eq:2:3.7}).  It is important to emphasize that these equations
characterize the tangent space of a manifold of dimensionality $K-K_c$,
where $K$ is the dimensionality of the complete decoupled manifold
and $K_c$ is the number of cyclic coordinates on this manifold.
 
Just as in Secs.~\ref{sec:2.2.1} and \ref{sec:2.2.2}
Eqs.~(\ref{eq:2:34.13}) lead to a generalized valley
algorithm, the analogue of Eq.~(\ref{eq:2:3.8}).  Superficially the problem
has been simplified compared to the situation without additional
constants of the motion owing to the reduction in the dimensionality
of the manifold specified by (\ref{eq:2:34.13}) to the value $K-K_c$.
On the other hand, actual computation
requires knowledge of elements previously absent from the calculation.
Thus we see from the definitions (\ref{eq:2:34.9})
that we need the quantities
\begin{equation}
\bar{B}^{lm} =f^l_{,\alpha}B^{\alpha\beta}f^m_{,\beta},
\label{eq:2:34.14}
\end{equation}
or, in other words, the basis vectors $f^l_{,\alpha}$ associated with
the conserved quantities, and these are not determined
by (\ref{eq:2:34.13}).  It is most convenient to carry forward the discussion
of this problem within the framework of the local harmonic approximation,
which is the subject matter of the remainder of this section.
For this continuation, we refer the reader to Sec.~\ref{sec:2.3.5}.

\subsection{Local harmonic formulations for decoupling
collective modes \label{sec:2.3}}
 
\subsubsection{Role of Frobenius' theorem\label{sec:2.3.1}}
 
Previously, we have sought conditions under which Eqs.~(\ref{eq:2:3.6}) or
(\ref{eq:2:3.7}), that catalogue the tangent vectors to the decoupled
manifold, determine candidates for decoupled surfaces.
To establish that there is a relationship with a local harmonic
approach, i.e., with an eigenvalue problem for small vibrations,
it is helpful to study the question of the existence
of surfaces in another light.
This involves the standard theory of the integrability of Eqs.
(\ref{eq:2:3.6}) or (\ref{eq:2:3.7}).  Consider for example Eq.~(\ref{eq:2:3.7})
applied to the case of a two-dimensional manifold, $K=2$,
where the point functions involved will be designated
as $V$ and $U$.  Provided the determinant
\begin{equation}
\left| \begin{array}{cc}
\bar{V}^{,1} & \bar{V}^{,2} \\
\bar{U}^{,1} & \bar{U}^{,2}
\end{array} \right|
\neq 0,                       \label{eq:2:6.1}
\end{equation}
these equations can be solved for the basis vectors (contravariant form)
\begin{equation}
g^{\alpha}_{,i}=\alpha_{i} V^{,\alpha} +
\beta_{i} U^{,\alpha},   \;\;i=1,2,
\label{eq:2:6.2}
\end{equation}
appropriate for the application of Frobenius's theorem \cite{35,36}.
The standard problem associated with this theorem is:  Given a
point in the underlying manifold, under what conditions do Eqs.~(\ref{eq:2:6.2})
determine a unique surface (tangent plane) through this point? In other words,
when are these equations integrable?
If these conditions (see below) are satisfied, then the totality of such
surfaces defines a ``foliation'' of the given manifold.  For our
purposes, we see that the basis vectors we seek are linear combinations
of the fundamental vector fields of the generalized valley formulation.  
 
The problem of interest to us is different.  To test decoupling, we
are not just seeking
a foliation of the full space,  but rather a special surface on
which a third vector field is tangent to the surface.  Nevertheless,
the special surface must be included in the foliation if there is to
be an exactly decoupled two-dimensional manifold.
The reason that we have not
emphasized Frobenius's theorem heretofore is first of all
that in most applications
we do not have an exactly decoupled surface, and second that it 
will play no role in any of the
algorithms suggested in this paper
for the construction of  an approximately decoupled
manifold.  The theorem enters the discussion for purely theoretical
reasons, as a tool for the transformation of the generalized valley
formulation of the exact decoupling conditions into a local harmonic
formulation, to be derived in this section.
 
For the moment we follow the standard mathematical custom of calling the
quantities
\begin{equation}
v^{(0)} = V^{,\alpha} (\partial/ \partial \xi^{\alpha}) ,
\;\;\;\;
v^{(1)} = U^{,\alpha} (\partial / \partial \xi^{\alpha})  ,
\label{eq:2:6.3}
\end{equation}
tangent vectors.  The integrability condition 
for Eq.~(\ref{eq:2:6.2}) is that the
two tangent vectors be in involution, i.e., closed under commutation,
\begin{equation}
[v^{(0)}, v^{(1)}]= c_{0} v^{(0)} + c_{1} v^{(1)} .
\label{eq:2:6.4}
\end{equation}
If we work out the commutator (\ref{eq:2:6.4}) for the vectors involved in
Eq.~(\ref{eq:2:6.2}), we find the conditions
\begin{equation}
V^{,\beta} U^{,\alpha} _{;\beta} - U^{,\beta} V^{,\alpha}
_{;\beta} = c_{0} V^{,\alpha} + c_{1} U^{,\alpha} .
\label{eq:2:6.5}
\end{equation}
Here the semicolon indicates the covariant derivative, but in
this case the curvature terms actually cancel between the two terms on
the left hand side of (\ref{eq:2:6.5}).  We study next an essential application of
this equation.
 
\subsubsection{Local harmonic formulation: first derivation including
curvature effects \label{sec:2.3.2}}
 
We continue to study in detail the case $K=2$.  To the point functions
$V$ and $U$ we adjoin the point function $T$, defined as the scalar
product of $\nabla V$ with $\nabla U$.  The corresponding gradient
vectors provide us with three examples of Eqs.~(\ref{eq:2:3.6}) or
(\ref{eq:2:3.7}). In the latter form, we have
\begin{subequations}
\begin{eqnarray}
V^{,\alpha}&=&\bar{V}^{,i}g^\alpha_{,i},  \label{eq:2:6.6a}   \\
U^{,\alpha}&=&\bar{U}^{,i}g^\alpha_{,i},  \label{eq:2:6.6b}  \\
T^{,\alpha}&=&\bar{T}^{,i}g^\alpha_{,i}.  \label{eq:2:6.6c}
\end{eqnarray}
\end{subequations}
For the transformation of these equations,
we require explicit expressions for the vector fields
$U^{,\alpha}$ and $T^{,\alpha}$ , calculated directly from
their definitions, namely,
\begin{eqnarray}
 U^{,\alpha}&=&2B^{\alpha\gamma} V_{;\delta\gamma}V^{,\delta},
\label{eq:2:6.7a}\\
 T^{,\alpha}&=&B^{\alpha \gamma} (U_{;\beta\gamma}
V^{,\beta} + U^{,\beta} V_{;\beta\gamma}) .
\label{eq:2:6.7b}
\end{eqnarray}
As a first step, notice that if we substitute (\ref{eq:2:6.7a}) and then
(\ref{eq:2:6.6a}) into (\ref{eq:2:6.6b}), the result is an equation of the form
\begin{equation}
2V^{,\alpha} _{;\delta} \bar{V}^{,i} g^{\delta}_{,i} =
\bar{U}^{,i} g^{\alpha} _{,i}  .
\label{eq:2:6.8}
\end{equation}
We note in passing that if $i$ is a single index, $i=1$,
as in the case $K=1$,
then this equation is already of the form,
\begin{equation}
V^{,\alpha}_{;\delta}g^{\delta}_{,i}=\omega^{2}g^{\alpha}_{,i},
\label{eq:2:6.9}
\end{equation}
where $\omega^2=\bar{U}^{,1}/\bar{V}^{,1}$,
which is the eigenvalue problem associated with the local random phase
approximation, or as it is usually called, the local harmonic
approximation (LHA).
Within the present context this name applies rather
to the combination of Eq.~(\ref{eq:2:6.6a}) and Eq.~(\ref{eq:2:6.9}), since the
simultaneous solution of both is required to determine a manifold,
as will become clear when we turn to applications of this method in later
sections.  Since Eq.~(\ref{eq:2:6.9}) has,
in general, N solutions, it is assumed that one has a criterion for
selecting the solution or solutions of interest (``collective modes").
(In general, the valley theory will display some corresponding
multiplicity of solutions.)  We remark that the involution condition was not
required for the case $K=1$; it becomes relevant for any larger value of $K$.
 
Returning to our main task, the case $K=2$, we have thus far replaced
Eq.~(\ref{eq:2:6.6b}) by (\ref{eq:2:6.8}).
We consider next the transformation of Eq.~(\ref{eq:2:6.6c}).
This is done in several steps.  First we substitute (\ref{eq:2:6.7b}) in order
to eliminate $T^\alpha$.  Next we introduce the involution
condition (\ref{eq:2:6.5}) in order to eliminate the covariant derivative
$U^{,\alpha} _{;\beta}$.  In the resulting equation, we finally insert
(\ref{eq:2:6.6a}) and (\ref{eq:2:6.6b})
in order to eliminate the vector fields $U^{,\alpha}$
and $V^{,\beta}$.   The equation thus obtained, which plays the role
of partner to (\ref{eq:2:6.8}), has the form
\begin{equation}
V^{,\alpha} _{;\beta} \bar{U}^{,j} g^{\beta} _{,j} =
(\bar{T}^{,i}-c_{0} \bar{V} ^{,i} - c_{1} \bar{U}^{,i})g^{\alpha}_{,i} .
\label{eq:2:6.10}
\end{equation}
Furthermore, if the determinant (\ref{eq:2:6.1}) does not vanish, (\ref{eq:2:6.8})
and (\ref{eq:2:6.10}) can be solved for $V^{,\alpha}_{;\beta}g^\beta_{,i}$,
\begin{equation}
V^{,\alpha}_{;\beta}g^{\beta}_{,i}=\Lambda_{ij}g^{\alpha} _{,j},
\label{eq:2:6.11}
\end{equation}
where $\Lambda$ is a real but not generally symmetric matrix.  If this matrix
has no degenerate eigenvalues, as we assume, (\ref{eq:2:6.11}) can,
by  a similarity transformation in the collective indices, be brought
to the form (\ref{eq:2:6.9}).  We have thus reached the LHA, where its
RPA component, Eq.~(\ref{eq:2:6.11}), must provide us with two
solutions, the two basis vectors for the tangent plane to $\Sigma$ at the
point under study.  These are the breve basis vectors defined in 
Sec.~\ref{sec:2.2}.  It should be
clear that the derivation of the local harmonic formulation given
above can be extended to any number of dimensions.

At this point, one might become curious to ask if any role can be assigned
to the remaining solutions of the LHA.  It will turn out that these
provide basis vectors for the non-collective manifold.

\subsubsection{Local harmonic formulation: second derivation including
curvature effects \label{sec:2.3.3}}
 
In this alternative approach, the introduction of Frobenius' theorem
is replaced by the application of the geodesic equation derived in
Sec.~\ref{sec:2.1.4}.  To apply this equation, we differentiate Eq.~(\ref{eq:2:6.6a})
with respect to $q^j$.  We thus obtain
\begin{equation}
V^{,\alpha}_{,\beta}g^\beta_{,j}=\bar{V}^{,i}_{,j}g^\alpha
_{,i}
+ \bar{V}^{,i}g^\alpha_{,ij}.
\label{eq:2:6.13}
\end{equation}
The form of the last term of (\ref{eq:2:6.13}) invites the application
of the geodesic Eq.~(\ref{eq:2:5.13}).  The substitution of this relation
and the
proper apportionment of the two resulting terms leads to the equation
\begin{equation}
V^{,\alpha}_{;\beta}g^\beta_{,i}=\bar{V}^{,j}_{;i}g^\alpha_{,j},
\label{eq:2:6.14}
\end{equation}
where the covariant derivatives are defined by the equations
\begin{subequations}
\begin{eqnarray}
V^{,\alpha}_{;\beta}&=&V^{,\alpha}_{,\beta}+\Gamma^\alpha_{\gamma\beta}
V^{,\gamma},      \label{eq:2:6.15}  \\
 \bar{V}^{,i}_{;j}&=&\bar{V}^{,i}_{,j}+\bar{\Gamma}^i_{lj}\bar{V}^{,l},
\label{eq:2:6.16}
\end{eqnarray}
\end{subequations}
and the affine connections are given by (2.41) and (2.42).  This equation is
of the same form as (\ref{eq:2:6.11}), and therefore the same additional
remarks apply.

It is convenient at this point and important in general to point out that
just as (\ref{eq:2:6.14}) exhibits the basis vectors $g^\alpha_i$ as right
eigenvectors of $V^{,\alpha}_{;\beta}$, the contragradient basis vectors
$f^i_{,\alpha}$ are left eigenvectors of the same matrix, namely
\begin{equation}
f^i_{,\beta}V^{,\beta}_{;\alpha} = \bar{V}^{,i}_{;j}f^j_{,\alpha}.
\label{eq:2:4aa.1}
\end{equation}
The proof requires only that we substitute into (\ref{eq:2:4aa.1}) the
decoupling condition (\ref{eq:2:2.18}) in the form
\begin{equation}
g^\alpha_{,j} =\bar{B}_{jk}f^k_{,\gamma}B_{\gamma\alpha}, \label{eq:2:4aa.2}
\end{equation}
and use the metric tensors to suitably raise and lower indices.

We summarize the results of this and the previous subsection as follows:  For
the case of exact decoupling, the generalized valley formulation is
fully equivalent to a local harmonic formulation.  In the generic case,
where decoupling is not exact, the two methods will yield results
that differ, that difference measured by the goodness-of-decoupling
criteria already described.  Checking one or more of these criteria
requires that we carry through calculations associated with both of
the formalisms.  Concretely, it means finding the solutions of the LHA
equation after doing a GVA calculation.

\subsubsection{Local harmonic formulation without a metric \label{sec:2.3.4}}

Given a Hamiltonian quadratic in the momenta, we have developed a
complete formalism for decoupling collective from
non-collective motion in the large amplitude adiabatic limit.  This
theory will cover the examples studied in Secs.~\ref{sec:3.2} and
\ref{sec:3.3}.  It will not serve, however, without further discussion
for the case of nuclear physics, where the initial Hamiltonian,
provided by a mean-field theory, is definitely not quadratic in the
momenta.  To deal with this case, which originally inspired these
investigations, we have a choice.  The first is to expand the given
Hamiltonian to quadratic terms in the momenta and to apply the
Riemannian theory to the result.  Here we shall develop an
alternative, which allows a general canonical transformation that is,
however, expanded in powers of the collective momenta, correct to
second order.  We pay for this increased generality of the canonical
transformation by having to deal with the curvature by successive
approximations instead of building it into the theory from the
beginning.

In this approach it is advantageous to use complex canonical coordinates,
\begin{eqnarray}
a_\alpha &=& \frac{1}{\sqrt{2}}(\xi^\alpha + i\pi_\alpha) \label{eq:2:61.1} 
\end{eqnarray}
(and complex conjugate) for the initial system.  We seek a canonical
transformation with the following requirement,
\begin{eqnarray}
H(a_\alpha,a_\alpha^{\ast}) &\rightarrow& \bar{H}(q^i,p_i,q^a,p_a)
\nonumber \\
&=&  \frac{1}{2}B^{ij}p_i p_j +V(q) + \;\;{\rm noncollective\; part}.
\label{eq:2:61.2}
\end{eqnarray}
We study the equation of motion
\begin{equation}
i\dot{a}_\alpha = \frac{\partial H}{\partial a_\alpha^{\ast}}, \label{eq:2:61.3}
\end{equation}
together with its complex-conjugate equation.
Introducing the definitions,
\begin{subequations}
\begin{eqnarray}
&&\partial_i a_\alpha \equiv\frac{\partial a_\alpha}{\partial q^i}, \;\;
\partial^i a_\alpha \equiv\frac{\partial a_\alpha}{\partial p_i},
\label{eq:2:61.5} \\
&&S_\alpha \equiv \frac{\partial H}{\partial a_\alpha^{\ast}}, \;\;
L_{\alpha\beta} \equiv\frac{\partial^2 \bar{H}}{\partial a_\alpha^{\ast}
\partial a_\beta^{\ast}}, \;\;
M_{\alpha\beta} \equiv\frac{\partial^2 \bar{H}}{\partial a_\alpha^{\ast}
\partial a_\beta}, \label{eq:2:61.7}
\end{eqnarray}
\end{subequations}
we then evaluate (\ref{eq:2:61.3}) under the assumption that we are describing a
decoupled motion ($q^a=p_a=0$).  This gives for the left hand side
\begin{eqnarray}
\dot{a}_\alpha &=& \partial_i a_\alpha \dot{q}^i
+\partial^i a_\alpha \dot{p}_i     \nonumber \\
&=& \partial_i a_\alpha(B^{ij}p_j)+\partial^i a_\alpha(-V_{,i}),
\label{eq:2:61.8}
\end{eqnarray}
where we have utilized the equations of motion on the decoupled manifold,
and assumed at the same time that we can drop a term containing
the rate of change of the
metric tensor, since it is second order in the collective momenta.
Since the resulting
expression contains terms of zero and first order in the collective momenta,
we expand the right hand side of (\ref{eq:2:61.3}) in powers of this momenta
and equate coefficients.

Before recording the two sets of equations that result from this
simple calculation, let us note, with $\vec{ q}$ representing the
collective variables, that what we are trying to calculate is
$a_\alpha(\vec{q})$ and $\partial^i a_\alpha(\vec{q})$.
However, in these equations, there also occur as variables the quantities
$\partial_i a_\alpha(\vec{q})$; to obtain a local harmonic formulation
that determines these quantities on a point by point basis
we need an additional set of equations.  We obtain these by
differentiating Eq.~(\ref{eq:2:61.3}) with respect to $q^i$, 
setting $p_i$ to zero.  This calculation brings in second derivatives of
$a_\alpha$, but these are dropped at this point as curvature
corrections.   Altogether we obtain the three sets of equations (and their
complex conjugates)
\begin{subequations}
\begin{eqnarray}
V_{,i}\partial^i a_\alpha &=& iS_\alpha, \label{eq:2:chf} \\
B^{ij}\partial_i a_\alpha &=& -i[M_{\alpha\beta}\partial^i a_\beta
      +L_{\alpha\beta}\partial^i a_\alpha^{\ast}],  \label{eq:2:lha1}\\
V_{,ij}\partial^j a_\alpha &=& i[M_{\alpha\beta}\partial_i a_\beta
      +L_{\alpha\beta}\partial_i a_\alpha^{\ast}].   \label{eq:2:lha2}
\end{eqnarray}
\end{subequations}
We emphasize the origin of these equations.  Equations (\ref{eq:2:chf})
and (\ref{eq:2:lha1}) are the terms of zero and first order in $p_i$ when
Eq.~(\ref{eq:2:61.3}) is combined with Eq.~(\ref{eq:2:61.8}).  Equation (\ref{eq:2:lha2})
is obtained by differentiating Eq.~(\ref{eq:2:chf}) with respect to $q^i$,
while suppressing the derivative of $\partial^j a_\alpha$.
The resulting set of equations constitute the present version of the
local harmonic approximation (LHA).  Of these, the first set is a
constrained Hartree-Fock set (CHF).  The second and third sets together
constitute the local RPA.

The system can be simplified further by the consistent assumption that
$M$ and $L$ are real symmetric matrices and that the partial derivatives
are either real or imaginary,
\begin{subequations}
\begin{eqnarray}
\partial_i a_\alpha^{\ast} &=& \partial_i a_\alpha, \label{eq:2:real} \\
\partial^i a_\alpha^{\ast} &=& -\partial^i a_\alpha. \label{eq:2:imag}
\end{eqnarray}
\end{subequations}
This allows us to eliminate the partials of $a_\alpha^{\ast}$.  The
formalism now consists of the CHF equations (\ref{eq:2:chf}) and two equivalent
eigenvalue equations obtained by combining (\ref{eq:2:lha1}) and (\ref{eq:2:lha2}),
of which one is
\begin{equation}
-(VB)^j_i \partial_j a_\alpha = [(L-M)(L+M)]_{\alpha\beta}
\partial_i a_\beta,   \label{eq:2:rpa1}
\end{equation}
and the other for $\partial^i a_\alpha$, is the transpose of (\ref{eq:2:rpa1}).
This implies that $VB$ and $BV$ have the same diagonal form (if there are no
degeneracies, as we assume again).
Orthonormalization conditions are provided
by the Lagrange bracket relations
\begin{equation}
\partial_i a_\alpha \partial^j a_\alpha^{\ast} - \partial^j a_\alpha
\partial_i a_\alpha^{\ast} = i \delta^j_i.  \label{eq:2:lb1}
\end{equation}

As opposed to the formalism based on point transformations, the present
formalism is poorer in not yielding at the first stage any internal checks
concerning the goodness of decoupling.  On the other hand, it does provide
the first approximation of a systematic expansion, which, with suitable
modifications can also be applied to anharmonic vibrations.
In previous work \cite{6}, also reviewed and extended \cite{22}, we have given
rather different (and more formally elaborate) treatments of the basic
content of this subsection, referring to it as the symplectic, as opposed
to the Riemannian, version of the local harmonic approximation.  We omit that
material here because it will not be applied in the remainder of this review.

It is worthwhile, in the event that we restrict ourselves to point
transformations, to make the connection of the present formalism with
the previous Riemannian treatment.  Using Eq.~(\ref{eq:2:61.1}) and the
reality conditions (\ref{eq:2:real}) and (\ref{eq:2:imag}), we have
\begin{subequations}
\begin{eqnarray}
\partial_i a_\alpha &=& \frac{1}{\sqrt{2}}\frac{\partial\xi^\alpha}
{\partial q^i} =\frac{1}{\sqrt{2}}\frac{\partial p_i}{\partial\pi_\alpha},
\label{eq:2:ng1} \\
\partial^i a_\alpha &=& \frac{i}{\sqrt{2}}\frac{\partial q^i}{\partial
\xi_\alpha}.   \label{eq:2:ng2}
\end{eqnarray}
\end{subequations}
Consequently, we can rewrite (\ref{eq:2:chf})--(\ref{eq:2:lha2}) as
\begin{subequations}
\begin{eqnarray}
V_{,i}\frac{\partial q^i}{\partial\xi^\alpha} &=& S_\alpha, \label{eq:2:ng3} \\
B^{ij}\frac{\partial\xi^\alpha}{\partial q^j} &=& (M-L)_{\alpha\beta}
\frac{\partial q^i}{\partial\xi^\beta} = B^{\alpha\beta}
\frac{\partial q^i}{\partial\xi^\beta},  \label{eq:2:ng4} \\
V_{,ij}\frac{\partial q^i}{\partial\xi^\alpha} &=& (M+L)_{\alpha\beta}
\frac{\partial\xi^\beta}{\partial q^i} =V_{,\alpha\beta}
\frac{\partial\xi^\beta}{\partial q^i}.   \label{eq:2:ng5}
\end{eqnarray}
\end{subequations}
In this form it is easy to show, by repetition of steps already carried
out in previous discussions of the Riemannian case, that in the event that
we retain the curvature term in (\ref{eq:2:ng5}), the final result is to
replace the second derivatives on both sides by covariant second derivatives.
Whereas previously we have confined our attention to the second order
form of the RPA equations, we now have a first-order form that will 
prove helpful in our future considerations.

\subsubsection{Further discussion  of the treatment of conserved quantities\label{sec:2.3.5}}

To continue the discussion of Sec.~\ref{sec:2.2.5}, we work with the quantity
\begin{equation}
Y \equiv V+ \frac{1}{2}p_l p_m \bar{B}_{lm}.   \label{eq:2:4aa.3}
\end{equation}
Following the reasoning of Sec.~\ref{sec:2.3.3}, we can derive the equations
\begin{eqnarray}
Y^{,\alpha} &=& Y^{,i}g^\alpha_{,i}, \label{eq:2:4aa.4} \\
Y^{,\alpha}_{;\beta}g^\beta_{,i} &=& Y^{,j}_{;i}g^\alpha_{,j}. \label{eq:2:4aa.5}
\end{eqnarray}
Since (\ref{eq:2:4aa.5}) is derived from (\ref{eq:2:4aa.4}) by differentiation and
subsequent use of the geodesic equation, one sees that it also
holds when we replace the index $i$ by one of the indices $l$ associated
with the conserved quantities.

The general problem posed by these equations is rendered more difficult
by the presence of the second term in (\ref{eq:2:4aa.3}), which is just the
part of the collective kinetic energy associated with the conserved
quantities.  As a first approximation, we may neglect this term in
(\ref{eq:2:4aa.4}) and (\ref{eq:2:4aa.5}).  The justification for this step is
as follows:  We have expressed interest in working only to second order
in the collective momenta.  The retention of this term implies that we are
determining the quantities $\bar{V},\,\bar{B}^{ij},\,\bar{B}^{lm}$, etc.
as functions not only of the collective variables $q^i$ but also as functions
of the constants $p_l$, thus implicitly carrying the calculation to higher
order in the momenta.  This may be of interest for some applications,
and therefore we shall return below to how the calculation to be
described can be extended to include the missing kinetic energy.

The local harmonic formulation contained in (\ref{eq:2:4aa.4}) and (\ref{eq:2:4aa.5})
was not applied to the non-nuclear applications contained in the sections
that follow immediately.  They have proved most useful, however, for our
applications to nuclear physics.  We shall therefore postpone a 
discussion of how the formalism is solved in detail as a local, point
by point evaluation of the point transformation $g^\alpha(q)$ and of the
associated basis vectors. Such calculations start at points of dynamic
equilibrium, where $V_{,i}=0$.  At such a point the two sets of equations are
uncoupled, allowing us to solve (\ref{eq:2:4aa.4}) and (\ref{eq:2:4aa.5}) in sequence.
At all other points the two sets of equations become and remain coupled,
but may be solved by iteration starting from a known neighboring solution.
In this procedure, we replace the matrix $V^{,i}_{;j}$ by a matrix of
eigenvalues $\Lambda_i$.

The one extra problem that we have in the case of conserved quantities is
that of identifying the solutions of (\ref{eq:2:4aa.5}) corresponding to
conserved quantities.  This identification is based on the fact that at
equilibrium points the corresponding eigenvalues are zero. This is because
at equilibrium points all components of the gradient of $V$ vanish,
and that in consequence
it follows from its definition that the covariant derivative $V_{;il}$
reduces to the ordinary derivative $V_{,il}$, which vanishes.  Thus
\begin{equation}
V^{,\alpha}_{;\beta}g^{\beta}_{,l} =0, \;\;\;({\rm equilibrium}).
\label{eq:2:4aa.6}
\end{equation}
Using a previously noted property of a canonical transformation,
\begin{equation}
\frac{\partial\xi^\alpha}{\partial q^l} =\frac{\partial p_l}{\partial
\pi_\alpha},   \label{eq:2:4aa.7}
\end{equation}
since in all cases of interest we have an explicit form for the conserved
quantity $p_l(\xi,\pi)$, it follows that we know the solutions of
(\ref{eq:2:4aa.6}).
As we move away from a point of equilibrium, the identification (\ref{eq:2:4aa.7})
is still correct, only the quantity in question satisfies an eigenvalue
equation with a non-zero eigenvalue.  Finally, solutions that take into
account the difference between $Y$ and $V$ can be obtained by slowly
``turning on'' non-zero values of $p_l$ and solving the equations by an iteration procedure based on
the solution for the previous value of the $p_l$ as starting point.

Further insight into the origin of the zero modes, as we shall refer to them,
can be obtained from the direct study of Eqs.~(\ref{eq:2:ng4}) and (\ref{eq:2:ng5}).
Since $B^{ij}$ and $V_{;ij}$ are both symmetric matrices, we choose to
study these quantities first in a representation in which the
collective matrix $V$ is
diagonal with eigenvalues ${\cal V}_i$, so that (\ref{eq:2:ng5}) is replaced
by the equation
\begin{equation}
{\cal V}_i f^i_{,\alpha} =V_{;\alpha\beta}g^\beta_{,i},  \label{eq:2:ng10}
\end{equation}
and (\ref{eq:2:ng4}) retains its previous form in the new coordinate system.
In this discussion it is understood that the index $i$ comprises all
collective coordinates, including those associated with conserved quantities.
If we think of the eigenvalues ${\cal V}_i$ as restoring forces, then we have
the usual physical picture associated with zero modes.  One occurs whenever
a restoring force vanishes.  From a {\em complete set} of solutions to
(\ref{eq:2:ng10}), we can then use (\ref{eq:2:ng4}) to find the reciprocal basis vectors
$f^i_{,\alpha}$,
\begin{equation}
f^i_{,\alpha} = B_{\alpha\beta}B^{ij}g^\beta_{,j}.   \label{eq:2:ng11}
\end{equation}

There is, however, another possibility for zero modes.  Now we consider the
representation in which the matrix $B$ in the collective space is diagonal
with eigenvalues ${\cal B}^i$.  If one of these eigenvalues, $i=l$, vanishes,
(corresponding to an infinite eigenmass), then again we have a zero mode
\begin{equation}
B^{\alpha\beta}f^i_{,\beta} =0.  \label{eq:2:ng12}
\end{equation}
{}From a complete set of solutions of (\ref{eq:2:ng4}) we can find a complete
set of solutions of (\ref{eq:2:ng5}) according to the equation
\begin{equation}
g^\alpha_{,i} = (V^{-1})_{;\alpha\beta}V_{;ij}f^j_{,\beta}.   \label{eq:2:ng13}
\end{equation}

{}From these two possibilities, we see that there is no special problem
from multiple zero modes, provided they are all of the potential or of
the mass type.  However, each of the procedures, to be completed, requires
that one of the two symmetric matrices involved be non-singular.

\newpage
\section{Some Simple Applications of the Generalized Valley Theory} \label{sec:3}

\subsection{The landscape model\label{sec:3.1}}

\subsubsection{Two-dimensional version\label{sec:3.1.1}}

In this section we shall study some examples of the decoupling theory
discussed in the previous section. The most straightforward case is
the decoupling of one degree of freedom in a model with two degrees of
freedom.  A very convenient two-dimensional model for this purpose is
the ``landscape model'' introduced by Goeke, Reinhard and Rowe
\cite{42}.
(The discussion in this subsection is
based on Ref.~\cite{22}.)

\begin{figure}
\centerline{\includegraphics[width=7cm]{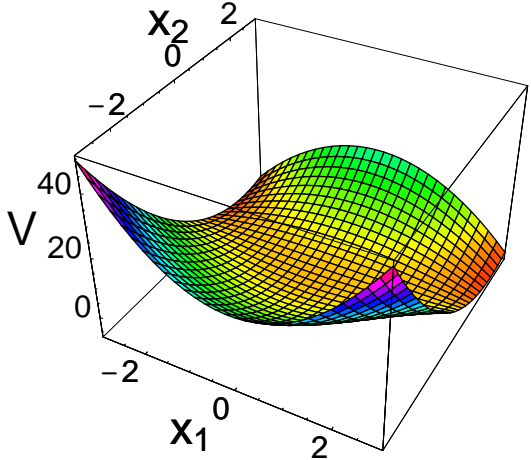}}
\caption{A three dimensional presentation of the potential energy
surface for the landscape model 
for  $\omega_1=1$, $\omega_2=2$, $\beta=-1$. \label{fig:3:lan1}}
\end{figure}

The Hamiltonian of this model is given by
\begin{equation}
H = \half (p_1^2 + p_2^2) + \half [(\omega_1 x^1)^2 + (\omega_2 x^2)^2]
+\beta(x^1)^2x^2.
\label{eq:3:landscape}
\end{equation}
The physics governing the model is most easily understood by looking
at the graphical representation of the potential energy surface shown in
Fig.~\ref{fig:3:lan1}.

The potential energy has a local minimum at
the origin where the energy is zero, and a pair of saddle points at
$(x^1,x^2) = (\pm \sqrt{2}\omega_1\omega_2/(2\beta),
                         -\omega_1^2/(2\beta))$, where the
potential energy is $\omega_1^4\omega_2^2/(8\beta^2)$.
The model derives its name from the landscape-like potential energy
surface. Note that, in contrast to a real landscape,
the potential energy decreases without bound beyond the saddle points. Quantum mechanically, this
allows a
particle to escape from the potential well, for instance, by tunneling
through in the neighborhood of the saddle point and
moving to infinity as, e.g., in the case of nuclear fission. This analogy
was one of the reasons for introducing the model.

We discuss this model with the help of the potential energy function,
$V$, and the square of its gradient, $U$.  The generalized valley equation,
\begin{equation}
V_{,\alpha} = \Omega U_{,\alpha},
\end{equation}
can be reduced to a single equation by eliminating $\Omega$,
\begin{eqnarray}
0 & = & V_{,1}U_{,2} - V_{,2} U_{,1} \nonumber\\
  & = &  -   x^{1}
\left\{4(x^{2})^{3}\beta ^{2}\omega_{2}^{2}
    +4(x^{2})^{2}\left[\beta\omega_{2}^{2}\omega_{1}^{2}-(x^{1})^{2}\beta
    ^{3}\right]
\right. \nonumber\\
&   &  \left. \qquad
+2x^{2}\left[ (x^{1})^{2}\beta^{2}(\omega_{2}^{2}
                                  -2\omega_{1}^{2})
            -\omega_{2}^{4}\omega_{1}^{2}
            +\omega_{2}^{2}\omega_{1}^{4}\right]
    +2(x^{1})^{4}\beta ^{3}
    -(x^{1})^{2}\beta\omega_{1}^{2}(\omega_{2}^{2}+\omega_1^2)
\right\}.
\end{eqnarray}
Here we have exhibited the generalized valley equation suggestively as
a cubic equation for $x^2$, with coefficients that are functions of $x^1$. 
Of the three solutions
of the cubic, one corresponds to a path that connects the minimum
with the
saddle point, following the direction of steepest descent near the saddle,
and the other two describe a closed path that corresponds
to the direction of steepest
ascent near the saddle. Of course we
restrict our attention to the first type of solution.

\begin{figure}
\centerline{\includegraphics[width=10cm]{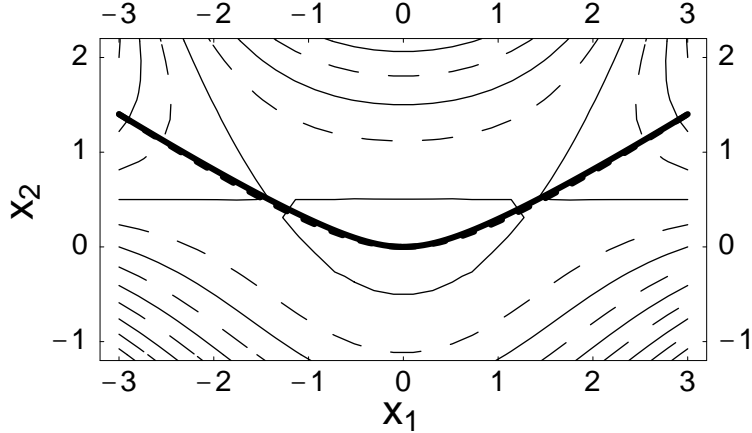}}
\caption{The solution of the valley equation (solid line)
for  $\omega_1=1$, $\omega_2=2$, $\beta=-1$. The thin solid and dashed lines
are contours of the potential energy.\label{fig:3:lan2}}
\end{figure}

In Fig.~\ref{fig:3:lan2}
we show the solution of the valley equation,
for the parameters $\omega_1=1$, $\omega_2=2$, $\beta=-1$,
as the solid line, drawn on a background of the contours of the potential
energy. As one can see, the valley passes through the minimum and
the saddle points.

\begin{figure}
\centerline{\includegraphics[width=10cm]{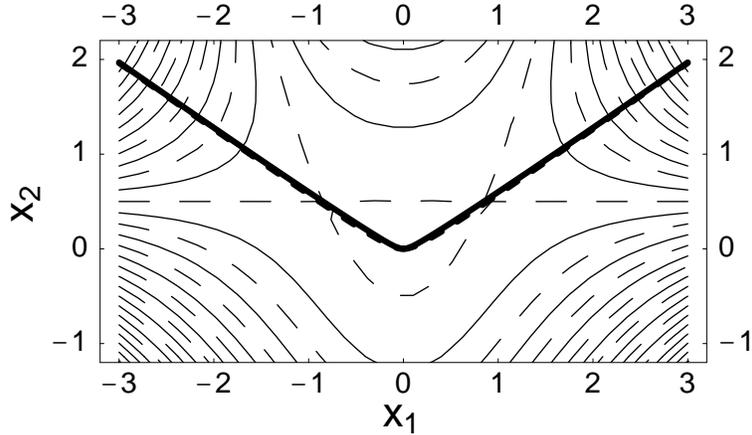}}
\caption{The solution of the valley equation (solid line)
for  $\omega_1=1$, $\omega_2=1.2$, $\beta=-1$. The thin solid and dashed lines
are contours of the potential energy.\label{fig:3:lan3}}
\end{figure}

We also show the same solution for a different $\omega_2$, $\omega_2=1.2$,
in Fig.~\ref{fig:3:lan3}
Although the solution looks similar, one should
notice the larger curvature of the valley near the origin. We shall
show that this indicates poorer decoupling.

Before considering the matter of decoupling, let us first discuss
alternative methods to determine an approximately decoupled manifold.
As described in Sec.~\ref{sec:2.3}, we can  use the local harmonic equation,
which takes a particularly simple form here because the metric tensor
is the unit matrix,
\begin{equation}
V_{,\alpha}^{\beta} f^{\mu}_{,\beta} = \Omega^2_\mu f^{\mu}_{,\alpha}.
\label{eq:3:LHA}
\end{equation}
It is supplemented by the force condition
\begin{equation}
V_{,\alpha} = \lambda f^{1}_{,\alpha}.
\label{eq:3:Visevec}
\end{equation}
Replacing $f^1$ by $V_{,\alpha}/\lambda$ in (\ref{eq:3:LHA}), we find
\begin{equation}
\Omega_1^2 V_{,\alpha} = V_{,\alpha}^{\beta} V_{,\beta} = U_{,\alpha},
\end{equation}
which shows that the local harmonic equation for the decoupling of one
coordinate is equivalent to the valley equation, a result already
known to us.

It is of some interest to consider an alternative to the subsidiary
condition Eq.~(\ref{eq:3:Visevec}). Because of the simple metric, given a
value of $f^1_{,\alpha}$, it and the associated value of
$g^\alpha_{,1}$ are proportional to each other; thus the latter may
replace the former in Eqs.~(\ref{eq:3:LHA}) and (\ref{eq:3:Visevec}) since
covariant and contravariant derivatives are now identical.  In the
case of exact decoupling we would find that $V_{,\alpha}$ is parallel
to the path $g^\alpha_{,1}$.  In the present model this can not be
true, since we do not have exact decoupling. Thus the condition that
$g_{,1}^\alpha$ is parallel to the path is different from
Eq.~(\ref{eq:3:Visevec}). We thus obtain a second algorithm by combining
Eq.~(\ref{eq:3:LHA}) for $g^\alpha_{,1}$ with the condition that the
latter be along the path, namely
\begin{equation}
g_{,1}^\alpha = dx^\alpha/dq,
\label{eq:3:parallel}
\end{equation}
which serves as replacement for Eq.~(\ref{eq:3:Visevec}).
For the model at hand, we know
from the solution of the GVE  that the collective path
can be parameterized as
\begin{equation}
x^1 = q,\;\;\;x^2 = h(q),
\end{equation}
which would give
\begin{equation}
g_{,1}^{\alpha} = (1,h'(q)).
\end{equation}
If we use this in the local harmonic equation, we find that
\begin{subequations}
\begin{eqnarray}
V_{,11} + V_{,12} h'&=&\Omega^2 ,\label{eq:3:firstLHA} \\
V_{,21} + V_{,22} h'&=&\Omega^2h'.\label{eq:3:secondLHA}
\end{eqnarray}
\end{subequations}
Using Eq.~(\ref{eq:3:firstLHA}) to eliminate $\Omega^2$ in the second
equation, we find a quadratic equation for $h'$ that can be converted
to an ordinary differential equation by solving for $h'$,
(note that the derivatives of $V$ are functions of $q$ and $h(q)$)
\begin{equation}
h' =
\frac{(V_{,22}-V_{,11}) \pm \sqrt{(V_{,22}-V_{,11})^2 + 4 V_{,12}V_{,21}}}
     {2V_{,12}}.
\label{eq:3:difE}
\end{equation}
This differential equation is solved as an initial value problem,
integrating outward from the point $x^1=0,x^2=0$, i.e., $h(0) =0$.
Since we would like to find that solution that is as close to the valley
as possible, we choose the minus sign in Eq.~(\ref{eq:3:difE}), so that
$h'(0)=0$ as well.

The solutions to this modified local harmonic equation are shown in
Figs.~\ref{fig:3:lan2} and \ref{fig:3:lan3} by the dashed lines.
They follow the valley closely, but not exactly. For that reason
they do not pass through the saddle point. This shows that the
GVA and the usual form of the LHA have some advantage over this form
of the local harmonic equation.

Now that we have calculated the valley we can easily calculate
the collective potential energy as
\begin{equation}
\bar{V}(q) = V(q,x^2(q)).
\end{equation}
If we wish to calculate the collective mass we need to have a set of
basis vectors.
As discussed in Sec.~\ref{sec:2.2.3}, there are at least 
two natural choices that can be made, which lead
to the same collective mass
in the case of exact
decoupling. In general, i.e., when decoupling is not exact, as just
stated above, these two sets will not lead to the same answer.
The first definition of the mass is based on the tangent vector,
\begin{equation}
g^\alpha_{,1} = dx^\alpha/dq = (1,dx^2/dx^1),
\end{equation}
so that the inverse mass can be evaluated as
\begin{equation}
\bar{B}_{11} = g^\alpha_{,1} B_{\alpha\beta} g^\beta_{,1}
             = 1 + (dx^2/dx^1)^2.
\label{eq:3:BBB}
\end{equation}
The other definition of the mass is based on the covariant form of the
gradient of $V$,
\begin{eqnarray}
V_{,\alpha}&=&\bar{V}_{,q} f^1_{,\alpha}     \nonumber\\
           &=& (V_{,\alpha}g^\alpha_{,1}) f^1_{,\alpha}     \nonumber\\
           &=& (V_{,1}+V_{,2}dx^2/dx^1) f^1_{,\alpha},
\end{eqnarray}
so that we find
\begin{eqnarray}
\breve{B}^{11} & = & f^1_{,\alpha} B^{\alpha\beta} f^1_{,\beta}
\nonumber\\
               & = & [(V_{,1})^2+(V_{,2})^2]/
               [V_{,1}+V_{,2}(dx^2/dx^1)]^2.
\end{eqnarray}
We use the inverse of Eq.~(\ref{eq:3:BBB}) as our definition of
the collective mass, i.e.,
\begin{equation}
\bar{B}^{11} = 1/[1+(dx^2/dx^1)^2].
\end{equation}
As stated before the difference between $\bar{B}$ and $\breve{B}$ is
related to the quality of decoupling. For that reason we evaluate
\begin{equation}
D = |\bar{B}_{11} \breve{B}^{11} - 1| .
\end{equation}
Let us now look at the behavior of $\bar{V}$, $\bar{B}$ and $D$ for the
two sets of parameters considered.

\begin{figure}
\centerline{\includegraphics[width=7cm]{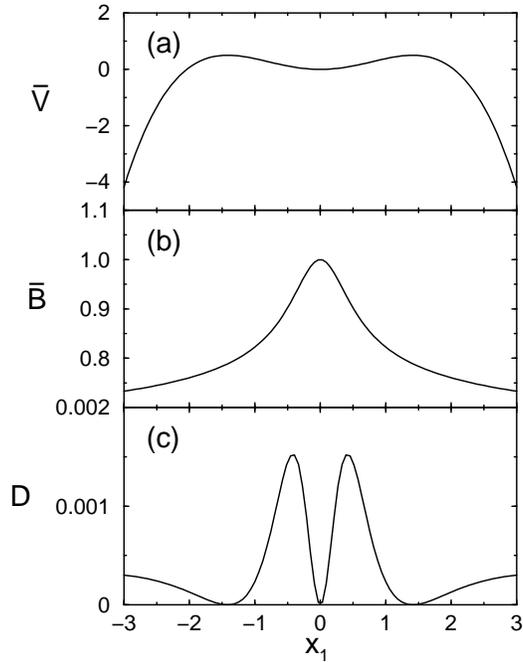}}
\caption{The solution to the collective path for the landscape model
for $\omega_1=1$,  $\omega_2=2$, $\beta=-1$. 
In panel (a) we show the collective potential energy, in panel (b) the
collective mass, and panel (c) contains the decoupling parameter $D$.
\label{fig:3:lan4}}
\end{figure}

We have plotted these three quantities
for  $\omega_1=1$, $\omega_2=2$, $\beta=-1$ in Fig.~\ref{fig:3:lan4}.
As can be seen from the value of
the decoupling measure $D$, Fig.~\ref{fig:3:lan4}c,
decoupling is very good.
In  Fig.~\ref{fig:3:lan4}b
we can see that the mass changes only very
slowly along the path, another indication that the decoupling is
excellent.

\begin{figure}
\centerline{\includegraphics[width=7cm]{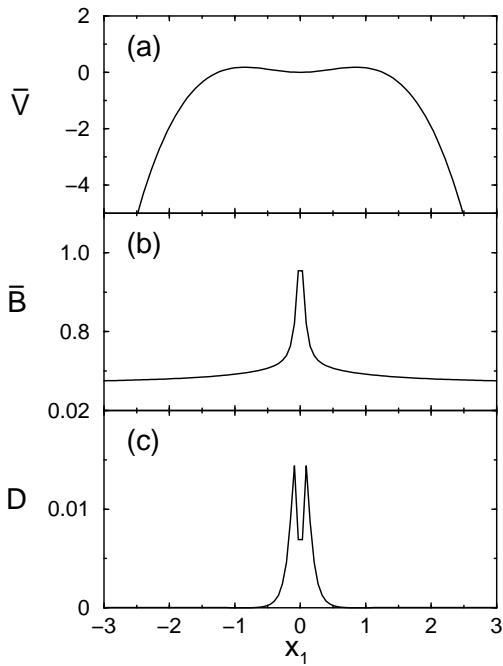}}
\caption{The solution to the collective path for the landscape model
for $\omega_1=1$,  $\omega_2=1.2$, $\beta=-1$. 
In panel (a) we show the collective potential energy, in panel (b) the
collective mass, and panel (c) contains the decoupling parameter $D$.
\label{fig:3:lan5}}
\end{figure}

If we once again reduce $\omega_2$ to $1.2$ (Figs.~\ref{fig:3:lan5}),
we find much larger values for $D$ as illustrated
in Fig.~\ref{fig:3:lan5}c.
The collective mass $\bar{B}$ also shows a
much more pronounced change. These are all indications of poorer
decoupling.

\subsubsection{Three-dimensional model\label{sec:3.1.2}}
In order to illustrate the process of decoupling two
coordinates (out of three) we use a model which
is a simple generalization of the model of
the previous section to one more dimension,
\begin{equation}
H              = \half[p_1^2+p_2^2+p_3^2]+
\half[\omega_1^2(x^1)^2 +
                       \omega_2^2(x^2)^2 +
                       \omega_3^2(x^3)^2 ] +
                  \beta[(x^1)^2+(x^2)^2]x^3.
\end{equation}
The potential energy has one minimum, at
${\vec x} = 0$, and two pairs of saddle points (at
${\vec x} = (\pm \omega_1\omega_3/[\beta\sqrt2],0,\omega_1^2/[2\beta])$
and
$\vec{x} = (0,\pm \omega_2\omega_3/[\beta\sqrt2],\omega_2^2/[2\beta])$.
These saddle points are connected through shallow one-dimensional valleys
with the minimum at the origin.

\begin{figure}
\centerline{\includegraphics{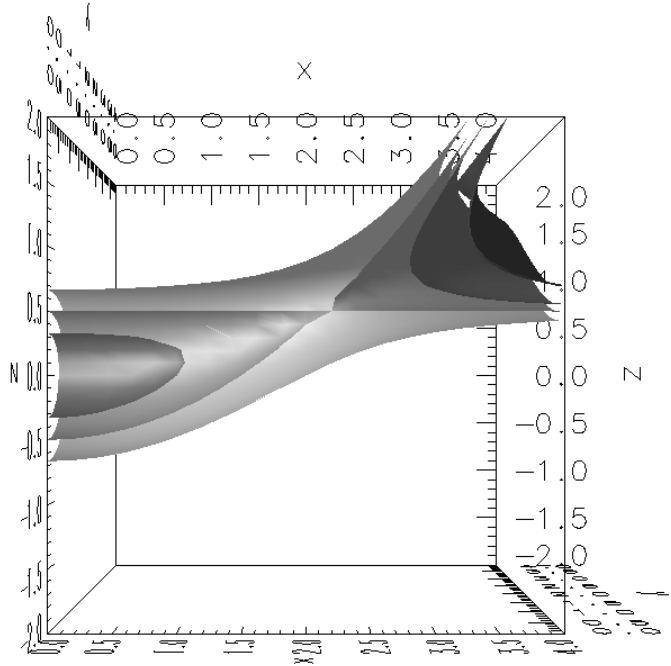}}
\caption{A set of isopotential surfaces for the 3D landscape model
($\omega_1=1$, $\omega_2=2$, $\omega_3=3$, $\beta=-1$).
{}From dark to light the energy values are -1, 0.5, 9/8 and 2. We clearly
see the saddle point for $x_2=0$ at $V=9/8$. 
For conciseness we have labeled the axes with $x$, $y$ and $z$, rather
than $x_1,x_2,x_3$. The range on the $y$ axis ranges from 0 to 4.
\label{fig:3:isopot}}
\end{figure}

This can clearly be seen in Fig.~\ref{fig:3:isopot}, where we show a set
of isopotential surfaces for the potential energy. These show clearly the minimum
at the origin, connected through a saddle point to the region of 
unbounded negative energy.

To decouple two coordinates out of three we need three scalar functions.
In this study we use the potential $V$, $U=\sum_i (\partial V/
\partial x^i)^2/2$ and
$W=\sum_i (\partial U/\partial x^i)^2/2$.
By elimination of the Lagrange multipliers
the generalized valley equation can be converted into an equation for
the vanishing of a determinant,
\begin{equation}
\left| \begin{array}{ccc}
         V_{,1} & U_{,1} & W_{,1} \\
         V_{,2} & U_{,2} & W_{,2} \\
         V_{,3} & U_{,3} & W_{,3}
\end{array} \right| = 0.
\end{equation}
By explicit evaluation we can show that, for fixed $x^1$ and $x^2$,
this equation is a sixth order polynomial equation for $x^3$, and
as such the solutions may be very complex. Furthermore, the equation
has the trivial solutions $x^2=0$ and  $x^1=0$. In the following
discussion we always discuss the determinant divided
by the factor $x^1x^2$.

We have studied this equation for $\omega_1=1$, $\omega_2=2$,
$\omega_3=3$ and $\beta=-1$. The potential energy is symmetric under the
transformation $x^1 \rightarrow -x^1$, as well as under $x^2 \rightarrow
-x^2$, so that we need only to consider the region $x^1 \geq 0$, $x^2
\geq 0$. Near the point $x^1=0$, $x^2=0$ we find that the GVE has only
two real solutions, only   one of which passes through $\vec{x}=0$. We
now wish to study this last solution. In order to be able to present clearly
what is happening, we have drawn graphical representations of
one-dimensional slices of the solutions to the GVE, $x^3(x^1,x^2)$.

\begin{figure}
\centerline{\includegraphics[width=5cm]{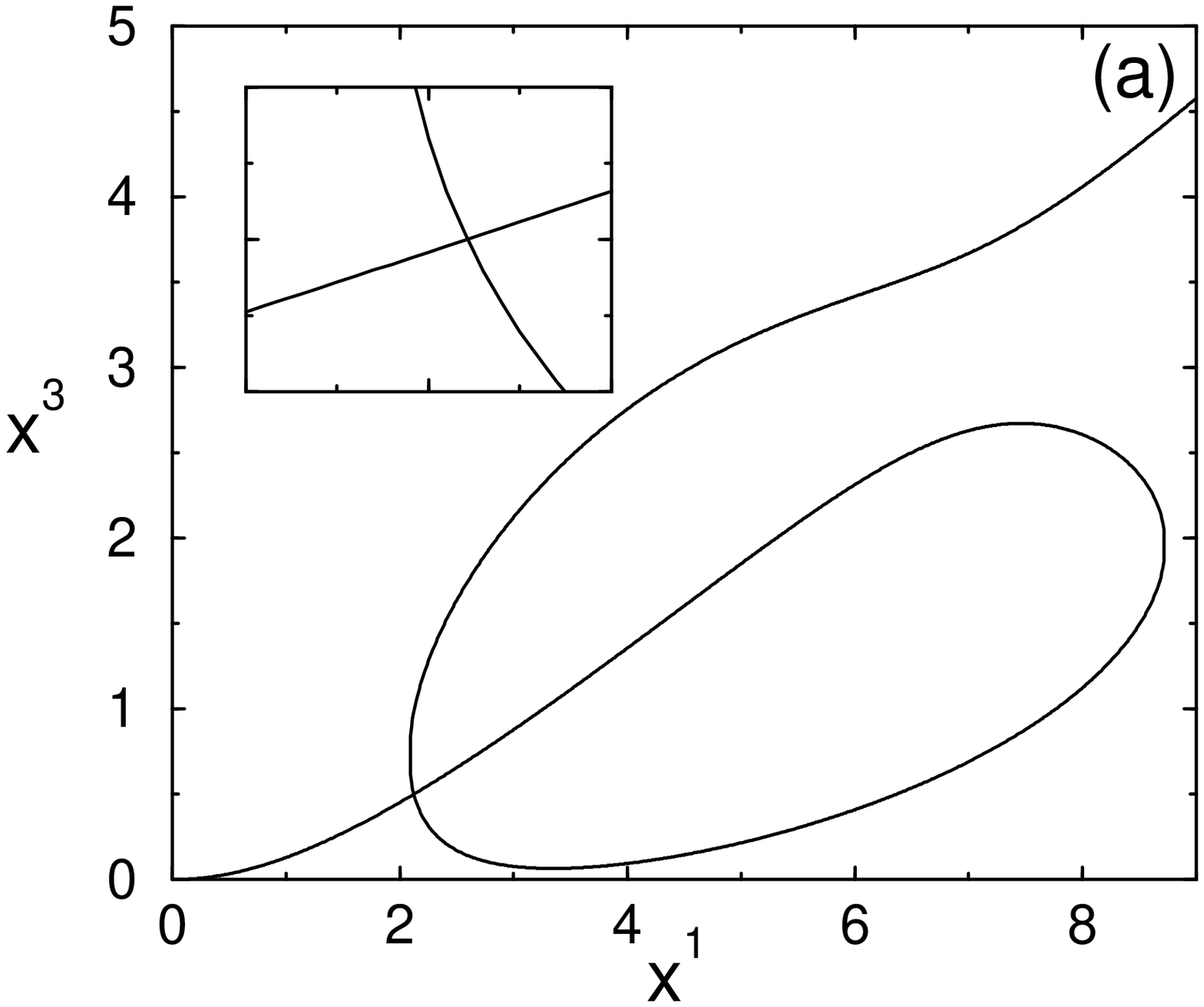}\qquad\includegraphics[width=5cm]{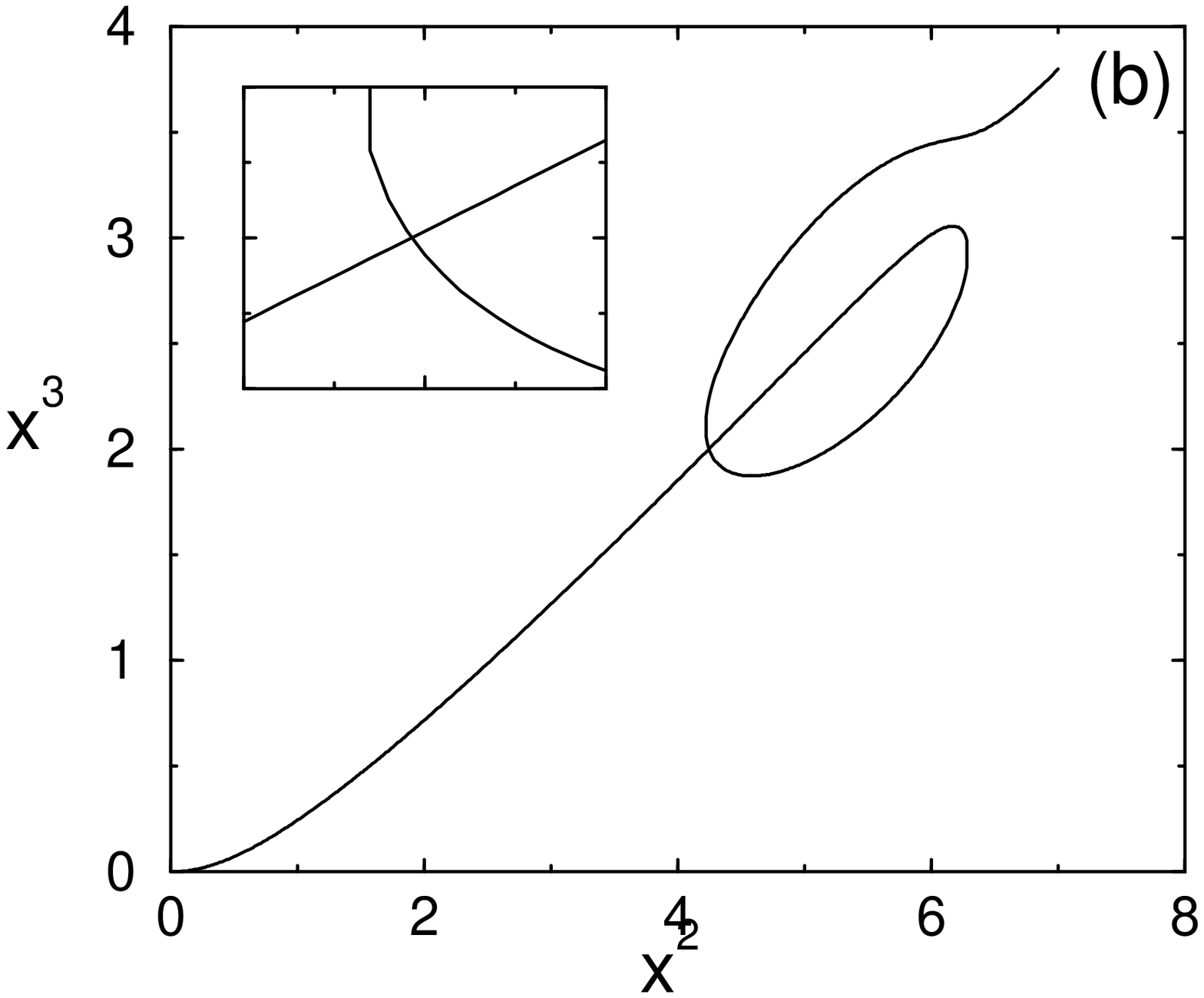}}
\centerline{\includegraphics[width=5cm]{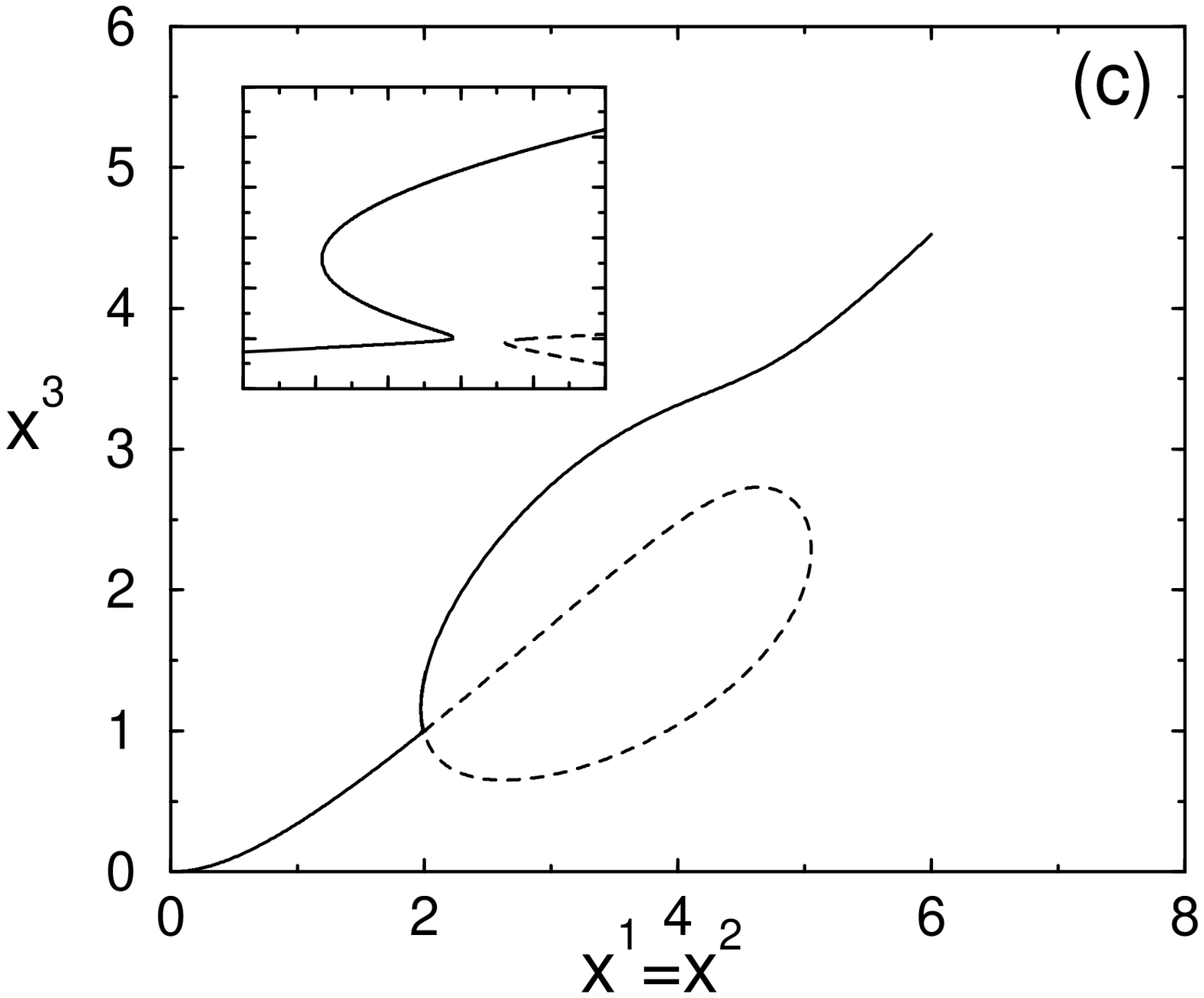}}
\caption{Three slices through the surface for the solution to the valley
equations for $\omega_1=1$,  $\omega_2=2$, $\omega_3=3$ and $\beta=-1$. 
In panel (a) we show the case $x^2=0$, in panel (b) $x^1=0$ and 
in panel (c) $x^1=x^2$.
\label{fig:3:lan6}}
\end{figure}

\begin{figure}
\centerline{\includegraphics{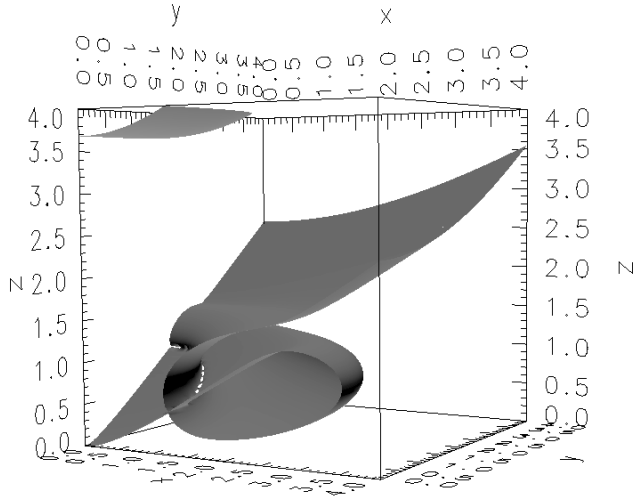}}
\caption{A three dimensional representation of the solution to the GVA
sketched in Fig.~\ref{fig:3:lan6}
\label{fig:3:lan6'}}
\end{figure}

The two slices shown in Figs.~\ref{fig:3:lan6}a and ~\ref{fig:3:lan6}b,
for $x^2=0$ and $x^1=0$, contain both the
minimum and a saddle point.  In that case we encounter a result
that is familiar from the case of decoupling
in two dimensional models: a second solution crosses the valley at the
saddle point. For the sake of clarity we have enlarged the region of the
crossing in the inset in Figs.~\ref{fig:3:lan6}.
The result is not too surprising once one realizes that we have
a one-dimensional valley solution for both $x^2=0$ and $x^1=0$. The two
cuts represented here are just these solutions.

The situation is very different, however, if we take any other cut
through the $x^1$--$x^2$ plane. A representative case, for $x^1=x^2$
is presented in Fig.~\ref{fig:3:lan6}c.
We find that on a course scale we do not
see any difference in behavior, i.e., we see a crossing, but if we
enlarge the region of the crossing we see that the crossing does not
really take place.
All the features illustated in Fig.~\ref{fig:3:lan6} can be obtained from
suitable cuts of the three-dimensional surface sketched in Fig.~\ref{fig:3:lan6'}.

The avoided crossing appears not to be due to a numerical artifact. In
an attempt to understand the situation we have just described, we have
studied the alternative procedure for defining an approximately
decoupled surface, based on the second form of the fundamental
decoupling theorem developed in Sec.~\ref{sec:2.2.1}. This is the form
that makes direct use of simplified vector fields rather than
proceeding from point functions and is more directly tied to the local
harmonic approximation when there is not exact decoupling.  In this
approach, we have already shown that $V_{,\alpha}$, $U_{,\alpha} =
V_{,\alpha}^{~\beta} V_{,\beta}$, $\tilde{W}_{,\alpha} =
V_{,\alpha}^{~\beta} V_{,\beta}^{~\gamma} V_{,\gamma}$ must be a set
of three dependent vectors, since they are all three linear
combinations of $f^1$ and $f^2$.  From this linear dependence we once
again obtain a determinant that should be zero,
\begin{equation}
\left| \begin{array}{ccc}
         V_{,1} & U_{,1} & \tilde W_{,1} \\
         V_{,2} & U_{,2} & \tilde W_{,2} \\
         V_{,3} & U_{,3} & \tilde W_{,3}
\end{array} \right| = 0.
\label{eq:3:LHAdet}
\end{equation}
This equation looks very similar to the GVE, but numerical study shows
that it does not suffer from the avoided crossing problems
noted for the GVA.

We shall now explain why the  GVA and the LHA give different results.
First consider the situation with respect to stationary paths in the
model under study.  These are of course solutions to the ``valley"
equation $V_{,\alpha}-\Omega_1 U_{,\alpha}=0$.  In addition to the two
valleys that bound the part of the decoupled surface under study, there
will be a ridge connecting the two associated saddle points. It is an
easy consequence of the equations of the LHA that because along such
an extremal path $V_{,\alpha}$ degenerates into proportionality to a
definite solution of the local harmonic equation (rather than being a
linear combination of such solutions, the general case), that not
only are $V_{,\alpha}$ and $U_{,\alpha}$ parallel to each other as
required by the valley equation but each is also parallel to
$\tilde{W}_{,\alpha}$.  Therefore along the ridge in question
the modified GVA equation (\ref{eq:3:LHAdet}) does not determine a local
tangent plane, but instead only a single direction common to the two
intersecting surfaces, an intersection that takes place along the ridge.
If, however, in the generalized valley equation we replace
$\tilde{W}_{,\alpha}$ by another vector field belonging to the set
named in the fundamental theorem, then as long as decoupling is not
exact, this alternate third vector field will not in general be parallel
to the first two, and a local tangent plane will be determined.
If the different choices
do not differ radically, however, we must expect the phenomenon
encountered, namely a near crossing.

\begin{figure}
\centerline{\includegraphics[width=5cm]{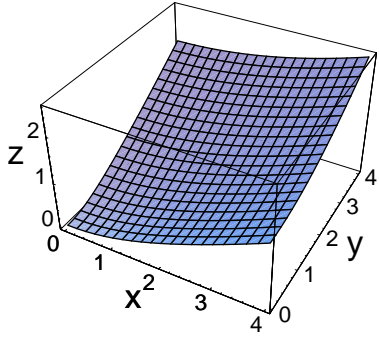}}
\caption{The solution to the LHA
equations for $\omega_1=1$,  $\omega_2=2$, $\omega_3=3$ and $\beta=-1$. 
\label{fig:3:lan7}}
\end{figure}

\begin{figure}
\centerline{\includegraphics[width=5cm]{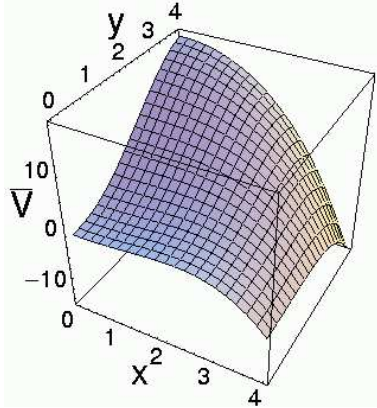}}
\caption{The collective potential energy for the LHA surface shown in
Fig.~\ref{fig:3:lan7}.
\label{fig:3:lan8}}
\end{figure}

\begin{figure}
\centerline{\includegraphics[width=5cm]{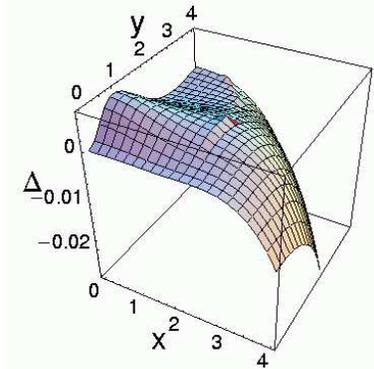}}
\caption{The difference between the LHA surface shown in
Fig.~\ref{fig:3:lan7} and the nearest solution to the GVA.
\label{fig:3:lan9}}
\end{figure}

It is simplest, therefore,
to restrict our attention to the solution $x^3(x^1,x^2)$
obtained from  the LHA. We have drawn that solution that runs through
both saddle points and the minimum in Fig.~\ref{fig:3:lan7}.
We also show the behavior of the collective potential energy in Fig.~\ref{fig:3:lan8}.
One can distinguish the presence of two shallow
valleys along the $x^1$ and $x^2$ axis.
That the surface $x^3(x^1,x^2)$ obtained from the local harmonic equation
 is very close to a solution of the GVA,
except in the region of the crossing can be seen in Fig.~\ref{fig:3:lan9}.
In that figure we plot
$\Delta$, the difference between the solution of the LHA for fixed $x_1,x_2$ 
and the nearest real solution to the GVA.
The difference is small, but grows as $x^1$ and $x^2$ both increase.

\begin{figure}
\centerline{\includegraphics[width=5cm]{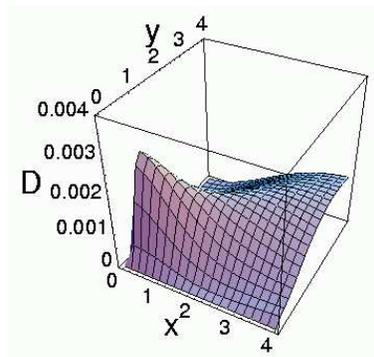}}
\caption{The decoupling measure $D$ for  the LHA surface shown in
Fig.~\ref{fig:3:lan7}.
\label{fig:3:lan10}}
\end{figure}

It now remains to discuss the quality of decoupling. With our
definition of the collective surface as $x^3(x^1,x^2)$ we easily obtain
\begin{subequations}
\begin{eqnarray}
g^\alpha_{,1}&=&(1,0,\partial x^3/\partial x^1),\\
g^\alpha_{,2}&=&(0,1,\partial x^3/\partial x^2).
\end{eqnarray}
\end{subequations}
The inverse collective mass now can be calculated as
\begin{equation}
\bar{B}_{ij} = \sum_\alpha g^\alpha_ig^\alpha_j.
\end{equation}
We can also derive a set of vectors from the gradients of $V$ and $U$,
\begin{eqnarray}
V_{,\alpha} & = & \bar{V}_{,1} f^1_{,\alpha} +
                  \bar{V}_{,2} f^2_{,\alpha},\nonumber\\
U_{,\alpha} & = & \bar{U}_{,1} f^1_{,\alpha} + \bar{U}_{,2}f^2_{,\alpha}.
\end{eqnarray}
[Here we use the chain rule, $\bar{X}_{,i} = X_{,\alpha} g^\alpha_{,i}$].
This can be inverted to give $f^1$ and $f^2$, with suitable limits taken
if the determinant
$\left|\begin{array}{cc} \bar{V}_{,1} & \bar{U}_{,1}\\
\bar{V}_{,2} & \bar{U}_{,2}
\end{array}\right|$
equals zero. We thus obtain
\begin{equation}
\breve{B}^{ij} = \sum_\alpha f_{,\alpha}^i f_{,\alpha}^j.
\end{equation}
We can now define the decoupling measure $D$:
\begin{equation}
D = \left|\Bigl[\half \sum_{ij} \bar{B}_{ij}\breve{B}^{ji}\Bigr]-1\right| .
\end{equation}
This quantity is displayed in Fig.~\ref{fig:3:lan7}.
As can be seen it is relatively small, showing its largest values for
$x^1=0$.  This is where the second and third eigenvalues of the local
harmonic equation are closer than for other points in the $x^1$--$x^2$
plane, so that the decoupling is poorer. Nevertheless, the results we
have found show excellent overall decoupling.

Apart from the problem with avoided crossings that we have discussed
in detail, there do not appear to be major differences between the
results obtained from the generalized valley equations and from the
local harmonic approximation. The fact that there are these avoided
crossings may lead to a slight prejudice in favor of the LHA, even
though we would like to point out that from a physical point of view,
in contrast to the mathematical discussion given above, we can easily
``span the gap'' by interpolation, making it less problematic to work
with the GVA.

\subsection{A simplified model for tunneling in many-particle
systems  \label{sec:3.2}}

\subsubsection{Description of the model \label{sec:3.2.1}}
As a more substantial example of the methods developed in the previous
sections, we consider a model of a system of $N$ oscillators, each
assigned as well a half unit of spin.  The Hamiltonian is chosen to be of
such a simple form that the model is equivalent to an oscillator
in interaction with a spin system that is almost classical.
This model was studied first in
\cite{37} by quite different methods.  The following discussion is
based on Ref.~\cite{14}.

   The system to be studied is described by the Hamiltonian,
\begin{equation}
H=\sum_{i=1}^{N}
\frac{1}{2}\left(p^{2}_{i}+q^{2}_{i}\right)+\kappa  \left(\sum _{i=1}^{N} q_{i} \right)
\left(\sum_{i=1}^{N}
\sigma _{zi} \right) -\lambda \left(\sum _{i=1}^{N} \sigma _{xi} \right)^{2}. \label{eq:3:ap1.1}
\end{equation}
One sees that only the center
of mass is coupled to the total spin and the internal motion is
unaffected by interaction. This
leads to a tremendous simplification, if we introduce the new variables
\begin{eqnarray}
q &=& \frac{1}{\sqrt{N}} \sum_{i=1}^{N} q_{i},  \;\;
p = \frac{1}{\sqrt{N}} \sum_{i=1}^{N} p_{i}, \label{eq:3:ap1.2} \\
S_{k} &=& \frac{1}{2}\sum_{i=1}^{N} \sigma_{ki},\; k=x,y,z, \label{eq:3:ap1.4}
\end{eqnarray}
as well as $N-1$  relative variables.
The Hamiltonian (\ref{eq:3:ap1.1}) then separates into an uninteresting piece,
representing the uncoupled relative oscillators, and the interesting part,
\begin{equation}
H^{\prime}=\frac{1}{2} p^{2}+\frac{1}{2} q^{2} +2
\kappa \sqrt{N} q S_{z} -4 \lambda S^{2}_{x}.  \label{eq:3:ap1.5}
\end{equation}
Henceforth we replace
 $H^{\prime}$ by $H$.  We shall also set $N=2J$ throughout this
discussion, where $J$ is the value of the ``angular momentum'' for the band of
states which 
can be shown to include
the ground state.

\subsubsection{Study of the classical limit
by the generalized valley algorithm \label{sec:3.2.2} }

In order to study the Hamiltonian (\ref{eq:3:ap1.5}) by the methods
of Secs.~\ref{sec:2.2} and \ref{sec:2.3},
it is convenient to introduce a mapping of the $su(2)$ algebra onto a
pair of quantum action-angle variables $\Pi$ and $\xi$
according to the formulas \cite{38,39}
(the non-standard choice of the quantization axis is related to the 
form of the Hamiltonian)
\begin{eqnarray}
   S_{+} &=& (S_{-})^{\dagger} =S_{z} + iS_{x}  \nonumber \\
 &=& \exp\left(\sfrac{1}{2}i\xi\right)\sqrt{(J+(1/2))^{2}-\Pi^{2}}
\exp\left(\sfrac{1}{2} i\xi\right)  \nonumber \\
  &=& \exp(i\xi)\sqrt{J(J+1)-\Pi(\Pi+1)}   , \label{eq:3:ap1.6} \\
 S_{y} &=& \Pi,                \label{eq:3:ap1.7}
\end{eqnarray}
where
\begin{equation}
[\Pi,\exp(i\xi)]=\exp(i\xi).   \label{eq:3:ap1.8}
\end{equation}
{}From these equations, we obtain approximate formulas for $S_{z}$ and
$S_{x}^{2}$,
required for the evaluation of Eq.~(\ref{eq:3:ap1.5}), by expanding in
powers of $(\Pi/J)$, assuming $\Pi$ to be no larger than of order unity. (This
assumption will 
have to be confirmed {\em a posteriori}
to be valid for the states of interest to us.)
These formulas are
\begin{subequations}
\begin{eqnarray}
S_{z} &=& \sqrt{J(J+1)}\cos(\xi)
-\frac{1}{4\sqrt{J(J+1)}}\{\cos(\xi),\Pi^{2}\},  \label{eq:3:ap1.9} \\
S_{x} &=& \sqrt{J(J+1)}\sin(\xi)-
\frac{1}{4\sqrt{J(J+1)}}\{\sin(\xi),\Pi^{2}\},   \label{eq:3:ap1.10} \\
S_{x}^{2} &=& J(J+1)\sin^{2}(\xi) -
\frac{1}{4}\{\sin(\xi),\{\sin(\xi),\Pi^{2}\}\},   \label{eq:3:ap1.11}
\end{eqnarray}
\end{subequations}
where the braces stand for the anticommutator of the corresponding
operators.
The Hamiltonian (\ref{eq:3:ap1.5}) thus becomes
\begin{subequations}
\begin{eqnarray}
H &=& V(q,\xi) + T ,  \label{eq:3:ap1.12}  \\
V(q,\xi) &=& \frac{1}{2}q^{2}+2\kappa q\sqrt{2J^{2}(J+1)}\cos(\xi)
 -4\lambda J(J+1)\sin^{2}(\xi)  ,    \label{eq:3:ap1.13}  \\
T &=& \frac{1}{2}p^{2}-\kappa q
\frac{1}{\sqrt{2(J+1)}}\{\cos(\xi),\Pi^{2}\}
+\lambda\{\sin(\xi),\{\sin(\xi),\Pi^{2}\}\}.   \label{eq:3:ap1.14}
\end{eqnarray}
\end{subequations}
The above expression for $H$ contains correctly the classical limit plus
the corrections of order $1/J$.  The choice of variables made in
Eqs.~(\ref{eq:3:ap1.6}) and (\ref{eq:3:ap1.7})
has the great advantage that most of the burden of variation of the spin
variables is carried by the trigonometric functions of
$\xi$ and 
that $\Pi$
remains small in the domain of interest to us.

Passing to the classical limit, we may rewrite the kinetic energy as
\begin{equation}
T_{{\rm class}} =\frac{1}{2}p^{2} +\frac{1}{2}B_\xi(q,\xi)\Pi^{2} ,\label{eq:3:ap1.15}
\end{equation}
where
\begin{equation}
B_\xi(q,\xi) = 8\lambda\sin^{2}(\xi)-
\kappa q\frac{4}{\sqrt{2(J+1)}}\cos(\xi),     \label{eq:3:ap1.16}
\end{equation}
and we find that the inverse mass matrix is diagonal
\begin{equation}
B = \left(\begin{array}{cc} 1 & 0 \\ 0 & B_\xi \end{array}\right).
\label{eq:3:Bspin}
\end{equation}

\begin{figure}
\centerline{\includegraphics[width=5cm]{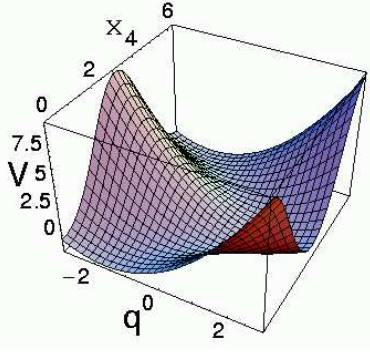}}
\caption{The profile of the potential energy surface $V(q,\xi)$.
\label{fig:3:5_1}}
\end{figure}

In the remainder of this section we shall use the choice of parameters
$\kappa =0.006403$, $\lambda =0.0005$ and $N=40$, the same as used in
\cite{37}. 
The profile of the potential energy function, $V(q,\xi)$, for this
choice of parameters is displayed in Fig.~\ref{fig:3:5_1}, where one
can easily see a valley, the precise course of which is defined
below by the solution of Eqs.~(\ref{eq:3:ap1.21}) and
(\ref{eq:3:ap1.22}).  We take $\xi$ in the range $0\leq \xi \leq
2\pi$.  One computes that $V$ is minimum at $\xi=n\pi$ and
$q=-2\kappa\sqrt{2J^{2}(J+1)}\cos(\xi)$, with
\begin{subequations}
\begin{equation}
V_{{\rm min}}= -4\kappa^{2}J^{2}(J+1) , \label{eq:3:ap1.17}
\end{equation}
and $V$ has saddle points at $q=0,\xi=(2n+1)\pi /2$, with
\begin{equation}
V_{{\rm saddle}} = -4\lambda J(J+1) .  \label{eq:3:ap1.18}
\end{equation}
We thus identify a potential barrier of height
\begin{equation}
V_{{\rm barrier}}
= 4J(J+1)(J\kappa^{2}-\lambda). \label{eq:3:ap1.19}
\end{equation}
\end{subequations}

We now turn to the machinery of our method.  To apply the valley algorithm,
one constructs the point function
\begin{equation}
U(q,\xi)=\frac{1}{2}(V_{,q}^{2}+B_\xi(q,\xi)V_{,\xi}^{2}) \label{eq:3:ap1.20}
\end{equation}
and solves the equations
\begin{eqnarray}
V_{,q}-\omega U_{,q} &=& 0, \label{eq:3:ap1.21} \\
V_{,\xi}-\omega U_{,\xi} &=& 0, \label{eq:3:ap1.22}
\end{eqnarray}
where by subscripts we denote the corresponding partial derivatives
and $\omega$ is a Lagrange multiplier.  These are the equations of a
valley on the potential energy surface as seen from the standpoint
of a metric defined by the classical kinetic energy (\ref{eq:3:ap1.16}).

For the case of two coordinates one can simplify these equations to
a single algebraic equation, the
determinantal condition
\begin{equation}
\left| \begin{array}{ll} V_{,q} & U_{,q} \\ V_{,\xi}&U_{,\xi}
\end{array}\right|
=0.
\label{eq:3:bu.22a}
\end{equation}

\begin{figure}
\centerline{\includegraphics[width=5cm]{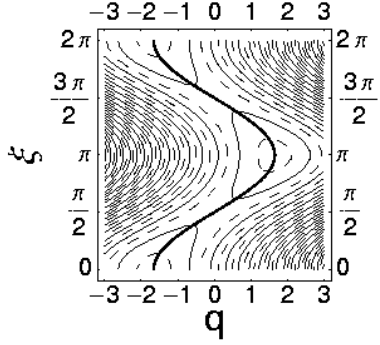}}
\caption{The collective path obtained as a solution of Eqs.~(\ref{eq:3:ap1.21})
and (\ref{eq:3:ap1.22}).\label{fig:3:5_2}}
\end{figure}

\begin{figure}
\centerline{\includegraphics[width=7cm]{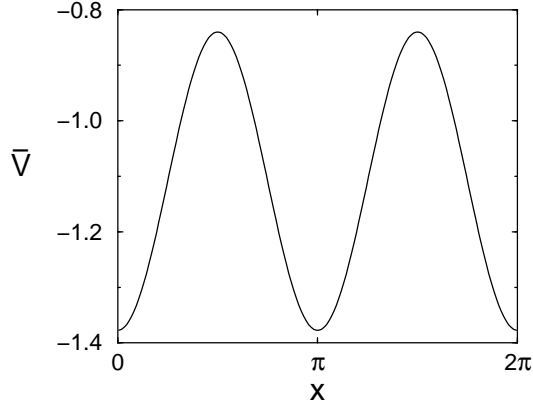}}
\caption{The collective potential energy, $V_{\text{coll}}=V(g(x),x)$.
\label{fig:3:5_3}}
\end{figure}

We shall parameterize the collective path by choosing a collective
coordinate $x$ defined by the equations
\begin{equation}
 \xi = x,  \;\;\;   q = g(x). \label{eq:3:ap1.23}
\end{equation}
The collective path and collective potential energy that result from
solving Eqs.~(\ref{eq:3:ap1.21}) and (\ref{eq:3:ap1.22}) are displayed
in Figs.~\ref{fig:3:5_2} and \ref{fig:3:5_3}, respectively.  The
latter is given by the formula  where we use the notation that a bar
denotes a function of collective variables only,
\begin{equation}
\bar{V}(x) = V(q(x),\xi(x)).  \label{eq:3:ap1.24}
\end{equation}

\begin{figure}
\centerline{\includegraphics[width=7cm]{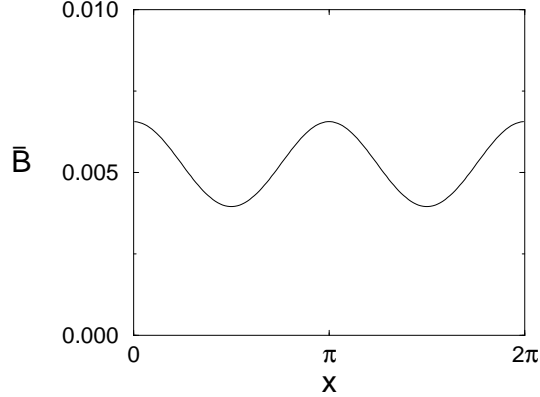}}
\caption{The inverse collective mass, $\bar{B}=B_\xi(1+g_{x}^{2}B_\xi)^{-1}$.
\label{fig:3:5_4}}
\end{figure}

Turning to the consideration of the collective mass, the theory
provides two distinct formulas, which agree only when there is exact
decoupling.  This can be understood physically as follows:  The
collective path (\ref{eq:3:ap1.23}) is the locus of points for which $\nabla V$ and
$\nabla U$ are
parallel, as seen from Eqs.~(\ref{eq:3:ap1.21}) and (\ref{eq:3:ap1.22}).
Exact decoupling occurs when the joint
direction of these vectors coincides with the tangent to the collective
path.  The formula for the mass that is determined by the tangent vector
to the collective path, which is an application of Eq.~(\ref{eq:3:BBB}), is
\begin{equation}
\bar{B}^{-1} =\frac{1}{B_\xi}\left(\frac{d\xi}{dx}\right)^2
+\left(\frac{dq}{dx}\right)^2
     =\frac{1 + B_\xi g_{,x}^{2}}{B_\xi} ,   \label{eq:3:ap1.25}
\end{equation}
where
\begin{equation}
\left( \begin{array}{cc}
         1  &  0 \\
         0 &  {1}/{B_\xi}
\end{array} \right),     \label{eq:3:ap1.26}
\end{equation}
is the inverse of the metric tensor, Eq.~(\ref{eq:3:Bspin}).  The
quantity $\bar{B}$ is plotted in Fig.~\ref{fig:3:5_4}.

\begin{figure}
\centerline{\includegraphics[width=7cm]{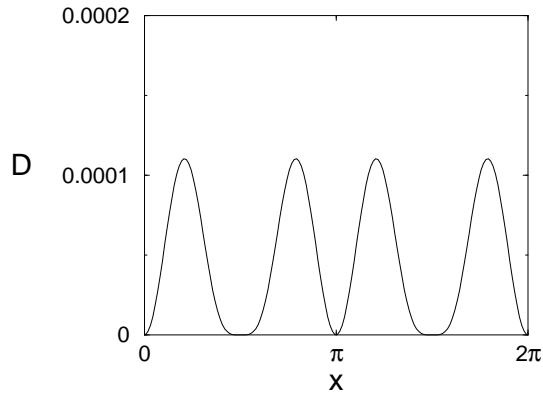}}
\caption{The invariant measure of the decoupling, Eq.~(\ref{eq:3:ap1.28}).
\label{fig:3:5_5}}
\end{figure}

If, however, the decoupling is not exact, we can derive an alternative
formula for the mass by replacing the components of the tangent
vector in Eq.~(\ref{eq:3:ap1.25}) by the components of a corresponding
vector parallel to $\vec \nabla V$.  This gives the formula
\begin{equation}
\breve{B}=\frac{V_{,q}^{2}+ B V_{,\xi}^{2}}{(V_{,q}g_{,x} + V_{,\xi})^{2}}.
\label{eq:3:ap1.27}
\end{equation}
In Sec.~\ref{sec:2.2.3} we have suggested an invariant measure, $D$,
of the goodness of decoupling, which is a point function on the
collective surface.  For the case of a collective path, it reduces to
the intuitive measure
\begin{equation}
D =|(\breve{B}-\bar{B})\bar{B}^{-1}|. \label{eq:3:ap1.28}
\end{equation}
This quantity is plotted in Fig.~\ref{fig:3:5_5}.  As one can see from this plot, the
collective mass computed using two independent methods is practically
the same, the differences being not larger than 0.012 percent. This simply
means that in this case the collective branch of the spectrum is
almost completely decoupled from the non-collective one.  This result will
be reflected when we assess the accuracy of our quantum results.

\begin{figure}
\centerline{\includegraphics[width=7cm]{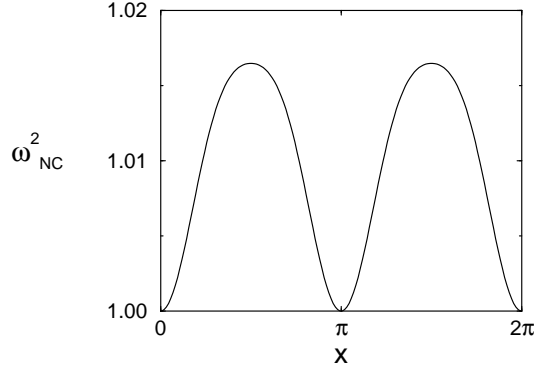}}
\caption{The square of the frequency of non-collective vibrations,
Eq.~(\ref{eq:3:ap1.29}).
\label{fig:3:5_6}}
\end{figure}

The theory discussed thus far does not pertain at all to the stability
of the motion in the neighborhood of the collective path.  This
is determined by the frequency for vibration along the local direction
perpendicular to the collective path, as given by a formula which
can be deduced from the considerations of Sec.~\ref{sec:2.2.4},
\begin{equation}
\omega^{2}_{\rm NC}=\frac{V_{,\xi\xi}(B_\xi g_{,x})^{2}-2V_{,\xi q}B_\xi g_{,x}+
V_{,qq}}{1+B_\xi g_{,x}^{2}}.      \label{eq:3:ap1.29}
\end{equation}
As seen from the plot in Fig.~\ref{fig:3:5_6}, the required stability is established.
This completes our discussion of the classical part of the problem.

\begin{aside}
 We provide a brief derivation of formula (\ref{eq:3:ap1.29}).  We use
index $1$ for the collective mode, $2$ for the non-collective mode.
Given the basis vector $g^\alpha_{,1}=(1,g_{,x})$ following from
Eq.~(\ref{eq:3:ap1.23}), we calculate $g^\alpha_{,2}$ by the procedure
described at the end of Sec.~\ref{sec:2.2.4}.  This yields the basis
vector
\begin{eqnarray}
g^\xi_{,2} &=& - \frac{g_x B_\xi}{\sqrt{1+B_\xi g_x^2}}, \label{eq:3:ncf1} \\
g^q_{,2}   &=& \frac{1}{\sqrt{1+B_\xi g_x^2}},   \label{eq:3:ncf2} \\
\bar{B}_{22} &=& 1.   \label{eq:3:ncf3}
\end{eqnarray}
Because of the choice (\ref{eq:3:ncf3}), the square of the non collective
frequency  has the value $\bar{V}_{22}$, whose value, in the light of
Eqs.~(\ref{eq:3:ncf1}) and (\ref{eq:3:ncf2}) is given by Eq.~(\ref{eq:3:ap1.29}).   
\end{aside}

\subsubsection{Requantization and comparison with exact diagonalization
\label{sec:3.2.3}}

The next step is the quantization of the collective Hamiltonian
in order to find the collective states.
Given the collective mass $ {\bar B}^{-1}$ and
the collective potential energy ${\bar V(\xi)}$, we choose the corresponding
quantized Hamiltonian in the form
\begin{equation}
H=\frac{1}{8}\{\Pi, \{\Pi,{\bar B}\}\}+{\bar V}(\xi), \label{eq:3:ap1.30}
\end{equation}
where the commutation relation between $\xi$ and $\Pi$ is to be taken in the
form (\ref{eq:3:ap1.8}),
and $ 0 \le \xi \le 2\pi $ with periodic boundary conditions.
There are always some ambiguities in
the quantization procedure, that arise from different orderings of the
coordinate and momentum operators, but for a well defined collective
motion they all differ by terms of relative order $1/J^{2}$
in the potential energy.  For the choice
made above, the corresponding classical Hamiltonian, obtained
after a Wigner transformation, is exactly that found in the previous
section.
The quantum results that follow for the ``ground-state-band'' are reported
in Table \ref{tab:3:spin}.
In the first two columns we list the exact and approximate (LACM)
mean excitation energies of the doublets.
The resulting decimal logarithms of the
energy splittings for the first five
levels are given in columns 3-5.  The exact values listed
in Table~\ref{tab:3:spin} were provided by the authors of \cite{37}.  The approximate
method used by them, imaginary time-dependent Hartree-Fock (ITDHF), could
be applied only to the doublet splittings, 
and can not be used to determine the individual eigenfunctions.
Our values, on the other hand,
were obtained as actual differences
of eigenvalues.

\begin{table}
\caption{
The mean excitation energies and the decimal logarithms of the
splittings of the energy levels below the barrier. Column 1
displays exact values for the mean energies of the doublets, column 2 our
results (LACM) for these values.  Columns 3-5 contain exact, ITDHF, and
LACM values for the doublet splittings.\label{tab:3:spin} }

\begin{tabular}{|r|r|r|r|r|} 
Energy (Ref.~\cite{37})&Energy(LACM)&Exact (Ref.~\cite{37})&
 ITDHF (Ref.~\cite{37}) & LACM  \\
        \tableline
G.S.   & G.S.  &  13.00 & 13.00 & 12.89 \\
0.154  & 0.154 &   8.96 &  8.57 &  8.89 \\
0.289  & 0.288 &   5.92 &  5.39 &  5.87 \\
0.403  & 0.401 &   3.51 &  2.85 &  3.51 \\
0.489  & 0.487 &   1.92 &  1.00 &  1.92 \\
\end{tabular}
\end{table}

 For all the levels below the
barrier, the agreement with the exact results are
very good. The fact that both the mean excitation energies
and the splittings are well reproduced means that the
formalism presented is accurate for the description of both
the classically allowed and forbidden regions. The values of the
 mean excitation energies are defined by the collective mass and
potential energy in the classically allowed region, while the
splittings depend on the properties of the collective
Hamiltonian in the classically forbidden region.

\subsection{A model of molecular isomerization \label{sec:3.3}}

\subsubsection{Introduction \label{sec:3.3.1}}

The next application, which also involves a tunneling process,
was stimulated by connection with an important molecular phenomenon known
as isomerization.
This refers to a situation where the ground state is (essentially) degenerate
with respect to the transfer of one Hydrogen bond to an equivalent position
by tunneling.  We shall study a simplified model for the influence of the
vibrational degrees of freedom on the tunneling rate.  We shall concentrate
first on the application of the GVA to a class of model Hamiltonians
which couple a double well to an oscillator degree
of freedom. We refer to our paper \cite{20} for a discussion of the
relationship of this effort to work done in the chemistry community on similar
models.  Aside from the specific problem of isomerization, there is an
important class of applications associated with the concept of reaction
path, based on the assumption that a chemical reaction proceeds along a
valley path of the potential energy of the composite system,
connecting neighboring minima through a saddle point.
For the models treated specifically, the
equations specifying the valley reduce to the problem of finding the
zeros at each point of the collective path of a relatively simple
determinant, the same  result as encountered in the previous applications in this section.  This
method is
not applicable to realistic reaction path problems.  In Ref.~\cite{20}
we have suggested an algorithm suitable for this case that combines
elements of both the GVA and the local harmonic approximation and bears some
relation to the methods we shall develop for the nuclear structure problem.

Before turning to a detailed study of the models, we take note of an
inherent simplification associated with molecular problems.  Though our
theory has been designed to deal with a general classical Hamiltonian of the
form
\begin{equation}
H(\pi,\xi) = \half\pi_\alpha B^{\alpha\beta}(\xi) \pi_\beta + V(\xi),
\label{eq:3:ap2.1}
\end{equation}
where $\xi$ and $\pi$ are an $N$-dimensional set of canonical
coordinates and momenta, respectively,
for the special case under study, the molecular
Born-Oppenheimer problem, the mass-tensor $B$ is
diagonal, and can even be made equal to the identity matrix by going to
so-called mass-scaled coordinates.
The most general form the inverse mass
$B^{\alpha\beta}$ can take, without this last rescaling is thus
\begin{equation}
B =
\left( \begin{array}{cccccc}
   m_1^{-1} & 0 & 0 & 0 & \ldots & 0 \\
   0 & m_1^{-1} & 0 & 0 & \ldots & 0 \\
   0 & 0 & m_1^{-1} & 0 & \ldots & 0 \\
   0&0&0&m_2^{-1} & \ldots & 0\\
   \vdots&\vdots&\vdots&\vdots&\ddots&\vdots\\
   0 & 0 &0 & 0&\ldots & m_N^{-1}
\end{array}\right).   \label{eq:3:ap2.2}
\end{equation}
In this case
the tensor notation may seem somewhat superfluous, but it
becomes more natural as soon as we go to other coordinate systems.

It is necessary to add a remark concerning quantization of the
models to be studied.  As discussed in Sec.~\ref{sec:2.2.4}, for motion {\em near}
a decoupled surface, the classical Hamiltonian takes the form
\begin{equation}
H = \half \bar{B}^{ij}p_ip_j  +  \bar{V} +
\half \left(\bar{B}^{ab}p_ap_b + \bar{V}_{,ab}q^aq^b \right),
\label{eq:3:ap2.3}
\end{equation}
where the motion orthogonal to
the collective surface is harmonic. For stability of the generalized
valley equation we must require that all frequencies be real, so that
it will cost energy to move away from the collective surface.
If the decoupling is not exact, we must add to the Hamiltonian (\ref{eq:3:ap2.3})
terms linear in $q^a$ and $p_a$.
Still we shall assume
that Eq.~(\ref{eq:3:ap2.3}) remains a good approximation to the problem
at hand, since the linear term should in general be small.
We check stability of the approximate Hamiltonian
by calculating the harmonic frequencies.

Of even greater weight in what follows is that the same local frequencies
can also be used to calculate corrections to the collective potential energy
when we requantize the problem.
If we assume that the motion in the non-collective coordinates
is much faster than the motion in the collective coordinates,
the non-collective coordinates contribute
only the zero-point energy of the (local) frequencies.
One thus writes,
\begin{equation}
H = \frac{1}{8}\{\{p_i,\bar{B}^{ij}\},p_j\} + \bar{V}
+\sum_{a=K+1}^N (\half \hbar \omega_a)
\label{eq:3:ap2.4}
\end{equation}
for the quantum Hamiltonian.
Here $\hbar\omega_a$ denote the local harmonic energies that depend
on the collective coordinates $q^i$.  Therefore the last term of
Eq.~(\ref{eq:3:ap2.4}) contributes to the collective potential energy.  Calculating
these frequencies is a standard small oscillations problem.

\subsubsection{Definition of and application to a class of model problems
\label{sec:3.3.2}}

We now apply the GVA to a specific model, to show
both the power and the drawbacks of our solution method.
The model in question describes a two-dimensional
tunneling phenomenon in a double-well
potential, which can be thought of as  a model of isomerization.
The quantum Hamiltonian to be studied belongs to a class having the general form
\begin{equation}
H=\half b_0\left(\frac{\partial^2}{\partial s^2 } + \frac{\partial^2}
{\partial Q^2 }\right)
+ V(s,Q),             \label{eq:3:ap2.5}
\end{equation}
with
\begin{equation}
V(s,Q) = V_0(s) - f(s)Q + \half m \omega^2 Q^2.   \label{eq:3:ap2.6}
\end{equation}
Here $s$ denotes a first guess for the ``reaction'' coordinate and $Q$ is
the coordinate of an orthogonal harmonic vibration.
Because the mass matrix is constant, the point function $U$ becomes
(to an overall scale)
\begin{equation}
U(s,Q) = \half \left(V_{,s}^2 +
V_{,Q}^2\right) .      \label{eq:3:ap2.7}
\end{equation}
The generalized valley equation states that the gradients
of $V$ and $U$ are parallel, which in this case leads to the
determinantal condition
\begin{equation}
V_{,s} U_{,Q} - V_{,Q} U_{,s} = 0.
\label{eq:3:ap2.8}
\end{equation}
We parameterize the path by a variable $t$,
\begin{eqnarray}
&& s =t,  \;\;
Q  = q(t)  \label{eq:3:ap2.10}
\end{eqnarray}
For the covariant collective mass, we thereby calculate, with dot indicating
derivative with respect to $t$,
\begin{eqnarray}
&&\bar{B}_{11}  = (1+\dot{q}^2)b_0^{-1},     \label{eq:3:ap2.11}
\end{eqnarray}
whose reciprocal is $\bar{B}^{11}$.
If we designate the basis vector defined by differentiating (\ref{eq:3:ap2.10})
as $g_{,1}=(1,\dot{q})$, then to study oscillations
in the non-collective coordinate, we need an orthogonal basis vector
$g_{,2}$, which can be chosen as
\begin{equation}
g_{,2} = (\dot{q},-1).     \label{eq:3:ap2.13}
\end{equation}
Thus we find
\begin{equation}
\bar{B}^{22} = \frac{b_0}{1+\dot{q}^2},   \label{eq:3:ap2.14}
\end{equation}
and, from Eq.~(\ref{eq:2:3.18})
\begin{eqnarray}
\bar{V}_{,22} & = & g_{,2} \left( \begin{array}{cc} V_{,ss} & V_{,sQ} \\
V_{,Qs} & V_{,QQ} \end{array}\right) g_{,2}\nonumber\\
& = &  (V_{,ss} \dot{q}^2 -2V_{,sQ} \dot{q} + V_{,QQ}).   \label{eq:3:ap2.15}
\end{eqnarray}
For the harmonic frequency, associated at each point
with the motion orthogonal to the path, we then find
\begin{equation}
\hbar \omega_2(t) = \sqrt{\frac{b_0}{1+\dot{q}^2}}
\left.
\sqrt{V_{,ss}\dot{q}^2-2V_{,sQ} \dot{q} + V_{,QQ}}
\right|_{s=t,Q=q(t)}.         \label{eq:3:ap2.16}
\end{equation}

Except for the decoupling measure $D$, we have now gathered all the
information needed to carry out a study of the models defined by
(\ref{eq:3:ap2.5}) and (\ref{eq:3:ap2.6}).
In order to calculate $D$ we need the basis vector $f^1$, designated as the
breve basis vector in Secs.~\ref{sec:2.2.2} and \ref{sec:2.2.3}. We use the conditions
\begin{equation}
(V_{,s},V_{,Q}) = \bar{V}_{,t} (f^1_{,s},f^1_{,Q})   \label{eq:3:ap2.17}
\end{equation}
to evaluate $f^1$ as
\begin{equation}
(f^1_{,s},f^1_{,Q}) = (V_{,s}+\dot{q}V_{,Q})^{-1}(V_{,s},V_{,Q}).
\label{eq:3:ap2.18}
\end{equation}
Thus we find
\begin{equation}
\breve{B}^{11} = b_0 (V_{,s}^2+V_{,Q}^2)(V_{,s}+\dot{q}V_{,Q})^{-2},
\label{eq:3:ap2.19}
\end{equation}
from which, in combination with (\ref{eq:3:ap2.11}), we can easily evaluate $D$.

Turning to the models, since according to Eq.~(\ref{eq:3:ap2.6}),
the potential is a simple
polynomial function of $Q$, we can readily derive, from
Eq.~(\ref{eq:3:ap2.8}), algebraic equations
for $Q$ as a function of $s=t$. In general these are cubic equations,
reducing to a quadratic equation if $f(s)$ is linear in $s$.

\subsubsection{Harmonic and Gaussian potential \label{sec:3.3.3}}

We have actually treated several distinct models \cite{20}
for the potential $V_0 (s)$, but shall report on the one that has the
most satisfactory physical properties.  It is patterned after a
choice \cite{40} that has been made for the inversion potential
of NH$_3$, and
consists of the combination of a harmonic and a Gaussian potential:
\begin{equation}
H = -\half \frac{\hbar^2}{m} \left(\frac{\partial^2}{\partial s^2}
+ \frac{\partial^2}{\partial Q^2}\right)
      +\half[A s^2  + B( \mbox{e}^{-Cs^2}-1)] +
       \half m \omega^2 (Q  -cs/m\omega^2)^2 .
\label{eq:3:ap2.27}
\end{equation}
Two of the three constants $A,B,C$ in the Hamiltonian were fixed
by choosing the
position of the minimum of the potential and the barrier height.
The third constant was determined by a requirement that the maximal
local frequency be relatively small
(${\rm max}\,(d^2V(s,0)/ds^2)\le\,100\,{\rm kcal/(mol\AA^2)}$).
We thus made the choice
\begin{eqnarray}
A & = & 27.87\;{\rm kcal/(mol\AA^2)},\\
B & = & 23.27\;{\rm kcal/mol},\\
C & = & 7.17\; {\rm \AA^{-2}}.
\end{eqnarray}

\begin{figure}
\centerline{\includegraphics[width=10cm]{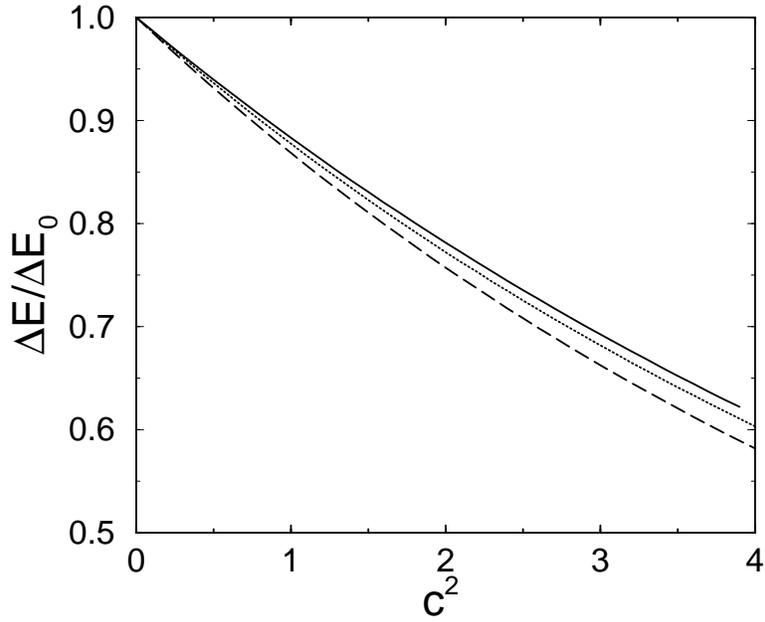}}
\caption{
  Values for the tunneling splitting for the Hamiltonian (\ref{eq:3:ap2.27}).
  The frequency is
  $\omega = 2980\; {\rm cm}^{-1}$.
  The continuous curve represents an exact result. The dashed curve is a
  calculation using the GVE approach, without zero-point corrections;
  the dotted curve is the result including quantum corrections.
\label{fig:3:3.1}}
\end{figure}

\begin{figure}
\centerline{\includegraphics[width=10cm]{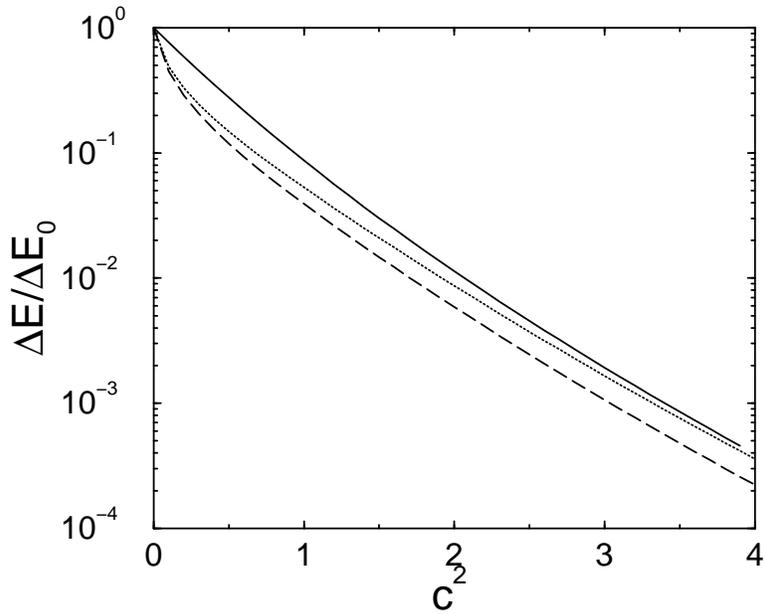}}
\caption{
  Values for the tunneling splitting for the Hamiltonian (\ref{eq:3:ap2.27}).
  The frequency is
  $\omega = 1333\;{\rm cm}^{-1}$.
  The continuous curve again represents an exact result. The dashed curve is a
  calculation using the GVA approach, without zero-point corrections;
  the dotted curve includes zero-point corrections.
}
\label{fig:3:3.2}
\end{figure}

\begin{figure}
\centerline{\includegraphics[width=10cm]{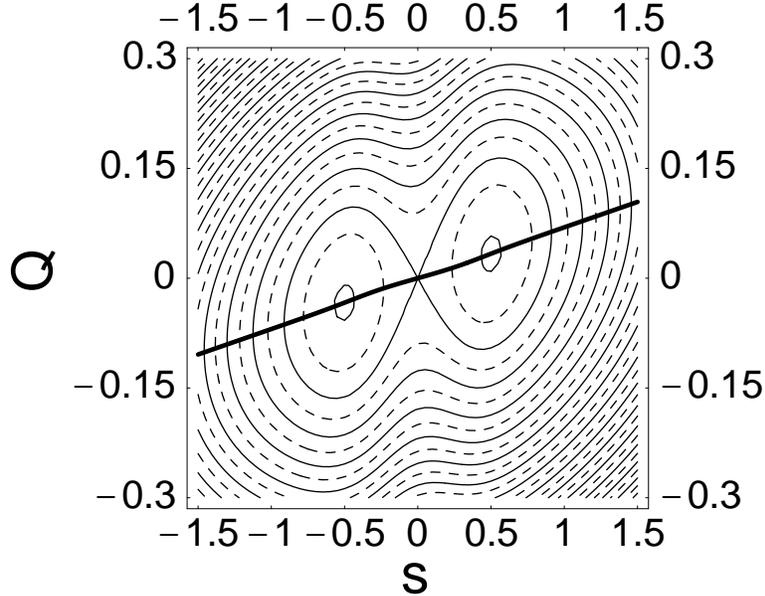}}
\caption{
  Energy surface for Eq.~(\ref{eq:3:ap2.27}), $c=1.05\;{\rm mdyn /\AA}$ and
  $\omega = 2980\;{\rm cm}^{-1}$.
  The solution to the GVA is represented by the continuous curve. Note that
  it is almost a straight line!
}
\label{fig:3:3.3}
\end{figure}

\begin{figure}
\centerline{\includegraphics[width=10cm]{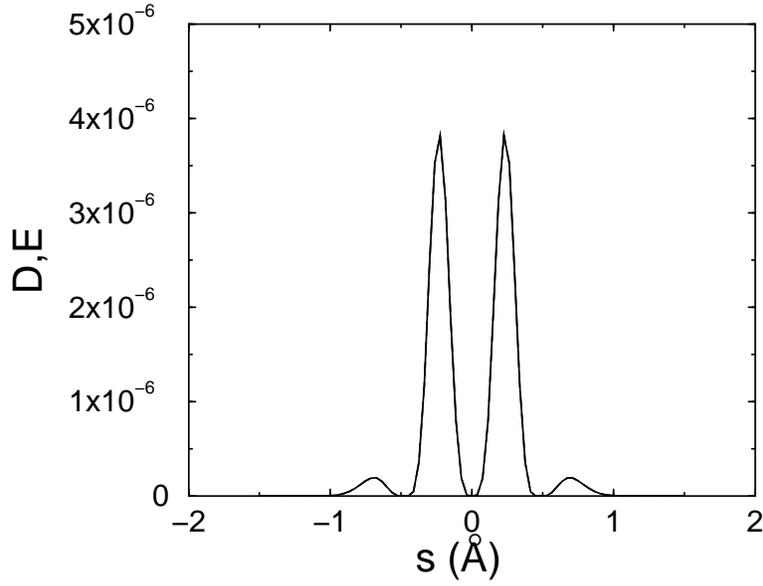}}
\caption{
  The decoupling parameters $D$ and $E$ (Eq.~(\ref{eq:3:E}))
  for the path given in Fig.~\ref{fig:3:3.3}. The difference between $D$ and $E$ can
  not be seen on the scale of the picture. The overall size of the
  parameters is also quite small, indicating excellent decoupling.
}
\label{fig:3:3.4}
\end{figure}

\begin{figure}
\centerline{\includegraphics[width=10cm]{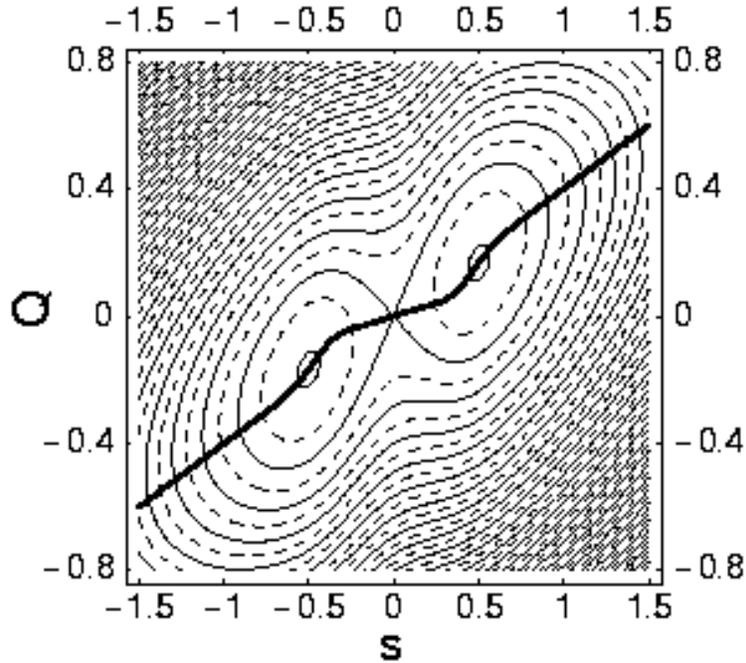}}
\caption{
  Energy surface for Eq.~(\ref{eq:3:ap2.27}), $c=1.05\;{\rm mdyn /\AA}$ and
  $\omega = 596\;{\rm cm}^{-1}$.
  The solution to the GVA is again
  represented by the continuous curve. Note that
  the oscillations are more pronounced than in Fig.~\ref{fig:3:3.3}.
}
\label{fig:3:3.5}
\end{figure}

\begin{figure}
\centerline{\includegraphics[width=10cm]{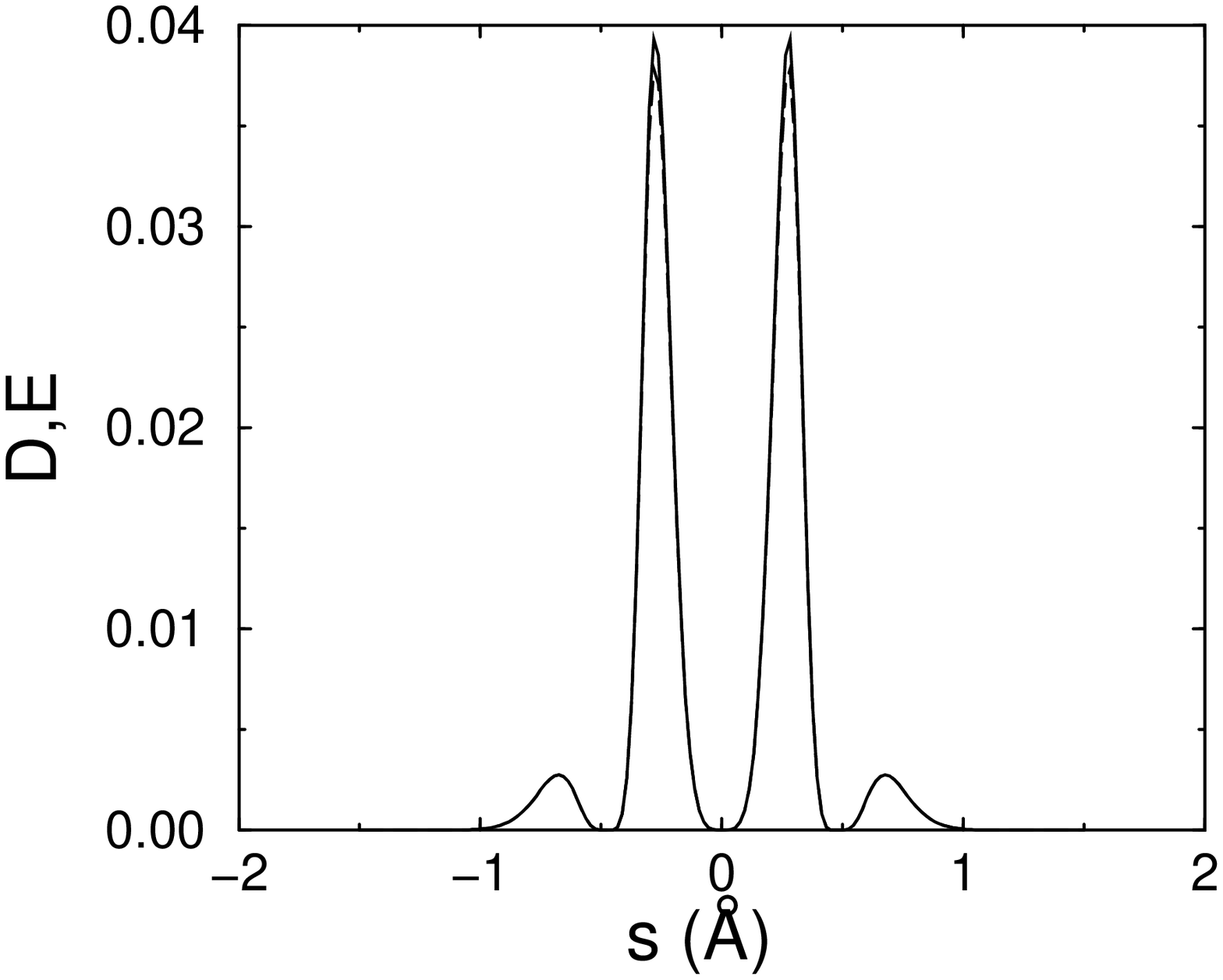}}
\caption{
  The decoupling parameters $D$ (dashed) and $E$ (continuous)
  (Eq.~(\ref{eq:3:E}))
  for the path given in Fig.~\ref{fig:3:3.4}. The difference between $D$ and $E$ can
  only be seen near the maxima. The size of $D$ and $E$ indicates
  relatively poor decoupling.
}
\label{fig:3:3.6}
\end{figure}

We performed two sets of calculations, one for $m\omega^2 = 750\; {\rm
kcal/(mol\AA^2)}$ ($\omega= 2980\;{\rm cm}^{-1}$) and another for
$m\omega^2 = 150\; {\rm kcal/(mol\AA^2)}$ ($\omega= 1333\, {\rm
cm}^{-1}$). We expect that adiabatic decoupling will be more accurate
for the larger of the two frequencies, and this will be verified by
our results.  In Fig.~\ref{fig:3:3.1} we show the results for the
tunneling splitting for the larger value of $\omega$ as a function of
$c$. The ``exact'' results were obtained by discretizing the
Hamiltonian (\ref{eq:3:ap2.27}) on a grid. As can be seen we find
perfect agreement with exact results up to fairly large $c$. There the
splitting becomes extremely small, and we can not trust our exact
results at that point.  In Fig.~\ref{fig:3:3.2} we show similar results for the
smaller of the two frequencies. Our results are not as convincing as
before, mainly due to the fact that the coupling is much
stronger. This can also be seen from the paths followed: The path for
$ c=0.35\;{\rm mdyn/\AA}$ is practically a straight line for the
largest of the two frequencies (Fig.~\ref{fig:3:3.3}), and has extremely good
decoupling (Fig.~\ref{fig:3:3.4}), but in the second case has large curvature
(Fig.~\ref{fig:3:3.5}) and poor quality of decoupling (Fig.~\ref{fig:3:3.6}).

In both Figs.~\ref{fig:3:3.4} and \ref{fig:3:3.6} we have plotted not
only the goodness of decoupling parameter $D$ but also another measure
$E$,
\begin{equation}
E = \bar{V}_{,a}\bar{B}^{ab}\bar{V}_{,b}/U,  \label{eq:3:E}
\end{equation}
which is the ratio of the square of the magnitude of the force orthogonal
to the collective surface to the square of the magnitude of the total force.
Clearly there is a great similarity between the two measures.

\subsubsection{Concluding remarks about simple applications\label{sec:3.3.4}}

In this section we have discussed how to apply the GVA and LHA to
simple models.  By solving the relevant equations we find a path (or,
in general, a surface) where we have optimal decoupling, in the sense
that we retain some residual coupling both in the kinetic and
potential energies.

Two of the examples chosen have involved decoupling one degree of
freedom from a Hamiltonian containing two degrees of freedom.  For
another fully-worked example of the same general type, the so-called
three-level Lipkin model, see \cite{9}.  We have also studied a case
of decoupling two degrees of freedom from a system with three degrees
of freedom \cite{13,22}, choosing a generalization of the so-called
landscape model \cite{42}.  The classical aspects of this model have
been studied fully, but a quantization has not been carried out, since
the potential energy is not bounded from below.  For this model we
placed special emphasis on the comparison between the GVA and LHA approaches.
The local harmonic
method may well be the one of choice for the many-body problem.
Finally, we may view the methods developed and applied thus far as
relevant to the problem of quantization of non-separable systems.

\newpage
\section{Applications to nuclear physics}\label{sec:4}

{}From the applications to date, it is apparent that the methods
developed are suitable for the decoupling of one and two degrees of
freedom from a parent system with a few more degrees of freedom. Thus,
as already remarked, they provide a useful approach to the approximate
quantization of non-separable systems.  The problem of major interest,
however, is that of decoupling one or a few degrees of freedom from a
system with many degrees of freedom, in particular, within the
framework of the nuclear many-body problem.  We were led to study this
problem within a classical Hamiltonian framework from the well-known
result \cite{43,43a,43b,44,44a} that the TDHF equations are equivalent
to Hamilton's equations.  
In this section we shall discuss the application of our ideas to
realistic nuclear applications. This report is mainly based on our
published work as referenced in the relevant sections. For some of our
older applications we have redone calculations in a manner that is better
suited to our current understanding of the correct approach. First we 
discuss how to transcribe the results
of our previous work into the language of nuclear physics and begin
the task of formulating and applying methods for solving the new
equations. In the GVA version the latter will be seen to assume the
form of a sequence of cranking equations, of which the first alone
defines the conventional cranking theory \cite{RS}.  The additional
equations fully constrain the cranking operator, rather than leave its
choice partly a matter of whim, as in existing theories. Indeed, the
unambiguous determination of this operator is tantamount to fixing the
approximately decoupled submanifold. An approximate method for solving
the generalized cranking formulation is then described that reduces
the many body problem to a few body problem of the type treated in our
previous work. The application of this method is illustrated for
several simplified models in Sec.~\ref{sec:4.2}.  In
Sec.~\ref{sec:4.3} we describe our study of the Silicon nucleus, and
in Sec.~\ref{sec:4.4} we recount our efforts to develop methods
applicable to heavy nuclei.

\subsection{Transcription of the time-dependent hartree-fock theory
to classical physics\label{sec:np}\label{sec:4.1}}

\subsubsection{Canonical variables and the adiabatic Hamiltonian\label{sec:4.1.2}}

The material presented in this section is largely based on 
Ref.~\cite{19}.  To utilize the theory described in the previous
sections we must recast the TDHF(B) equations into the form of Hamilton's
equations.  We shall utilize a method of our own devising. (In
Appendix A, we describe various choices that have been suggested for
carrying out this transformation.) 
As the simplest example we  consider the TDHF
equations,
\begin{equation}
i\dot{\rho}_{ab} = [{\cal H},\rho]_{ab}, \label{eq:4:7.1}
\end{equation}
where
\begin{equation}
 {\cal H}_{ab} = h_{ab} + V_{acbd}\rho_{dc}, \label{eq:4:7.2}
\end{equation}
and $h_{ab}=h_{ba}^{\ast}$, $V_{abcd}=-V_{bacd}=-V_{abdc} 
=V_{cdab}^{\ast}$ are the matrix elements of the single-particle
($h$) and two-body ($V$) parts of a non-relativistic 
nuclear many-body 
Hamiltonian. The labels $a,b,\ldots$ refer to a complete orthonormal set 
of single particle functions $\phi_{a}$; this set will be further 
subdivided into a set $h$ occupied in the reference Slater determinant 
and an unoccupied set $p$. We have
\begin{equation}
\rho = \sum_{h}\phi_{h}^{\ast}\phi_{h},\;\;\;(\rho^{2})_{ab} 
= \rho_{ab}. \label{eq:4:7.3}
\end{equation}
 
The most convenient choice of basis for exhibiting the canonical 
structure of (\ref{eq:4:7.1}) is a dynamical basis  in which $\rho$ is 
instantaneously diagonal, i.e., where $\rho$ in (\ref{eq:4:7.1})
is diagonal. In this basis (\ref{eq:4:7.1}) is equivalent to the 
pair of equations
\begin{eqnarray}
i\dot{\rho}_{ph} &=& {\cal H}_{ph} = \delta W_{\mathrm{HF}}/\delta \rho_{hp}, \nonumber\\ 
i\dot{\rho}_{hp} &=& -{\cal H}_{hp} = -(\delta W_{\mathrm{HF}}/\delta \rho_{ph}) ,
\label{eq:4:7.4}
\end{eqnarray} 
where
\begin{equation}
 W_{\mathrm{HF}}[\rho] = h_{ab}\rho_{ba} + \frac{1}{2}V_{abcd} 
\rho_{ca}\rho_{db}, \label{eq:4:7.5}
\end{equation}
the Hartree Fock functional, serves as Hamiltonian. 
Equations (\ref{eq:4:7.4}) are already in Hamiltonian form, and we identify 
$\rho_{ph}$ and $\rho_{hp}=\rho_{ph}^{\ast}$ as a conjugate pair of complex canonical 
variables. We may also introduce real canonical coordinates $\xi$ and $\pi$,
\begin{eqnarray}
\rho_{ph} &=& \frac{1}{\sqrt{2}}(\xi^{ph} + i\pi_{ph}), \nonumber \\ 
\rho_{hp} &=& \frac{1}{\sqrt{2}}(\xi^{ph} - i\pi_{ph}) 
= \frac{1}{\sqrt{2}}(\xi^{hp} + i\pi_{hp}). \label{eq:4:7.6}
\end{eqnarray}

In order to utilize the theory of point transformations developed in such
detail in Sec.~\ref{sec:2}, we can construct an adiabatic classical
Hamiltonian 
by expanding $W_{\mathrm{HF}}$
in powers of $\pi$. With the help of (\ref{eq:4:7.6}), we find
the standard Hamiltonian form,
\begin{eqnarray}
H(\xi,\pi) &\simeq& W[\rho(\xi, 0)]+ \frac{1}{2}\pi_{ph}\pi_{p'h'}
(\delta^{2} W/\delta\pi_{ph}\delta\pi_{p'h'}) \nonumber \\ &\equiv&
V(\xi) + \frac{1}{2}\pi_{\alpha}B^{\alpha\beta}(\xi)
\pi_{\beta}, \label{eq:4:7.24}
\end{eqnarray}
if we replace the $ph$ labels by $\alpha$. There is no term linear in
momenta, as follows from
\begin{equation}
\frac{\delta W}{\delta\pi_{ph}} = \frac{i}{\sqrt{2}}\left( 
\frac{\delta W}{\delta\rho_{ph}} - \frac{\delta W}{\delta\rho_{hp}} 
\right) = \frac{i}{\sqrt{2}}({\cal{H}}_{hp}-{\cal{H}}_{ph}) = 0,
\label{eq:4:7.25}
\end{equation}
since the matrix elements of ${\cal H}$ can be chosen real if the
system under study is time reversal invariant.  Furthermore
\begin{equation}
\frac{\delta^{2}W}{\delta\pi_{ph}\delta\pi_{p'h'}} = 
B^{php'h'} = -\frac{1}{2}\left(\frac{\delta^{2} W}{\delta\rho_{ph}
\delta\rho_{p'h'}} - \frac{\delta^{2} W}{\delta\rho_{ph} 
\delta\rho_{h'p'}} -\frac{\delta^{2} W}{\delta\rho_{hp}\delta\rho_{p'h'}} + 
\frac{\delta^{2} W}{\delta\rho_{hp}\delta\rho_{h'p'}}\right). \label{eq:4:7.26} 
\end{equation} 
The techniques necessary to evaluate Eq.~(\ref{eq:4:7.26}) are reviewed at the
end of the present discussion. The result is
\begin{eqnarray}
B^{php'h'} &=& \frac{1}{2}\delta_{hh'}({\cal{H}}_{pp'} + 
{\cal{H}}_{p'p}) - \frac{1}{2}\delta_{pp'}({\cal{H}}_{hh'} + 
{\cal{H}}_{h'h}) \nonumber \\ 
&&+ \frac{1}{2}(V_{ph'hp'} + V_{hp'ph'} - V_{pp'hh'} - V_{hh'pp'}). 
 \label{eq:4:7.27}
\end{eqnarray}

The preceding formula simplifies if we consider either 
separable interactions in the Hartree approximation or Skyrme 
interactions in conjunction with spin and isospin saturated systems, 
for in those cases the last set of terms depending explicitly on the 
two-body matrix elements cancel. Since our initial applications all 
conform to one or the other of these approximations, the remaining 
formulas of the transcription will apply only to these cases. It is 
straightforward, if more tedious, to elaborate formulas
corresponding to the general case.
 
We also restrict further transcription to the case of a single collective 
coordinate. Again when these will be needed, there will be no essential 
difficulty in adding the formulas applicable in more general instances.
To apply the GVA, we need in addition to the elements of the Hamiltonian,
the second point function, U, defined by
\begin{equation}
U = {\cal{H}}_{ph}B^{php'h'}{\cal{H}}_{p'h'}. \label{eq:4:7.28}
\end{equation}

\begin{aside}
We interject a derivation of the formulas needed for the evaluation of
(\ref{eq:4:7.26}) and similar formulas. 
If $\Theta_{ab}$ is a matrix element of an arbitrary single particle 
operator in a basis in which the density matrix $\rho$ is diagonal, 
we want to know how these matrix elements vary when we vary the 
density matrix. The basic formulas are
\begin{eqnarray}
\delta|a) &=& \delta_{ah}\sum_p |p)\delta\rho_{ph}
 +\delta_{ap}\sum_h |h)\delta\rho_{hp}, \label{eq:4:B.1}
\end{eqnarray}
and its complex conjugate, taking into account that $\delta\rho_{ph}^{\ast}
=-\delta\rho_{hp}$. From Eq.~(\ref{eq:4:B.1}) and associated statements,
one then verifies that
\begin{subequations}
\begin{eqnarray}
\partial\Theta_{ab}/\partial\rho_{ph} &=& -\delta_{ap}\Theta_{hb} 
+ \Theta_{ap}\delta_{bh} , \label{eq:4:B.2} \\ 
\partial\Theta_{ab}/\partial\rho_{hp} &=& \Theta_{pb}\delta_{ah} 
- \Theta_{ah}\delta_{pb} . \label{eq:4:B.3}
\end{eqnarray}
\end{subequations}
By the same technique we derive, for example,
\begin{equation}
\partial V_{abcd}/\partial\rho_{ph} = -\delta_{ap}V_{hbcd} 
+ \delta_{ch}V_{abpd} 
 - \delta_{bp}V_{ahcd} + \delta_{dh}V_{abcp}. \label{eq:4:B.4}
\end{equation}
Finally by combining (\ref{eq:4:B.2})-(\ref{eq:4:B.4}), we can derive the formulas
\begin{subequations}
\begin{eqnarray}
\partial{\cal H}_{ab}/\partial\rho_{ph} &=& -{\cal H}_{hb}\delta_{ap} 
+ {\cal H}_{ap}\delta_{bh} + V_{ahbp}, \label{eq:4:B.5} \\ 
\partial{\cal H}_{ab}/\partial\rho_{hp} &=& {\cal H}_{pb}\delta_{ah} 
- {\cal H}_{ah}\delta_{bp} + V_{apbh}, \label{eq:4:B.6}
\end{eqnarray}
\end{subequations}
that have proved useful above.
\end{aside}
 
\subsubsection{Equation for the collective path\label{sec:4.1.3}}

We consider now a point transformation from the variables $\xi^{ph}$
to the variables $q^\mu$, which are once again divided into the
collective ones $q^i$, $i=1,\ldots,K$, and noncollective ones $q^a$,
$a=K+1,\ldots,N_{\mathrm{ph}}$.  For ease of distinction we shall
write $Q^i$ for the collective variables, and confine our attention to
the decoupled manifold $q^a=0$. We thus write
\begin{equation}
\xi^{ph} = \xi^{ph}(Q^{i}), \;\;\; Q^{i} = Q^{i}({\vec{\xi}}). \label{eq:4:7.30}
\end{equation}
For a collective path, to which we restrict present considerations,
the superscript $i$ becomes superfluous. We
also set
\begin{equation}
\delta V/\delta\rho_{hp} = {\cal H}_{ph}, \;\; 
\delta U/\delta\rho_{hp} \equiv {\cal H}_{ph}^{(1)},\;\; 
\delta Q/\delta\rho_{hp} \equiv f_{ph}=\sqrt{2} f_{,ph}. \label{eq:4:7.31}
\end{equation}
With this nomenclature, the equations specifying that the gradients of the
potentials must be parallel for a path to be a valley,
 and that therefore they
must be proportional to a common vector $f_{ph}$, take the concise forms
 \begin{equation}
 {\cal H}_{ph} = \lambda f_{ph}, \;\;\; 
  {\cal H}_{ph}^{(1)} = \mu f_{ph}, \label{eq:4:7.32}
 \end{equation}
 where $\lambda = d\bar{V}/dQ$ and $\mu = d\bar{U}/dQ$. Each of these 
 equations is of the cranking form, differing in the 
 structure of the cranking Hamiltonians and in the definition 
 of the cranking parameters (Lagrange multipliers), but both driven by the same cranking 
 operator $f$. The cranking operator which accomplishes this heavy 
 burden is no longer freely at our disposal, but must be a self 
 consistent solution of the two sets of conditions. 

 Since the quantity $f_{ph}$ is symmetric, $f_{ph}=f_{hp}$,
 and is equal to the derivative
 $\delta Q/\delta \rho_{ph}$, we find that
 \begin{equation}
Q = \int \trace(f \delta \rho)
 = 2 \int f_{ph} \delta \rho_{ph}. \label{eq:4:qfromf}
 \end{equation}
 The relation (\ref{eq:4:qfromf}) will allow us to evaluate the collective
 coordinate along the path, and thus plays a very important role in the
 following discussion.

Instead of the pair of equations (\ref{eq:4:7.32}), we shall often
use the equivalent local harmonic formulation,
\begin{eqnarray}
{\cal H}_{ph} &=& \lambda f_{ph}, \nonumber\\
M_{ph}^{p''h''}f_{p''h''} & = & \Omega^2 f_{ph},\nonumber\\ 
M_{ph}^{p''h''} & = & \bar{V}_{;p'h'ph}{B}^{p'h'p''h''}. 
\label{eq:4:LHA2}
\end{eqnarray}

\subsubsection{Calculation of the collective mass\label{sec:4.1.4}}

Using Eq.~(\ref{eq:4:7.31}) the generalized valley
equation takes the form
\begin{equation}
{\cal H}_{ph} - \Lambda{\cal H}_{ph}^{(1)} =0. \label{eq:4:7.35}
\end{equation}
Within the nuclear context we shall refer to this equation as the {\em
generalized} cranking equation. For the simple examples studied in
Sec.~\ref{sec:3}, we solved the equivalent of (\ref{eq:4:7.35}) as a determinantal
condition. In the true many-body problem, we solve this equation
as a cranked Hartree-Fock equation with cranking operator
${\cal H}_{ph}^{(1)}$. The solution yields a density matrix
$\rho(Q)$, which depends 
parametrically on the collective coordinate $Q$. This density 
matrix specifies the collective path and is thus equivalent to the 
determination of the form of the first of Eqs.~(\ref{eq:4:7.30}). From
this result we can compute the covariant form of the collective mass
(written formally below for any number of collective coordinates),
\begin{equation}
\bar{B}_{ij} = \frac{\partial\xi^{ph}}{\partial Q^{i}} 
B_{php'h'}\frac{\partial\xi^{p'h'}}{\partial Q^{j}}. \label{eq:4:7.36}
\end{equation}
It can be shown that it 
is this form that is related directly to the usual cranking formula for 
the mass parameters found in traditional nuclear treatments. The difficulty
with this calculation is that it requires the inversion of the matrix,
$B^{php'h'}$, which may be a formidable problem for heavy nuclei when many
shells have to be included for a realistic calculation.

On the other hand, the formula which uses the given mass matrix directly,
\begin{equation}
\breve{B}^{ij} =\breve{f}^i_{,ph}B^{php'h'}\breve{f}^j_{,p'h'}, \label{eq:4:7.37}
\end{equation}
is calculated from the covariant basis vectors $\breve{f}^i_{ph}$, which
result usually from the solution of (\ref{eq:4:7.32}), in practice
either from the method developed in Sec.~\ref{sec:4.2} or from the
LHA.  These ``physical'' basis vectors, as we have previously
explained, are not equivalent to the basis vectors that appear in
(\ref{eq:4:7.36}).  Thus the proposed construction carries with it two
candidates for the collective mass tensor, as we have distinguished by
the breve notation. This naturally leads to the decoupling measure $D$
defined in Eq.~(\ref{eq:2:3.13}).

We conclude this section with an alternative (but numerically 
equivalent) method of calculating the  mass tensor. We consider only
the case of one collective coordinate, but the calculation can be
generalized. Indeed, it follows directly from the definition of the point
function $U$ that
\begin{equation}
\bar{B} = U/(d\bar{V}/dQ)^{2}. \label{eq:4:7.39}
\end{equation}
 
\subsubsection{Condition for local stability\label{sec:4.1.5}}

We begin by quoting the formulas of Sec.~\ref{sec:2.2.4} with only
completely obvious changes of notation.
We wish to study the non-collective energy,
\begin{eqnarray}
H_{NC} &=& \half \bar{V}_{,ab}\delta q^a\delta q^b
 +\half p_a\bar{B}^{ab}p_b \nonumber \\
&=& V_{NC} + T_{NC}, \label{eq:4:7.40}
\end{eqnarray}
since wherever it is positive, we have local stability.
Here
\begin{equation}
\bar{V}_{,ab} = V_{,\alpha\beta}\xi^{\alpha}_{,a}\xi^{\beta}_{,b}, \;\;
\bar{B}^{ab} = f^{a}_{,\alpha}B^{\alpha\beta}f^{b}_{,\beta}. \label{eq:4:7.41} 
\end{equation}
We see
from this equation that solution of the small oscillation problem posed by
Eq.~(\ref{eq:4:7.40}) requires the specification at
each point of the surface of a coordinate system spanning the space 
orthogonal to the collective space. This is, in principle, 
an elementary geometrical problem, that was illustrated for the non-nuclear
problem, e.g., in Sec.~\ref{sec:3.3}.
 
To complete the transcription to nuclear physics, it remains only
to specify the
formula for $V_{,\alpha\beta}$, which is needed to evaluate the first
equation of (\ref{eq:4:7.41}). Using the formulas (\ref{eq:4:B.2})-(\ref{eq:4:B.6}),
we calculate
\begin{eqnarray}
V_{,\alpha\beta} &=& \frac{1}{2}\left[\frac{\delta^{2}V} 
{\delta\rho_{ph}\delta\rho_{p'h'}} + \frac{\delta^{2}V}{\delta\rho_{ph} 
\delta\rho_{h'p'}} + \frac{\delta^{2}V}{\delta\rho_{hp}\delta\rho_{p'h'}} 
+ \frac{\delta^{2}V}{\delta\rho_{hp}\delta\rho_{h'p'}}\right] \nonumber\\ 
&=& \frac{1}{2}\delta_{hh'}({\cal H}_{pp'} + {\cal H}_{p'p}) 
- \frac{1}{2}\delta_{pp'}({\cal H}_{hh'} + {\cal H}_{h'h}) \nonumber\\ 
&& + \frac{1}{2}(V_{ph'hp'}+ V_{hp'ph'} + V_{pp'hh'}+ V_{hh'pp'}).\label{eq:4:7.42} 
\end{eqnarray} 

\subsubsection{Method of solution including extended adiabatic
approximation \label{sec:mos}\label{sec:4.1.6}}

We are finally ready to consider the major problem with which Secs.~\ref{sec:4.2},
\ref{sec:4.3}, and \ref{sec:4.4} will be largely concerned, namely how to actually
solve the generalized cranking equations (\ref{eq:4:7.32}) (and their
multi-coordinate generalizations) for specific applications. (The
material presented in this section is drawn largely from
Ref.~\cite{19}.) We shall initially restrict the discussion to
interactions of the form 
 \begin{equation}
  V_{abcd} = \sum_{\sigma}\kappa_{\sigma}(q_{\sigma})_{ac} 
 (q_{\sigma})_{bd}. \label{eq:4:7.33}
 \end{equation}
Remark first that
we are dealing with a vector space of high or even infinite dimension,
with the components of a given vector labeled by the combination
$(ph)$. The quantities ${\cal H}_{ph}$ and ${\cal H}_{ph}^{(1)}$ are
the components of two vectors in this space. We are looking for a
basis, or rather (as we change the collective variable) a continuous,
differentiable one-dimensional manifold of bases, such that the two
vectors are everywhere parallel to each other, and in so far as
possible, also parallel to the tangent to the manifold.
 
We now describe a method which yields the exact solution for a
restricted class of models considered in Sec.~\ref{sec:4.2}, and that
we believe should serve as a reasonable approximation otherwise. It
will be seen, essentially for reasons of symmetry, that for these
models, ${\cal H}_{ph}$ and all the remaining physical vectors can be
expressed in terms of the $ph$ matrix elements of a small number of
one-body operators. Building on this lead, we consider a set of
one-body observables, $q^{i},\, (i=1...L)$,
\begin{equation}
q^{i} =\trace (\rho \hat{o}^{(i)}), \label{eq:4:7.43}
\end{equation}
where the $\hat{o}^{(i)}$ are fixed one-particle operators,
with matrix elements $o^i_{ph}$.
We then assume that the ``true'' collective coordinate, $Q$, is a function of 
the $q^{i}$, 
$Q = Q(q^{i})$. The cranking operator, $f$, is then 
given by the expression
\begin{equation}
f_{ph}=(\delta Q/\delta \rho_{hp})=\sum_{i=1}^{L}a_{i}o^{(i)}_{ph},\;\; 
a_{i}=(\delta Q/\delta q^{i}),\;\;o^{i}_{ph}=(\delta q^{i}/\delta\rho_{hp}).
\label{eq:4:7.44}
\end{equation}
 
With these assumptions, the first of Eqs.~(\ref{eq:4:7.32}) takes the form
\begin{equation}
\bar{{\cal H}}_{ph}=\biggl({\cal H}-\lambda\sum_{i}a_{i}\hat{o}^{(i)}\biggr)_{ph}=0,\label{eq:4:7.45}
\end{equation}
which is recognized as a constrained 
Hartree problem of a special type. The assumption $Q=Q(q^i)$ implies that
the cranking operator $f$ is a vector in a space of prescribed
dimensionality equal to the number of $o^i$. This establishes an analogy
with the problems studied in Secs.~\ref{sec:2} and \ref{sec:3}.
 For the present discussion
(\ref{eq:4:7.45}) is a cranking equation with a single cranking operator 
defined
by fixed values of $a_2...a_L$. As will be demonstrated by examples in
Sec.~\ref{sec:4.2}, the solution of the cranking equation in concert
with the second of
Eqs.~(\ref{eq:4:7.32}) will actually tie down self-consistent values 
of the $a_i$.
It is, moreover, perfectly feasible to extend this idea to finding more
than a single collective coordinate within the L dimensional manifold
of the $o^i$. In this sense $L$ plays the role of the effective number
of dimensions to which the problem has been reduced by the special
assumptions made.

Next we must take into account the discovery in the course of this work
that in order to reproduce the exact results of the Suzuki model
studied in Sec.~\ref{sec:4.2}, we need to consider extended point
transformations as described in Sec.~\ref{sec:4.1.3}. For present
purposes the essential modification of previous formulas
is that in all the point functions
beyond the potential $V$, the mass tensor $B^{\alpha\beta}$ is replaced
by the tensor defined in Eqs.~(\ref{eq:2:2.32}) and (\ref{eq:2:2.33}),
\begin{equation}
\tilde{B}^{\alpha\beta} = B^{\alpha\beta} -\bar{V}_{,\lambda}
f^{(1)\lambda\alpha\beta}. \label{eq:4:7.45'}
\end{equation}
Recalling the definition of the matrix $f^{(1)}$,
\begin{equation}
Q^\mu = Q^\mu(\xi, \pi) = f^{(0)\mu}(\xi) 
+ \frac{1}{2} f^{(1)\mu\alpha\beta}(\xi)\pi_\alpha\pi_\beta 
 + {\cal O}(\pi^4), \label{eq:4:7.46}
\end{equation}
we thus recognize $f^{(1)\mu}$ as a second derivative of $Q^\mu$ with
respect to the components of $\pi$.

Returning to the calculation of the tilde mass, we write $\tilde{B}
=B +\Delta B$, where the first term is given by Eq.~(\ref{eq:4:7.27}), and where,
as previously asserted,
for separable interactions in the Hartree approximation, the case presently
under consideration, the set of terms depending explicitly on the 
two body matrix elements cancel. For the
calculation of $\Delta B$,
on the other hand, we have, according to Eqs.~(\ref{eq:4:7.45'}) and (\ref{eq:4:7.46}),
\begin{eqnarray} 
\Delta B^{php'h'} & = & - \deriv{\bar{V}}{Q} \frac{\partial^2 Q} 
 {\partial\pi_{ph}\,\partial\pi_{p'h'}} 
 = - \deriv{\bar{V}}{Q} \frac{\partial^2 Q} 
 {\partial\rho_\alpha\,\partial\rho_\beta} 
\frac{\partial \rho_{\alpha}}{\partial \pi_{ph}}
\frac{\partial \rho_{\beta}}{\partial \pi_{p'h'}}
\nonumber\\ 
 & = & \frac{1}{2}\deriv{\bar{V}}{Q} \frac{\partial^2 Q} 
 {\partial\rho_\alpha\,\partial\rho_\beta} 
 (\delta_{\alpha,ph}-\delta_{\alpha,hp}) 
 (\delta_{\beta,p'h'}-\delta_{\beta,h'p'}). \label{eq:4:delB} 
\end{eqnarray}
In the special case that $Q=\trace(f\rho)$, where $f$ is a one 
body operator, using the differentiation rules
(\ref{eq:4:B.2})-(\ref{eq:4:B.6}), we find the simple result
\begin{equation} 
\Delta B^{php'h'} = - \deriv{\bar{V}}{Q} (f_{pp'}\delta_{hh'}- 
 f_{hh'}\delta_{pp'}). 
\label{eq:4:DeltaB} 
\end{equation}

In general, however, $f$ will also be a function of the density. The most 
general assumption regarding this density dependence that we shall 
entertain in this work is 
that $Q$ is an arbitrary functional of a basis of one-body operators, as
described in connection with (\ref{eq:4:7.43}) and (\ref{eq:4:7.44}), 
\begin{equation} 
Q = Q[q^i], \;\;\; 
q^i = \trace(\rho \hat{o}^i), \label{eq:4:Qqi} 
\end{equation} 
and $\hat{o}^i$ is a set of one-body operators. We then find that
Eq.~(\ref{eq:4:DeltaB}) is still valid, but $f$ is now given by 
\begin{equation} 
f = \sum_i Q_{,q_i} o^i. 
\label{eq:4:fhat} 
\end{equation} 
{}From Eq.~(\ref{eq:4:Qqi}) it can be shown that 
the derivative $\delta Q/\delta \rho_{ph}$ that occurs in the cranking 
equation is given by the $ph$-matrix elements of (\ref{eq:4:fhat}). 
The point function U is defined in Eq.~(\ref{eq:4:7.28}), except for the
replacement of $B$ by $\tilde{B}$.
Substituting the explicit value for $\tilde{B}$, this can 
be expressed as 
\begin{eqnarray} 
 U 
&=& {\cal H}_{hp} \bar{\cal H}_{pp'}{\cal H}_{p'h} - 
{\cal H}_{ph} \bar{\cal H}_{hh'}{\cal H}_{h'p}. \label{eq:4:formu} 
\end{eqnarray} 

Finally, the formula for the  collective mass can be read off from 
Eqs.~(\ref{eq:4:7.28}) and (\ref{eq:4:formu}), namely,
\begin{eqnarray}
\frac{1}{2}\bar{B}&=& f_{hp}\bar{\cal H}_{pp'}f_{p'h}
-f_{ph}\bar{\cal H}_{hh'}f_{h'p}. \nonumber\\
&=& \frac{1}{2}\trace \rho[f,[\bar{{\cal H}},f]]. \label{eq:4:newmass}
\end{eqnarray}
Below, in Sec.~III C, this formula will be applied to a model for monopole
vibrations, including an exactly solvable limiting case.

\subsection{Algorithms\label{sec:4:alg}}
\subsubsection{Algorithm in the local harmonic approximation\label{sec:alg:LHA}}
In our approach to tackling the solution of the decoupling problem for the nuclear case, we
have usually preferred to construct an algorithm based on the local
harmonic approximation.  Other options are available, however, and will be discussed in the next
section (\ref{sec:alg:GVA}).  Many aspects of the
algorithms are similar, so that it is useful to spend some time here
discussing the LHA  approach.  It is based on a self-consistent
iteration between the generalized cranking problem and the local
harmonic equation which determines the cranking operator, and will be
formulated here in terms of Hartree-Fock, though it generalizes
trivially to HFB.

First we solve the Hartree-Fock equation at the HF minimum, where
$\lambda$ in Eq.~(\ref{eq:4:7.32}) is zero. We then solve the RPA equation to obtain $f_{ph}$.
For this starting point, there is no problem of self-consistency,
since $\lambda=0$.  We next use this initial $f$ as input to solve the
cranking equation at a ``nearby point'', characterized more precisely
below, which gives a slightly different $ph$-basis. This means that we
have to solve the RPA again, leading to a new $f$. We continue to
cycle until convergence has been achieved at the given point. We can
then move on the next point.

We can move to a neighboring point either by making a small change in
$\lambda$, the Lagrange multiplier, and later finding the change in the collective coordinate
$Q$, or else make a small fixed change in $Q$, afterwards finding
the new value of $\lambda$. The latter gives a
more stable algorithm and is therefore usually the technique adopted.
Using Eq.~(\ref{eq:4:qfromf})
we find that for small real $\Delta \rho_{ph}=\rho_1-\rho_0$,
the change in the collective coordinate is given by
\begin{eqnarray}
\Delta Q &=& 2\int_{\rho_0}^{\rho_1} \trace(f\delta\rho)
\approx \sum_{ph}[ f_{ph}(\rho_0) + f_{ph}(\rho_1)]\Delta \rho_{ph}
\nonumber \\
&=& [f(0) +f(1)]\cdot \Delta\rho 
\label{eq:4:DeltaQ},
\end{eqnarray}
which is the simplest (trapezoidal) approximation to the area
under the curve, an approximation that should suffice
for sufficiently small changes.
By summing expressions of the form (\ref{eq:4:DeltaQ}), we can assign
values of $Q$ at any point along the collective path, relative to an
arbitrarily chosen initial value.

The structure of the algorithm the algorithm used to solve the LHA 
is as follows:
\begin{enumerate} 
\item 
Start at the HF minimum $(\lambda=0)$ and diagonalize the Hartree-Fock 
Hamiltonian $\cal H$. 
\item 
At this point solve the RPA equation, and use this to define $f_{ph}$
(there is no problem of self-consistency,
since $\lambda=0$).
\item 
We use this initial $f$ as input to solve the cranking equation
at a ``nearby point'', characterized more precisely below.
\item
Diagonalizing this equation leads to a slightly different $ph$-basis.
Solve the RPA equation, and thus obtain a new $f$. 
\item
Use this as input to the cranking equation, and perform
an RPA step until
selfconsistency is achieved, i.e., $f$ and the $ph$-basis no longer
change.
\item We now again move to a nearby point, and start 
the whole process again from step 4. 
\end{enumerate}

\begin{aside}
The outline above omits essential details concerning the exact manner
in which the iteration should be carried out in order to guarantee that
$\Delta Q$ retains a fixed value during the procedure.  Our aim is to
find a solution, under this constraint, of the cranking condition ${\cal H}[\rho(1)]-\lambda
f(1)=0$, at the same time that
$f(1)$ is a solution of the local RPA equation evaluated for $\rho(1)$.
The procedure involves a double iteration sequence:  For a fixed, n{\it th}
approximation, $f^{(n)}$ to the cranking operator $f(1)$, we must find a
solution of the associated cranking equation.  This will involve an iteration
with index $i$, that is closely related to the method of steepest descent.
When this iteration is completed, we can then use the local RPA to compute
$f^{(n+1)}$ and start the next cycle of iteration.

At any stage of the double sequence labeled by $n$ and $i$, we shall approximate
Eq.~(\ref{eq:4:DeltaQ}) by the formula
\begin{equation}
\Delta Q = [f(0) + f^{(n)}(1)]\cdot\Delta_i^{(n)}\rho, \label{larev4:1}
\end{equation}
which converges to the correct limit.  Explicitly
$f^{(n)}(1)$ is the approximation to the cranking operator obtained by
solving the local RPA equation with the density matrix that is the solution of the cranking equation 
\begin{equation}
{\cal H}[\rho^{(n-1)}]-\lambda^{(n-1)}f^{(n-1)}=0. \label{larev4:2}
\end{equation}
Further, $\Delta_i^{(n)}\rho=\rho^{(n)}_i-\rho(0)$ is an i{\it th} approximation to
$\Delta^{(n)}\rho$ to which it must approach as $i\rightarrow \infty$.  In turn, in the limit
$n\rightarrow \infty$, $\Delta^{(n)}\rho$ must approach
$\Delta\rho = \rho(1)-\rho(0)$ and $f^{(n)}\rightarrow f(1)$.  It follows from
Eq.~(\ref{larev4:2}) that $\lambda^{(n-1)})={\cal H}\cdot f^{(n-1)}/
f^{(n-1)}\cdot f^{(n-1)}$. 

We suppose that $f^{(n)}$ is known 
(we specify below how to start the iteration), and we wish to calculate $\Delta^{(n)}\rho$.  We
choose for the i{\it th} iteration
\begin{eqnarray}
\Delta_i^{(n)}\rho &=& \Delta_{i-1}^{(n)}\rho +\epsilon^{(n)}_i\frac{f^{(n)}}
{f^{(n)}\cdot f^{(n)}} +\Delta_{\perp i}^{(n)}\rho,  \label{larev4:3} \\
\Delta_{\perp i}\rho &=& \delta_i^{(n)}\{{\cal H}[\rho_i^{(n)}]-
\lambda^{(n)}_i f^{(n)}\}.    \label{larev4:4}
\end{eqnarray}
Here $\lambda^{(n)}_i$ is determined by the condition
 $\Delta_{\perp i}^{(n)}\rho\cdot f^{(n)} =0$, and $\delta_i^{(n)}$ 
is chosen small for small
$i$, so as to prevent the initial iteration steps from diverging and
can be set to unity for sufficiently large $i$, since it multiplies
a factor which itself tends to zero as we approach a solution of the 
cranking equation.  The meaning of
Eq.~(\ref{larev4:3}) is that ultimately we represent $\Delta^{(n)}\rho$
as a linear combination of a part proportional to $f^{(n)}$ and a part 
orthogonal to the latter. We start the present iteration procedure with 
the choice
\begin{equation}
\Delta^{(n)}_1\rho = \epsilon^{(n)}_1 f^{(n)}/f^{(n)}\cdot f^{(n)}, 
\label{larev4:5}
\end{equation}
where $\epsilon_1^{(n)}$ is adjusted to give the preassigned value of $\Delta Q$.  However the
component of the change of the density matrix
along the direction of $f^{(n)}$ must be adjusted in order to retain the value
of $\Delta Q$ as we iterate, and this has been allowed for in 
Eq.~(\ref{larev4:3}).  

When the iteration on $i$ converges, we arrive at a value of $\Delta^{(n)}
\rho$, and this in turn allows us to calculate $f^{(n+1)}$ from the local
RPA equation and thus to start the next cycle of the double iteration procedure.  We can initiate
the entire cycle with the choice $f^{(1)}
=f(0)$.
\end{aside}

As stated above, the definition of the collective coordinate is obtained 
by adding the
small increments $\Delta Q$ (\ref{eq:4:DeltaQ}) as we move along the
path. This is clearly related to the choice of normalization of
$f_{ph}$, as a different choice leads to a different value of $\Delta
Q$. This freedom, corresponding to a point transformation in the
collective coordinate, is reflected by the dependence of the
collective mass on $f$, 
\begin{equation}
\bar{B}=f_{ph} B^{php'h'} f_{p'h'}.
\label{eq:4:BBar}
\end{equation}
The equations (\ref{eq:4:DeltaQ}) and (\ref{eq:4:BBar}) suggest a few
obvious choices of normalization: We could either choose to take
$\bar{B}=1$, which leads to a very attractive Hamiltonian without a
position dependent mass, or require 
\begin{equation}
f_{ph}(\rho_0) \Delta \rho^{ph}=
f_{ph}(\rho_1) \Delta \rho^{ph}.
\end{equation}
We shall generally use the first choice, apart from a model discussed
in the next section, where the second choice is more natural.

\subsubsection{Path following algorithms\label{sec:alg:GVA}}

The algorithm discussed in the previous section functions reasonably well
in many cases, but was found to be somewhat unstable. The double
iterations do not always converge, and in the case of crossing paths it is
more likely to jump to the other path than follow the current one
beyond the crossing.
It is not hard to find heuristic remedies that
work in some cases, but we prefer a more robust and stable algorithm.

A search through the relevant literature show that such algorithms exist,
but they require a minor reinterpretation of the equations. 
Both the LHA and GVA based approaches are,  for the case of
one collective coordinate, fully equivalent attempts to find a
one-dimensional path in a multi-dimensional space. 
In the equations for this case (here written in the GVA form)
\begin{equation}
V_{,\alpha}=\lambda U_{,\alpha},
\end{equation}
we interpret both $\lambda$ and the particle-hole coordinates as
unknowns.  We thus have $N_{\mathrm{ph}}$ non-linear equations in
$N_{\mathrm{ph}}+1$ unknowns ($\rho_{ph}$ and $\lambda$). Solutions to
such an underdetermined set are obviously one-dimensional manifolds
$\lambda(q),\xi_{ph}(q)$.  Similar sets of non-linear equations arise
in many contexts, and special techniques have been developed to deal
with these \cite{Rheinboldta,Rheinboldtb}. These algorithms are based
on a path-following algorithm in the $(\lambda,\xi_{ph})$
coordinates. As in the previous algorithm a step along the path is
made by adding a small number times the tangent vector
$\partial_Q(\lambda,\xi_{ph})$ to an existing solution; a point on the
line is then found by a Newton iteration orthogonal to the tangent. 
Substantial use is also made
of curvature information. Unlike the  algorithm discussed in the
previous subsection, the step-size
along the path is chosen based on the (speed of) convergence of 
the Newton iteration. If we wish to find points where one of
the coordinates reaches a certain value, this can be done using interpolation
on the points on the line. Such algorithms normally follow the path
across crossings, and are well capable to detect crossing solutions (usually
called ``bifurcations'' in the relevant literature).

We apply the method to the LHA or GVA equations by following the path
using the $ph$ coordinates (and, making a slight error, also the
adiabatic truncation in these coordinates) corresponding to the
starting point, the point labelled $0$ in
Eq.~(\ref{eq:4:DeltaQ}). When we reach a point that is a chosen
distance away from the starting point we change our coordinates to the
local basis at the new point. This unfortunately also induces a non-trivial
change in the inertia matrix due to the difference in the
adiabatic truncation at each point (our adiabatic truncation is determined
locally). If this change is small, as is usually the case, a quick Newton
iteration typically fixes this problem. In principle, this could be a source
of problems.

In implementing this method we have based our programs on the PITCON
routines developed by Rheinboldt and collaborators
\cite{PITCON}. These are specifically designed to deal with the
one-parameter case (lines). Some limited work exists on
multi-dimensional problems, but that is of little relevance for our
approach. 

Nevertheless, we would like to be able to study two- and possibly even
higher-dimensional collective surfaces.  Clearly to generate a surface
we must solve a generalized valley equation containing three
functions, and two Lagrange multipliers. As discussed in
Sec.~\ref{sec:3.1.2}, the GVA suffers from problems with avoided
crossings, that we might be able to deal with, but would blur the
picture somewhat.  In the only application we have made up to date, we
have therefore chosen an approach based on the second form of the
decoupling theorem of Sec.~\ref{sec:2.2.1}, similar to the one 
mentioned in the discussion leading up to Eq.~\ref{eq:3:LHAdet},
\begin{subequations}
\begin{equation}
V_{,ph}=\lambda
V_{,ph}^{~p'h'} V_{,p'h'}+\mu V_{,ph}^{~p''h''}V_{,p''h''}^{~p'h'} V_{,p'h'}.
\end{equation}
These equations, which contain two free parameters, are supplemented with
one additional equation containing no additional parameters,
thus ``slicing'' through the solution surface.  This last 
equation was chosen to state that $Q_{20}$ and $Q_{22}$ lie on a
given line,
\begin{equation}
\alpha_0  Q_{20}+\alpha_2Q_{22}=\phi,
\end{equation}
\end{subequations}
where $\alpha_0$ and $\alpha_2$ are some given numbers, and a set of
$\phi$ is chosen such that we get a set of equidistant slices. A
starting point for each slice is constructed by following the valley
until a solution satisfying the chosen value of $\phi$ is found.  One
thus sees that the this approach will work as long as the slices do not
 become tangential to the
surface.

Additional complications arise that are not so much of a
numerical as of a topological nature: two-dimensional topology is much
richer than one-dimensional, leading to a host of problems.

In principle we need not solve any RPA equation, which leads to a
considerable speed-up in the calculation. On the other hand, for the
routines to operate in a stable fashion the derivative of the tangent
vector needs to be known at least approximately, which is a
calculation related to that of the RPA matrix, and requires a similar
amount of time. The time spend in this calculation
and the related matrix inversion are limiting factors
in the application of the algorithm to large-scale nuclear physics problems.

\subsection{ The Suzuki model and extensions\label{sec:4.2}}

\subsubsection{The exactly solvable model\label{sec:4.2.1}}
 
As first applications of the machinery developed in
Sec.~\ref{sec:4.1}, we first re-studied the tunneling problem
described in Sec.~\ref{sec:3.2} and then studied an exactly solvable
model of spin-less fermions in one-dimension that is also a model for
monopole vibrations of nuclei \cite{19}.  (This model has been studied
previously by several authors \cite{45,46,47,48,51}.)  We shall
recount only the results of the second investigation. In
Ref.~\cite{26}, of which we shall not give an account in the present
review because it takes us too far from our main path, we have shown
that this model is the simplest of a class of exactly solvable models
of non-negligible physical interest, in particular in connection with
application to the symplectic and pseudo-symplectic models of
structure of deformed nuclei \cite{49,50,52}. The method of solution
developed in Ref.~\cite{26} are very special to the algebraic
structure of the particular models.  The aim of the present
investigation is to show that the extended adiabatic approximation 
works for the same
problems. The model of interest is described by a Hamiltonian that in
second quantized form is written
\begin{equation}
\hat{H} =\int \left[ \psi^{\dagger}(x)\frac{1}{2} 
(p^{2}+x ^{2})\psi(x)\right]dx + \frac{1}{2}\kappa\hat{Q}^{2}, \label{eq:4:suz1}
\end{equation}
where
\begin{equation}
\hat{Q}=\int \psi^{\dagger}(x)x^{2}\psi(x)\,dx \label{eq:4:suz2}
\end{equation}
is a monopole operator.
For this model \cite{45,46,47,48}, it is known that the
time-dependent Hartree equation has a decoupled solution
governed by the collective Hamiltonian
\begin{equation}
H = 4 Q P^2 +
\frac{1}{2}Q +\frac{1}{2}\kappa Q^{2} +\frac{N^4}{8Q}.
\end{equation}

In order to apply our LACM techniques, we need to study the Hartree
approximation corresponding to the Hamiltonian (\ref{eq:4:suz1}),
which is
\begin{equation}
{\cal H}=\frac{1}{2}(p^{2}+x^{2}) + \kappa Qx^{2},
\;\;Q = \trace (\rho x^{2}). \label{eq:4:suz3}
\end{equation}
The variable $Q$ is the ``natural'' first choice of collective coordinate
in a cranking treatment. We may anticipate that if there is to be an exactly 
decoupled coordinate then it must be $Q$, i.e., the cranking operator must 
be $x^{2}$, and the associated cranking Hamiltonian, also of harmonic
form, will then be
\begin{equation}
\bar{\cal H}={\cal H}-\lambda f =\frac{1}{2}(p^{2}+\omega^{2}x^{2}), \;\; 
\omega^{2}=1+2(\kappa Q-\lambda). \label{eq:4:suz4}
\end{equation}
The $N$ lowest energy eigenfunctions of $\bar{\cal H}$, $\phi_{h}$, (energy 
$\epsilon_{h}$), will provide a density matrix $\rho(Q)$ which specifies a 
submanifold of dimension one. 
 
{}From the solutions $\phi_{h}$ we can find the collective potential
energy and the collective mass. We have first 
\begin{equation}
\bar{V}(Q)=\trace \left[\frac{1}{2}(p^{2}+x^{2})\rho \right] 
+\frac{1}{2}\kappa Q^{2} 
=\sum_{h}\epsilon_{h}(Q) - \frac{1}{2}\kappa Q^{2}+Q\bar{V}_{,Q},
\label{eq:4:suz5}
\end{equation}
where we have recalled the definition 
$\lambda=(d\bar{V}/dQ)\equiv \bar{V}_{,Q}$. 
The sum over single particle energies of $\bar{{\cal H}}$ is eliminated
by means of the virial theorem,
\begin{equation}
\sum\epsilon_{h}=\omega^{2}\trace(x^{2}\rho)=\omega^{2}Q. \label{eq:4:suz6}
\end{equation}
We thus arrive at the first order differential equation
\begin{subequations}
\begin{equation}
\bar{V}(Q)=Q[\omega^{2} + \bar{V}_{,Q}] -\frac{1}{2}\kappa Q^{2}, \label{eq:4:suz7}
\end{equation}
of which the solution is
\begin{equation}
\bar{V}(Q)= \frac{1}{2}Q +\frac{1}{2}\kappa Q^{2} +\frac{\beta}{Q},
\label{eq:4:suz8}
\end{equation}
\end{subequations}
with the value of the constant $\beta$ yet to be determined. 
 
To calculate $\beta$, 
let $Q_{0}$ be the equilibrium value of $Q$, as fixed in part from the
equation
\begin{equation}
0=V_{,Q}=\frac{1}{2}+\kappa Q_{0} -\frac{\beta}{Q_{0}^{2}}.
\label{eq:4:suz9}
\end{equation}
{}From the virial theorem (\ref{eq:4:suz6}), by computing $\trace
(x^2\rho)$, we find $ Q_{0}=N^{2}/2\omega_{0},$ where $\omega_{0}$ is
the equilibrium value of $\omega$, namely, $\omega_{0}^{2} = 1+2\kappa
Q_{0}$.  Combining the various results, we find that
$\beta=(1/8)N^{4}$.  This agrees with the exact result derived in
Ref.~\cite{26}.
 It is easy to see that the relation $Q=N^2/2\omega$ between $Q$
and $\omega$ holds
for any point on the collective path. Combining this observation with 
Eq.~(\ref{eq:4:suz4}) leads to another derivation of the value of $\beta$. 
 
We remark parenthetically that the second term of (\ref{eq:4:suz8}) 
is associated in an obvious 
way with the interaction term of the original many-body Hamiltonian, and that 
if in the latter we were to make the replacement, 
$\frac{1}{2}\kappa \hat{Q}^{2} \rightarrow {\cal V}(\hat{Q}),$ 
then the corresponding replacement would take place in (\ref{eq:4:suz8}).
The value of
$Q_{0}$ would change, but assuming the system to be stable, the value of 
$\beta$ would be unaffected. 
 
We consider next the calculation of the collective mass, utilizing 
Eq.~(\ref{eq:4:newmass}), which for the present example takes the form
\begin{equation}
B=\trace\rho[x^2,[\bar{{\cal H}},x^{2}]]= 8Q. \label{eq:4:suz11}
\end{equation}
As shown independently in Ref.~\cite{26}, this is  the correct
value.  We thus see that to obtain this value and thus the exact value
for the collective Hamiltonian requires the modified theory of
extended point transformations. We verify as well that all the conditions for
exact decoupling are now satisfied: 
\begin{enumerate}
\item 
first of all, as the
collective mass depends only on $Q$, one has the geodesic decoupling
condition $B_{,a} = 0$; 
\item the force condition $\bar{V}_{,a}=0$ has
previously been satisfied and 
\item  finally the block-diagonality of
the mass tensor can always be imposed by the proper choice of
coordinate system. 
\end{enumerate}
This will be contrasted below with the non-exactly solvable extensions
of the present model, where the value of $B$ will be a function
of coordinates additional to those assumed to be collective.

It is of interest for future purposes to delve more deeply into the
properties of the current model and its relation to the GVA. We
consider the function $U$, given in Eq.~(\ref{eq:4:formu}) and its
first derivative $U_{,ph}$. If in (\ref{eq:4:formu}), we use the cranked
Hartree condition $\bar{{\cal H}}_{ph}=0$ and the formula
(\ref{eq:4:suz11}) for the collective mass, we obtain the projection
of $U$ onto the collective submanifold,
\begin{equation}
\bar{U} = 4\bar{V}_{,Q}^2 Q, \label{eq:4:suz12}
\end{equation}
whence
\begin{equation}
U_{,ph} = [8\bar{V}_{,QQ}\bar{V}_{,Q}Q+ 4\bar{V}_{,Q}^2](x^2)_{ph}
\equiv \mu (x^2)_{ph}. \label{eq:4:suz13}
\end{equation}
This result is of the form expected for exact decoupling, and
determines the Lagrange multiplier $\mu$.

\subsubsection{A local harmonic approximation applicable to the conventional
and extended Suzuki models}

When a model is not exactly solvable, we must apply some version of
the general theory, either the GVA or the LHA. For application to
nuclear models, we have so far found it more convenient to utilize the
LHA. In this method, we adjoin the constrained Hartree-Fock condition
(\ref{eq:4:7.45}) to the local harmonic eigenvalue equation
appropriate to extended adiabatic approximation
\begin{eqnarray} 
M_{ph}^{p''h''}f'_{p''h''} & = & \Omega^2 f'_{ph},\nonumber\\ 
M_{ph}^{p''h''} & = & 
\bar{V}_{;p'h'ph}\tilde{B}^{p'h'p''h''}. 
\label{eq:4:eigv} 
\end{eqnarray} 
This equation has the structure of a local RPA equation with the
characteristic differences already encountered in the section on
the extended GVA, namely that we must use the modified mass, $\tilde{B}$.

The algorithm we have used in this case is a slight variation of the
one discussed in Sec.~\ref{sec:alg:LHA}.  
We find it more convenient to approximate $\hat{f}$ in a basis of
suitably chosen one body operators $\hat{o}^{(i)}$,
\begin{equation} 
\hat{f} = \sum_i c_i \hat{o}^{(i)}, 
\end{equation} 
and to solve the associated projected eigenvalue equation 
\begin{subequations}
\begin{equation} 
M_{ij} e_j^{(k)} = \Omega_k O_{ij} e_j^{(k)}, 
\label{eq:4:RPAapprox} 
\end{equation} 
where 
\begin{equation} 
M_{ij} = o^{(i)}_{ph} M^{ph}_{p'h'}B^{p'h'p''h''}o^{(j)}_{p''h''} 
\label{eq:4:Mij} 
\end{equation} 
and 
\begin{equation} 
O_{ij} = o^{(i)}_{p'h'} B^{p'h'p''h''} o^{(j)}_{p''h''}. 
\label{eq:4:Oij} 
\end{equation} 
\end{subequations}
Clearly this is an approximation, but it has the practical advantage that, 
for the model we shall consider in the next section, it is much more 
stable. It also helps to reduce the dimensionality of the RPA matrix. 
Since the RPA matrix is diagonalized many times during the calculation 
this gives a sizeable reduction of computation time. We may add, 
parenthetically, that this method has the conceptual advantage of 
tying our work, in a clear manner, to previous, non-selfconsistent 
cranking calculations, where the form of $f$ is fixed {\em \`a priori}. 

It should also be noted that in this calculation
we normalize $\hat{f}$, and thus $c_i$ with the
second method discussed in Sec.~\ref{sec:alg:LHA}, where we chose
$\Delta\rho_{ph} f_{ph}^{(0)}=\Delta\rho_{ph} f_{ph}^{(1)}$.  This choice of
normalization condition  for the Suzuki model leads to the
collective operator being the quadrupole operator, and the inverse
mass taking the value $8Q$, so that the numerical calculation ties
right on to the analytic one.

\subsubsection{Practical calculations for an extended monopole model}
 
To test the applicability of the algorithm,
we study a family of models that contains the Suzuki model as a 
special case. Though we have already solved this special case exactly, it is
important to remark that it is not strictly an example of adiabatic
large amplitude collective motion. When we examine the local RPA, we find
that the eigenvalue corresponding to the exact cranking operator (proportional
to $x^2$) is the second rather than the lowest eigenvalue, and this is the
one that we therefore follow as we turn up the extra interaction included in
the general model.

For present purposes we have modified the Hamiltonian of the
Suzuki model by adding an $x^4\cdot x^4$ interaction, 
\begin{equation} 
\hat{H} = \int \psi(x)^{\dag}\frac{1}{2}(p^2+x^2)\psi(x) dx + 
\frac{1}{2}\kappa_2 \left(\int \psi(x)^\dagger x^2 \psi(x) dx\right)^2 + 
\frac{1}{2}\kappa_4 \left(\int \psi(x)^\dagger x^4\psi(x) dx\right)^2. 
\label{eq:4:modSuzi} 
\end{equation} 
The Hartree Hamiltonian for this Hamiltonian is 
\begin{equation} 
\bar{\cal H}_{\lambda} = \frac{1}{2} p^2 + 
 \frac{1}{2} (1+2\kappa_2 Q_2) x^2 + 
 (\kappa_4Q_4) x^4, \label{eq:4:modHart} 
\end{equation} 
where 
\begin{equation} 
Q_i = \trace(\rho x^i). 
\end{equation} 
 
For the calculations reported here we used a Slater determinant of ten
fermion states ($N=10$). The interaction strength is fixed at the
value $\kappa_2=0.01412 (\approx \sqrt{2}/N^2)$. All equations were
solved in a finite dimensional harmonic oscillator basis (100
functions, $\omega=1$). This of course introduces an error due to the
limited size of the basis, but for $\kappa_4=0$ we checked numerically
that this has only a small effect. It would also be better to change
the oscillator frequency as we move along the path, but this is too
cumbersome to implement for the simple problem at hand.  To solve the
RPA equation, we applied the method described above using
$x^2$ and $x^4$ as basic operators.

Our method does not fix an origin for the value of the collective
coordinate, in particular its value $Q_0$ at the Hartree minimum, This
is closely related to the fact that our formalism does not change if
we add a constant to all diagonal matrix elements of the cranking
operator $f$. For this reason we will give all our results as a
function of $Q-Q_0$.
\begin{figure}
\centerline{\includegraphics[width=7cm]{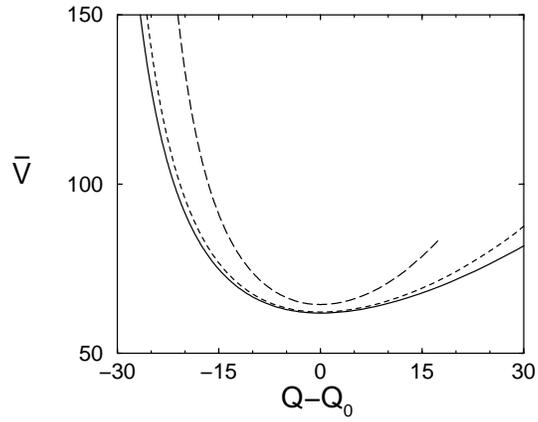}}
\caption{The collective potential energy for the modified Suzuki model.
The meaning of the lines is discussed in the main text.\label{fig:4:suz2}}
\end{figure}
\begin{figure}
\centerline{\includegraphics[width=7cm]{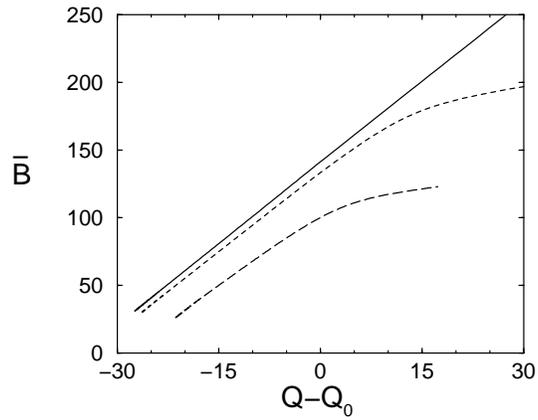}}
\caption{The collective mass for the modified Suzuki model.
The meaning of the lines is discussed in the main text.\label{fig:4:suz3}}
\end{figure}
\begin{figure}
\centerline{\includegraphics[width=7cm]{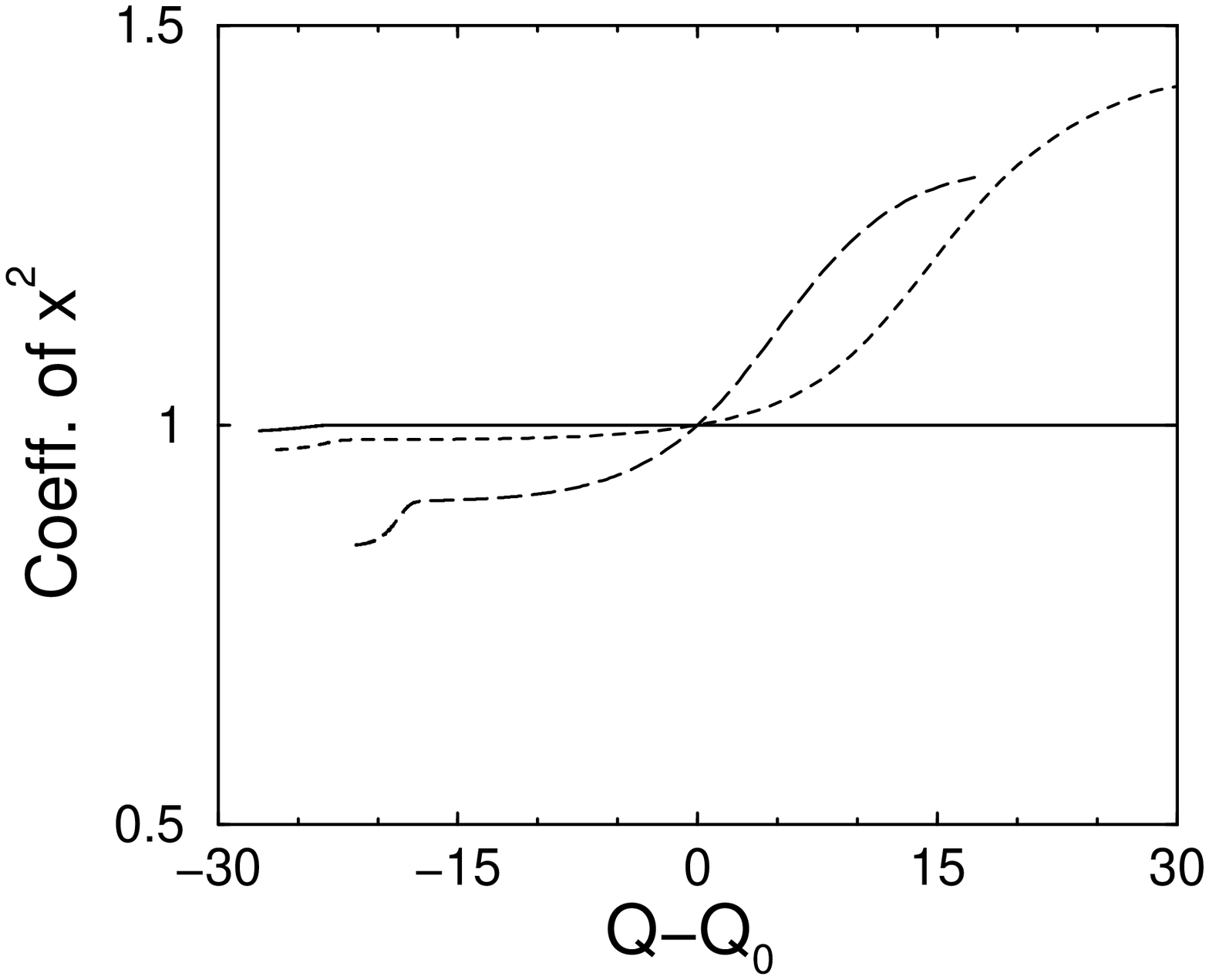}}
\caption{The coefficients of $x^2$ in the collective operator.
The meaning of the lines is discussed in the main text.\label{fig:4:suz4}}
\end{figure}
\begin{figure}
\centerline{\includegraphics[width=7cm]{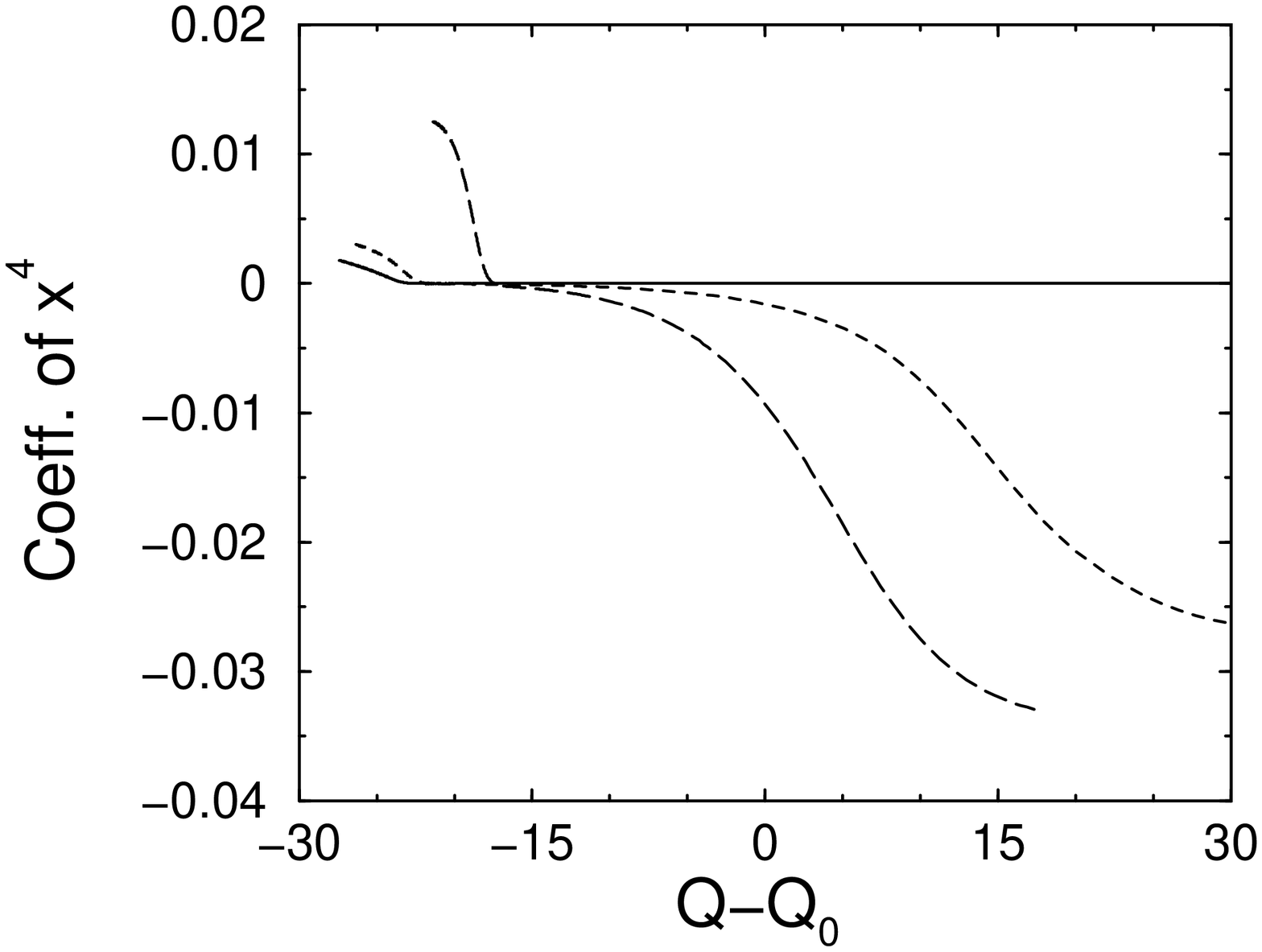}}
\caption{The coefficients of $x^4$ in the collective operator.
The meaning of the lines is discussed in the main text.\label{fig:4:suz5}}
\end{figure}

In a series of figures, we compare results for three values of
$\kappa_4$ (solid line $\kappa_4=0$, short dashes 0.00001 and long dashes
0.0001).  In Fig.~\ref{fig:4:suz2} we give the potential energy for
these models, the curve for the Suzuki model agreeing very closely
with the exact result.  Clearly an extra repulsive interaction
increases the potential energy at the minimum, and also results in a
larger second derivative.  The mass is a much more sensitive parameter
since it depends on the particle-particle and hole-hole matrix
elements of the cranking operator, that are only indirectly determined
from the RPA equation.  As can be seen in Fig.~\ref{fig:4:suz3}, the
mass changes drastically from one value of $\kappa_4$ to the other;
the numerical calculation for the Suzuki model again coincides with
the exact result.

We have not yet discussed the dependence of the cranking 
operators on the collective coordinate.
{}From the solution of the RPA equation we get, at each point of the 
collective path, a collective operator of the form 
\begin{equation} 
\hat{f}_{ph} = c_2 (x^2)_{ph} + c_4 (x^4)_{ph}. 
\label{eq:4:fromfit} 
\end{equation} 
In Fig.~\ref{fig:4:suz4} we give the dependence of the coefficient of $x^2$ 
in the 
collective operator as a function of the value of the collective coordinate, 
and of the coefficient of $x^4$ in Fig.~\ref{fig:4:suz5}.
Apart from a change of the cranking operator along the 
collective path, we can also see some effects of the limited model space 
used in the calculations. One such effect is the sharp turn of the 
$\kappa_4=0.0001$ curves at large negative $Q-Q_0$. This will disappear 
if we increase the number of basis functions.

\subsection{Application to the study of the Silicon nucleus \label{sec:si1}\label{sec:4.3}}

\subsubsection{Introduction\label{sec:4.3.1}}

In this subsection, we describe a calculation carried out using our
methods where an attempt was made to fit actual nuclear data. We can
characterize the outcome of the efforts described in these two
sections as a moderately successful (but still incomplete) description
of the low-lying spectrum of $^{28}$Si.  In our current opinion, this
was not exactly the right step to take immediately after our
investigation of monopole models. In the course of the latter
investigation, we developed approximation methods especially suited to
the application to Hamiltonians with separable interactions.  For such
interactions, the expansion of the self-consistent cranking operators
in a small basis of operators suggested by the Hamiltonian itself
would certainly yield improvement with respect to cranking
calculations of the type pioneered by Kumar and Baranger
\cite{1a,1b,1c,1d,1e} and, at the same time, permit the inclusion of
curvature effects within the framework of extended point
transformations. Results of these studies will be given in
Sec.~\ref{sec:4.5} below.

The reason for choosing application to the $sd$ shell
nucleus $^{28}$Si was that it is known to exhibit shape
coexistence, since bands of both prolate and oblate shape can
be identified, and probably do mix. Staying within the $sd$ shell,
one can also compare the results of our collective approach with those of
an exact shell-model calculation for the chosen Hamiltonian. In turn,
the latter should have some reasonable relation to experiment.
Unfortunately this appears to rule out simple separable interactions (such
as the quadrupole-quadrupole interaction), at least for $^{28}$Si. 
Furthermore, for this nucleus we can compare our calculation 
with the only
existing application to a bound system \cite{58,59} 
of a ``competing'' self-consistent
theory of large amplitude collective motion \cite{60}.

The main consequences of utilizing a so-called realistic interaction is that
it is more difficult than for separable interactions to include the effects
associated with the extended adiabatic approximation. 
We have, however, included
the effects of curvature and found that they are small. Since
we expect the effects of the extended adiabatic approximation 
in general, to be of the same order of magnitude as 
those due to curvature, we feel that we may proceed with some confidence.
The material that follows is based largely on Refs.~\cite{23,24}.

\subsubsection{Equations of the local harmonic formulation\label{sec:4.3.2}}
 
The local harmonic form of the generalized valley theory, that we
apply in this section, consists of the two equations contained in 
Eq.~(\ref{eq:4:LHA2}), namely the cranking equation and the associated
RPA eigenvalue equation.

These equations have to be solved
self-consistently:
The cranking equation 
determines, for given
$f_{ph}$, a set of single-particle wave functions, using as before the label
 $p$ for empty orbits and $h$ for filled orbits. In turn, given this basis,
the local harmonic equation determines $f_{ph}$.
If we neglect the affine connection we obtain a set of equations that
is very close to the set derived in the Holzwarth-Yukawa formalism \cite{60}
by Pelet and Letourneux, \cite{58,59}, the only difference
being that we use RPA where they use a further approximation, the
TDA (Tamm-Dancoff approximation)
\cite{RS}.
 
The RPA equation in (\ref{eq:4:LHA2}) is a linear eigenvalue problem
and thus does not fix the
scale of the eigenvectors. We propose to make the very convenient
choice of normalization
\begin{equation}
f^\mu_{,\alpha} B^{\alpha\beta} f^\nu_{,\beta} =
\delta^{\mu\nu}. \label{eq:4:cnorm}
\end{equation}
The normalization is arbitrary, whereas the orthogonality follows from 
(\ref{eq:4:LHA2}).
For this special choice the inverse $B_{\alpha\beta}$ also takes a very
simple form,
\begin{equation}
B_{\alpha\beta} = f^\mu_{,\alpha} \delta_{\mu\nu} f^\nu_{,\beta}. \label{eq:4:Bl}
\end{equation}

Finally,
we want to measure the quality of decoupling along the
path. This can be done by comparing two different forms
of the collective mass that can be calculated in the theory, following 
the procedure described in Sec.~\ref{sec:2.2.3}.
 
\subsubsection{Symmetries\label{sec:4.3.3}}
In this section we shall discuss the consequences of two symmetries 
that we impose on the Slater determinants.
Taking advantage of the fact that isospin is a good symmetry for light
nuclei, we require that proton and neutron orbits be occupied with equal
probability. This corresponds to considering only the manifold
of $T=0$ states, which constitute the low-energy part of the
spectrum of $^{28}$Si.
This symmetry reduces the number of active particles
we have to consider by a factor of two, so that
the effective number of single-particle
degrees of freedom is reduced from 12 to 6. 
 
Furthermore, following Pelet \cite{58,59}
we impose ``ellipsoidal'' symmetry, i.e.,
we require the intrinsic nuclear shape to be invariant under a rotation of
180 degrees about any of the three symmetry axes.
It is well known \cite{61} 
that such a
symmetry requires that $K$, the projection of the angular momentum on the
intrinsic $z$-axis is even and also relates wavefunction components with
positive $K$ to those with negative $K$. In the latter regard, it duplicates
the function of time reversal invariance, which
for static solutions of the Hartree-Fock problem for even
nuclei already implies that any pair of time-reversed orbits is either
occupied or unoccupied.
The requirement of ellipsoidal symmetry is indeed a strong limitation
in the implementation of our algorithm, since any reasonable interaction
\cite{62,63,64} gives that for $^{28}$Si
the lowest-energy RPA mode at the Hartree-Fock
minimum is a $\Delta K=3$ state.
We thus reject this solution as a possible choice of
cranking operator, and choose the lowest even-$\Delta K$
solution for this purpose.
 
The reduction to ellipsoidally symmetric Slater determinants leads to an
extra reduction by a factor of two in the number of single-particle
degrees of freedom, so that
we finally end up with three active ``particles''.
Taking these reductions into account
we can easily rewrite the Hartree-Fock energy $W$
in terms of a density matrix within the restricted space only
(so that $\trace(\rho) = 3$). The Hartree-Fock Hamiltonian becomes
\begin{equation}
{\cal H}_{\alpha\beta} = 4\delta_{\alpha\beta} \epsilon_a
+ \sum_{\gamma\delta} \tilde{V}_{\beta\delta\alpha\gamma}
\rho_{\gamma\delta}.
\end{equation}
(Here $\tilde{V}$ is a suitable symmetrized form of the two-body
interaction, and the factor 4 originates from the fact that each state
is occupied four times.)  We can evaluate the mass and the second
derivative of the potential using methods described in
Sec.~\ref{sec:4.1}, and find (the Roman letters $p$,$h$ are used to
denote all the quantum numbers of the particle-hole orbits except the
isospin projection)
\begin{subequations}
\begin{eqnarray}
B^{php'h'} & = & {\cal H}_{pp'}\delta_{hh'}
 - {\cal H}_{hh'}\delta_{pp'}
 + \tilde{V}_{hp'ph'} -\tilde{V}_{pp'hh'}, \label{eq:4:Bexplicit}\\
V_{,php'h'} & = &
 {\cal H}_{pp'}\delta_{hh'}
 - {\cal H}_{hh'}\delta_{pp'}
 + \tilde{V}_{hp'ph'} +\tilde{V}_{pp'hh'}.
\label{eq:4:Vexplicit}
\end{eqnarray}
\end{subequations}
The derivative of $B$ is also needed for the calculation of the RPA matrix.
We refer to Appendix B of \cite{23} for its evaluation.
At the Hartree-Fock minimum the resulting local harmonic equation
is indeed the standard RPA equation, e.g.,
Eq.~(8.83) in \cite{RS}, as can be seen from
Eq.~(8.71) in the same reference.

In order to understand how to calculate the quantities (\ref{eq:4:Bexplicit})
and (\ref{eq:4:Vexplicit}), we back up one step.
Instead of rewriting the energy density in terms of the reduced density
matrix that takes into account all symmetries, we consider
the general RPA (taking into account the proton-neutron symmetry, though)
at a point of ellipsoidal symmetry.
Consider, for example, the kinetic energy matrix $B$, that can be obtained
as a second derivative of the energy functional. We separate the 
single-particle basis
into two disjoint sets that are mutually conjugate under time reversal,
and restrict the indices $p$ and $h$ to label only states in one of these
sets, so that the time reversed states $\bar{p}$ and $\bar{h}$ are
members of the other set. (In case of a summation we shall later
write $\sum'$ to indicate this limitation.)
We now can order the states in such a way that $B$ takes the form
\begin{equation}
B^{\alpha\beta} =
\left( \begin{array}{llll}
 B^{php'h'} & B^{ph\bar{p}'\bar{h}'}&
 B^{ph\bar{p}'h'} & B^{php'\bar{h}'} \\
 B^{\bar{p}\bar{h}p'h'} & B^{\bar{p}\bar{h}\bar{p}'\bar{h}'}&
 B^{\bar{p}\bar{h}\bar{p}'h'} & B^{\bar{p}\bar{h}p'\bar{h}'} \\
 B^{\bar{p}hp'h'} & B^{\bar{p}h\bar{p}'\bar{h}'}&
 B^{\bar{p}h\bar{p}'h'} & B^{\bar{p}hp'\bar{h}'} \\
 B^{p\bar{h}p'h'} & B^{p\bar{h}\bar{p}'\bar{h}'}&
 B^{p\bar{h}\bar{p}'h'} & B^{p\bar{h}p'\bar{h}'}
\end{array}\right).
\end{equation}
 
{}From the explicit expression for $B$, Eq.~(\ref{eq:4:Bexplicit}),
we can easily
derive that all the entries in this matrix with an odd number of barred
indices are zero, as a consequence of time reversal invariance of the
underlying shell-model Hamiltonian.
A further consequence of this symmetry is that all entries of $B$
are invariant
under the interchange of all unbarred with all barred indices.
If we now introduce the orthogonal transformation
\begin{equation}
O = \frac{1}{\sqrt{2}}
\left(\begin{array}{rrrr}I & -I&0&0 \\ I & I&0&0\\
 0&0&I & -I \\ 0&0&I & I
\end{array} \right) ,
\end{equation}
we obtain the following block-diagonal form for the transform of $B$
\begin{equation}
 O^T B O =
\frac{1}{2}
\left( \begin{array}{llll}
 B^{php'h'} +B^{\bar{p}\bar{h}p'h'} & 0 & 0 & 0 \\
 0&B^{php'h'} -B^{\bar{p}\bar{h}p'h'} & 0 & 0 \\
 0&0& B^{\bar{p}h\bar{p}'h'} + B^{\bar{p}hp'\bar{h}'}&0\\
 0&0&0& B^{\bar{p}h\bar{p}'h'} - B^{\bar{p}hp'\bar{h}'}
\end{array}\right).
\end{equation}
The second derivative of the potential energy takes the same form.
We thus have split the total RPA problem into four disjoint subproblems.
{}From the structure of the matrix $O$ we can infer that the basis vectors
that span the upper left block have equal $ph$ and $\bar{p}\bar{h}$
matrix elements (and zero matrix elements between a state and a time
reversed one).
The eigenvectors of this part of the RPA matrix can serve
as cranking operators, since they conserve the ellipsoidal symmetry.
The next block to the lower right
 still has entries in the same space, but the
$ph$ entries are minus the $\bar{p}\bar{h}$ entries.
The basis vectors for the
next two blocks only have entries $\bar{p}h$ and $p\bar{h}$, the first
with equal sign, and the last (on the lower right) with opposite sign.
As will be discussed in the next subsection the lower three blocks play an
essential role in the calculation of the moments of inertia.
 
\subsubsection{Calculation of the inertia tensor\label{sec:4.3.4}}
 
If we study a real nucleus, we know that there exist two (for an axially
deformed nucleus) or three (the general triaxial case) cyclic
coordinates, that can be identified as
the angles conjugate to the angular momenta.
On the classical level the momenta are
\begin{equation}
J_i = \trace(\rho^{\tau_z} \hat{J}^{\tau_z}_i) = 2\trace(\rho \hat{J_i}).
\end{equation}
Here we have taken the proton-neutron degeneracy into account in the last
term. We shall not yet use ellipsoidal symmetry.
At an axial minimum we can easily find these momenta, since they correspond
to zero eigenvalues of the RPA matrix. In the covariant RPA they no
longer correspond to zero modes at all points on the collective path,
but it is still correct to expand the energy
to second order in the components of the angular momenta to obtain the
inertia tensor and we should still be able to calculate these momenta by
following the appropriate solutions of the local RPA.

First let us analyze what form the momenta take in the local ($ph$)
basis. From the matrix elements ($C^i_\alpha$ are the expansion coefficients
of the state $i$ in shell model states $\alpha$)
\begin{eqnarray}
\langle p |\hat{J}_m | h \rangle & = &
C^p_\alpha C^h_\beta
\langle (nljm)_a |\hat{J}_m | (nljm)_b \rangle \nonumber \\
&=&\delta_{j_aj_b} \sqrt{j_a(j_a+1)} C^{j_a\;\;1\;\;j_b}_{m_a\;m\;m_b}
C^p_\alpha C^h_\beta
\end{eqnarray}
and the symmetry properties of the Clebsch-Gordan coefficients we find
\begin{equation}
\langle p | \hat J_m | h \rangle =
\langle h | \hat J_{-m} | p \rangle (-1)^m.
\end{equation}
If we now use
\begin{eqnarray}
\hat J_x & = & (\hat J_{-1}-\hat J_{+1}) /\sqrt{2},\nonumber\\
\hat J_y & = & i(\hat J_{+1}+\hat J_{-1}) /\sqrt{2},
\end{eqnarray}
we find
$\langle p | \hat J_x | h \rangle = \langle h | \hat J_x | p \rangle$
is real and
$\langle p | \hat J_y | h \rangle = -\langle h | \hat J_y | p \rangle$
is purely imaginary. To first order in the coordinates and momenta we
now have
\begin{eqnarray}
J_x & = &2\sqrt{2}\sum_{ph}
 \xi_{ph} \langle p | \hat J_x | h \rangle,\nonumber\\
J_y & = &-2\sqrt{2}\sum_{ph}
\pi_{ph}i \langle p | \hat J_y | h \rangle,\nonumber\\
J_z & = &2\sqrt{2}\sum_{ph}
 \xi_{ph} \langle p | \hat J_z | h \rangle.
\label{eq:4:35}
\end{eqnarray}
It is not hard to show that $\hat{J}_x$ and $\hat{J}_z$ have zero
 matrix elements
between a state and a time reversed one, whereas
$\langle p | \hat J_y | {h} \rangle = -
 \langle \bar{p} | \hat J_y|\bar{h}\rangle $. Similarly $J_x$ and $J_y$
belong to the last two symmetry classes discussed in the previous
subsection.

It is of interest to note that of the components of the angular momentum,
only $J_y$ is ``momentum-like'', whereas the other two components are
``coordinate-like''. According to the discussion in Sec.~\ref{sec:2.3.5}, this
could result in technical difficulties if this occurred in the same
symmetry subspace of the RPA matrix. As we have just explained, this is
not the case here, and we therefore expect the considerations of \ref{sec:2.3.5} to
apply separately to each component.

To calculate the inverse moments of inertia, which corresponds to the
calculation of the inverse mass for the kinetic energy,
we now have to evaluate second derivatives of the mean-field energy,
$W$, namely, $\partial^2 W/\partial J_i \partial J_j$.
This quantity can be reexpressed by the chain rule in terms of
$B^{php'h'}$ and $V_{,php'h'}$, e.g.,
\begin{equation}
{\cal I}_y^{-1} =
\frac{\partial\pi_\alpha}{\partial J_y}
B^{\alpha\beta}
\frac{\partial\pi_\beta}{\partial J_y}.
\end{equation}
Using the canonicity conditions, Eqs.~(\ref{eq:2:2.25}),
for $J_i$ and the conjugate angles $\theta_i$, we find that the quantities
$\partial\pi_\beta/\partial J_y$ and
$\partial\xi_\beta/\partial J_i$, $i=x,z$, can be replaced by
$\partial \theta_y/\partial \xi^\beta$ and
$-\partial \theta_i/\partial \pi_\beta$, respectively.
Since we do not have explicit expressions for the matrix elements of
the angles it does not appear feasible to use the resulting expressions,
but there exists an alternative.
Consider the moments of inertia themselves, instead of their inverses.
In analogy with Eq.~(\ref{eq:4:7.36}) we can calculate, for
${\cal I}_y$,
\begin{equation}
{\cal I}_y = \frac{\partial J_y}{\partial\pi_{ph}} B_{php'h'}
\frac{\partial J_y}{\partial\pi_{p'h'}}, \label{eq:4:momin2}
\end{equation}
where we have once more used the canonicity conditions to equate
\begin{equation}
\frac{\partial \xi^\alpha}{\partial\theta_y} =
\frac{\partial J_y}{\partial\pi_{\alpha}}.
\end{equation}
The expression (\ref{eq:4:momin2}) can also be obtained through expansion of the
energy in terms of $\dot\theta_y$. If we apply the same analysis to
the other two moments of inertia we find that they
can be evaluated by a similar
expression involving the second derivative of the potential energy,
\begin{equation}
{\cal I}_j = \frac{\partial J_j}{\partial\xi^{ph}} V^{,php'h'}
\frac{\partial J_j}{\partial\xi^{p'h'}}.
\label{eq:4:I_j}
\end{equation}
Even though these equations may look unfamiliar, at the Hartree-Fock
minimum they correspond to the usual RPA equations for the moments
of inertia (see e.g.\ Eq.~(8.113) in \cite{RS}).
 
It may not appear obvious that the inertia tensor is diagonal. As
noted before, however, each of the vectors of particle-hole matrix
elements of the components of $J$ belongs to a different symmetry
class discussed in Sec.~\ref{sec:4.3.3}. Since both the mass and
potential energy matrix are block-diagonal, we thus can not have
off-diagonal matrix elements in the inertia tensor. This is a
consequence of ellipsoidal symmetry, and will no longer hold if we
would allow, say, octupole components in the cranking operator.
 
Note that we have not used the covariant derivative of $V$ in
(\ref{eq:4:I_j}) for the sake of a consistent treatment of the
different moments
of inertia. In the general non-adiabatic theory we would expect complete
symmetry between the two expressions, which can only be reached for the
present case by disregarding the covariant derivative.
 
{}From the explicit expressions for the derivatives of the different $J$'s
obtained from Eq.~(\ref{eq:4:35}) we find that, using the
symmetries to restrict the summations over half the single particle
states (as indicated by a prime),
\begin{subequations}
\begin{eqnarray}
{\cal I}_z & = &
{\sum_{php'h'}\!}'\,
\left(2\sqrt{2}\langle p | \hat J_z |{h} \rangle\right)
\sqrt{2}\left(V_{,php'h'}-V_{,\bar{p}\bar{h}p'h'}\right)\sqrt{2}
\left(\langle p'| \hat J_z | h'\rangle 2\sqrt{2}\right)\nonumber\\
& = &
{\sum_{php'h'}\!}'\,
16 \langle p | \hat J_z |{h} \rangle
\left(V_{,php'h'}-V_{,\bar{p}\bar{h}p'h'}\right)
\langle p'| \hat J_z | h'\rangle,\\
{\cal I}_y & = &
{\sum_{php'h'}\!}'\,
16 i\langle p | \hat J_y |\bar{h} \rangle
\left(B^{p\bar{h}p'\bar{h}'}-B^{\bar{p}hp'\bar{h}'}\right)
i\langle p'| \hat J_y | \bar{h}'\rangle,\\
{\cal I}_x & = &
{\sum_{php'h'}\!}'\,
16 \langle p | \hat J_x |\bar{h} \rangle
\left(V_{,p\bar{h}p'\bar{h}'}+V_{,\bar{p}hp'\bar{h}'}\right)
\langle p'| \hat J_x | \bar{h}'\rangle.
\end{eqnarray}
\end{subequations}
 
\subsubsection{Numerical algorithm\label{sec:4.3.5}}
 
We have applied the numerical algorithm discussed in
Sec.~\ref{sec:alg:GVA}. At each point on the path we solve a local
RPA, of which the eigenvectors were normalized according to
$f^{\mu}_{\,\alpha} B^{\alpha\beta} f^{\nu}_{,\alpha}=\delta_{\mu\nu}$.
One of the important areas where these eigenvectors find application
is in the definition of the decoupling measure $D$.
Due to the choice of normalization we find that $\bar{B}^{11} =1$.
We therefore only need to calculate
\begin{equation}
\breve{B}_{11} = \frac{d\xi^\alpha}{dq} B_{\alpha\beta}
 \frac{d\xi^\beta}{dq}, \label{eq:4:brB}
\end{equation}
where $B_{\alpha\beta}$ is given by Eq.~(\ref{eq:4:Bl}).
If we furthermore approximate the derivatives in (\ref{eq:4:brB})
by finite differences,
\begin{equation}
\frac{d\xi^\alpha}{dQ} \approx
\sqrt{2} \frac{\Delta\rho^\alpha} {\Delta Q},
\end{equation}
the evaluation of $D$ yields
\begin{eqnarray}
D & = & \breve{B}_{11} -1 \nonumber\\
&=& \sum_\mu (\Delta q^\mu/\Delta Q)^2 -1.
\label{eq:4:Deq}
\end{eqnarray}
where we have defined
\begin{equation}
\Delta q^\mu = \sqrt{2} \sum_{ph} \Delta\rho_{ph} f^\mu_{,ph}.
\end{equation}
Thus if $\Delta q^1 \equiv \Delta Q$ is the only non-zero number
in the sequence
$D =0$. If any of the other coordinates is comparable to $\Delta q^1$
we do not have good decoupling. As emphasized below, the result of this
calculation suggests how to choose additional collective coordinates to
improve the decoupling.

\subsubsection{Quantum corrections\label{sec:4.3.6}}

Before attempting to compare the results found from the application
of the algorithm
described in Sec.~\ref{sec:2.3.5} with an exact diagonalization, it is appropriate
to include the first quantum corrections. In practice this means the
quantum corrections to the classical potential energy, which is 
the largest contribution to the classical collective Hamiltonian.
(By comparison, the kinetic energy is
of relative order $(1/N)$, where $N$ is the number of degrees of freedom
that participate in the motion.) Among previous attempts to found a
theory of large amplitude collective motion on a quantum basis such as
the method of
generator coordinates \cite{60,66}, the Born-Oppenheimer method \cite{67},
a generalized coherent state method \cite{69},
and the equations of motion method \cite{48,4},
none provided a {\em systematic}
expansion in $(1/N)$. (The work mentioned here for the first time
will be reviewed in Sec.~\ref{sec:7}, the final section of this review.)
 
The basic problem is that of computing quantum corrections about
mean field solutions describing {\em non-equilibrium} configurations.
Thus in the simplest case of one collective coordinate, that we choose
as an example, the fluctuations must be studied at an arbitrary point
on a ``collective path''.
 
The interest of studying quantum fluctuations about non-equilibrium
mean fields has certainly not escaped the attention of previous
workers in the field, including an extensive study of rotating
nuclei \cite{71a,71b} using a boson expansion method, a preliminary study of
quantum fluctuations about an {\em arbitrary} time-dependent mean
field \cite{72}, and semiclassical quantization of periodic solutions
\cite{73} using a path integral method. In the last
work it is implied that the adiabatic case of interest to us
had been disposed of earlier \cite{74}, but a study of this last reference
does not sustain the claim. It appears
that Reinhard, Goecke and their co-workers are the only authors who
have evaluated and included a part of this correction in their calculations
\cite{66}. They too have not evaluated the full $(1/N)$ corrections, however.
We have developed a new method for studying quantum
fluctuations about mean fields, that has also been applied to
the study of translational motion \cite{75} and rotational motion at high
spin \cite{76}.

We have decided that it is unwarranted to reproduce a full account
of this somewhat complex theory, since we have nothing to add to the
detailed exposition given in
Ref.~\cite{24}. We shall therefore confine ourselves to quoting the result
of applying this theory to the collective potential energy. We find
\begin{equation}
V(Q)=V_0(Q) + \mbox{$\frac{1}{2}$}\left[\sum_a\omega_a(Q) -\trace
{\cal A}\right] -\left[(H_{11})^{(0)} +(H_{04} +
H_{40})^{(0)}\right]. \label{eq:4:Energy}
\end{equation}
Here $V_0(Q)$ is our previous classical result for the potential energy,
$\omega_a(Q)$ are the solutions of the local RPA, other than the solution(s)
associated with the large amplitude collective motion,
\begin{eqnarray}
(H_{11})^{(0)} &=& \mbox{$\frac{1}{2}$}\sum_{php'h'a}
 ({\cal H}_{pp'}\delta_{hh'}-
 {\cal H}_{hh'}\delta_{pp'})
{\cal Y}^{\ast}_a(ph){\cal Y}_a(p'h'), \label{eq:4:sum6} \\
(H_{04})^{(0)} &=& (H_{40})^{(0)\ast} =\frac{1}{4}{\cal B}_{php'h'}
{\cal X}^{\ast}_a(p'h'){\cal Y}_a(ph), \label{eq:4:sum3}
\end{eqnarray}
and ${\cal A}$ and ${\cal B}$ are the standard RPA matrices when the RPA is
written in terms of the Tamm-Dancoff amplitude ${\cal X}_a(ph)$ and the
ground-state correlation amplitude ${\cal Y}_a(ph)$,
\begin{eqnarray}
{\cal A}_{php'h'} &=& {\cal H}_{pp'}\delta_{hh'}
 -{\cal H}_{hh'}\delta_{pp'} +V_{p'hh'p}, \\
{\cal B}_{php'h'} &=& V_{hh'pp'}.
\end{eqnarray}

At the point of equilibrium, $d\bar{V}/dQ=0$, this expression reduces to a
value of the ground state
energy that agrees in the limit of weak residual interaction with
the result of perturbation theory \cite{79}
(after addition of the term $\frac{1}{2}\omega_1$,
associated with the collective coordinate $Q$). It also contains
the zero-point energy calculated by Reinhard and Goeke \cite{66}
as a special case: If we discard the $\omega_a$ terms, and approximate the
trace by the contribution along the collective path (see below for more
detail), we obtain the result of these authors.

\subsubsection{Results\label{sec:4.3.7}}

We shall first discuss results obtained based on one collective mode
without inclusion of quantum corrections. We shall then see that substantial
improvement can be achieved by including the latter and by making
approximate provision for a second collective coordinate.

\begin{table}
\caption{
The single particle energies used in the calculation \label{tab:4:1}}
\begin{center} 
\begin{tabular}{l|r}
shell & s.p.e. (MeV) \\
\hline
1d5/2 &-3.9478\\
2s1/2 & -3.1635\\
1d3/2 & 1.6466\\
\end{tabular}
\end{center} 
 \end{table}

We have applied the algorithm discussed in Sec.~\ref{sec:4.3.5} to a
description of $^{28}$Si in the $sd$-shell, using Kuo's interaction \cite{62},
with single-particle energies as given in Table \ref{tab:4:1}.
This interaction is not the best shell-model interaction available (it
is known that Wildenthal's W interaction \cite{62,63}
gives a much better description of the $sd$-shell nuclei), but this
interaction is as close as possible to the unpublished interaction
used by Pelet and Letourneux.
Note that we use ellipsoidal symmetry, which is a limitation for the
present case, where the lowest RPA mode at the HF minimum is
a $|\Delta K|=3$ mode, so that it may not be totally correct
to consider only even multipoles.
 
The model exhibits a deep deformed minimum on the oblate side
built from a Slater
determinant in which orbits with $m=5/2,3/2,1/2,-5/2,-3/2,-1/2$ are
occupied. There is no stable prolate solution with the same orbits
occupied.
Further study reveals, however, that there is another
minimum with positive quadrupole moment when the orbits
$m=3/2,1/2,1/2,-3/2,-1/2,-1/2$ are occupied. Since all axially deformed
states within this
manifold are orthogonal to all states considered previously, we find
that there is no path through the subspace of axially deformed Slater
determinants from one state to the other. The lowest mode of even $K$ of the
RPA is not an axial ($K=0$) mode, however, but a $K=2$ mode. It therefore
seems plausible that there exists a path going through triaxial shapes
from one to the other. Actually this point
has already been studied by Pelet
and Letourneux \cite{58,59}, and such a path has been found. 
 
\begin{figure}
\centerline{\includegraphics[width=7cm]{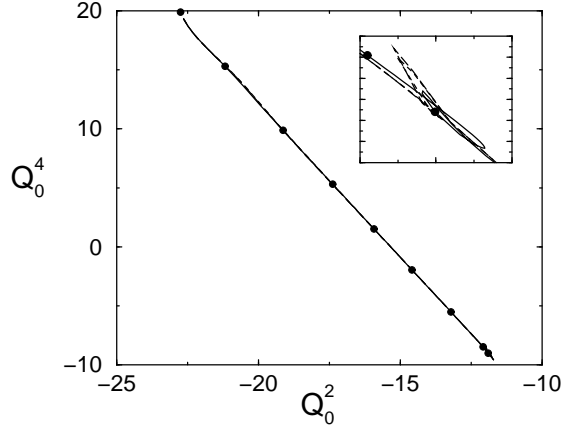}}
\caption{
The value of the hexadecapole moment $\langle r^4 Y^4_0\rangle$ as a
function of the quadrupole moment $\langle r^2 Y^2_0 \rangle $. The
oblate minimum is located at the upper left corner and the prolate one
at the lower right. At the scale shown, we cannot distinguish the path
determined using covariant derivatives from that obtained neglecting
curvature corrections. The markers are drawn at those positions where
the collective coordinate is a multiple of $1/4$, $Q=i/4$, starting
from the prolate minimum.  See the text for a discussion of the insert.
\label{fig:4:q2-q4}}
\end{figure}
We have applied our algorithm to the same calculation and found the
corresponding path. Since each point of the collective path
corresponds to a
Slater determinant, or equivalently a set of occupied orbits,
it is hard to visualize this path.
We need projections of the manifold of Slater determinants on
some two-dimensional surface in order to represent the path
graphically.
Guided by Pelet and Letourneux we give
the values of the hexadecapole moment
$\langle\sqrt{4\pi/9}\, r^4 Y^4_0\rangle$ as a function of the quadrupole
moment $\langle \sqrt{4\pi/5}\,r^2Y^2_0\rangle$ along the path in
Fig.~\ref{fig:4:q2-q4}.
The prolate minimum does not have positive quadrupole moment
since the symmetry axis of the prolate solution is the $x$ and not the
$z$-axis.
The oblate minimum is located at the
upper left corner and the prolate one at the lower right. The solid
line has been calculated using the covariant derivatives, the dashed
line using ordinary derivatives and the dotted line corresponds to
the TDA calculation of Pelet. Clearly there is a smooth and
continuous change of the two parameters along the path.
The lower right of this plot, however, is not as smooth as it seems.
For that reason we give an enlargement in the inset.
The region of this large curvature will be shown below to correspond
to a region of bad decoupling, a behavior that has also been noted for
simple models.

\begin{figure}
\centerline{\includegraphics[width=7cm]{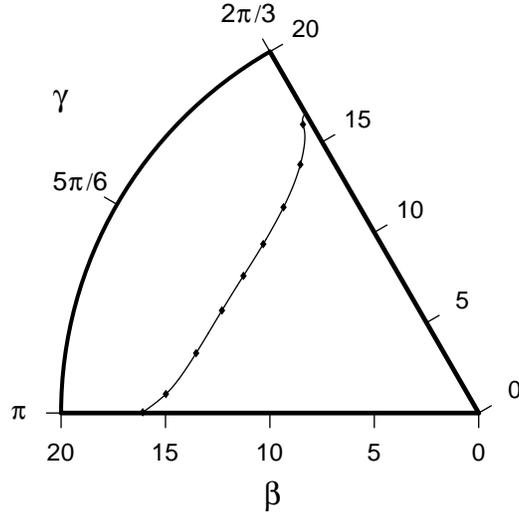}}
\caption{
The one-dimensional collective path projected onto the quadrupole
deformation $\beta-\gamma$ plane. The oblate minimum is attained for
$\gamma=\pi$, corresponding to symmetry around the z axis.  The dots
indicate the points where the collective coordinate takes the values 
$Q = i/4$, starting from the oblate minimum where
$Q=i=0$.
\label{fig:4:beta-gamma}}
\end{figure}

\begin{figure}
\centerline{\includegraphics[width=7cm]{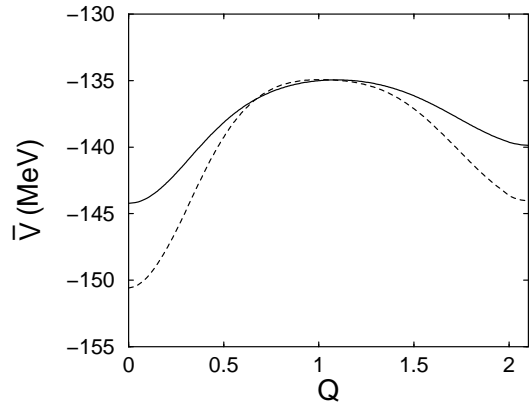}}
\caption{ 
The potential energy $V_0(Q)$ along the collective path (solid line)
as well as the quantum corrected potential energy $V(Q)$ (dashed line).
$Q=0$ corresponds
to the oblate minimum, whereas $Q=2.15$ for the prolate minimum.
\label{fig:4:q-V}}
\end{figure}

In Fig.~\ref{fig:4:beta-gamma} 
we have drawn a different representation
of the path, where
we plot the quadrupole $\beta$ and $\gamma$ parameters along the path,
defined as
\begin{equation}
\langle \sqrt{4\pi/5}\,r^2Y^2_0\rangle = \beta \cos\gamma,\;\;\;
\langle \sqrt{4\pi/5}\,r^2Y^2_2\rangle = \beta/\sqrt{2}\sin\gamma.
\end{equation}
It is difficult to distinguish results obtained by the three methods
described previously.  The behavior near the
prolate minimum is again not as smooth as at all other points.
As can be seen from Figs.~\ref{fig:4:q2-q4} and \ref{fig:4:beta-gamma}
the difference between the covariant and non-covariant approaches
is surprisingly small. This is even more surprising if one notes that
the affine connection does not have small matrix elements.
Furthermore, as we have argued before, the difference between TDA and
RPA is small as well. With the results of our calculation, we evaluate
the collective potential $\bar{V}$, the solid line in Fig.~\ref{fig:4:q-V}.
(The dashed line includes quantum corrections to be discussed below.)
To the accuracy of the figure the curves obtained using
the covariant and the non-covariant approach coincide.

\begin{figure}
\centerline{\includegraphics[width=7cm]{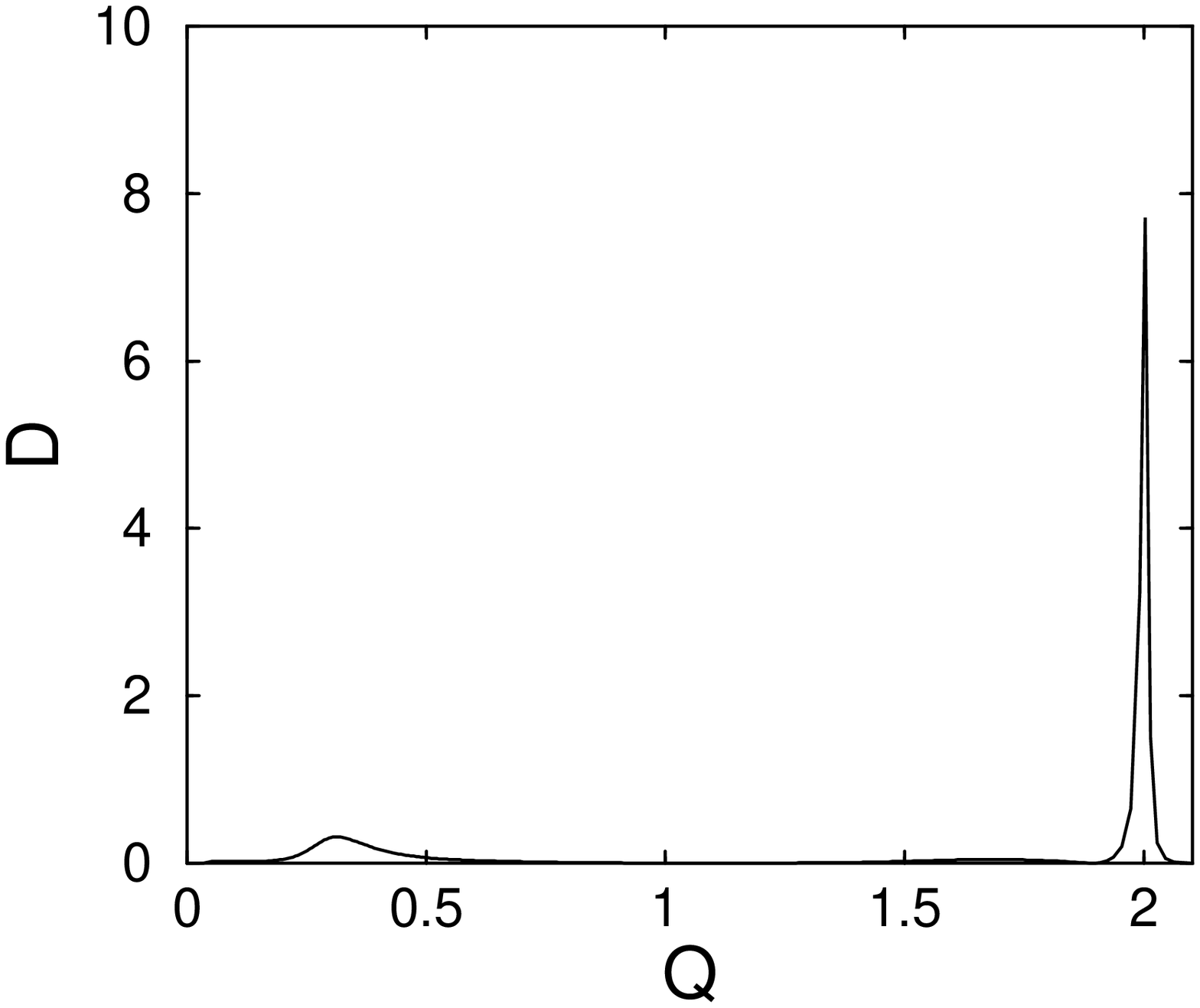}}
\caption{ 
The decoupling measure $D$ along the collective path. Note the extremely
large value near $Q=2$.
\label{fig:4:q-D}}
\end{figure}

\begin{figure}
\centerline{\includegraphics[width=7cm]{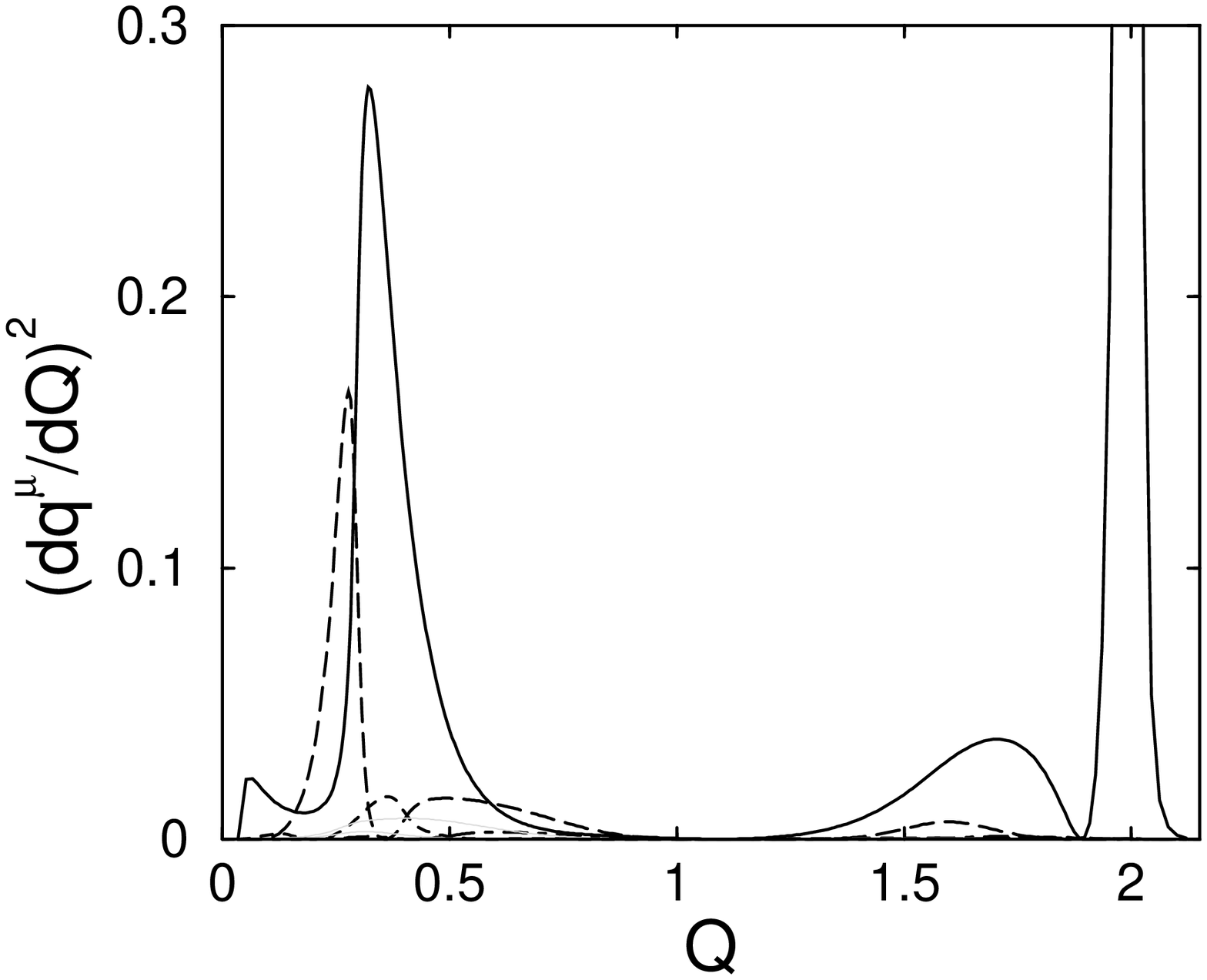}}
\caption{ 
The change of the non-collective coordinates with the collective
coordinate, which are related to $D$ as discussed in the text.
The solid line represents the second coordinate (in harmonic
frequency), the dashed the third.\label{fig:4:q-coors}}
\end{figure}

Having obtained the path we can now calculate the decoupling measure
$D$. Using Eq.~(\ref{eq:4:Deq}) we find in Fig.~\ref{fig:4:q-D} that
the quantity $D$ is not small everywhere along the path. The largest
value of $D$ is found near the prolate minimum, but $D$ also exhibits
another ``bump'' not too far from the oblate minimum.  This explains
the rapid change in properties of the path in Figs.~\ref{fig:4:q2-q4}
and \ref{fig:4:beta-gamma}, since the region of rapid change (large
curvature) occurs exactly where decoupling is bad.  These features
seem to indicate that we need to include more than one collective
coordinate.  To see whether an approach with two coordinates may be
able to solve this problem we have separated the contributions
$(\Delta q^\mu/\Delta Q)^2$ for the four choices of $\mu$, ordered
according to the eigenvalues of the RPA.  We clearly see in
Fig.~\ref{fig:4:q-coors}
that the second coordinate (the solid line) gives the
most important contribution to this quantity almost everywhere. This
seems to hold promise for the calculation of a two-dimensional potential
energy surface.

\begin{figure}
\centerline{\includegraphics[width=7cm]{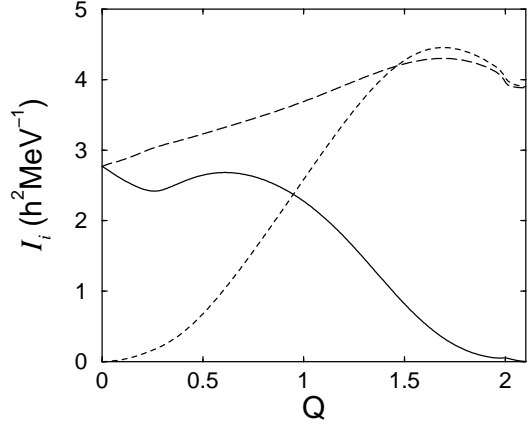}}
\caption{ 
The change of the moments of inertia along the collective path.
The solid line represents the moment of inertia around the
$z$-axis, the long-dashed around the $y$ one, and the dashed around
the $x$.\label{fig:4:q-momi}}
\end{figure}

To complete the discussion of parameters of the Hamiltonian we have also
calculated the moments of inertia, using the approach in Sec.~\ref{sec:4.3.4}.
These are displayed in Fig.~\ref{fig:4:q-momi}.
The moments of inertia behave relatively smoothly (although they
do not follow the irrotational pattern). Where $D$ has its largest value
the moments of inertia seem to change least smoothly.

\begin{figure}
\centerline{\includegraphics[width=7cm]{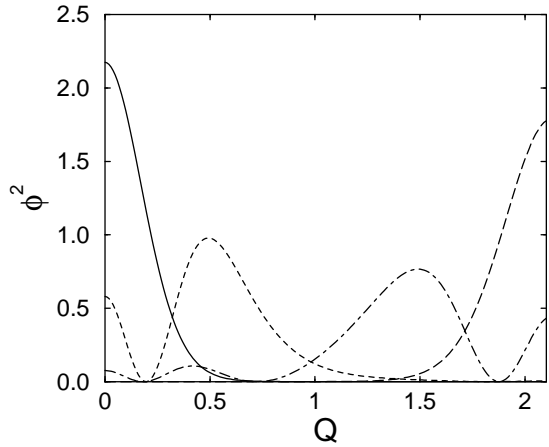}}
\caption{The square of the wave function for the four lowest $0^+$
eigenstates. The solid line is the lowest, the long-dashed the second,
the short-dashed the third and the dashed dotted the fourth state.
\label{fig:4:q-eigv}}
\end{figure}

\begin{figure}
\centerline{\includegraphics[width=7cm]{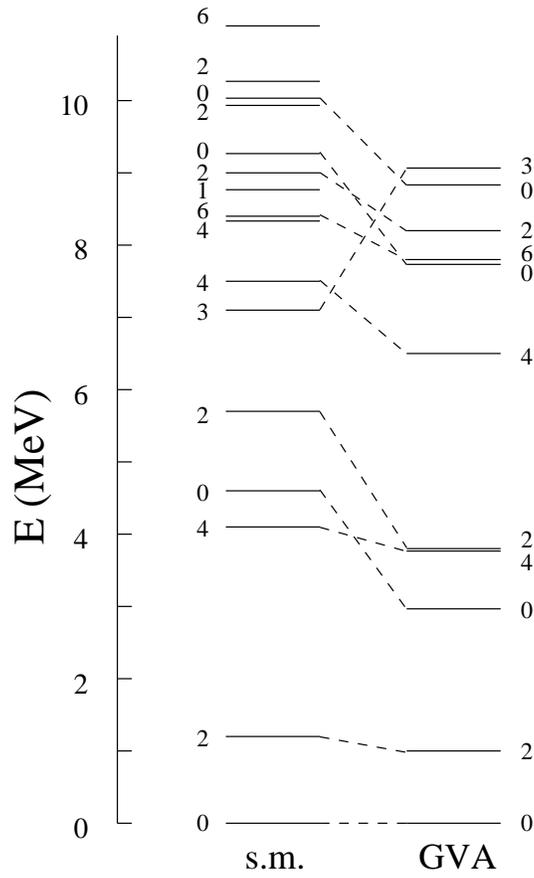}}
\caption{The spectrum calculated from a requantization of the collective Hamiltonian, 
\label{fig:4:spectrum}}
\end{figure}

Though we have found that decoupling is not good everywhere,
we have nevertheless used the Hamiltonian we have constructed,
\begin{equation}
{\cal H}=\half \sum_i \frac{J_i^2}{{\cal I}_i} + \half P^2 + \bar{V}(Q),
\label{eq:4:Hcl}
\end{equation}
to generate spectra and compare to the full shell-model calculation in
the $sd$ model space. To that end we need to requantize the Hamiltonian
(\ref{eq:4:Hcl}). Since we are working in
the intrinsic system we have to be very careful. The necessary symmetry
properties were, in effect, set out a long time ago by Bohr \cite{65}
and are discussed in Appendix C of Ref.~\cite{23}, where details
of the numerical solution of the Schr\"odinger equation for the Hamiltonian
(\ref{eq:4:Hcl}) are also given. The only thing that needs to concern us
here is that we decompose the eigenfunctions as
\begin{equation}
\phi(Q\Omega)_{IMa}
=
\sum_K \phi_{IKa}(Q) \braket{\Omega}{IMK},
\end{equation}
where $\braket{\Omega}{IMK}$ are appropriately symmetrized angular
functions,
the $K$ summation only runs over positive and even values, and $K=0$
is excluded for odd $I$. Using these ideas it is not very
hard to write down a finite difference representation
for the Hamiltonian matrix which in its turn can be
diagonalized in order to obtain eigenvalues and eigenfunctions.
In Fig.~\ref{fig:4:q-eigv}
we give the eigenfunctions
for the states with zero angular
momentum. The ground state can not be very sensitive to the badness of
decoupling, since it is very small for $Q>1$,
but the first and other excited states have sizable values in
the region of bad decoupling. This means that we do not believe that
the corresponding
eigenvalues are very good approximations to the shell model values.
If we look at the spectrum as given in Fig.~\ref{fig:4:spectrum},
we indeed see that
the ground state rotational band is reproduced quite well. The band
built on the oblate minimum (the second $0^+$ state), comes at much
too low an excitation energy. Again the moment of inertia of that band is
more or less correct. The other $0^+$-states also come out at a lower energy
than their shell-model counterparts. Further note the $3^+$ state that is
found to lie at too high an excitation energy. This is probably
partially due to the neglect of odd multipoles in our calculation, which
certainly would lower the energy of such a state, but this question
requires further study.

Let us now study
how quantum corrections to the potential energy alter this picture. We write
\begin{equation}
\bar{V}(Q) = \bar{V}_0(Q) + 
\mbox{$\frac{1}{2}$}\left[\sum_a \omega_a - \trace(A)\right]
 -\left[(H_{11})^{(0)} + (H_{04}+H_{40})^{(0)}\right] -
 \sum_i \frac{\langle J_i^2 \rangle}{2{\cal I}_i} .
\label{eq:4:correction}
\end{equation}
The first two corrections terms are the ones described in the previous
section and do not need further explanation apart from the statement
that we use only the ellipsoidal excitations to calculate these two terms;
the last term, which is a subtraction of the expectation value of the
rotational energy in the local Hartree-Fock state, is just
a part of the total expression for the quantum corrections obtained
from the remaining three quarters of the space
(see, e.g., Ref.~\cite{RS}, Eq.~(8.111)).
 
\begin{figure}
\centerline{\includegraphics[width=7cm]{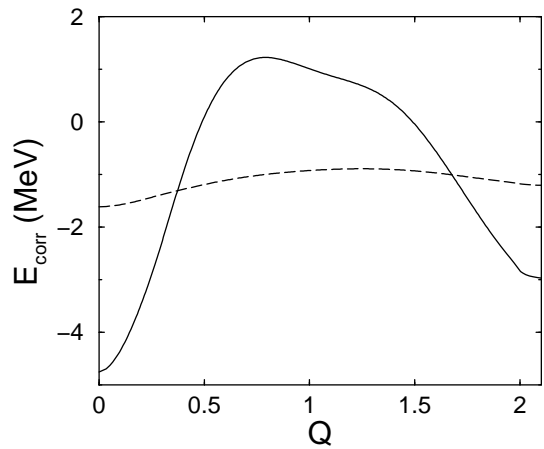}}
\caption{The quantum corrections to the potential. The solid line is the RPA correction, and the
dashed line the rotational energy correction.
\label{fig:4:qcoor}}
\end{figure}

The quantum corrected potential energy is given in Fig.~\ref{fig:4:q-V} as
the dashed line. Clearly we have deeper minima and a higher barrier,
so that the wave functions will become more strongly localized on the
minima.
In Fig.~\ref{fig:4:qcoor} we give each of the
 corrections to the potential energy separately.
The solid line represents the quasiboson result
 $\mbox{$\frac{1}{2}$}(\sum_a \omega_a - \trace({\cal A}))$,
the long-dashed line represent the rotational energy
$- \sum_\i {\langle J_i^2 \rangle}/{2{\cal I}_i}$. We have not shown
the ``overcounting correction''
$-[(H_{11})^{(0)} + (H_{04}+H_{40})^{(0)}]$.
We find that to good approximation this equals zero,
a result that can
be shown to become exact in the perturbation limit.
We do not understand why this relation holds so well for our calculation,
but it may be due to a weak residual interaction.
In any event, in the following we shall
disregard this term because of its small contribution,
and concentrate on the two remaining terms.

\begin{table}
\caption{ A comparison of various approximate calculations of the
$0^+$ energies. $E_\mu$ gives absolute energies, whereas
$E^x_\mu$ represents the excitation energies. The first column gives the
exact shell-model results. The next two column represent the results
of quantization of the one-dimensional collective Hamiltonian; the
first without quantum corrections, and the second with inclusion of
these corrections. The column labeled ``2D'' gives the results of
the approximate two-dimensional calculation discussed in the text.
The last column gives the results obtained from the RPA on either
minimum.\label{tab:4:2}}

\begin{center}
\begin{tabular}{l|r|r|r|r|rl}
&shell model & \multicolumn{2}{c}{1D}& 2D & RPA \\
\hline
& & no zero pt & zero pt\\
\hline
$E_{0^+_1}$ & -149.638 & -140.535 & -146.941
 &-146.904 & -147.556\\
$E_{0^+_2}$ & -145.121 & -137.580 & -141.987
&-142.163 & -142.381\\
$E_{0^+_3}$ & -140.409 & -132.822 & -134.929
&-138.134 & -138.256& (138.058)\\
$E_{0^+_4}$ & -139.634 & -127.841 & -130.858
& -137.347& -136.618& (136.187) \\
\hline
$E^x_{0^+_2}$& 4.517 & 2.954 & 4.953
&4.741& 5.175\\
$E^x_{0^+_3}$& 9.229& 7.712 & 12.012
&8.670& 9.300 & (9.498) \\
$E^x_{0^+_4}$& 10.004& 12.693 & 16.091
&9.557&10.937 & (11.369) \\
\end{tabular}
\end{center}
\end{table}

In Table \ref{tab:4:2} we give the numerical values we have calculated for some
low-lying $0^+$ states. We both give absolute and excitation energies
to show the effects of various approximation schemes in the best possible
way. The first column gives the exact shell model result, as a benchmark
to measure the quality of various approximations.
The second and the third column give the results of a requantization of
the one-dimensional Hamiltonian without (second column) and with
(third column) the inclusion of quantum corrections. Not surprisingly
the absolute value for the energies becomes closer to the true value
if we include quantum corrections, even though we apparently still need
more to attain the correct result. It is gratifying to see
that the excitation energy of the first excited $0^+$-state, the
band head of the first excited band, is much closer to the true value.
 
For the sake of comparison we give the RPA value for the oblate and prolate
minima, which are $V(Q) + \frac{1}{2} \omega_1$, as well as the
excitations
obtained by adding $\omega_1$ (or $\omega_2$) to this value. As can
be seen $\omega_1$ and $\omega_2$ are quite close. This has to do with
the fact that we should really use two collective coordinates.
Stopping short of that approach we have done a poor man's
calculation along the collective path.
To each point we assign
two coordinates $(Q_1,q_2)$ where $Q_1$ is the $Q$ used
previously and $q_2$ is the result of summing the numbers $\Delta Q_2$
defined in Sec.~\ref{sec:4.3.5} in our discussion of the goodness of decoupling.
We then
requantize an approximation to the
Hamiltonian (valid for $0^+$-states alone), that includes the second
collective coordinate in an harmonic approximation only,
\begin{eqnarray}
H&=&\mbox{$\frac{1}{2}$} (P_1^2+P_2^2) +
E_{\rm HF}(Q^1) + \mbox{$\frac{1}{2}$}\omega_2^2(Q^2-q^2(Q^1))^2
\nonumber \\ &&
+\mbox{$\frac{1}{2}$}\sum_{a>2}\omega_a - \mbox{$\frac{1}{2}$} \trace[{\cal
A}(Q^1)]
-\mbox{$\frac{1}{2}$} \sum_j\langle J^2_j\rangle/{\cal I}_j(Q^1).
\end{eqnarray}
The results of this calculation are given in the column labeled
`2D' and can be seen to give a much better overall result for the
excitation spectrum.

\begin{figure}
\centerline{\includegraphics[angle=-90,width=7cm]{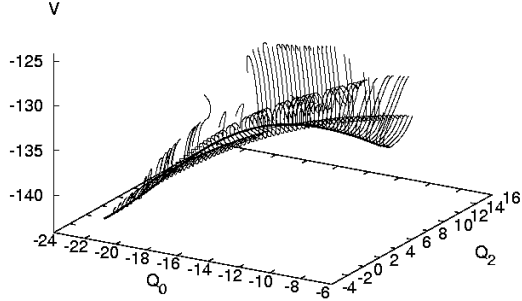}}
\caption{A representation of the two-dimensional potential energy
surface for $^{28}$Si using lines that ``slice'' through the surface.
\label{fig:4:si2D}}
\end{figure}

\begin{figure}
\centerline{\includegraphics[width=7cm]{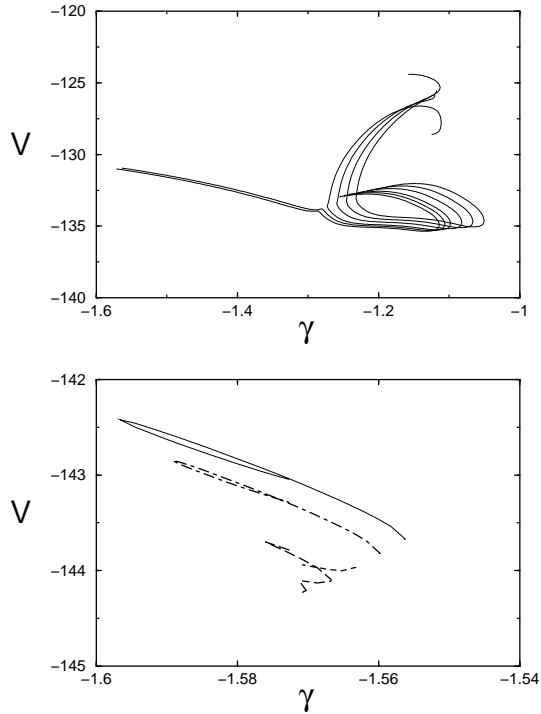}}
\caption{A combination of a few slices through the potential energy
surface. The lower panel is near the HF minimum, the upper panel show
a reconnection phenomenon about midway through the valley.
\label{fig:4:si2D2}}
\end{figure}

A full 2D calculation has proven to be a major problem. The first
reason for this is that it was not obvious how to construct a stable
algorithm. Using the path-following approach that we have applied to
find the valley, it was realized that the best approach we could
imagine was to calculate slices through the two-dimensional surface,
starting from points on the valley. We chose to use the constraint
that $-1/2(Q_0-q_0)=\sqrt{3/8}(Q_2-q_2)$, where $q_0$ and $q_2$ are
the quadrupole coordinates on the valley floor. Unfortunately the
path-following algorithm using local coordinates is not as stable as
the standard form of PITCON, and we encounter some failures, but these can
probably be patched up. The real problem is the structure of the
surface. As can be glanced from Fig.~\ref{fig:4:si2D}, the structure
is quite complex. An even more quantitative feel can be obtained from
Fig.~\ref{fig:4:si2D2}, where we show both the first few slices (lower
panel), and a bifurcation-like change in the character of the surface
(upper panel). We have stopped each line calculation when the density
matrix becomes axial, since then we can continue the potential surface
by reflection around the $z$-axis. Note that this is stronger than the
condition $\gamma=n\pi/3$, since a state can be triaxial with an axial
quadrupole moment. Indeed, in the calculations near the HF minimum we
see, apart from lack of convergence in some cases, that we do find
states that have axial quadrupole moments, but are not axial.
We also note that these slices end up on two different axial
points. This suggests, and this is also seen in a calculation
for axial deformation only, that the bottom of the potential surface
looks somewhat like the bottom of a paraboloid, very different from what we
would expect.

Equally worrisome is the behavior in the middle of the path, where we
find what looks like an avoided crossing between two intersection
surfaces becoming a funnel (i.e., the surfaces touch), after which the
surfaces reconnect in a different way. This is a difficult problem to
deal with computationally.

Clearly the potential energy surface found does not lend itself easily
to a quantization procedure, even if we fix the convergence problems
with the algorithm. This points at the weakest point in this model
study, the fact that effective interactions in small model spaces must
be extremely complicated to describe the full dynamics. Such
interactions should not naturally lead themselves to a collective
approach, and that is a large source of the difficulties.

We have, however, shown that we can construct a one-dimensional path that
gives a reasonable 
description of the lowest states in $^{28}$Si.  
Inclusion of the quantum corrections
in our calculation for $^{28}$Si improves the calculation of both
binding and excitation energies. We have shown that a quasi
two-dimensional calculation improves the results even more. 

\subsection{A basis for self-consistent cranking operators\label{eq:4:Sec.ssc}\label{sec:4.4}}

In the usual Hartree-Fock or Hartree-Fock-Bogoliubov calculations which are
being used to study deformed and superdeformed nuclei as well as fission
\cite{no96,no97,no114}, the multidimensional surface is 
generated by cranking, using some one-body operators to move away from
the local minima. The most commonly employed constraint operators are
just the multipole spherical harmonics. Despite some apparent
successes of such calculations, the procedure is far from being
satisfactory \cite{no115}, both from the theoretical and practical
points of view. One may indeed raise the following questions:

\begin{enumerate}
\item Even though the quadrupole $L=2$ mode seems to play a dominant
role in the deformation of nuclei, there is no reason to limit oneself
to this or other space-dependent cranking operators. Spin-dependent components
 may not be negligible. Furthermore, the radial structure may turn out to 
be much more complicated than the simple $r^L$ form of the electric multipole
operators.

\item In order to cover the space of the collective motion, it is 
most probable that one is required to use a large number of these operators,
 should they not be well chosen. For heavy nuclei, a calculation with many
(three or more) cranking operators is impracticable. One therefore must
 wonder whether it is possible to cover the space of the collective motion 
by choosing a small set of operators that are well-defined linear 
combinations of elementary ones. A positive answer to this question would be 
of significant value for the study of heavy nuclei.
\end{enumerate}

In this section we analyze a few aspects of these questions.
Section \ref{sec:4.4.1} shows what we can learn from the case of the Silicon nucleus
\cite{27}. The more interesting
study of the Baranger-Kumar pairing-plus-quadrupole model is described in
Sec.~\ref{sec:4.4.2} \cite{31a,31b}. The account of a model study that follows in 
Sec.~\ref{sec:4.5} also bears on this subject.

\subsubsection{Self-consistent cranking operators for the $^{28}Si$ nucleus\label{sec:4.4.1}}

The Kuo interaction used for the study of this nucleus is very complex.
Aside from the space structure, it also depends on spin and isospin.
It is therefore reasonable to expect that the cranking operator reflects 
these degrees of freedom as well. As a first study, let us neglect the
isospin, which is reasonable for this light $N=Z$ nucleus. We shall choose 
the following two sets of one-body operators: the usual set of
spin-independent multipole operators
\begin{equation}
\hat{o}^{S=0}_{J\Sigma} = F_J(r) Y_{J\Sigma}(\hat{r})
\end{equation}
and the spin-dependent operators
\begin{equation}
\hat{o}^{S=1}_{LJ\Sigma}=F_{LJ}(r) [ S\times Y_L ]_{J\Sigma}
\end{equation}
where $S$ is the spin operator and $F(r)$ represents a radial dependence.

    To see whether the cranking operator, the chosen collective solution 
of the RPA equation (\ref{eq:4:LHA2}) can be expressed in terms of the above 
sets of operators, we shall calculate it in two ways:

\begin{enumerate}
\item We first solve the original p-h RPA equation to get a set of p-h matrix 
elements $f_{ph}$. These are properly normalized.

\item Next, we solve the projected RPA equation according to the method 
given in Sec.~\ref{sec:4.1.6} over the above sets of operators. 
This gives a new set of p-h matrix elements
$\bar{f}_{ph}$. 
\end{enumerate}
The criteria for good projection is the smallness of the 
following quantity
\begin{eqnarray}
\delta &=& \sum_{ph,p'h'} f_{ph} B^{ph,p'h'} ( f_{p'h'} - \bar{f}_{p'h'} )
\nonumber \\
&=& 1 -f_{ph}B^{php'h'}\bar{f}_{p'h'},\label{eq:4:delta}
\end{eqnarray}
where we have used the normalization imposed in Eq.~(\ref{eq:4:cnorm}).
In view of the chosen normalization, $ 0 < \delta < 1 $ with 
$\delta = 0$ corresponding to exact projection and $\delta=1$ to the case where
the two solutions are orthogonal.

 Let us limit our study to just one point on the collective path, 
namely at the HF minimum: It is at this point that the number of operators
required is the smallest, while at the same time it is sufficient for the stated purpose.

\begin{table}
\caption{ The energy of the collective state and the value of the 
representation measure $\delta$ when more and more elementary operators
are added.\label{tab:4:3}}

\begin{center} 
\begin{tabular}{c|c|c|r|l} 
$J$ & $S$ & $L$ & $\Omega$ & $\delta$ \\ 
\hline
2 & 0 & 2 & 10.531 & 0.7383 \\
4 & 0 & 4 & 10.592 & 0.6852 \\
2 & 1 & 2 & 6.860 & 0.1572 \\
3 & 1 & 2 & 6.631 & 0.1226 \\
3 & 1 & 4 & 6.631 & 0.1226 \\
4 & 1 & 4 & 6.034 & 0.0000 \\
\end{tabular}
\end{center}
\end{table}

    First of all, we assume that the model space consists of just the
$sd$-shell, as given in Sec.~\ref{sec:4.3} above. In this case, the radial dependence is
irrelevant, as explained below. Table \ref{tab:4:3} gives the results for the 
quantity $\delta$ defined above together with the energy $\Omega$ of the 
collective state obtained when more and more operators are added to the
operator basis (in going from one line to the next in the table).
It is seen that the spin-independent operators alone are far from being
sufficient to represent the RPA cranking operator. Even after having 
exhausted all the available values of $J$ compatible with the model space,
 $\delta$ is still found to have a large value of $0.6852$ and $\Omega =
10.59$ MeV is still much larger than the value obtained by direct 
diagonalization of the RPA equation which is $\Omega = 6.034 $.
It turns out that the spin-dependent part ( $ S=1 $ ) is very important,
much more important than the spin-independent one. As a matter of fact,
by including all the given operators, one can show that the basis is complete 
and thus gives an exact ($\delta=0$) representation of the self-consistent 
cranking operator. The possibility of an exact representation of $f$ in this
case also means that there is no need to consider a
more complicated radial form
factor $F(r)$. This is so because the average value of $r^k$ is the same for 
any state of the model space and changing $k$ just adds an overall
factor which can be absorbed in the normalization.

\begin{table}
\caption{ The smallest value $\delta_{min}$ obtained when the model
space is enlarged sequentially for three choices of the radial form
factor $F(r)$.\label{tab:4:4}}
\begin{center}
\begin{tabular}{l|l|l|l}
Model & $F(r)=r^2$ & $F(r)= $ & $\rho(r) r^2$ \\ 
space & & $r^2 + r^4$ & \\
\hline
+ $2s_{1/2}$ & 0.00 & 0.00 & 0.00 \\
+ $1d_{3/2}$ & 0.0345 & 0.00 & 0.00 \\
+ $1d_{5/2}$ & 0.1986 & 0.0011 & 0.1375\\
+ $0g_{7/2}$ & 0.2473 & 0.0932 & 0.1206 \\
+ $0g_{9/2}$ & 0.5146 & 0.4132 & 0.3644 \\
\end{tabular}
\end{center}
\end{table}

To see whether more complicated form factors may be necessary, we shall
enlarge the model space to include, in addition to the $sd$-shell, the next
positive parity $sdg$-shell. By adding the different subshells one by one 
into the model space and trying to get the best representation, it is found 
that a single r-dependence of the form $r^k$ is no longer sufficient. Table \ref{tab:4:4}
shows the best ( the smallest ) $\delta$ , denoted $\delta_{min}$, obtained
when the model space is enlarged successively: $\delta_{min}$ increases from
zero to $0.5146$ when all the $sdg$-shell is included. Recalling that 
$\delta_{min}$ corresponds to including {\em all} the available angular momenta
and spins so that, to improve the result, the only remaining possibility 
is an additional radial dependence. As a matter of fact, adding {\em another}
radial dependence of the form
\begin{equation}
F(r) = r^k G(r)
\end{equation}
leads to much better results, as shown in the third column of Table
\ref{tab:4:4}. These results have been obtained with $k=2$ and $
G(r)=r^2$, that is with {\em two} radial form factors $r^2$ and
$r^4$. It looks as if an additional $r$ dependence is needed every time
one adds a new shell, with the consequence that the number of
elementary operators is doubled each time. Of course, this is
undesirable. One would like to limit as much as possible the number of
elementary operators for computational reasons. It has been found that
the radial form factor
\begin{equation}
F(r) = \rho(r) r^2
\end{equation}
where $\rho(r)$ is the average density, allows a better fit of the 
cranking operator than $r^2$ alone and gives results comparable to those
found by using two independent radial form factors (see the last column 
of Table \ref{tab:4:4}). The attractive feature of the above form is that it allows an 
automatic cutoff of the unbounded nature of $r^2$ and its wild behavior 
when a large number of harmonic oscillator states is used in the
calculation.

    In conclusion, the theory allows the determination of a self-consistent
cranking operator that is expressible in terms of linear combination of
elementary one-body operators. For the study of heavy nuclei for which the
construction of the complete particle-hole RPA matrix and its diagonalization
may become prohibitive, the knowledge of a limited set of basis operators
provides a practical way to solve the problem. The result constitutes a 
considerable improvement over the current practice based on fixed cranking 
operators.

\subsubsection{Cranking operators in the Pairing-Plus-Quadrupole Model\label{sec:4.4.2}} 

    The P+Q model is probably one of the most simple and successful
nuclear Hamiltonians that allows us to discuss realistic problems
involving pairing and quadrupole degrees of freedom. Baranger and
Kumar analyzed in great detail the (adiabatic) collective motion in
the P+Q model assuming that the collective variables are the mass
quadrupole operators \cite{1a,1b,1c,1d,1e}. Thus, they reduced the large number
of two-quasiparticle (2qp) degrees of freedom (of the order of
thousands) into only two collective coordinates, $\beta$ and
$\gamma$. However, a previous study of the $O(4)$ model \cite{30a}
suggests that even for this simple Hamiltonian, the self-consistent
collective coordinate is not as trivial as it seems to be. 
We show below that the
normal-mode coordinate of the random-phase approximation (RPA) is
quite different from the mass quadrupole operator. This is
especially true when the system is deformed.

    For simplicity, let us assume that we are at the HFB minimum. At this
point, the local RPA is equivalent to the quasiparticle RPA with the
constrained Hamiltonian
\begin{eqnarray}
H' &=& H - \sum_{\tau=n,p} \lambda_\tau N_\tau ,\\
H &=& \sum_k \epsilon_k c_k^\dagger c_k
 -\sum_{\tau=n,p} \frac{G_\tau}{2}
 \left( P_\tau^\dagger P_\tau + P_\tau P_\tau^\dagger \right)
 -\frac{\chi}{2} \sum_{K=-2}^2 Q_{2K}^\dagger Q_{2K} \nonumber\\
 &=& \sum_k \epsilon_k c_k^\dagger c_k
 -\frac{1}{2} \sum_\sigma \kappa_\sigma R_\sigma R_\sigma
 +\frac{1}{2} \sum_\sigma \kappa_\sigma S_\sigma S_\sigma ,
\end{eqnarray}
where $\epsilon_k$ are spherical single-particle energies and
$N_\tau = \sum_{k\in\tau} c_k^\dagger c_k$ are the number operators
for neutrons ($\tau=n$) and protons ($\tau=p$).
The operators
$R_\sigma$ and $S_\sigma$ are the hermitian and anti-hermitian
components, respectively, of the pairing operators,
$P_\tau^\dagger=\sum_{k\in\tau, k>0} c_k^\dagger c_{\bar k}^\dagger$,
and the dimensionless quadrupole operators,
$Q_{2K} = b_0^{-2} \sum_{kl} \bra{k} r^2 Y_{2K} \ket{l} c_k^\dagger c_l$,
where $b_0=(\hbar/m\omega_0)^{1/2}$ is the harmonic oscillator length.
The Hamiltonian contains five operators of $R_\sigma$-type and four of
$S_\sigma$-type.
Together with the corresponding coupling constants $\kappa_\sigma$,
\begin{equation}
\label{eq:4:PQ_table}
\begin{array}{ccccccc}
R_\sigma &=& (P_+^{(+)})_n, & (P_+^{(+)})_p,
 &Q^{(+)}_{20}, &Q^{(-)}_{21}, &Q^{(+)}_{22},\\
S_\sigma &=& (P_-^{(+)})_n, & (P_-^{(+)})_p,& &Q^{(+)}_{21}, &Q^{(-)}_{22},\\
\kappa_\sigma &=& G_n, & G_p, &\chi, &\chi, &\chi,
\end{array}
\end{equation}
where
\begin{eqnarray}
(P_\pm^{(+)})_\tau &=& \frac{1}{\sqrt{2}} ( P_\tau \pm P_\tau^\dagger ),
 \quad \mbox{ for } \tau=n,p, \nonumber \\
\label{eq:4:sig_good_Q}
Q^{(\pm)}_{2K} &=& \frac{1}{\sqrt{2}} ( Q_{2K} \pm Q_{2-K} ) , \quad
\mbox{ for } K=0,1,2.
\end{eqnarray}
The $(\pm)$ superscripts indicate the signature quantum number,
$e^{-i\pi J_x} o^{(\pm)} e^{i\pi J_x} = \pm o^{(\pm)}$.
Following the standard formulation of the model,
we shall neglect the Fock terms,
the contributions of the pairing force to the Hartree potential
and those of the quadrupole force to the pairing potential.

 Utilizing the procedure outlined in Sec.~\ref{sec:4.1}, and illustrated
in Sec.~\ref{sec:4.4.1}, one arrives at the
classical Hamiltonian\begin{equation}
\Ham \equiv \bra{\Psi} H \ket{\Psi} \approx
 E_0 + \frac{1}{2} B^{\alpha\beta} \pi_\alpha \pi_\beta
 + \frac{1}{2} V_{\alpha\beta} \xi^\alpha \xi^\beta \ ,
\label{eq:4:H_HA}
\end{equation}
in terms of the canonical variables $(\xi,\pi)$.
Here, each of the indices ($\alpha,\beta,\cdots$) indicates a pair of
2qp indices ($ij,kl,\cdots$).
The mass and curvature parameters are explicitly given by
\begin{subequations}
\begin{eqnarray}
B^{\alpha\beta} &=&
 E_\alpha \delta_{\alpha\beta}
 + 2\sum_\rho \chi_\rho S^{(\rho)}_\alpha {S^{(\rho)}_\beta} \ ,\\
V_{\alpha\beta} &=&
 E_\alpha \delta_{\alpha\beta}
 - 2\sum_\rho \chi_\rho R^{(\rho)}_\alpha {R^{(\rho)}_\beta} \ ,
\end{eqnarray}
\end{subequations}
where $E_\alpha$ are the Hartree-Bogoliubov quasi-particle energies.
Following Ref.~\cite{1a,1b,1c,1d,1e}, we multiply the quadrupole operators by a factor
 $\alpha_\tau^2$
with $\alpha_n=(2N/A)^{1/3}$ and $\alpha_p=(2Z/A)^{1/3}$,
and also reduce
the quadrupole matrix elements between the states of the upper
shell by a factor
$\zeta = ({\cal N}_{\rm L}+\frac{3}{2})/({\cal N}+\frac{3}{2})$,
where ${\cal N}$ is the oscillator quantum number operator and
${\cal N}_{\rm L}$ is the number of quanta in the lower shell.
Thus, the modified quadrupole operators are defined as
$Q_{2K} \equiv (Q_{2K})_n + (Q_{2K})_p$, with
$(Q_{2K})_n=\alpha_n^2\zeta (r^2 Y_{2K})_n$
and $(Q_{2K})_p=\alpha_p^2\zeta (r^2 Y_{2K})_p$
(which we shall refer to as ``the quadrupole operators'').

Recalling that we are studying only the HFB minimum, the solution of t
he RPA equation
\begin{equation}
V_{\alpha\gamma}B^{\gamma\beta} f^\mu_{,\beta}
 = (\Omega^\mu)^2 f^\mu_{,\alpha} \ ,
\label{eq:4:full_RPA_eq}
\end{equation}
which involves the diagonalization of the RPA matrix 
$V_{\alpha\gamma}B^{\gamma\beta}$
whose dimension is equal to the number of active 2qp degrees of
freedom.
For separable forces, this can be simplified by solving a dispersion relation,
which facilitates the numerical calculations for heavy nuclei.
In general, however, the RPA diagonalization requires extensive computational
resources.
Now let us approximate an eigenvector using a selected 
set of one-body operators
$\{ \hat{o}^{(i)} \}$:
\begin{equation}
{f}_{,\alpha} = \sum_i c_i o^{(i)}_\alpha \ ,
\end{equation}
where $o^{(i)}_\alpha$ indicate the 2qp matrix elements of operator $\hat{o}^{(i)}$. 
Then, instead of the full RPA equation (\ref{eq:4:full_RPA_eq}),
we obtain a projected RPA equation
\begin{equation}
{M}^{ij} C^n_j = \left( \bar{\Omega}^n \right)^2
 \bar{O}^{ij} C^n_j \ ,
\label{eq:4:projected_RPA_eq}
\end{equation}
see Eqs.~(\ref{eq:4:RPAapprox}-\ref{eq:4:Oij}).
The dimension of  ${M}^{ij}$ and
${O}^{ij}$ is equal to the number of selected one-body
operators $\{\hat{o}^{(i)}\}$. Therefore, if we can approximate the RPA
eigenvectors by using a small number of operators, it will
significantly reduce the computational task.

A criterion for good projection may be given by the closeness of the
projected RPA frequencies $\bar{\Omega}$ to the real
RPA frequencies $\Omega$.
Another criterion is the smallness of the quantity $\delta$, defined
as in Eq.~(\ref{eq:4:delta}).

The theory has been applied to several heavy isotopes. Here we
report the numerical results for even-even Sm isotopes
(A=146$\sim$154). The form of the P+Q model is that discussed in the
second and third of the series of papers by Baranger and Kumar \cite{1c,1d}.
The model space and the parameters, such as the spherical
single-particle energies, the pairing and quadrupole force strengths,
are taken from Table 1 in the third paper. The equilibrium
parameters ($\beta$, $\gamma$, $\Delta$, $\lambda$) are found to agree
with Table 2 of the same paper. The ground states of $^{146,148}$Sm
are spherical ($\beta=0$) and the others have prolate shapes
($\beta>0, \gamma=0$).

\begin{figure}
\centerline{\includegraphics[width=10cm]{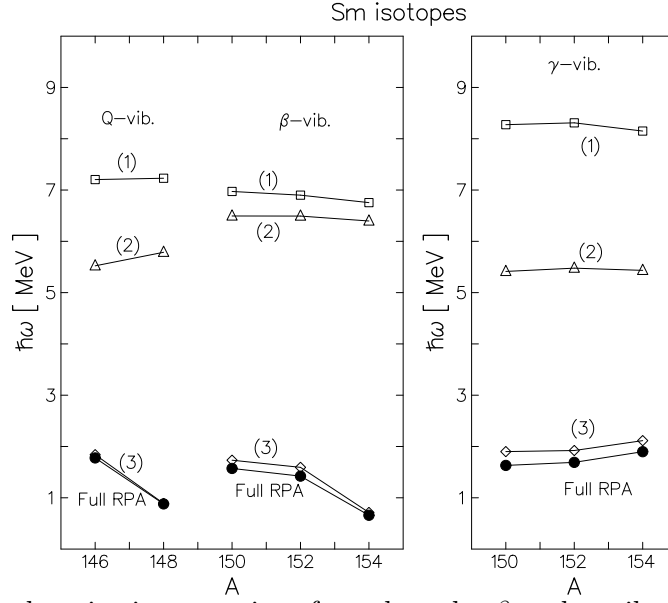}}
\caption{Calculated excitation energies of quadrupole, $\beta$ and
$\gamma$ vibrations for even-even Sm isotopes.
Note that the ground states of $^{146,148}$Sm are spherical.
The closed circles indicate the RPA results while the open symbols 
are the results of projected RPA calculations.
See the main text for the difference between (1), (2) and (3).
}
\label{fig:4:RPA-PRPA}
\end{figure}

Figure \ref{fig:4:RPA-PRPA} 
shows the excitation energies (RPA
frequencies) of $\beta$ and $\gamma$ vibrations, obtained by the RPA
and projected RPA calculations. For the projected RPA calculations,
we have adopted three different sets of one-body operators. The first
and simplest choice is to use the operators appearing in the separable
forces, the pairing and quadrupole operators, $P_\tau$,
$P_\tau^\dagger$, $(Q_{2K})_\tau$ ($\tau=n,p$). This choice is denoted
as (1) in the figure. In this case the projected RPA matrices of
Eq.~(\ref{eq:4:RPAapprox}) are two dimensional for spherical
nuclei and for $\gamma$ vibrations, while they are six dimensional for
the $\beta$ vibrations. The calculated frequencies are $7\sim 8$ MeV
which are $5\sim 6$ MeV larger than the corresponding RPA
frequencies. In the second set, labeled as (2), we increase the
number of operators. We keep both the pairing and quadrupole operators,
but include two additional quadrupole operators with ``monopole'' radial
dependence, $(r^0Y_{2K})_\tau$. We also include the
hexadecapole operators, $(r^4 Y_{4K})_\tau$, and the rank-2
spin-dependent operators, $([r^0Y_2 \times \vec{s} ]^{(2)}_K)_\tau$,
$([r^2Y_2 \times \vec{s} ]^{(2)}_K)_\tau$. As far as the frequencies
are concerned, we can see some improvement over the case (1) for
spherical and the $\gamma$ vibrations in deformed nuclei, though they
are still much higher than the real RPA frequencies. For the $\beta$
vibrations, the inclusion of the additional rank-2 (and higher rank)
operators seems not so important. Actually we see that the $\beta$
vibrations are found to have a significant amount of monopole
components. For the last set, denoted as (3), we adopt the same operators
as (1) but each 2qp matrix element is weighted with a factor $(E_{\rm
2qp})^{-2}$. This means that we employ a set of {\it state-dependent}
one-body operators $\{ \widetilde{o}^{(i)}\}$ defined by
\begin{equation}
\widetilde{o} \equiv \sum_\alpha
 \frac{o_\alpha}{(E_\alpha)^2} (a^\dagger a^\dagger)_\alpha
 + \mbox{h.c.} \ .
\label{eq:4:scale}
\end{equation}
The result of this projection is now almost identical to that of the full RPA.

\begin{table}
\caption{Calculated values of $\delta$, Eq.~(\ref{eq:4:delta}),
of the projected RPA solutions for Sm isotopes.
The columns (1), (2), (3), (1-a) and (1-b)
represent the different projections (see text).
For the spherical nuclei ($^{146,148}$Sm),
there is no distinction between $\beta$ and $\gamma$ vibrations.
\label{tab:4:5}}

\begin{tabular}{c|ccccc|ccccc}
 & \multicolumn{5}{c}{$\beta$ vibration} &
 \multicolumn{5}{c}{$\gamma$ vibration} \\
A & (1) & (1-a) & (1-b) & (2) & (3) & (1) & (1-a) & (1-b) & (2) & (3)\\ \hline
146 & 0.271 & 0.132 & 0.225 & 0.421 & 0.009 & & & & &\\
148 & 0.243 & 0.131 & 0.184 & 0.314 & $3\times 10^{-4}$ & & & & &\\
150 & 0.602 & 0.499 & 0.519 & 0.632 & 0.026
 & 0.610 & 0.342 & 0.507 & 0.685 & 0.092\\
152 & 0.497 & 0.346 & 0.433 & 0.526 & 0.020
 & 0.616 & 0.279 & 0.472 & 0.691 & 0.081\\
154 & 0.513 & 0.117 & 0.437 & 0.534 & 0.002
 & 0.636 & 0.208 & 0.426 & 0.679 & 0.052\\ 
\end{tabular}
\end{table}

In Table \ref{tab:4:5}, the quality of projection $\delta$,
Eq.~(\ref{eq:4:delta}), is listed. In the cases (1) and (2), where the RPA
vectors are projected on the elementary operators, $\delta \gtrsim
0.25$ for $^{146,148}$Sm and $\delta \gtrsim 0.5$ for the others.
Therefore, roughly speaking, selected one-body operators possess at
most 75\% (50\%) of overlap with the real eigenvectors in spherical
(deformed) nuclei. On the other hand, the projection (3) exhausts
more than 90\% of real eigenvectors even for the worst case. At first sight
it may
look strange that $\delta$ is larger for (2) than for (1), while the energy
for (2) is lower. This is due to the fact that case (2)
is dominated by certain neutron components. Since the relevant
neutron 2qp energies are lower than those of protons, this
proton-neutron asymmetry leads to a decrease in the frequency
$\bar{\Omega}$ and at the same time an increase in $\delta$. This is
also a reflection of the poor quality of the approximation.

Figure  \ref{fig:4:RPA-PRPA}  and Table \ref{tab:4:5} indicate
that it is very difficult to obtain sensible results
by using elementary one-body operators (i.e., of the form (1) or (2)).
This is mainly due to the fact that the RPA eigenvectors, when being projected
onto elementary one-body operators, have
unrealistically large amplitudes for high-lying 2qp components.
In order to demonstrate this, we introduce a cut-off energy $\Lambda_{\rm cut}$
for the 2qp matrix elements,
i.e., $o^{(i)}_\alpha = 0$ for $E_\alpha > \Lambda_{\rm cut}$.
We then perform the projected RPA calculation
with the truncated one-body operators from set (1).
The resulting values $\delta$ are listed in Table \ref{tab:4:5} 
for $\Lambda_{\rm cut}=5$ MeV (1-a) and for 10 MeV (1-b).
We see that the major contributions to the RPA modes come
from the 2qp components
with $E_{\rm 2qp} < 5$ MeV. We thus conclude that
the superiority of the projection (3) simply comes from its being capable of
suppressing
the unnecessary high-energy components by the factor $(E_{\rm 2qp})^{-2}$.

This suppression factor is not arbitrary, but can be derived
from the following simple argument.
If we have a single-mode separable force $H=-(1/2)\chi R R$
(assuming no coupling among different modes),
we can determine the RPA eigenvectors analytically,
$f^\mu_{,\alpha} \propto R_\alpha/((E_\alpha)^2-(\Omega^\mu)^2)$.
In the limit that $\Omega^\mu \ll E_{\rm 2qp}$,
the projection on $f_{,\alpha} \propto R_\alpha/(E_\alpha)^2$
gives the exact answer.
It is worth noting that taking $R=Q_{22}$,
we find exactly the formula for $\gamma$
vibrations as discussed in the first paper of Baranger and Kumar \cite{1a}.

Let us examine the projection (3) in more detail. For spherical Sm
nuclei, the RPA eigenvector has a good isoscalar character and can be
approximated as $\bar{f} \approx (\widetilde{Q}_2)_n +
(\widetilde{Q}_2)_p$ where the tilde indicates that the matrix elements
include the suppression factor as in Eq.~(\ref{eq:4:scale}).
For deformed nuclei, where the collectivity of the
vibrational states is smaller than for spherical nuclei and the pairing
modes can mix with the quadrupole ones, the situation is more complex.
Taking $^{154}$Sm as an example, the
eigenvectors of $\beta$ and $\gamma$ vibrations are
\begin{subequations}
\begin{eqnarray}
\bar{f}^{\beta{\rm-vib}}&=& (\widetilde{Q}_{20})_n
 + 0.91 (\widetilde{Q}_{20})_p
 - 0.48 \widetilde{P}_n - 0.44 \widetilde{P}_p
 + 0.085 \widetilde{P}_n^\dagger - 0.14 \widetilde{P}_p^\dagger\ ,\\
\bar{f}^{\gamma{\rm-vib}}&=& (\widetilde{Q}_{22})_n
 + 0.87 (\widetilde{Q}_{22})_p \ .
\end{eqnarray}
\end{subequations}
For the $\beta$ vibration, we find a significant mixing with the
monopole (pairing) modes.

In conclusion, we have examined the possibility of expressing
the self-consistent
cranking operator in terms of a limited set of one-body operators.
It seems very difficult to approximate the normal-mode vectors
with use of elementary one-body operators. This difficulty disappears, however,
when we use a small number of {\it state-dependent} one-body operators.
This may reflect the importance of the self-consistent determination of
the collective coordinates for large amplitude collective motion,
because the coordinates now have a strong state-dependence as well.
The structure of the self-consistent cranking operators is clearly
changing when we move from spherical to deformed nuclei.
For the study of large amplitude
collective motion in heavy nuclei for which
the diagonalization of the RPA matrix becomes too time-consuming,
the results obtained here may give a hint for a correct
choice of a state-dependent basis of operators.
The choice of a limited set of (state dependent) basis operators
provides a practical way to solve the LHA through the projection.
With the self-consistent cranking operators,
the LHA should provide a significant improvement over the conventional
CHFB calculation based on fixed cranking operators.

\subsection{A model study of shape transitions and shape coexistence\label{sec:4.5}}

\subsubsection{Introduction\label{sec:4.5.1}}

Clearly one of the most important aspects of heavy nuclear systems is
the role of the pairing interaction, which needs to be tackled within
the framework of our theory.  In a first application \cite{30}, that
we shall not reproduce here, we applied the LHA to analyze properties
of collective motion in a semi-microscopic model of nucleons
interacting through a pairing force, coupled to a single harmonic
variable giving a macroscopic description of the remaining degrees of
freedom. It turned out that with this method the system automatically
selects either diabatic (shape coexistence) or adiabatic collective
surfaces according to the strength of the pairing
interaction. However, we felt that it would be beneficial to study a
fully microscopic Hamiltonian.

To this end we investigate the collective motion in a model which
describes a system of nucleons interacting through a simplified
version of the pairing-plus-quadruple force \cite{PSS80}.  Although
the Hamiltonian has a very simple form, we shall see that the model
can reproduce the qualitative features of many kinds of interesting
situations observed in real nuclei, such as a spherical-to-deformed
transition, and nuclei with shape coexistence, where more than one
equilibrium shape play a role.

In the case of a single-$j$ shell the model Hamiltonian is built from
the generators of an $o(4)$ algebra, which makes exact diagonalization
feasible. The model has been originally developed to describe
$K^\pi=0^+$ excitations in deformed nuclei \cite{PSS80}, and has also
been used as a test-bed for various methods applied to the calculation of
collective excitations such as the boson expansion method
\cite{Mat82a,Mat82b}, the selfconsistent collective coordinate method
\cite{Mat86}, and a semiclassical method \cite{SM88a,SM88b}. The model can
be generalized to multiple shells, where it has been used to investigate
shape-coexistence phenomena \cite{FMM91}. Finally, a similar model has
been used to study the collective mass parameter in finite
superconducting systems \cite{AB88}.

Although the low-lying spectra in nuclei are mostly dominated by the
quadrupole phonon ($J^\pi=2^+$) excitations, the anharmonicity 
of this mode is very
important for many nuclei, especially in a shape-transition region,
where the nature of the ground-state changes rapidly with particle
number. For instance, the even-even Sm isotopes show a typical
example of the spherical to deformed shape transition in which the
spectrum shows a strong anharmonicity between the spherical ($N\leq
84$) and deformed ($N\geq 90$) nuclei, especially for $^{148,150}$Sm.
These phenomena are primarily related to the competition between the
monopole and quadrupole interactions among the valence particles
outside a closed core. The pairing-plus-quadrupole model
was designed to describe this
competition and is well able to reproduce the most important
aspects of the experimental data (see \cite{PPQQ} and references
therein). Later the boson expansion method has been applied to the
same model (with an additional quadrupole pairing interaction) for the
description of the shape transition in the Sm isotopes \cite{KT72a,KT72b},
and shows excellent agreement with the experimental data. Since the
$O(4)$ model is very similar to the pairing-plus-quadrupole model, it
would be of significant interest to see whether our method of large
amplitude collective motion can properly describe the shape
transition phenomena in this exactly solvable model.

The importance of shape-coexistence in nuclear physics can be seen
from the multitude of theoretical approaches and the amount of
experimental data as compiled in the review paper Ref.~\cite{Woo92}.
An important example can be found in even semi-magic Sn and Pb
isotopes, where the ground states are spherical. However, deformed
excited $J^\pi = 0^+$ states have been observed at low-excitation
energies in many of these nuclei. These excited states are regarded
as states associated with proton two-particle-two-hole (2p-2h)
excitations across the closed shell. Using the Nilsson picture, which
shows down-sloping single-particle levels above the proton closed
shell, and up-sloping levels below it, it is possible to assign a
configuration of two particles lying on down-sloping levels and two
holes on up-sloping levels. The configuration-constrained
Nilsson-Strutinsky calculations as performed by Bengtsson and
Nazarewicz \cite{BN89} have suggested that the diabatic
potential-energy curve obtained by switching off the interaction
between the 2p-2h and the ground-state (0p-0h) configuration gives a
more accurate picture than the conventional adiabatic potential
energy. This question, whether the nuclear collective potential is
adiabatic or diabatic, is quite old, and was originally raised by Hill
and Wheeler \cite{HW53}.
It is our aim to investigate in the $O(4)$ model
whether the method is able to provide us with useful information about
shape mixing, and to test whether it makes useful predictions 
concerning whether
the collective potential energy is diabatic or adiabatic.

\subsubsection{Formalism: removal of spurious modes\label{sec:4.5.2}
\label{sec:formalism}}

We have discussed most of the important aspects of our approach in
previous sections. The only point of principle that requires special
discussion is the treatment of spurious modes.  A typical example is
given by the Nambu-Goldstone (NG) mode associated with the violation
of particle-number conservation. The general problem was first
discussed in Sec.~\ref{sec:2.2.5}, where it was pointed out that NG
modes can have their origin either in the absence of restoring forces,
or (less frequently considered) the occurrence of infinite
eigenmasses. In the application to the rotational motion in $^{28}$Si,
we encountered both cases, and found no basic difficulty dealing with
them because they occurred in different symmetry spaces of the local
RPA. In the present example, we shall see that the NG mode associated
with particle-number violation is of the infinite eigenmass
type. Therefore, we first describe a general technique for handling
such a case.

For the models to be discussed in the following sections,
the modes associated with a change in average particle number
are given by a linear combination of coordinates:
\begin{equation}
\tilde{f}^{ng}(\xi) = \sum_{\alpha=1}^n c_\alpha \xi^\alpha \ ,
\end{equation}
where $c_\alpha$ is a constant. The problem is that this
mode leads to a zero eigenvalue of the mass, \begin{equation}
B^{\alpha\beta} \tilde{f}^{ng}_{,\beta}
 = B^{\alpha\beta} c_\beta = 0 \ .
\end{equation}
This means that we cannot invert the mass matrix. The only sensible
way to deal with this is to remove these degrees of freedom from our
space, by defining a new set of coordinates,
$\tilde{\xi}^{\mu} = \tilde{f}^\mu(\xi)$.
These are required to satisfy
\begin{eqnarray}
\label{eq:4:new_co_1}
B^{\alpha\beta} \tilde{f}^\mu_{,\beta} \neq 0 \ ,\quad&\mbox{ for }&
\forall\alpha\mbox{ and }\mu=1,\cdots,n-M \ ,\\
\label{eq:4:new_co_2}
B^{\alpha\beta} \tilde{f}^\mu_{,\beta} = 0 \ ,\quad&\mbox{ for }&
\forall\alpha\mbox{ and }\mu=n-M+1,\cdots,n \ ,
\end{eqnarray}
where we assume that there are $M$ Nambu-Goldstone modes ($\mu > n-M$).
Then, we may formulate the LHA
in the space of $n-M$ dimension, $\{\tilde{\xi}^\mu\}_{\mu=1,\cdots,n-M}$,
in which $\det(B^{\mu\nu}) \neq 0$.
\begin{eqnarray}
{\cal M}^\nu_\mu f^i_{,\nu} = (\omega^i)^2 f^i_{,\mu} \ ,\\
{\cal M}^\nu_\mu \equiv \tilde{V}^{,\nu}_{;\mu}
 = \tilde{B}^{\nu\nu'}\tilde{V}_{;\nu'\mu} \ ,
\end{eqnarray}
where indices $\mu,\nu,\cdots$ run only from 1 to $n-M$.
Our aim is to provide a feasible method to calculate this LHA,
namely to calculate the mass parameter $\tilde{B}^{\nu\nu'}$,
potential $\tilde{V}(\tilde{\xi})$, and their derivatives.

The second equation (\ref{eq:4:new_co_2}) determines
tangent vectors of the NG modes.
The rest of coordinates $\tilde{f}^\mu$ for $\mu=1,\cdots,n-M$
are arbitrary as long as
their derivatives are linearly independent of the others.
The full Jacobian matrix $\tilde{f}^\mu_{,\alpha}$
allows us to define the derivatives of the inverse transformation,
$\tilde{g}^\alpha_{,\mu}$ as the inverse of $\tilde f$,
\begin{equation}
\tilde{f}^\mu_{,\beta} \tilde{g}^\beta_{,\nu} = \delta^\mu_\nu
\ ,\quad
\tilde{f}^\mu_{,\beta} \tilde{g}^\alpha_{,\mu} = \delta^\alpha_\beta \ .
\end{equation}
Since all $\tilde{f}^\mu_{,\alpha}$ are constant
(independent of coordinates),
all $\tilde{g}^\alpha_{,\mu}$ are also constant and
the derivatives $\tilde{f}^\mu_{,\alpha\beta}$
(or $\tilde{g}^\mu_{,\alpha\beta}$) 
all vanish.
This implies that within the NG subspace the connection vanishes,
$\Gamma = 0$,
and the geometric character of the transformation of any tensor 
is fully determined in the subspace that does not contain the NG modes.
One can use this to calculate the new mass parameter and its derivatives as
\begin{subequations}
\begin{eqnarray}
\tilde{B}^{\mu\nu} &=&
 \tilde{f}^\mu_{,\alpha} B^{\alpha\beta} \tilde{f}^\nu_{,\beta} \ ,\\
\tilde{B}^{\mu\nu}_{,\lambda} &=&
 \tilde{f}^\mu_{,\alpha} B^{\alpha\beta}_{,\gamma}
 \tilde{f}^\nu_{,\beta} \tilde{g}^\gamma_{,\lambda} \ ,
\end{eqnarray}
\end{subequations}
and the derivatives of the potential as
\begin{subequations}
\begin{eqnarray}
\tilde{V}_{,\mu} &=& \tilde{g}^\alpha_{,\mu} V_{,\alpha}\ ,\\
\tilde{V}_{,\mu\nu} &=& \tilde{g}^\alpha_{,\mu} \tilde{g}^\beta_{,\nu}
 V_{,\alpha\beta}\ .
\end{eqnarray}
\end{subequations}

\subsubsection{The $O(4)$ model
\label{sec:O4_model}\label{sec:4.5.3}}

We shall first study the properties of the single-shell $O(4)$ model.
We define fermionic operators
$c^\dagger_{jm}$ and $c_{jm}$ that create or annihilate a particle in
the $J_z=m$ sub-state. In terms of these operators we define four pairing 
($P$, $P^\dagger$, $\tilde P$, and $\tilde P^\dagger$) and two 
multipole operators ($N$ and $Q$) that close under commutation, and 
generate the algebra $o(4)$,
\begin{subequations}
\begin{eqnarray}
P^\dagger &=& \sum_{m>0} c_{jm}^\dagger c_{j\bar m}^\dagger \ , \quad
\tilde{P}^\dagger = \sum_{m>0} \sigma_{jm} c_{jm}^\dagger c_{j\bar m}^\dagger \ ,\\
N &=& \sum_m c_{jm}^\dagger c_{jm}\ , \quad
Q = \sum_m \sigma_{jm} c_{jm}^\dagger c_{jm}\ , \quad\\
\sigma_{jm} &=& \left\{
 \begin{array}{ll}
 +1 & \mbox{for } |m| < \Omega/2\\
 -1 & \mbox{for } |m| \geq \Omega/2
 \end{array} \right. .
\end{eqnarray}
\end{subequations}
Here we need to require that the pair multiplicity $\Omega = j+1/2$
is an {\em even} integer in order for the algebra to close. The sign
of $\sigma_{jm}$ is chosen so as to mimic the behavior of the matrix
elements of the axial quadrupole operator $\bra{jm} r^2 Y_{20}
\ket{jm}$, and we shall call $Q$ the quadrupole operator in the
remainder of this work, even though it does not carry the correct
multipolarity.

As is well-known, the algebra $o(4)$ is isomorphic to $su(2)\oplus
su(2)$. This can be made explicit in terms of the quasi-spin operators
\begin{subequations}
\begin{eqnarray}
\label{eq:4:su2_gen_1}
A_+ &= \frac{1}{2} \left( P^\dagger + \tilde{P}^\dagger \right)
 = A_-^\dagger\ ,
 \quad A_0 =& \frac{1}{4} \left( N + Q - \Omega \right) \ ,\\
\label{eq:4:su2_gen_2}
B_+ &= \frac{1}{2} \left( P^\dagger - \tilde{P}^\dagger \right)
 = B_-^\dagger\ ,
 \quad B_0 =& \frac{1}{4} \left( N - Q - \Omega \right) \ ,
\end{eqnarray}
\end{subequations}
which generate two independent $su(2)$ algebras,
\begin{subequations}
\begin{eqnarray}
\left[ A_+ , A_- \right] &=& 2 A_0 \ , \quad \left[ B_+ , B_- \right] = 2 B_0 \ ,\\
\left[ A_0 , A_\pm \right] &=& \pm A_\pm \ , \quad \left[ B_0 , B_\pm \right] = \pm B_\pm \ ,\\
\left[ { A}_\mu , { B}_{\mu'} \right] &=& 0 \ .
\end{eqnarray}
\end{subequations}

The Hamiltonian of the model is chosen as a simple quadratic form in (some of)
the generators of $o(4)$,
\begin{equation}
H = -G P^\dagger P - \frac{1}{2} \kappa Q^2 \ ,
\label{eq:4:O4_Hamiltonian}
\end{equation}
which mimics the pairing-plus-quadrupole model. Even
though the Hamiltonian looks simple, it does not have a closed-form
solution (it does not have $O(4)$ dynamical symmetry). Nevertheless a
numerically exact solution for the Hamiltonian (\ref{eq:4:O4_Hamiltonian})
can be obtained by simple diagonalization. To this end one rewrites
the Hamiltonian in terms of the quasi-spin
operators $\vec{A}$ and $\vec{B}$,
\begin{equation}
\label{eq:4:H_O4}
H = -G \left( A_+ + B_+ \right) \left( A_- + B_- \right)
 - 2 \kappa \left( A_0 - B_0 \right)^2 \ .
\end{equation}
This Hamiltonian commutes with the total particle number $N=2(A_0
+ B_0) + \Omega$, and there are no further constants of the motion.
The pairing force tends to align the two quasi-spin vectors $\vec{A}$
and $\vec{B}$, so as to obtain the maximal pairing matrix elements,
while the quadrupole force tends to de-align them (to maximize $(A_0 -
B_0)^2$). In this picture, the non-integrability of the model, as
well as the physics described, is related to the competition between
the pairing and the quadrupole force. This is identical to a
competition between alignment and de-alignment of the quasi-spins.

For a fixed number of particles $N=2n_0$,
we construct from the vacuum state $\ket{0}$ all states with
a constant number of generators $A_+$ and $B_+$,
\begin{equation}
\ket{n_0,k_a} = 
\left[ \frac{ \left(\frac{\Omega}{2} - k_a \right)!
 \left(\frac{\Omega}{2} - n_0 + k_a \right)! }
 { \left\{\left(\frac{\Omega}{2}\right)!\right\}^2
 k_a ! \left( n_0 - k_a \right)! } \right]^{1/2}
 A_+^{k_a} B_+^{n_0 - k_a} \ket{0} \ , 
\end{equation}
where $0 \leq k_a \leq n_0$.
Finding the eigenvectors of the Hamiltonian now involves a trivial
matrix diagonalization in this basis of dimension ($n_0 + 1$).

The mean-field description of the Hamiltonian (\ref{eq:4:H_O4}) is most
easily based on the use of a product of $su(2)$ coherent states, one for
the $A_\mu$ sub-algebra, and another for the $B_\mu$ one. (This is the
transcription of the TDHFB formalism described in general terms in
Sec.~\ref{sec:A.5}, and is our only explicit use of this method in
the review.) Each of these
states is characterized by a complex variable, $z_a$ and $z_b$
\cite{Per86}. The time-dependent mean-field dynamics in this
parameterization defines the classical Hamiltonian problem to which
we shall apply
our methodology. We can also parameterize the coherent state with
four real angles \cite{SM88a,SM88b,Per86}, 
\begin{eqnarray}
\ket{z_a,z_b} &=& \exp\left[z_a A_+ - z_a^* A_-
 + z_b B_+ - z_b^* B_- \right] \ket{0} \ ,\nonumber \\
 &=& \left( \cos\frac{\theta}{2} \cos\frac{\chi}{2} \right)^{\Omega/2}
 \exp\left[ \tan\frac{\theta}{2} \exp(-i\phi) A_+
 + \tan\frac{\chi}{2} \exp(-i\psi) B_+ \right] \ket{0}\ ,
\label{eq:4:coh1}
\end{eqnarray}
where we have used
\begin{equation}
z_a = \frac{\theta}{2} \exp(-i\phi) \ ,\quad
z_b = \frac{\chi}{2} \exp(-i\psi) \ .
\end{equation}

The time-dependent Hartree-Fock Bogoliubov (TDHFB) equations are in
this case the classical equations of motion obtained from the
stationary condition of the coherent-state action $\delta S = 0$,
where
\begin{eqnarray}
S[z] &=& \int^t dt\, \bra{z_a,z_b} i \partial_t - H \ket{z_a,z_b} \ ,\nonumber\\
 &=& \int^t dt\,
 \frac{\Omega}{2} \left( \dot{\phi} \sin^2\frac{\theta}{2}
 + \dot{\psi} \sin^2\frac{\chi}{2} \right)
 - \int^t dt\, \Ham (\theta, \chi; \phi,\psi) \ ,
\end{eqnarray}
and
\begin{equation}
\label{eq:4:clH_0}
\Ham = \bra{z_a,z_b} H \ket{z_a,z_b} \ .
\end{equation}
In order to facilitate our work we introduce
real canonical variables $\xi^\alpha$ and $\pi_\alpha$,
\begin{eqnarray}
\xi^1 &=& \frac{\Omega}{2} \sin^2 |z_a|
 = \frac{\Omega}{2} \sin^2 \frac{\theta}{2}\ , \quad
 \xi^2 = \frac{\Omega}{2} \sin^2 |z_b|
 = \frac{\Omega}{2} \sin^2 \frac{\chi}{2}\ , \\
\pi_1 &=& \arg(z_a) = -\phi\ , \quad 
 \pi_2 = \arg(z_b) = -\psi\ .
\end{eqnarray}
Since these variables are canonical, the equations of motion are of
Hamiltonian form
\begin{equation}
\dot{\pi}_\alpha = -\frac{\partial \Ham }{\partial \xi^\alpha} \ , \quad
\dot{\xi}^\alpha = \frac{\partial \Ham }{\partial \pi_\alpha} \ ,
\end{equation}
where the classical Hamiltonian (\ref{eq:4:clH_0}) is the coherent state
expectation value of the Hamiltonian rewritten in terms of canonical
variables (the explicit form can be found from that of the more general
Hamiltonian discussed in the following section,
Eqs.~(\ref{eq:4:Hamfirst}-\ref{eq:4:Hamlast}), upon substitution of
$q_\alpha=1$).
The adiabatic Hamiltonian is then found by
expanding the full Hamiltonian with respect to $\pi$ up to second order,
and is defined in Eqs.~(\ref{eq:4:HamADfirst}-\ref{eq:4:HamADlast}).

We discuss next the problem of defining a requantization
procedure and the consequences of the adiabatic truncation with
respect to momentum. The classical limit of the single-$j$
Hamiltonian has two constants of motion: $\Ham = E$ and $\langle N
\rangle = 2 \sum_\alpha \xi^\alpha = N_0$. Since the phase space is
four dimensional, this implies the complete integrability of the
system, and there is a two dimensional torus on which all classical
orbits lie. Due to this special feature of this model, one can apply
the Einstein-Brillouin-Keller (EBK) quantization condition. This has
been done in Refs.~\cite{SM88a,SM88b} and good agreement with the exact
results has been obtained for both energy spectra and transition
amplitudes. However, it is impossible to extend this
quantization method to non-integrable systems like the ones we will
discuss in the following sections. We wish to use the same quantization
procedure for the simplest form of the model and the more
complicated cases discussed later on, and shall turn to our favorite
technique first.

After truncation of the Hamiltonian up to second order in momentum, we
can define a collective Hamiltonian by evaluating its value 
for points on the collective space $\Sigma$ which is
parameterized by $x^1$ and $p^1$, since we have chosen $x^a=p^a=0;
a=1,\cdots,n-1$,
\begin{subequations}
\begin{eqnarray}
\bar{\Ham}_{\rm col} &=& \left.\bar{\Ham}_{\rm ad}\right|_{\Sigma}
 = \frac{1}{2} \bar{B}^{11}(x^1) p_1^2
 + \bar{V}(x^1) \ ,\\
\bar{B}^{11}(x^1) &=& \sum_{\alpha\beta}
 f^1_{,\alpha} f^1_{,\beta}
 B^{\alpha\beta}\left(\xi^\alpha = g^\alpha(x^1, x^a=0) \right) \ ,\\
\bar{V}(x^1) &=& V\left(\xi^\alpha = g^\alpha(x^1, x^a=0) \right)\ .
\end{eqnarray}
\end{subequations}
Since the scale of the collective coordinate $x^1$ is arbitrary,
we choose to normalize $f^1_{,\alpha}$ so as to make $\bar{B}^{11} = 1$.
Subsequently, the Hamiltonian $\bar{\Ham}_{\rm col}$
is quantized in this flat space as \cite{30}
\begin{equation}
\label{eq:4:H_col_1}
\hat{H}_{\rm col} = - \frac{1}{2} \frac{d^2}{dx^2} + \bar{V}(x) \ ,
\end{equation}
where we have replaced $(x^1,p_1)$ by $(x,p)$.

In order to evaluate the matrix elements of a one-body operator $F$
(either diagonal or transition matrix elements), we first obtain the
collective classical representation of the operator $F$, which in
keeping with the adiabatic approximation is expanded in powers of
momentum,
\begin{equation}
\bar{\cal F}(x,p) =
\left. {\cal F}(\xi,\pi)\right|_{\Sigma}
 = \left. \bra{z} F \ket{z}\right|_\Sigma
 = \sum_{i=0}^\infty \left. {\cal F}^{(i)}(\xi,\pi) \right|_\Sigma
 = \sum_{i=0}^\infty \bar{\cal F}^{(i)}(x,p) \ .
\end{equation}
Here ${\cal F}^{(i)}$ is the term of $i$-th order in $\pi$.
The function $\bar{\cal F}$ is requantized, by making the replacement
$\bar{\cal F}(x,p) \rightarrow \bar{F}(x,\frac{d}{dx})$,
at which point one will have to confront the problem of operator
ordering between $x$ and $p$. We shall avoid this problem by
keeping,
invoking once again the assumption of slow collective motion,
only the zeroth order term $\bar{\cal F}^{(0)}$.
It is an interesting question what the effect of higher order
terms will be. This is clearly outside the scope of the present work, and
requires further investigation.
Fortunately, in the current model,
we have no ambiguity for the quadrupole operator $Q$
because ${\cal Q}^{(i)} = 0$ for $i\neq 0$. For convenience we denote the
classical limit of the quadrupole operator by $q$.
The transition matrix elements can thus be calculated by the
one-dimensional integral,
\begin{equation}
\bra{n'} F \ket{n} = \int dx \Psi_{n'}^*(x) \bar{F}^{(0)}(x)
 \Psi_n(x) \ ,
\end{equation}
where $\Psi_n$ are the eigenfunctions of the collective Hamiltonian
(\ref{eq:4:H_col_1}),

\begin{figure}
\centerline{\includegraphics[width=0.9\textwidth]{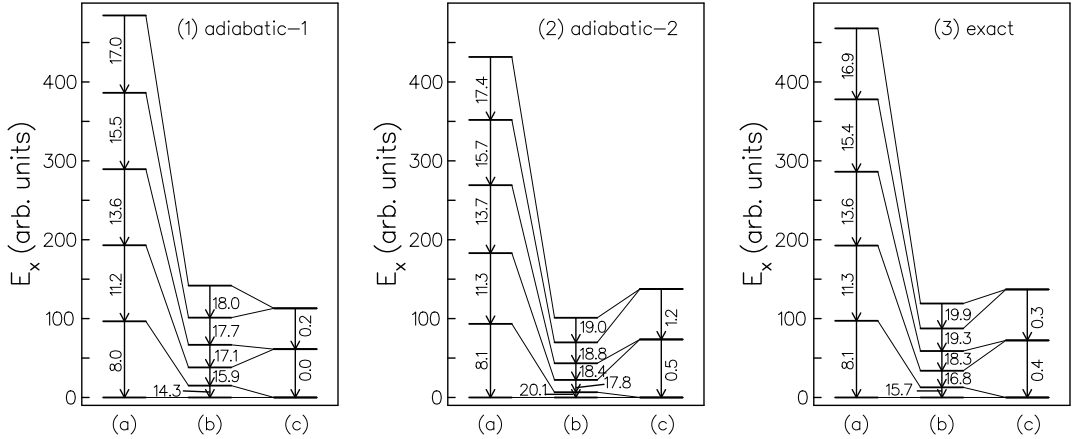}}
\caption{The excitation energies and transition matrix elements
$|\bra{n-1} Q \ket{n}|$, all in arbitrary units, in the single-$j$
shell $O(4)$ model as a function of the ratio between the strength of
the pairing and that of the quadrupole force. The case (a) is a weak
quadrupole force, $2\kappa/G=0.079$, (b) a medium sized one,
$2\kappa/G=1.63$ and (c) a very strong one, $2\kappa/G=12.7$. We have
constrained $G^2+(2\chi)^2=1$. The three panels give our standard
adiabatic quantization (1), the results from the ``adiabatic in
coordinates'' method (2), and the exact results (3).
}\label{fig:4:singlej-1}
\end{figure}

\begin{figure}
\centerline{\includegraphics[width=0.5\textwidth]{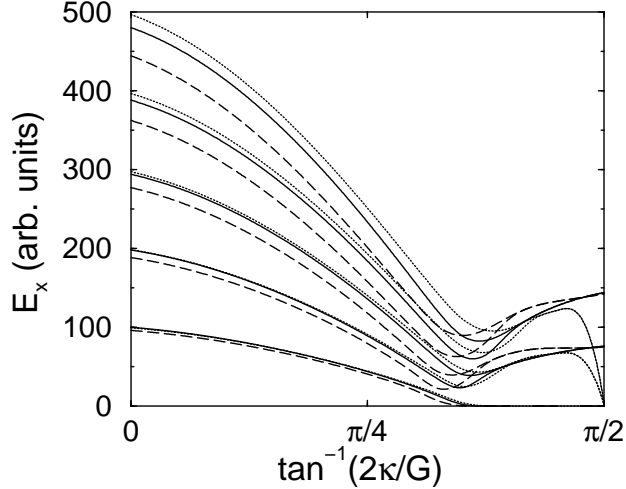}}
\caption{The excitation energies in the single-$j$ shell $O(4)$ model
for 40 particles in a shell with $\Omega=100$ as a function of the
parameter essentially the ratio between the strength of the pairing
and that of the quadrupole force. The left end corresponds to the
case of pure pairing force and the right end to the pure quadrupole
force. The solid lines are the exact results, and the dotted
lines represent the standard requantization of the adiabatic
collective Hamiltonian. The dashed lines represent the expansion in
terms of coordinates discussed in the text. }\label{fig:4:singlej-2}
\end{figure}

{}From the number of coordinates and momenta found (2+2) we see that the
configuration space of the single-$j$ shell model is two-dimensional.
Since there is a zero mode ($\det(B^{\alpha\beta}) = 0$) corresponding
to the NG mode associated with the particle number violation, one may
obtain a one-dimensional path $\Sigma$
without the application
of the LHA. Rather than plotting this path we have chosen to represent
the results of requantization for energies and transition strengths.
These results are presented in panel (1) of Fig.~\ref{fig:4:singlej-1},
and as the dotted lines in Fig.~\ref{fig:4:singlej-2}. We
obtain good agreement with the exact results over a wide range of
parameters except very close to a pure quadrupole force. Due to the peculiar
nature of the quadrupole operator the mass parameter goes to zero in this
limit ($G=0$), and there is no
kinetic term. Thus an eigenstate of Hamiltonian is a coordinate
eigenstate $\ket{x}$ at the same time. Then, the periodic nature of
the momentum becomes important, which we have ignored in our calculations.
Taking account of the periodicity of momentum,
one finds that the
coordinate operator $x$ should have discrete eigenvalues.
In order to check that it is possible to deal with this problem,
we have expanded the Hamiltonian
up to second order with respect to the coordinates rather than momenta,
keeping all order in momenta. We 
have also imposed periodic boundary conditions on the wave function
$\Psi(p) = \Psi(p + \pi/2)$.
The result of this quantization is shown in panel (2) of Fig.~\ref{fig:4:singlej-1},
and as the dashed lines in Fig.~\ref{fig:4:singlej-2}.
The agreement is good in the no-pairing limit $G=0$, while it is not
as good as the standard quantization anywhere else.
Since we are not really interested in the (integrable) pure quadrupole
model, but rather in competition between the pairing and quadrupole
forces, we shall ignore the $G=0$ limit in the rest of this work.
We shall thus follow the conventional adiabatic quantization procedure
in coordinate space as described above.

\subsubsection{The Multi-O(4) Model
\label{sec:multi_O4}\label{sec:4.5.4}}


It is a greater challenge to our approach to study the multi-shell
case. There exists a straightforward extension of the model, by
addition of the individual pairing generators, and summing the
quadrupole operators of each shell with a weight factor (we shall thus
not have a direct coupling between the different shells). The
operators are, for $j=j_1,j_2,\cdots,j_\Lambda$, (for each shell $j_i$
we take the pair degeneracy $\Omega_i = j_i + 1/2$ to be even)
\begin{eqnarray}
P^\dagger &=& \sum_{i,m_i>0} c_{j_im}^\dagger c_{j_i\bar m_i}^\dagger \ ,\quad
\tilde{P}^\dagger =
 \sum_{i,m_i>0} \sigma_{j_im_i} c_{j_im_i}^\dagger c_{j_i\bar m_i}^\dagger \ ,\\
N &=& \sum_{im_i} c_{j_im_i}^\dagger c_{j_im_i} \ ,\quad
Q =
 \sum_{im_i} q_i \sigma_{j_im_i} c_{j_im_i}^\dagger c_{j_im_i} \,
\end{eqnarray}
where $q_i$ represents the magnitude of quadrupole moment carried by
single-particle states.
For each shell we can define quasi-spin
$[su(2)\oplus su(2)]$ generators $\vec{A}^i$ and $\vec{B}^i$
in a manner similar to Eqs.~(\ref{eq:4:su2_gen_1}) and (\ref{eq:4:su2_gen_2}).
We choose a slightly more general Hamiltonian than in the previous chapter
by adding a term containing spherical single-particle energies,
\begin{equation}
H = \sum_{jm} \epsilon_j c_{jm}^\dagger c_{jm}
 -G P^\dagger P - \frac{1}{2} \kappa Q^2 \ .
\end{equation}
The exact solution can be also obtained by diagonalization in a basis set
\begin{equation}
\bigotimes_{i=1}^\Lambda \ket{k_a^{i}, k_b^{i}}
= \prod_{i=1}^\Lambda (A_+^i)^{k_a^i} (B_+^i)^{k_b^i}
 \ket{0} \ ,
\end{equation}
where $0 \leq k_a^i,k_b^i \leq \Omega_i/2$ and $\sum_i (k_a^i + k_b^i)
= n_0 = N_0/2$. This is no longer as trivial a calculation as before,
since the dimension of the basis increases rapidly with the number of
shells $\Lambda$, but can still be done, provided that one chooses
$\Lambda$ sufficiently small. On the other hand, since the
dimension of TDH(F)B configuration space increases only linearly with
$\Lambda$, the amount of effort required for the ALACM calculation is
still rather small. The time-dependent mean-field state is obtained through
the use of the coherent-state representation as before,
and is given by the product of states (\ref{eq:4:coh1})
\begin{eqnarray}
\ket{\vec{z}} &=& \prod_i^\Lambda \ket{z_i}\ .
\end{eqnarray}
We choose the canonical variables as
\begin{subequations}
\begin{eqnarray}
\xi^\alpha &=& \left\{ \begin{array}{ll}
 \frac{\Omega_i}{2} \sin^2 \frac{\theta_i}{2}\ , &
 \mbox{for } \alpha = i = 1,\cdots,\Lambda \ ,\\
 \frac{\Omega_i}{2} \sin^2 \frac{\chi_i}{2}\ , &
 \mbox{for } \alpha = \Lambda+i = \Lambda+1,\cdots,2\Lambda \ ,
 \end{array} \right. \\
\pi_\alpha &=& \left\{ \begin{array}{ll}
 -\phi_i\ , & \mbox{for } \alpha = i = 1,\cdots,\Lambda \ ,\\
 -\psi_i\ , & \mbox{for } \alpha = \Lambda+i = \Lambda+1,\cdots,2\Lambda \ .
 \end{array}\right.
\end{eqnarray}
\end{subequations}
It is convenient to allow the indices of $e$, $q$ and $\Omega$ to range
from 1 to $2\Lambda$ by copying the original list
of parameters, e.g.,
\begin{equation}
e_\alpha = \left\{ \begin{array}{ll}
 e_i\quad(i=\alpha) &
 \mbox{for } \alpha = 1,\cdots,\Lambda \ ,\\
 e_i\quad(i=\alpha-\Lambda) &
 \mbox{for } \alpha = \Lambda+1,\cdots,2\Lambda \ .
 \end{array} \right.
\end{equation}
Using these definitions, the classical Hamiltonian can be given in the compact
form
\begin{subequations}
\begin{eqnarray}
\label{eq:4:Hamfirst}
\Ham &=& {h}_{\rm sp} + {\cal H}_P + {\cal H}_Q \ ,\\
{h}_{\rm sp} &=&2\sum_\alpha e_\alpha \xi^\alpha \ ,\\
\label{eq:4:clH_P_2}
\Ham_P(\xi,\pi) &=& -\frac{G}{16} \left\{
 \left| \sum_\alpha e^{i\pi_\alpha} S_\alpha \right|^2
 + 32 \sum_\alpha \Omega_\alpha^{-1} (\xi^\alpha)^2 \right\} \ ,\\
\label{eq:4:clH_Q_2}
\Ham_Q(\xi) &=& -2\kappa \left\{
 \left( \sum_\alpha \sigma_\alpha q_\alpha \xi_\alpha \right)^2
 + \sum_\alpha \Omega_\alpha^{-1} q_\alpha \xi^\alpha
 (\Omega_\alpha - \xi^\alpha ) \right\} \ ,\\
S_\alpha &=& 2 \sqrt{2\xi^\alpha(\Omega_\alpha - 2\xi^\alpha)} \ ,\\
\sigma_\alpha &=& \left\{ \begin{array}{l}
 +1 \quad \mbox{for } \alpha = 1,\cdots,\Lambda \ ,\\
 -1 \quad \mbox{for } \alpha = \Lambda+1,\cdots,2\Lambda \ .
 \end{array} \right.
\label{eq:4:Hamlast}
\end{eqnarray}
\end{subequations}
The adiabatic limit of this Hamiltonian is
\begin{subequations}
\begin{eqnarray}
\label{eq:4:HamADfirst}
\Ham_{\rm ad} &=& \frac{1}{2} \sum_{\alpha\beta}
 B^{\alpha\beta} \pi_\alpha \pi_\beta
 + V(\xi) \ ,\\
B^{\alpha\beta} &=& \frac{G}{8} 
\left[\delta_{\alpha\beta}
 \left( S_\alpha\sum_\gamma S_\gamma\right) -S_\alpha S_\beta \right] \ ,\\
V(\xi) &=& V_P(\xi) + V_Q(\xi) \ ,\\
V_P(\xi) &=& \Ham_P(\xi, \pi=0) \ , \quad V_Q(\xi) = \Ham_Q (\xi) \ .
\label{eq:4:HamADlast}
\end{eqnarray}
\end{subequations}
The terms in Eqs.~(\ref{eq:4:clH_P_2}) and (\ref{eq:4:clH_Q_2})
proportional to $\Omega_\alpha^{-1}$ arise from the exchange contributions, and
will be neglected.


Using the LHA we identify the collective degree
of freedom amongst the $2\Lambda$ coordinates. As before we first must 
remove the NG mode corresponding to a change in particle number explicitly,
due to the zero mass parameter associated with this mode.
The particle number is simply given by the sum of the numbers for the 
individual shells, ${\cal N} = 2 \sum_\alpha \xi^\alpha$. 
It is easy to show that
\begin{equation}
\sum_\beta B^{\alpha\beta} {\cal N}_{,\beta}
 = 2 \sum_\beta B^{\alpha\beta} = 0 \ .
\end{equation}
Thus we apply the prescription for removing the spurious mode
and use the LHA to
determine the collective path in the remaining
($2\Lambda-1$)-dimensional coordinate space. Since we shall mainly
investigate how the LHA can deal with the transition spherical to
deformed, it is sufficient to study only two shells. We take the size
of these shells to be equal, $\Omega_1 = \Omega_2 = 10$, and put $16$
particles in the available space. We split the degeneracy by taking
$e_1 = 0$ and $e_2=1$, and we use a different value of the ``quadrupole
moment'' for each shell as well, $q_1 = 3$ and $q_2=1$. The pairing
force is fixed at $G=0.3$, and we only vary the quadrupole force
strength $\kappa$.

\begin{figure}
\centerline{\includegraphics[width=0.8\textwidth]{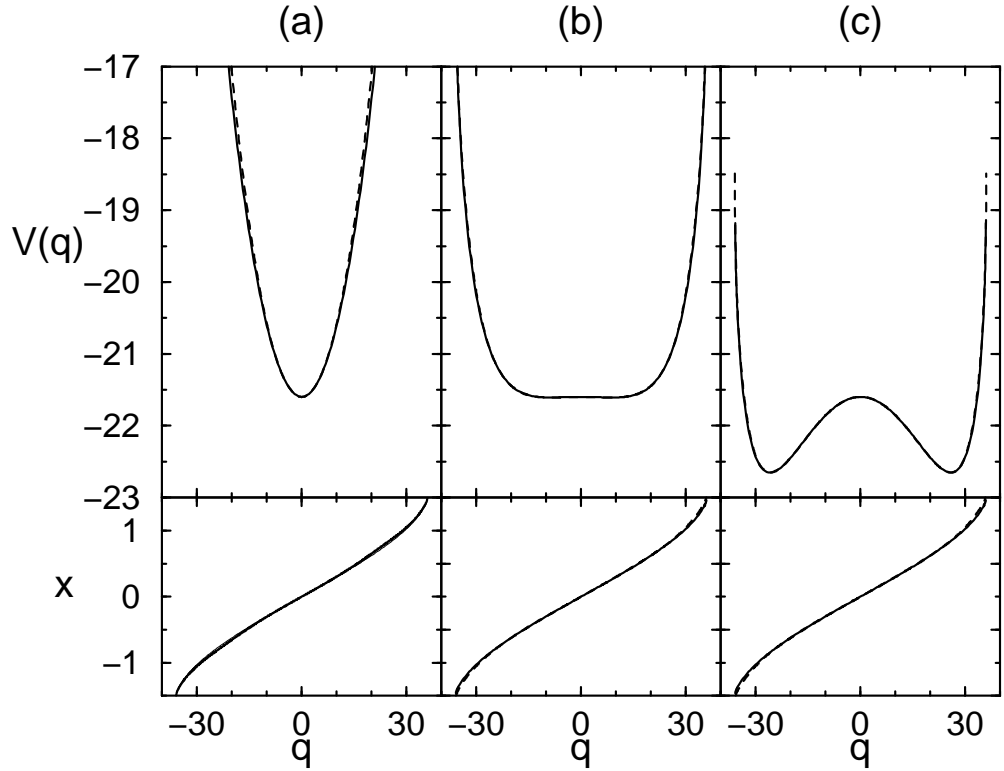}}
\caption{The collective potential energy $V(q)$ and the collective
coordinate $x$ (normalized to unit mass) as a function of the
quadrupole moment $q=\langle \hat Q \rangle$. We show both the LHA 
(dashed line) and the
CHB (solid line) results in each figure. These results
are, for this model, essentially indistinguishable. The case (a)
corresponds to a weak quadrupole force ($\kappa=0.01$), (b) to a
slightly stronger force ($\kappa=0.03$) and (c) to the strongest,
$\kappa=0.035$. The units of all displayed quantities are arbitrary.}
\label{fig:4:2shell1}
\end{figure}

We show a representation of the collective potential energy and the
collective coordinate in Fig.~\ref{fig:4:2shell1}. We represent these
quantities as a function of the expectation value of the quadrupole
operator. As an alternative to the LHA approach, we have also
performed a simple constrained Hartree-Bogoliubov (CHB) calculation,
where one imposes a value for the expectation value of the quadrupole
operator. We determine the mass for this case by replacing the RPA
eigenvector by the coordinate derivative of the quadrupole expectation
value. We then renormalize the coordinate to obtain a constant mass.

\begin{figure}
\centerline{\includegraphics[width=0.9\textwidth]{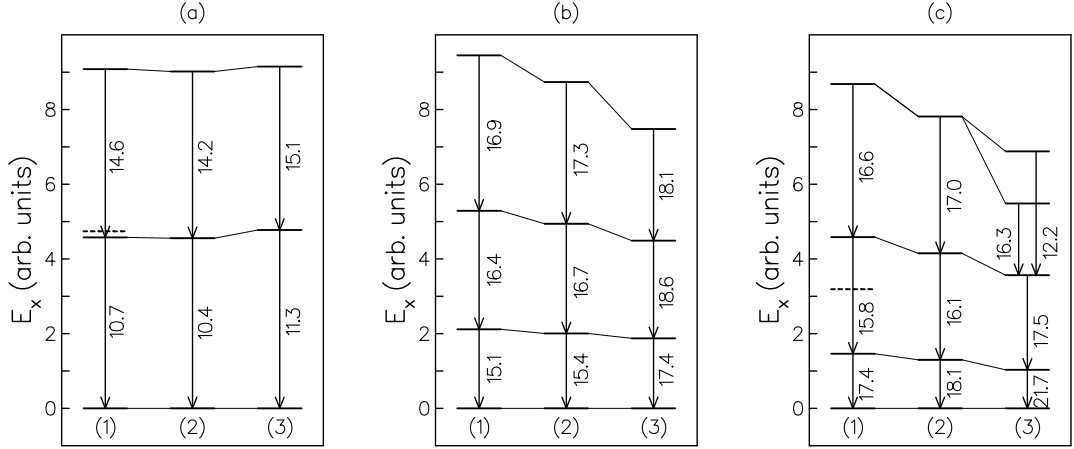}}
\caption{The excitation energies $E_x$ and transition matrix elements
$|\bra{n'} Q \ket{n}|$ (numbers next to arrows) in the two-shell case
discussed in the text. The three cases (a), (b) and (c) correspond to
those discussed in Fig.~\ref{fig:4:2shell1}. In each of the three panels
the left one (1) is obtained from requantizing the CHB, the middle one
(2) is obtained from requantizing the LHA result, and the right one
corresponds to exact diagonalization. The dashed line shows the lowest
RPA eigenvalue.}
\label{fig:4:2shell2}
\end{figure}

We have investigated a full shape transition scenario, where we have
changed the strength of the quadrupole interaction so that the collective
Hamiltonian changes from spherical and harmonic for case (a) to flat
for case (b) to deformed for case (c). We see that the difference
between the LHA and CHB calculations is relatively small. This is also
borne out by the spectra and transition strength in
Fig.~\ref{fig:4:2shell2}. We can see the similarity between the two
approximate calculations. If anything, the LHA gives slightly better
results than the CHB based calculations. We seem to be unable to
reproduce the large density of states found in the exact calculation
for ``deformed'' nuclei, where there are indications from the
transition strengths that some of the states in the approximate
calculations are split into several of the exact states. Note,
however, that at an excitation energy of 6, we are 5 units above the
barrier, so this may just be due to our choice of parameters. The
shape mixing in the low-lying excited states appears to be described
sensibly, however. We would have liked to be able to choose an even
large value of $\kappa$, but if we do that the system collapses to the
largest possible quadrupole moment in the model space, which leads to
all kinds of unphysical complications.

\subsubsection{A multi-shell $O(4)$ model with neutrons and protons
\label{sec:two_comp_O4}\label{sec:4.5.5}}

In heavy nuclei, the number of neutrons is normally (much) 
larger than that of protons, which often leads to a radically
different shell structure at the Fermi surface for neutrons and protons.
In order to perform a model study of such phenomena, where we can
still perform an exact calculation, we adapt 
the multi-shell $O(4)$ model introduced in the previous section
to one describing systems with both neutrons and protons
\cite{FMM91}.
 We shall
then use this model to analyze the collective dynamics of 
shape-coexistence nuclei, as observed for instance in semi-magic nuclei.
At the same time we shall concentrate on the diabatic/adiabatic dichotomy
already mentioned in the introduction.

The model is a simple extension of the multi-shell $O(4)$ model
in the previous section, with the main difference that we do not have
pairing between proton and neutron orbitals,
\begin{eqnarray}
H &=& H_n + H_p + H_{np}\ ,\\
H_n &=& \sum_{i\in n,m_i} \epsilon_i c_{j_im_i}^\dagger c_{j_im_i}
 -G_n P_n^\dagger P_n -\frac{1}{2}\kappa Q_n^2 \ ,\\
H_p &=& \sum_{i\in p,m_i} \epsilon_i c_{j_im_i}^\dagger c_{j_im_i}
 -G_p P_p^\dagger P_p -\frac{1}{2}\kappa Q_p^2 \ ,\\
H_{np} &=& - \kappa Q_n Q_p \ ,
\end{eqnarray}
where $P_{n(p)}$ and $Q_{n(p)}$ are the pairing and quadrupole operators
for neutrons (protons) (see the definitions in Sec.~\ref{sec:multi_O4}).
In this model there are two trivial NG modes associated with
the change of neutron and of proton number, which can both be removed 
explicitly in the manner discussed before. 

We study this model for a single-shell for neutrons, with pair
degeneracy $\Omega_n=50$, containing 40 particles. We take the
single-particle quadrupole matrix element $q_n=1$, and use a pairing
strength $G_n=0.3$, and assume zero single-particle energy. For
protons we take two shells, both with $\Omega_{p1}=\Omega_{p2}=2$,
$q_{p1}= q_{p2}=2$, having single-particle energies
$e_{p1}=-e_{p2}=5$. We study two different set of interaction
parameters, both with $\kappa=0.1$. The first is $G_n=G_p=0.3$, the
second has the same neutron pairing strength, but $G_p=10$.

\begin{figure}
\centerline{\includegraphics[width=0.8\textwidth]{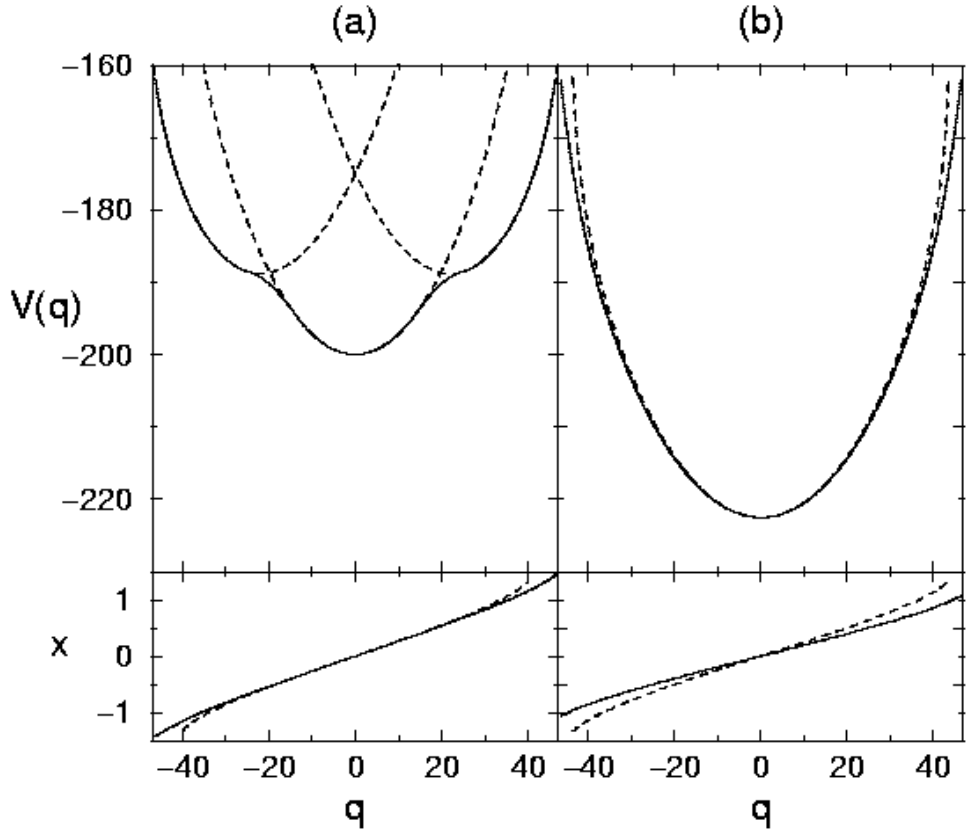}}
\caption{The collective potential energy $V(q)$ and the collective coordinate
$x$ as a function of the quadrupole moment $q= q_p+q_n $. We
show both the LHA (dashed lines) and the CHB
results (solid line) in
each figure. The case (a) corresponds to a weak proton
pairing force ($G_p=0.3$), (b) to a strong pairing force ($G_p=10$).
The rest of the parameters are given in the main text. The units of all 
quantities displayed are arbitrary.
}\label{fig:4:pnshell1}
\end{figure}

As displayed in Fig.~\ref{fig:4:pnshell1}, the collective potential
energy for the weaker pairing strength shows a very interesting
behavior, with two shoulders appearing in the CHB collective
potential energy. This is what is normally called the adiabatic
potential energy, and the shoulders arise from an avoided crossing. As
in our previous work \cite{30} the adiabatic LHA method chooses a
diabatic (crossing) set of potential energy curves. These
shape-coexistence minima are related to 2p-2h excitation in the proton
model space, promoting two particles from the lowest Nilsson orbitals
to the down-sloping ones. This is of course very similar to the
phenomena observed in shape coexistence in semi-magic nuclei.

We get another surprise for the strong pairing case. Here the
collective potentials look very similar, and rather structureless, but
the collective coordinates are different. This can be traced back to
the fact that the collective coordinate in the LHA is not $q_p+q_n$,
but a different combination. At the minimum, the lowest RPA mode
correspond approximately to $q_n+\frac{1}{10}q_p$. This is similar to
the situation analyzed in great detail in our study of $^{28}$Si, and once
again shows the importance of self-consistency in the selection of the
collective coordinate. The real collective coordinate is {\em not} the
mass-quadrupole operator, with consequences that produce discernible
differences in the predictions of the two methods.

\begin{figure}
\centerline{\includegraphics[width=0.9\textwidth]{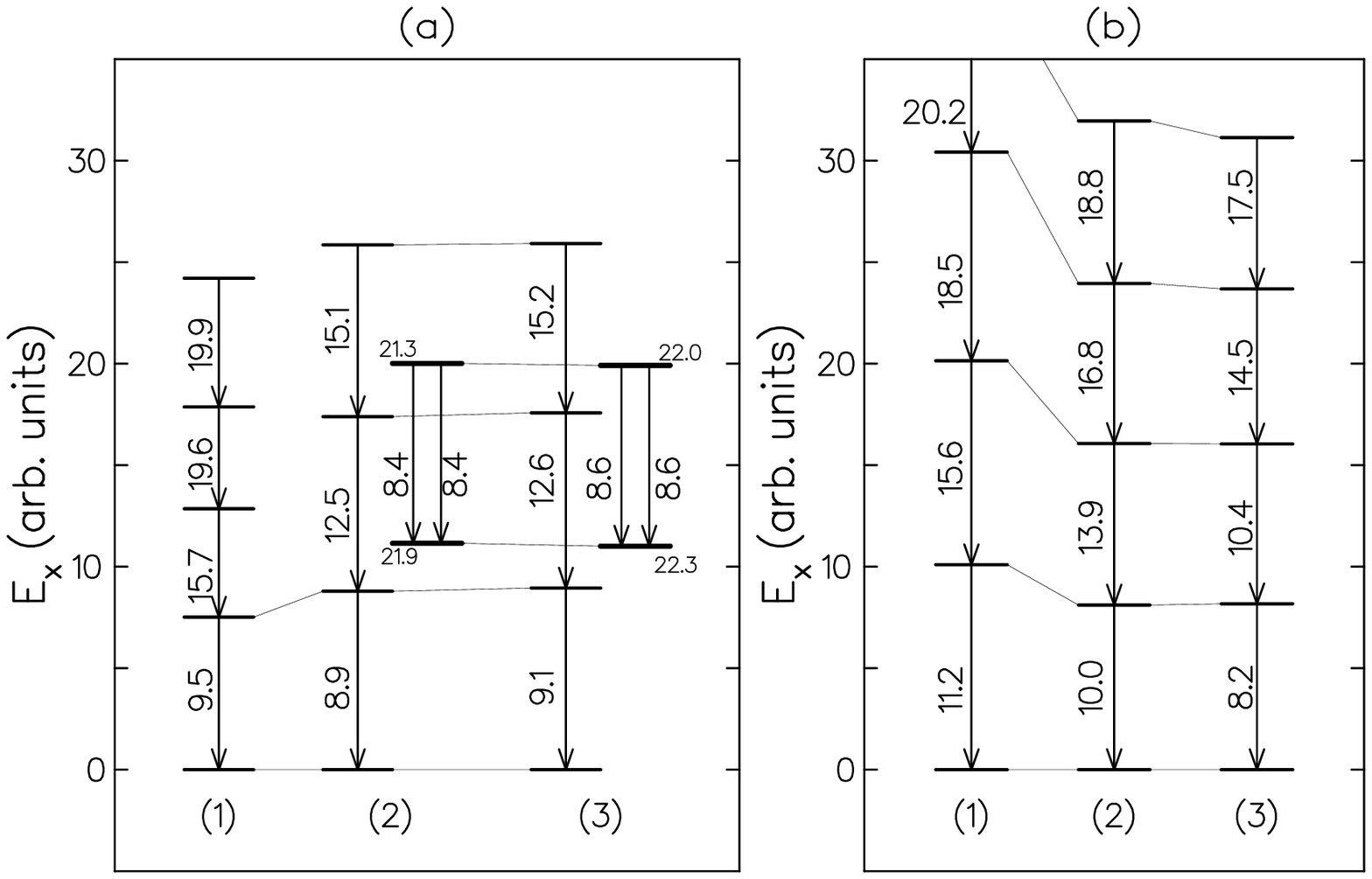}}
\caption{The excitation energies $E_x$ and transition matrix elements
$|\bra{n'} Q \ket{n}|$ (numbers next to arrows), in arbitrary units,
for the proton-neutron
model discussed in the text. The two cases (a) and (b) correspond to
those discussed in Fig.~\ref{fig:4:pnshell1}. In each of the panels the
left one (1) is obtained from requantizing the CHB, the middle one (2)
is obtained from requantizing the LHA result, and the right one (3)
corresponds to the exact diagonalization. Levels denoted by thick
lines are doubly degenerate. The numbers next to the arrows denote
the size of the transition matrix elements.}
\label{fig:4:pnshell2}
\end{figure}

In Fig.~\ref{fig:4:pnshell2} we show the consequences of these results 
for the requantization. For the weak pairing case the diabatic picture
obtained through the LHA gives almost perfect results, whereas the CHB
potential energy curve fails to provide the correct answer. In the case of
the strong pairing the incorrect choice of the collective coordinate
leads to too large a level spacing in the CHB calculations, whereas the LHA
and exact calculations agree again.

For the weak pairing case the exact calculation almost exhibits the doublet
structure found from the diabatic potential energy curves; the splitting
is less than one part in $10^5$ in the exact calculation. Nevertheless,
the exact calculation consists of symmetric and antisymmetric states, which 
leads to the transition matrix element 22.3 between these two states. Since
in the LHA calculation the states do not mix, we have printed the
{\em diagonal} matrix element instead. For very weak mixing this is
the correct comparison, as is borne out by the results.

The decoupling measure $D$, Eq.~(\ref{eq:4:Deq}), is small for all
these cases. The worst case is the strong pairing case, where it rises
from $0$ for $q=0$ to $3\times10^{-3}$ for $q=40$. For this reason we
also believe that the ``scalar Berry potential'' \cite{30} will be
small, and we have not included this, or any other quantum corrections,
in our calculations.

\subsection{Concluding remarks}

We have shown that the LHA and/or the GVA give workable tools in
situations that might be considered as a guide towards realistic
nuclear physics problems.

In the area of shape coexistence, as studied by the model calculations
discussed in the last section, and a bit less so for $^{28}$Si, it
appears that our approaches contain excellent methods to obtain
sensible results, with reasonable effort. 
The problems with the LHA is that even though the size of the
space is much smaller than that which enters into a straightforward
diagonalization, the application of the local RPA, which needs to be 
repeated many times at every point along the collective path, may be
prohibitive for realistic calculations.  Fortunately, many problems
can be studied with simplified separable forces, as in the model
discussed here. We are at the moment considering the old
pairing-plus-quadrupole model, that has been applied so successfully
in the physics of heavy nuclei. Such a model can be dealt with much
more straightforwardly than more microscopic Skyrme or Gogny-force
based approaches. This will allow us to shed light on a good treatment
of shape-coexistence, as well on the interesting question of the
choice of the collective (cranking) operator, which was already found
to be non-trivial in certain limits of the $O(4)$ model.

One might argue that even that is not enough, and we should really
adopt a fully microscopic quantum many-body approach. We
believe that we can address this problem, and are actively considering
the approaches available to us. From the discussion
given above, it should be clear that an efficient calculation hinges
on efficient solution of
the RPA. We are investigating two approaches to this problem, the use
of iterative diagonalization of the RPA using Lanczos procedures, or
the approximation of the RPA by using separable forces
\cite{sep:RPA}, which can be diagonalized much more easily.

In the end such an approach may be more efficient than the
path-following version of the GVA, which doesn't require an RPA
diagonalization, but rather depends on a matrix inversion of similar
dimensionality, which makes such calculations also very time consuming.

\newpage
\section{Quantum theory of LACM and Berry phase}\label{sec:5}

\subsection{The Born-Oppenheimer approximation\label{sec:5.1}}

\subsubsection{Introduction\label{sec:5.1.1}}

Having completed our account of the applications to nuclear physics
carried out to date, we turn next to work that is concerned with the
further development of the theory. Thus far we have studied large
amplitude collective motion on a classical basis and then quantized
the emergent collective Hamiltonian. We also developed and applied a
theory for computing further corrections arising from the quantum
oscillations of the non-collective degrees of freedom. The combined
theoretical structure, applied in Secs.~\ref{sec:4.3} and
\ref{sec:4.4} may well suffice for most applications. 
Nevertheless, in this section we return to the fundamentals of the
theory of large amplitude collective motion by seeking to establish a
quantum theory from the beginning. We attempted such an approach in
our earliest efforts in this field
\cite{3,4}, but met with only limited success.

The renewed effort, which we now report, has been rendered more
satisfactory by an adaptation of the fundamental notions of the
Born-Oppenheimer approximation, including the improvements brought
about by the inclusion of the contributions associated with the Berry
phase \cite{81,82}. In this subsection, we deal with the problem
without Fermi-Dirac statistics, an account drawn from Ref.~\cite{25}.
In subsection \ref{sec:5.2}, we present some applications, and in
subsection \ref{sec:5.3}, we develop the theory including Fermi-Dirac
statistics.

\subsubsection{Summary of previous viewpoint and its limitations\label{sec:5.1.2}}

In contrast to the previous sections, 
we base our study on a {\it quantum} Hamiltonian, that with the help
of the summation convention, takes the form
\begin{equation}
H = {H}(\xi^\alpha,\pi_\alpha)=\frac{1}{8} \{\pi_\alpha,
\{\pi_\beta,{B}^{\alpha\beta}(\xi)\}\} + {V}(\xi)
={T} + {V},
\label{eq:5:12.2.1} 
\end{equation}
that describes $N$ coordinate and momentum pairs,
$\xi^\alpha$ and $\pi_\alpha$, $\alpha=1,...,N$, which satisfy the canonical
commutation relations, using units where $\hbar=1$, 
\begin{equation}
[\xi^{\alpha},\pi_{\beta}] = i\delta^\alpha_\beta. \label{eq:5:2.100}
\end{equation}
Associated with this
Hamiltonian is a form for the scalar product in Hilbert space
that we describe at the end of the current discussion.

Within the quantum mechanical framework, we limit ourselves to
locally invertible point transformations,
\begin{eqnarray}
 \xi^\alpha &=& g^\alpha(q),\;\; q^\mu = f^\mu(\xi), \nonumber \\
 q&= & \{Q^i ,q^a\}, \;\; i=1...K, \;\; a=K+1...N. \label{eq:5:12.2.2}
\end{eqnarray}
For a more concise notation we denote the set of collective coordinates by $Q$,
and use $q$ for the non-collective ones. We can now 
find the Hamiltonian for the new variables. For the potential energy, we have
\begin{equation}
{ V}(\xi)={ V}(g^\alpha(Q,q))\equiv \bar{V}(Q,q). \label{eq:5:12.1.1}
\end{equation}
We transform the kinetic energy recognizing the tensor character of
the mass matrix and with the help of the relation
\begin{eqnarray}
\pi_\alpha&=& \frac{1}{2}\{f^\mu_{,\alpha},p_\mu\}. \label{eq:5:12.2.17} 
\end{eqnarray}
The justification for (\ref{eq:5:12.2.17}) is also discussed below.
In terms of the new variables, the kinetic energy
takes the form
\begin{equation}
T=\frac{1}{8}\{p_\mu ,\{p_\nu ,\bar{B}^{\mu\nu}(Q,q)\}\} + \bar{V}_{\mathrm{quant}}(Q,q),
\label{eq:5:13.2.19}
\end{equation}
where the second term, which is specifically a quantum potential arising from
the non-commutativity of coordinates and momenta, has the form
\begin{equation}
8\bar{V}_{\mathrm{quant}}(Q,q)=[f^\mu_{,\alpha\gamma}g^\gamma_{,\nu}g^\alpha_{,\lambda}
\bar{B}^{\nu\lambda}]_{,\mu} - [f^\mu_{,\alpha\gamma}g^\gamma_{,\mu}]_{,\nu}
g^\alpha_{,\lambda}\bar{B}^{\nu\lambda}. \label{eq:5:13.1.2}
\end{equation}
For further work, the momenta are also divided
into collective and non-collective subsets,
$p_\mu = \{P_i,p_a\}$. In order to understand how to decompose the
Hamiltonian into collective and non-collective parts on the quantum level,
we turn in the next section to a discussion of the Born-Oppenheimer
approximation.

\begin{aside}
We define a scalar product and provide the proof of Eq.~(\ref{eq:5:12.2.17}),
which specifies how the momentum operator transforms under a general point
transformation. It is essential to recognize that this result
is tied to a choice of scalar product. We suppose that the Hamiltonian
(\ref{eq:5:12.2.1}) is to be used in conjunction with the metric
\begin{equation}
\langle\Psi_a|\Psi_b\rangle\equiv\int d\xi^1\cdots d\xi^N\Psi^{\ast}_a(\xi)
\Psi_b(\xi). \label{eq:5:A.1} \end{equation}
Thus \begin{equation}
\pi_\alpha\rightarrow -i(\partial/\partial\xi^\alpha).
\label{eq:5:A.2} \end{equation}
Now carry out the point transformation (\ref{eq:5:12.2.2}) with Jacobian $J$,
\begin{equation} J=|\partial\xi^\alpha/\partial x^\beta|.
\label{eq:5:A.3} \end{equation}
If we introduce a new wave function \begin{equation}
\psi_a = J^{\frac{1}{2}}\Psi_a , \label{eq:5:A.4} \end {equation}
the metric is preserved in the sense \begin{equation}
(\Psi_a,\Psi_b)=(\psi_a,\psi_b)=\int dx^1\cdots dx^N\psi^{\ast}_a(x)
\psi_b(x). \label{eq:5:A.5} \end{equation}
This is the metric that is associated with Eq.~(\ref{eq:5:12.2.17}).
The proof is given in appendix A of Ref.~\cite{25}. 
\end{aside}

\subsubsection{Application of the Born-Oppenheimer approximation\label{sec:5.1.3}}

We set ourselves two tasks. We shall first find the
leading corrections to the form of the collective Hamiltonian
given above, arising
from the coupling of the fast to the slow variables. This will be done
with the help of the standard Born-Oppenheimer (BO) approximation.
In order to evaluate these corrections explicitly, however, we shall
then commit ourselves to further approximations for the
dependence of the full Hamiltonian
on the fast variables, that yield a normal
mode description of the latter.

To go beyond the restricted picture presented in Sec.~\ref{sec:5.1.2},
we introduce the BO picture into the theory of large
amplitude collective motion. To carry this out, we adopt a
representation of the wave functions of the collective states
$|n\rangle$ in the coordinate space $|Q^i,q_a\rangle \equiv |Q,q\rangle$,
of the form
\begin{eqnarray}
\langle Q^i q^a|n\rangle &\equiv & \langle Q,q|n\rangle 
 =\sum_\nu(Q|n\nu)[q|\nu\!:\!Q],
\label{eq:5:13.2.4}
\end{eqnarray}
Here $\nu$ are quantum numbers that describe the state of excitation
of the intrinsic degrees of freedom, so that the sum in
(\ref{eq:5:13.2.4}) is over various collective bands, each term
representing a product of a wave function for the collective motion
and a wave function for the intrinsic motion, the latter tied to the
instantaneous value of the collective coordinates. This is the essence
of the BO picture, though it is not an approximation until we restrict
the sum.  We utilize a notation where a ket with an angular bracket
indicates a state in the full Hilbert space, a square bracket denotes
a state in the space of fast variables (which depends parametrically
on the slow variables, as indicated by the notation``$:\!Q$'' in the
state vector), and a ket with a parenthesis is part of the collective space.  We
suppose that for fixed $Q$, the states $[q|\nu\!:\!Q]$ are a complete
set of functions for the fast coordinates,
\begin{equation}
 \sum_\nu [q|\nu\!:\!Q][\nu\!:\!Q|q']
 = \delta(q-q'). \label{eq:5:13.2.5a} \end{equation}
Though for the moment we have not specified the equation
of which they are the solutions, we shall be able to do so, at least 
approximately, as a consequence of the developments to be carried out
 in this section.

For the remainder of the current discussion, we shall consider the
simplest case in which the fast variables occupy, for any given value
of the slow variables, the state of lowest energy, which is assumed to
be non-degenerate. For the case to be studied here,
$\nu$ thus takes a single value chosen to be zero. 
We suppress this zero in the coefficient function
that appears in (\ref{eq:5:13.2.4}), and is now written as 
$(Q|n)$. This can 
be identified as the wave function for the collective motion.
This identification agrees with the one that was made in the previous sections,
where the description of large amplitude collective motion was
not tied to a BO approximation. There we emphasized the connection
between the decoupled Hilbert space of the collective coordinates
and the crucial existence of a $K$-dimensional decoupled
coordinate manifold described by the functions $g^\alpha(Q,q=0)$.
The approximate separability of the associated
Hilbert space is based on the assumption that the functions
$[q|\nu\!:\!Q]$ describing the fast variables are confined
to a narrow region in $q$ space in the neighborhood of the collective
surface. If this picture holds, we may expect the mathematical
details to work out reasonably.

Adopting the BO approximation, we set ourselves the task of finding
an effective Hamiltonian to describe the motion of the collective
variables. This operator is defined, though not yet operationally,
by means of the equation, \begin{equation}
(n'|H_{{\rm eff}}(Q,P)|n)=\langle n'|H(\xi,\pi)|n\rangle.
\label{eq:5:13.2.6} \end{equation}
This relation will assume the status of a definition of the collective
Hamiltonian within the space of slow variables and $H_{{\rm eff}}$ 
recognized as a generalization of $H_C$ only after we
specify how to eliminate the fast variables from the right hand side.
The procedure that we shall follow is closely akin to the
traditional BO approach, with characteristic differences arising
from the facts that at the beginning we cannot specify which are the
slow and which the fast variables, and that the treatment of the
fast variables comes as a kind of afterthought, dependent in detail
on the prior treatment of the collective variables.

Let us start with the
potential energy, \begin{equation}
{V}(\xi)={V}(g^\alpha(Q,q))\equiv \bar{V}(Q,q),
\label{eq:5:13.2.7} \end{equation}
and evaluate the associated piece of (\ref{eq:5:13.2.6}) 
\begin{eqnarray}
\langle n'|{V}(\xi)|n\rangle &\equiv &(n'|\bar{V}_{{\rm eff}}|n)\nonumber \\
&\cong&\int dQdq\,(n'|Q)[ 0\!:\!Q|q] {V}(
g^\alpha(Q,q))[q|0\!:\!Q]( Q|n),
\label{eq:5:13.2.8} 
\end{eqnarray}
in the BO approximation. Within this framework,
the only feasible way
of integrating out the fast variables would appear to be to expand
$\bar{V}$ in powers of $q^a$, \begin{equation}
\bar{V}(Q,q)=\bar{V}(Q) +\bar{V}_{,a}q^a+\frac{1}{2} \bar{V}_{,ab}q^a q^b +... \;,
\label{eq:5:13.2.9} \end{equation}
leading to 
\begin{equation}
\bar{V}_{{\rm eff}}(Q) = \bar{V}(Q) + \bar{V}^{(1)}(Q) + \bar{V}^{(2)}(Q) +... \;,
\label{eq:5:13.2.10} 
\end{equation}
where, 
if the function $[q|0\!:\!Q]$ is chosen to be normalized,
the various pieces take the form
\begin{subequations}
\begin{eqnarray}
\bar{V}(Q)&=&\bar{V}(Q,0), \label{eq:5:2.11} \\
\bar{V}^{(1)}(Q)&=&\bar{V}_{,a}\int|[q|0\!:\!Q] |^2 q^a
\equiv \bar{V}_{,a}(Q)\langle q^a\rangle_Q, \label{eq:5:13.2.12} \\
\bar{V}^{(2)}(Q) &=& \frac{1}{2} \bar{V}_{,ab}(Q)\langle q^aq^b\rangle_Q.
\label{eq:5:13.2.13}
\end{eqnarray}
\end{subequations}
(In Sec.~\ref{sec:5.3}, we shall adopt an alternative method for the nuclear case.)

We thus see that the leading term is independent of the
wave function for the fast variables, coinciding with
the classical result for the potential energy of large amplitude
collective motion.
The computation of this term requires only the form of the collective
submanifold, $\xi^\alpha =g^\alpha(Q,0)$, which can be determined
by the general theory presented in Sec.~\ref{sec:2}.
This is not immediately obvious,
since we are dealing with a quantum theory, and the so-called decoupling
conditions are operator conditions. These conditions, which follow
straightforwardly from decoupling requirements imposed on the transformation
(\ref{eq:5:12.2.2}), are
\begin{eqnarray}
\{p_i, \bar{B}^{ia}\} &=& 0, \label{eq:5:DC1} \\
\bar{V}_{,a} +\frac{1}{8}\{p_i,\{p_j, \bar{B}^{ij}_{,a}\}\} &=& 0. \label{eq:5:DC2}
\end{eqnarray}
It is apparent that they will be satisfied provided the classical
decoupling conditions (\ref{eq:2:2.15}-\ref{eq:2:2.17}) are.

To go beyond this lowest order, we need, besides the wave function of the
fast variables, to rearrange the expansion of $V$ so that each term in the
expansion is a scalar under coordinate transformations.
At the conclusion of the present discussion,
it is shown that if we wish to interpret the small
quantities $q^a$ as components of a vector, we must replace the
ordinary second derivative by a covariant second derivative,
$\bar{V}_{;ab}$, that can be computed from known or calculable quantities
according to the equations 
\begin{subequations}
\begin{eqnarray}
\bar{V}_{,a}(Q) &=& {V}_{,\alpha}g^\alpha_{,a}, \label{eq:5:13.2.14} \\
\bar{V}_{;ab}(Q)&=&{V}_{;\alpha\beta}g^\alpha_{,a}g^\beta_{,b}.
 \label{eq:5:13.2.15} 
\end{eqnarray}
\end{subequations}

Turning now to the quantum corrections, we assume that the decoupling
condition, $\bar{V}_{,a}=0$ is sufficiently well satisfied that
we can neglect the term $\bar{V}^{(1)}(Q)$.  The third term
$\bar{V}^{(2)}(Q)$, is the first, $\Delta^{(1)}\bar{V}(Q)$, of a sequence of
contributions that we shall identify as the leading corrections to the
potential energy.

The remaining terms
will arise from the study of the kinetic energy, given after
transformation of coordinates by Eqs.~(\ref{eq:5:13.2.19})
and (\ref{eq:5:13.1.2}).
In order to integrate out the fast variables in the contribution that
these terms make to $H_{{\rm eff}}$, consider the first term
of the kinetic energy. We expand the mass
tensor in powers of $q$, and keep initially only the leading term
$\bar{B}^{\mu\nu}(Q,0)\equiv \bar{B}^{\mu\nu}(Q)$. 
In this approximation, we shall first
restrict the study to the contribution of
those terms where the indices $\mu,\nu$ take on values $i,j$
in the collective subset.

As a preliminary to this calculation, we study the simpler object
\begin{eqnarray}
(n'|(P_i)_{{\rm eff}}|n)&=&\langle n'|P_i|n\rangle 
 =(n'|(P_i -A_i)|n) 
 \equiv (n'|D_i|n). \label{eq:5:12.3.50} \end{eqnarray}
Here $P_i$ is the momentum operator in the collective subspace, and
\begin{equation}
A_i \equiv i\int dq\, [0\!:\!Q|q]\left(\partial_{Q^i} [q|0\!:\!Q]\right).
\label{eq:5:12.3.51} \end{equation}
Notice, however, that if we calculate straightforwardly, the expectation value
of $P_iP_j$ contains an additional contribution,
\begin{eqnarray}
(n'|(P_iP_j)_{{\rm eff}}|n)&=&\langle n'|P_iP_j|n\rangle 
= (n'|(P_iP_j -A_iP_j -A_jP_i + S_{ij})|n), \label{eq:5:3.52} \\
S_{ij} &=& -\int dq\, [0\!:\!Q|q](\partial_{Q^i}\partial_{ Q^j}
 [q|0\!:\!Q]. \label{eq:5:12.3.53} 
\end{eqnarray}
Since $S_{ij}$ is symmetric, 
it is straightforward to see that the
canonical commutation relations among the collective variables projects
without change.

It is useful to rewrite (\ref{eq:5:3.52}) in a form that makes contact
with standard results. We have as a definition and as a generalization of
(\ref{eq:5:12.3.51})
\begin{eqnarray}
(i\partial_j)[q|\nu\!:\!Q]& =&\sum_{\nu'}
{}[q|\nu'\!:\!Q](A_j(Q))_{\nu'\nu}, \label{eq:5:13.3.55} \\
(A_j)_{\nu'\nu}&=&i\int dq\,[\nu'\!:\!Q|q]\left(\partial_{Q^j}[q|\nu\!:\!Q]\right).
\label{eq:5:3.550}
\end{eqnarray}
Because of the $Q$ dependence of the matrix element, it now follows that
\begin{eqnarray}
S_{ij}&=& \sum_\nu (A_i)_{0\nu}(A_j)_{\nu 0} +(i\partial_i)A_j 
= (i\partial_i)A_j +A_iA_j +S^{\prime}_{ij}, \label{eq:5:3.56} \\
(A_j)_{00} &=&A_j , \;\;
S^{\prime}_{ij} = \sum_{\nu\neq 0}(A_i)_{0\nu}(A_j)_{\nu 0} \label{eq:5:3.58}
\end{eqnarray}
Consequently, we may also rewrite Eq.~(\ref{eq:5:3.52}) as
 \begin{equation}
(n'|(P_iP_j)_{{\rm eff}}|n) = (n'|(D_iD_j + S^{\prime}_{ij})|n).
 \label{eq:5:3.59} \end{equation}
In the simplest case, where $[0\!:\!Q|q]$ is a real
wave function, it follows that $A_i$ vanishes, but the contribution
$S^{\prime}_{ij}$ remains to be taken into account.

We are now in a position to apply
to the computation of the collective kinetic energy
the same reasoning as just carried out for a product of momentum
operators. Making use
of the analogue of (\ref{eq:5:3.59}), the result is
 \begin{equation}
\frac{1}{8}\{D_i ,\{D_j ,B^{ij}(Q)\}\} +\frac{1}{2} S^{\prime}_{ij}B^{ij}.
\label{eq:5:13.2.21} \end{equation}
where the second term can be incorporated into the
collective potential energy as a second such contribution,
$\Delta^{(2)}\bar{V}(Q)$.
An additional contribution of this type is
obtained by setting $q^a=0$ in Eq.~(\ref{eq:5:13.1.2}), 
\begin{equation}
\Delta^{(3)}\bar{V}(Q)= 
\bar{V}_{\mathrm{quant}}(Q,0), \label{eq:5:13.1.20} 
\end{equation}
where $\bar{V}_{\mathrm{quant}}$ is the quantum potential defined in
Eqs.~(\ref{eq:5:13.2.19}) and (\ref{eq:5:13.1.2}).

It remains for us to discuss the contributions from $T$ that depend
on $\bar{B}^{ai}(Q)$ that ``mix" the collective and non-collective indices
and those that depend on the non-collective mass tensor $\bar{B}^{ab}(Q)$.
The former can be neglected because of the decoupling condition
that the transformed mass tensor does not mix collective and non-collective
spaces. The leading contribution of the latter is seen
to be another contribution to the potential energy, 
\begin{equation}
\Delta^{(4)} \bar{V} =
\frac{1}{2} \bar{B}^{ab}(Q)\langle p_a p_b\rangle_Q,
\label{eq:5:13.2.26} \end{equation} 
where the average is that defined in Eq.~(\ref{eq:5:13.2.13}).

To summarize our findings, we have derived the following effective
Hamiltonian, \begin{equation}
H_{{\rm eff}} = \frac{1}{8}\{D_i ,\{D_j ,\bar{B}^{ij}(Q)\}\} + \bar{V}(Q)
+\Delta \bar{V}(Q), \label{eq:5:13.2.27} \end{equation}
where $\Delta \bar{V}(Q)$ is the sum of four terms that summarize the leading
quantum corrections including the coupling to the fast variables,
\begin{equation}
\Delta \bar{V} = \sum_{i=1}^4 \Delta^{(i)} \bar{V}, \label{eq:5:13.2.28} 
\end{equation}
given respectively in or in relation to Eqs.~(\ref{eq:5:13.2.13}),
(\ref{eq:5:13.2.21}), (\ref{eq:5:13.1.20}), and (\ref{eq:5:13.2.26}).

Let us contrast this result with the corresponding form appropriate to
the more familiar adiabatic treatment of the structure of molecules.
The introduction of a curved metric aside, the main difference is that
in the molecular case, the ground state wave function of the fast
variables, for a fixed value of $Q$, may be assumed known, and its
eigenvalue, $\epsilon_0 (Q)$, together with $\Delta^{(2)} \bar{V}$,
contained in Eq.~(\ref{eq:5:13.2.21}) constitutes the collective
potential energy \cite {81,82}. In the present instance we cannot
assume that we know the Hamiltonian of the fast variables, except in
an approximate sense that we now discuss.  Instead of specializing the
transformed Hamiltonian operator to the values $q^a=p_a=0$, as in the
classical limit, we now retain terms up to second order in these
variables. In this treatment, we may replace
$\bar{V}_{\mathrm{quant}}(Q,q)$ by $\bar{V}_{\mathrm{quant}}(Q,0)$,
since this term is already a small correction, and the remaining terms
linear in the fast variables may be dropped because of the decoupling
conditions. To the specified accuracy, we obtain the following
approximate quantum Hamiltonian for the full space, 
\begin{subequations}
\begin{eqnarray}
H&=&H_{\mathrm{C}} +\bar{V}_{\mathrm{quant}}(Q) +H_{\mathrm{NC}}, 
\label{eq:5:13.2.50} \\ 
H_{\mathrm C} &=&\frac{1}{8}\{P_i,\{P_j,\bar{B}^{ij}(Q,0)\}\} + \bar{V}(Q,0),
\label{eq:5:2.50a}\\ 
H_{\mathrm{NC}}&=&\frac{1}{2} p_a p_b \bar{B}^{ab}(Q)+
\frac{1}{2} q^a q^b \bar{V}_{ab}(Q).
\label{eq:5:13.2.51} 
\end{eqnarray}
\end{subequations}
The justification for this form is that upon application of the BO
procedure, it results in the $H_{{\rm eff}}$ that has been derived.

For each value of $Q$, the Hamiltonian $H_{\mathrm{NC}}$ represents a standard
normal mode problem. Let $c_\alpha,c^{\dag}_\alpha$ be normal mode
destruction and creation operators, $\Omega_\alpha$ the corresponding
frequencies, and $\hat{n}_\alpha=c^{\dag}_\alpha c_\alpha$. Assuming
local stability, i.e., $\Omega_\alpha$ real and positive, we have
\begin{equation}
H_{\mathrm{NC}}(Q)=\sum_\alpha \left(\hat{n}_\alpha(Q) +\half\right)\Omega_\alpha(Q).
\label{eq:5:13.2.60}
\end{equation}
The practical importance of the contribution of (\ref{eq:5:13.2.60}) has already
been noted in the applications described in Secs.~\ref{sec:3.2} and \ref{sec:3.3},
especially the latter, and also in Sec.~\ref{sec:4.3.6}.

The quantum Hamiltonian, expressed in terms of the optimum choice of
variables, but in a restricted approximation, has thus been found.
Except for the term $\bar{V}_{\mathrm{quant}}(Q)$, it has been
expressed in terms of elements that can be readily calculated within
the GVA or local RPA.  Evaluation of $\bar{V}_{\mathrm{quant}}(Q)$
requires calculation of the curvature of the basic vectors (quantities
such as $f^\mu_{,\alpha\beta}$). For good decoupling, this term should
be small and will be omitted from further consideration.

We return to a discussion of the correction terms, $\Delta \bar{V}$ by
which $H_{{\rm eff}}$, Eq.~(\ref{eq:5:13.2.27}), differs from
$H_{\mathrm C}$.  We have just discarded $\Delta^{(3)} \bar V$. The
sum $\Delta^{(1)} \bar V + \Delta^{(4)} \bar V$ has been seen in
(\ref{eq:5:13.2.60}) to be a sum of oscillator terms in the
approximation considered, its contribution then depending, naturally,
on the state of motion of the fast variables. In this section we have
considered only the lowest energy state for these variables, so that
the contribution of this term is just the zero-point energy.

It remains for us to discuss the contribution of the term
$\Delta^{(2)} \bar V$, the Berry scalar potential, associated with the
projection of the kinetic energy onto the collective subspace. This
contribution goes together with that from the ``vector potential",
$A_i$.  The manner in which both the vector and scalar potentials
contribute is best studied within the context of the examples worked
out in Sec.~\ref{sec:5.3} that follows.

\begin{aside}
In terms of the original coordinates, let us consider the change in the
potential energy between two neighboring points, $\xi$ and $\xi +\delta
\xi$. To second order in $\delta\xi$ we have the usual terms of a
Taylor expansion, \begin{eqnarray}
\Delta V &=& V(\xi +\delta\xi) - V(\xi) \nonumber \\
&=& V_{,\alpha}\delta\xi^\alpha +\frac{1}{2} V_{,\alpha\beta}\delta\xi^\alpha
\delta\xi^\beta, \label{eq:5:B.1} \end{eqnarray}
that now appears most unsatisfactory, since $\Delta V$ is a scalar,
but the second term of (\ref{eq:5:B.1}) contains the ordinary rather than
the covariant second derivative. This defect is removed by replacing
the quantity $\delta\xi^\alpha$ by substitution from the relation
\begin{equation}
d\xi^\alpha = \delta\xi^\alpha +\frac{1}{2}\Gamma^\alpha_{\beta\gamma}
\delta\xi^\beta\delta\xi^\gamma, \label{eq:5:B.2} 
\end{equation}
that contains the Christoffel symbol, 
\begin{equation}
\Gamma^\alpha_{\beta\gamma} =\frac{1}{2} {B}^{\alpha\delta}
({B}_{\delta\beta,\gamma} + {B}_{\delta\gamma,\beta}
-{B}_{\beta\gamma,\delta}). \label{eq:5:B.3}
\end{equation}
We thus obtain the form
\begin{equation}
\Delta \bar{V} = V_{,\alpha}d\xi^\alpha +\frac{1}{2} V_{;\alpha\beta}
d\xi^\alpha d\xi^\beta, \label{eq:5:B.4}
\end{equation}
where the second term contains the covariant derivative
\begin{equation}
V_{;\alpha\beta} = V_{,\alpha\beta} -\Gamma^\delta_{\alpha\beta}V_{,\delta}.
 \label{eq:5:B.5}
\end{equation}
It is now apparent from (\ref{eq:5:B.4}) that $d\xi^\alpha$ are the
components of a vector, and therefore transformation to any alternative
set of coordinates such as the $q^\mu$ is standard.

This allows us to calculate the quantity $\bar{V}^{(2)}$ of
Eq.~(\ref{eq:5:13.2.13}) from given dynamical quantities, provided we
can define a complete set of coordinate axes at each point of the
decoupled manifold. This is done in two stages.  In the first stage,
the basis vectors at each point of the tangent space to the collective
submanifold, for example the set $f^i_\alpha$, are determined by the
algorithm that discovers the collective submanifold. A set of basis
vectors $f^a_\alpha$, orthogonal to the tangent space is then
determined (non-uniquely) by the requirement that these be orthogonal
to the $f^i_\alpha$ and to each other with respect to the metric
$\tilde{B}^{\alpha\beta}$.  The basis vectors orthogonal to the
tangent space are precisely the elements needed to compute the
Hamiltonian of the fast variables.
\end{aside}

\subsection{Examples with Berry phase\label{sec:5.2}}

In this subsection we shall discuss a few applications of our formalism.
First we shall show the role of the Berry phase in a model
where we have degenerate excited states in the non-collective subspace.
In the next subsection we shall extend the model to one where the ground state
is influenced by the Berry potentials. We shall compare the spectrum
of the collective Hamiltonian with that of an exact diagonalization.

Another application of interest is based on a model of Letourneux and
Vinet \cite{84,85} that contains two interacting modes, one representing
the low energy quadrupole mode and the other the higher energy dipole mode.
We have carried through our own discussion of this model in Ref.~\cite{29}
but have chosen not to include it in this review.

\subsubsection{Model with Berry phase in excited states\label{sec:5.2.1}}

It is simplest to illustrate the main points by choosing models in
which the collective coordinates have already been identified, so that
we need not enter into the intricacies of the theory of large amplitude
collective motion {\em per se}. For instance, in the first model studied 
below, there are no terms in the Hamiltonian linear in the fast variables. 
Thus the model satisfies the decoupling conditions exactly. The 
models worked out in this section come from Ref.~\cite{25}.
We study the Hamiltonian 
\begin{subequations}
\begin{eqnarray}
H&=& H_{{\rm core}} + H_{{\rm sp}} + H_{{\rm int}} 
=H_{{\rm core}} + H_{{\rm NC}} \label{eq:5:5.1} \\
H_{{\rm core}}&=& \frac{\vec{P}^2}{2M} + V(Q), \label{eq:5:5.2} \\
H_{{\rm sp}}&=& \omega(Q)(a_1^{\dag}a_1 +a_2^{\dag}a_2 +1), \label{eq:5:5.3}\\
H_{{\rm int}}&=& -G(Q_-\, a_1^{\dag}a_2 +Q_+\, a_2^{\dag}a_1). \label{eq:5:5.4}
\end{eqnarray}
\end{subequations}
Here the coordinates $\vec{Q}$ and the canonical momenta $\vec{P}$ 
are both two-dimensional
vectors. We use the notation $\vec{Q}=(Q_1,Q_2)$, $Q_{\pm}=Q_1\pm iQ_2$, and
$Q=(Q_1^2 +Q_2^2)^{1/2}$. (In this section, we are not maintaining the 
distinction between upper and lower indices.)
Furthermore, the $a_i,a_i^{\dag},i=1,2$
are boson destruction and creation operators, $G$ is a coupling strength,
and the frequency, $\omega(Q)$, of the uncoupled boson modes has been given
a so far unspecified dependence on $Q$ that will be chosen for analytic and
numerical convenience. Note that the
Hamiltonian, (\ref{eq:5:5.1}) conserves the boson number,
\begin{equation}
N = a_1^{\dag}a_1+ a_2^{\dag}a_2 = {\rm constant}. \label{eq:5:5.5}
\end{equation}

Since the $N=0$
problem is completely trivial for this model, the first interesting case
is $N=1$. Here, the state vectors may be written exactly as a superposition 
\begin{equation}
|n\rangle = \int d\vec{Q}\,\{(\vec{Q}|n1)a_1^{\dag}|0]
+ (\vec{Q}|n2)a_2^{\dag}|0]\}, \label{eq:5:5.6} 
\end{equation}
where $|0]$ is the vacuum state. The use of square brackets for the
vacuum state of the fast variables is consistent with the notation
introduced earlier. The resulting eigenvalue equation in the space of
the collective variables, $\vec{Q}$, is determined by the two-by-two
effective Hamiltonian, $H_{\mathrm{eff}}$, with matrix elements
\begin{subequations}
\begin{eqnarray}
(H_{{\rm eff}})_{11}&=&H_{{\rm core}} +2\omega(Q)=(H_{{\rm eff}})_{22}, \\
(H_{{\rm eff}})_{12}&=&-GQ_-=(H_{{\rm eff}})_{21}^{\ast}. \label{eq:5:5.7}
\end{eqnarray}
\end{subequations}
Below we shall describe exact solutions of the associated Schr\"{o}dinger
equation.

These exact solutions are to be compared with the adiabatic approximation,
with and without the Berry potential terms. For this approximation, we
require the normal modes of $H_{\mathrm{NC}}$, which we calculate in a standard way
from the equations of motion, \begin{eqnarray}
[a_1,H_{\mathrm{NC}}]&=& \omega a_1 - GQ_- a_2, \nonumber \\
{}[a_2,H_{\mathrm{NC}}]&=& \omega a_2 - GQ_+ a_1, \label{eq:5:5.8} \end{eqnarray}
by forming the matrix elements,
\begin{equation}
\psi_i = [0|a_i|\Psi]. \label{eq:5:5.9}
\end{equation}
With $\Omega$ representing the energy of the state $|\Psi]$, ,
we obtain the equations \begin{eqnarray}
\Omega \psi_1 &=& \omega \psi_1 -GQ_-\psi_2, \nonumber \\
\Omega \psi_2 &=& -GQ_+\psi_1 +\omega\psi_2, \label{eq:5:5.10} \end{eqnarray}
that yield the eigenvalues \begin{eqnarray}
\Omega^{(1)}(Q) &=& \omega(Q) - GQ, \nonumber \\
\Omega^{(2)}(Q) &=& \omega(Q) + GQ, \label{eq:5:5.11} \end{eqnarray}
that are degenerate when $Q=0$.
The associated normalized solutions of (\ref{eq:5:5.10}) are represented most
conveniently by introducing the normal-mode creation operators 
\begin{equation}
b_i^{\dag} = \psi^{(i)}_j a_j^{\dag}. \label{eq:5:5.12}
\end{equation}
We make the explicit choice
\begin{subequations}
\begin{eqnarray}
b_1^{\dag} &=& \frac{1}{\sqrt{2}}a_1^{\dag}+
\frac{1}{\sqrt{2}}\exp{i\phi(\vec{Q})}a_2^{\dag}, \label{eq:5:5.13} \\
b_2^{\dag} &=& \frac{1}{\sqrt{2}}\exp{i\phi(\vec{Q})}a_1^{\dag} -
\frac{1}{\sqrt{2}}a_2^{\dag}, \label{eq:5:5.14} \\
\tan\phi(\vec{Q})&=&Q_2/Q_1. \label{eq:5:5.15}
\end{eqnarray}
\end{subequations}

We apply these elementary results to the adiabatic approximation.
In this case we represent a suitable subset of the eigenfunctions
(\ref{eq:5:5.6}) in the form
\begin{equation}
|n\rangle = \int d\vec{Q}\, (\vec{Q}|n)b_1^{\dag}|0\!:\!Q], \label{eq:5:5.16}
\end{equation}
where the notation $|0\!:\!Q]$ refers to the vacuum for the normal modes.
(For the current model, it coincides with the uncoupled vacuum.)
The considerations of Sec.~\ref{sec:5.1.3} now apply to this 
class of state vectors and, in particular, we apply 
Eq.~(\ref{eq:5:3.59}). As a special case of this equation, we have 
\begin{eqnarray}
\sum_i (P_i)^2 &\rightarrow& \sum_i (P_i -A_i)^2 +\sum_i |(A_i)_{21}|^2,
\label{eq:5:5.160} 
\end{eqnarray}
\begin{subequations}
\begin{eqnarray}
A_i &=& i[0\!:\!Q|b_1\partial_i b_1^{\dag}|0\!:\!Q], \label{eq:5:5.17} \\
(A_i)_{21} &=& i[0\!:\!Q|b_2\partial_i b_1^{\dag}|0\!:\!Q], \label{eq:5:5.18}
\end{eqnarray}
\end{subequations}
With the help of Eqs.~(\ref{eq:5:5.13}) and (\ref{eq:5:5.14}), the quantities
of interest are found to take the values \begin{eqnarray}
\vec{A} &=& \frac{1}{2Q^2}(Q_2,-Q_1), \label{eq:5:5.19} \\
\sum_i|(A_i)_{21}|^2 &=& (1/4Q^2). \label{eq:5:5.20} 
\end{eqnarray}
As expected these results are singular for $Q=0$, which is where the two states 
become degenerate and the adiabatic approximation breaks down.
The collective or adiabatic Hamiltonian, $H_{\mathrm{C}}$, that thus emerges from
the assumption that the state vectors of interest can be written in the
form (\ref{eq:5:5.16}), has the structure 
\begin{equation}
H_{\mathrm{C}} = (1/2M)[(\vec{P}-\vec{A})^2 +(\vec{A}_{21})^2] +{V}(Q)
+(3/2)\Omega^{(1)}(Q)+(1/2)\Omega^{(2)}(Q). \label{eq:5:5.21} 
\end{equation}

We turn to the problem of solving the associated eigenvalue problem.
We wish to compare the results of exact (numerical) calculations
with the eigenvalues of the collective Hamiltonian. It is useful
to choose $\omega(Q)$ such that the latter is exactly solvable.
One such choice is
\begin{equation}
\omega(Q) = \frac{1}{2} GQ .
\end{equation}
In this case the additional contribution to the potential
is zero,
\begin{equation}
\frac{3}{2}\Omega^{(1)}(Q)+\frac{1}{2} \Omega^{(2)}(Q)
= 0.
\end{equation}
We further take $M=1$ and $V(Q)=\frac{1}{2}Q^2$.
We use the definition of the two-dimensional angular momentum,
\begin{eqnarray}
L& =&- i \frac{\partial}{\partial \phi} = -i (Q_1 P_2-Q_2P_1), 
\end{eqnarray}
to simplify the ${\vec P}\cdot{\vec A}$ term. 
We can further simplify the equation
$H_c \psi(\vec Q) = E \psi(\vec Q)$
by substituting
\begin{equation}
\psi(\vec Q) = Q^{-1/2} \chi(Q) {\rm e}^{im\phi}
\end{equation}
and find
\begin{equation}
\left\{
-\frac{1}{2} \frac{d^2}{d Q^2}
+ \frac{1}{2Q^2}\left[m^2+m+\frac{1}{4}\right]
+\frac{1}{2} Q^2\right\} \chi(Q) =
E\chi(Q).
\end{equation}
Even though the centrifugal term is different, this is still
very similar to the radial equation for the two-dimensional harmonic
oscillator. It can be solved by the substitution
\begin{equation}
\chi(q) = q^{\alpha+1/2} {\rm e}^{-q^2/2} L_n^{(\alpha)}(q^2),
\end{equation}
As is well known, the Laguerre function $L_n^{(\alpha)}$ satisfies the equation 
\begin{equation}
- \frac{d^2}{d q^2} L_n^{(\alpha)}
+(q^2+\frac{1-4\alpha^2}{4q^2})L_n^{(\alpha)} = (4n+2\alpha+2)L_n^{(\alpha)}.
\end{equation}
We thus find that
\begin{eqnarray}
\alpha(m) &=& \sqrt{1/4+(m+1/2)^2} , \\
 E_{nm} &=& (2n+\alpha(m)+1). \label{eq:5:EnmB}
\end{eqnarray}
Without the Berry's phase terms we would have found
\begin{equation}
 E_{nm} = (2n+|m|+1).
\label{eq:5:EnmnoB}
\end{equation}
When $G=0$ the solution (\ref{eq:5:EnmnoB}) is exact.
Thus Eq.~(\ref{eq:5:EnmB}), which is independent
of $G$, cannot be valid for all $G$. It should be
valid in the adiabatic limit, which means that the two
frequencies $\Omega$ must be very different. This
occurs when $G$ is very large.

The exact solution of the problem can be calculated using
a harmonic oscillator basis in circular coordinates for the $Q$ degrees
of freedom.
This is coupled to a state containing either one $a^\dagger_1$ or
one $a^\dagger_2$ boson,
\begin{equation}
|n,m,n_1,n_2\rangle = |n,m) (a^\dagger_1)^{(n_1)}(a^\dagger_2)^{(n_2)}|0],
\;\;n_1+n_2=1.
\end{equation}
The interaction Hamiltonian only couples states with $n_1=1,m=m_1$
to states with $n_2=1,m=m_1+1$, but also changes the radial quantum
number of the oscillator. We can thus choose the value of $m_1$.
Since we
wish to obtain accurate results for large $G$, we allow the number of $Q$
harmonic oscillator quanta to be fairly large. We have used
up to 200 harmonic oscillator states in each block (which leads
to a 400 by 400 matrix eigenvalue problem).

In Table \ref{tab:5:1} we give a selected set of results
for $m_1=0$ and a number of values of $G$. 
Similar results for $m_1=1$ are listed in Table \ref{tab:5:2}.
We clearly see the
convergence to the collective model results including the
Berry phase for large $G$.

\begin{table}
\caption{The lowest 10 eigenvalues of the coupled problem for
$m_1=0$ as a function of $G$. The column labeled Berry lists the
eigenenergies of the collective Hamiltonian (\ref{eq:5:5.21}).
\label{tab:5:1}}
\begin{center} 
\begin{tabular}{rrrrrrrr}
G &
0.01 & 0.1 & 1 & 10 & 100 & 1000 &Berry\\
\hline
$E_1$ &
1.00875 & 1.07849 & 1.36229 & 1.58463 & 1.66532 & 1.69302 & 1.7071\\
$E_2$ &
2.01328 & 2.13134 & 2.92007 & 3.48844 & 3.63505 & 3.68298 & 3.7071\\
$E_3$ &
3.01540 & 3.14492 & 4.09426 & 5.39473 & 5.60870 & 5.67438 & 5.7071\\
$E_4$ &
4.01827 & 4.18125 & 5.30157 & 7.29413 & 7.58443 & 7.66658 & 7.7071\\
$E_5$ &
5.01997 & 5.19061 & 6.57000 & 9.17666 & 9.56145 & 9.65932 & 9.7071\\
$E_6$ &
6.02221 & 6.22068 & 7.66964 & 11.0262 & 11.5393 & 11.6525 & 11.7071\\
$E_7$ &
7.02367 & 7.22768 & 8.90918 & 12.8133 & 13.5177 & 13.6459 & 13.7071\\
$E_8$ &
8.02558 & 8.25425 & 10.0230 & 14.4981 & 15.4965 & 15.6396 & 15.7071\\
$E_9$ &
9.02687 & 9.25972 & 11.1883 & 16.0926 & 17.4754 & 17.6335 & 17.7071\\
$E_{10}$ &
10.0286 & 10.2840 & 12.3410 & 17.7119 & 19.4544 & 19.6276 & 19.7071\\
\end{tabular}
\end{center}
\end{table}

\begin{table}
\caption{The lowest 10 eigenvalues of the coupled problem for
$m_1=1$ as a function of $G$. The column labeled Berry lists the
eigenenergies of the collective Hamiltonian \protect{(\ref{eq:5:5.21})}.
\label{tab:5:2}}
\begin{center} 
\begin{tabular}{rrrrrrrr}
\hline
G &
0.01 & 0.1 & 1 & 10 & 100 & 1000 &Berry\\
\hline
$E_1$ &
 2.01308 & 2.11333& 2.41566& 2.54921& 2.57601& 2.58042& 2.58114\\
$E_2$ &
 3.01670 & 3.17369& 4.20360& 4.51926& 4.57057 & 4.57959& 4.58114\\
$E_3$ &
 4.01807 & 4.16340& 5.44213& 6.48766& 6.56479& 6.57866& 6.58114\\
$E_4$ &
 5.02086 & 5.21512& 6.43129& 8.45380& 8.55880& 8.57766 & 8.58114\\
$E_5$ &
 6.02202 & 6.20307& 7.86356& 10.4165& 10.5527& 10.5766& 10.58114\\
$E_6$ &
 7.02436 & 7.25000& 8.79958& 12.3739& 12.5464& 12.5755& 12.58114\\
$E_7$ &
 8.02538 & 8.23691& 10.1070& 14.3218& 14.5400& 14.5744 & 14.58114\\
$E_8$ &
 9.02745 & 9.28063& 11.1975& 16.2512& 16.5336& 16.5733 & 16.58114\\
$E_9$ &
 10.0284 & 10.2669& 12.3141& 18.1346& 18.5270& 18.5721& 18.58114\\
$E_{10}$ &
 11.0302 & 11.3082& 13.5364& 19.8627& 20.5204& 20.5709 & 20.58114\\
\hline
\end{tabular}
\end{center} 
\end{table}

\subsubsection{Model with ground-state correlations\label{sec:5.2.2}}

We next study a model differing from the one just investigated only in
the form of the interaction. In this model Eqs.~(\ref{eq:5:5.1})-(\ref{eq:5:5.3})
stand as given, but Eq.~(\ref{eq:5:5.4}) is modified to \cite{25} 
\begin{eqnarray}
H_{{\rm int}} &=& -G_0 Q(a_1^{\dag}a_2 + a_2^{\dag}a_1) 
 -G_1(Q_+ a_1 a_2 +Q_- a_1^{\dag}a_2^{\dag}). \label{eq:5:5.22} 
\end{eqnarray}
The second term of this interaction
spoils the conservation of boson number used to simplify the solution of the 
previous problem. For the present problem, it is already interesting to study 
the spectrum when the fast variables are in their ground levels, since there
 are now non-trivial ground-state correlations and associated non-trivial 
Berry potentials, that we shall calculate.

We first turn to the study of the BO approximation. As before we need 
     the normal modes of the fast variables, as determined from the
equations of motion 
\begin{eqnarray}
[a_1,H_{\mathrm{NC}}] &=& \omega a_1 -G_0Q a_2 -G_1Q_- a_2^{\dag}, \nonumber \\
{}[a_2,H_{\mathrm{NC}}] &=& \omega a_2 -G_0Q a_1 -G_1Q_- a_1^{\dag}, \label{eq:5:5.24}
\end{eqnarray}
and their hermitian conjugates. In terms of the definitions
\begin{eqnarray} b_i^{\dag}&=&\psi_j^{(i)}a_j^{\dag}-\chi_j^{(i)}a_j,
\label{eq:5:5.25} \\ \psi_j^{(i)}&=&[0\!:\!Q|a_j|\Psi^{(i)}],
\label{eq:5:5.26}\\ \chi_j^{(i)}&=&[0\!:\!Q|a_j^{\dag}|\Psi^{(i)}],
\label{eq:5:5.27}\\ |\Psi^{(i)}] &=&b_i^{\dag}|0\!:\!Q], \label{eq:5:5.28}
\end{eqnarray}
we obtain the eigenvalue equations \begin{eqnarray}
\Omega\psi_1&=&\omega\psi_1 -G_0Q\psi_2 -G_1Q_-\chi_2, \nonumber \\
\Omega\psi_2&=&\omega\psi_2 -G_0Q\psi_1 -G_1Q_-\chi_1, \nonumber \\
-\Omega\chi_1&=&\omega\chi_1 -G_0Q\chi_2 -G_1Q_+\psi_1,\nonumber \\
-\Omega\chi_2&=&\omega\chi_2 -G_0Q\chi_1 -G_1Q_+\psi_1.\label{eq:5:5.29}
\end{eqnarray} 
The physical eigenvalues are the positive roots of
\begin{equation}
\Omega^2(Q) = (\omega \mp G_0Q)^2 -G_1^2Q^2, \label{eq:5:5.30} 
\end{equation}
which are again degenerate at $Q=0$.
As for the first model studied,
we label these solutions $\Omega^{(i)},i=1,2$.
The corresponding amplitudes are determined from the equations of motion and
the normalization conditions 
\begin{equation}
|\psi_1|^2 +|\psi_2|^2 -|\chi_1|^2 -|\chi_2|^2 =1. \label{eq:5:5.31}
\end{equation} 
The simplest forms for the amplitudes are achieved by
repeated use of the eigenvalue condition (\ref{eq:5:5.30}). We thus find
\begin{eqnarray}
\psi_2^{(j)}&=&(-1)^{j+1}\psi_1^{(j)}={\rm real}, \nonumber\\
\chi_i^{(j)}&=&\exp(i\phi)\tilde{\chi}_i^{(j)}, \nonumber\\
\tilde{\chi}_2^{(j)}&=&(-1)^{j+1}\tilde{\chi}_1^{(j)}={\rm real},
\nonumber\\
\psi_1^{(j)}&=&
\frac{(\Omega^{(j)}+\omega)+(-1)^jG_0Q}
{\sqrt{2}[(\omega+\Omega^{(j)})^2+(-1)^j2G_0Q(\omega+\Omega^{(j)})
+(G_0^2-G_1^2)Q^2]^{1/2}},
\nonumber\\
\chi_1^{(j)}&=&
 (-1)^{j+1}f^{(j)} (Q)\psi_1^{(j)},
\nonumber\\
f^{(j)}&=&\frac{G_1Q}{(\Omega^{(j)}+\omega)+(-1)^jG_0Q}. \label{eq:5:5.37}
\end{eqnarray} 
Here $\phi$ is the polar angle for $Q$, $\phi ={\rm arctan}(Q_2/Q_1)$.

These results are thus available for application to the BO
approximation studied by means of the assumption 
\begin{equation}
|n\rangle =\int d\vec{Q}\,|\vec{Q}) (\vec{Q}|n)|0\!:\!Q]. 
\label{eq:5:5.39} 
\end{equation}
It follows that the Berry vector potential is given by the formula 
\begin{equation} 
A_i = i[0\!:\!Q|\partial_i|0\!:\!Q].
\label{eq:5:5.40} 
\end{equation}
To carry out this calculation, we need the form of the correlated vacuum state,
as given by the equation, see Ref.~\cite{83},
\begin{equation}
|0\!:\!Q] = {\cal N} \exp[\frac{1}{2} Z_{ij}a_i^{\dag}a_j^{\dag}]|0],
\label{eq:5:5.41} 
\end{equation}
where ${\cal N}$ is a real normalization factor, whose value we
shall not need,
$a_i |0]=0$, and $Z_{ij}$ is a generally complex array whose role will
be considered in more detail below. Using the constancy of
the norm of the correlated ground
state as a function of the values of $\vec{Q}$, we can rewrite (\ref{eq:5:5.40})
as 
\begin{equation}
A_i =\frac{1}{2} i \left\{[ 0\!:\!Q|\partial_i|0\!:\!Q]
- [\partial_i 0\!:\!Q|0\!:\!Q]\right\}, \label{eq:5:5.42}
\end{equation} 
which shows that the derivative of the (real) normalization factor
${\cal N}$ does not contribute. This leaves the result
\begin{equation}
A_i = \frac{1}{4}i\left[\langle a_k^{\dag}a_l^{\dag}\rangle\partial_i
Z_{kl} - \langle a_k a_l\rangle\partial_i Z_{kl}^{\ast}\right],
\label{eq:5:5.43} 
\end{equation}
where the averages, indicated by angular brackets,
are with respect to the correlated vacuum.

We next indicate how (\ref{eq:5:5.43}) can be evaluated in terms of the solutions
we have found above for the equations of motion. The quantities $Z_{kl}$
are solutions of the equations 
\begin{equation}
\psi_j^{(i)}Z_{jk}^{\ast} = \chi_k^{(i)}, \label{eq:5:5.44} 
\end{equation}
that follow from the definition $b_i |0\!:\!Q] =0.$ The expectation values
can also be evaluated with the help of the formulas 
\begin{equation}
a_i = \psi_i^{(j)}b_j +\chi_i^{(j)\ast}b_j^{\dag},\label{eq:5:5.45} 
\end{equation}
and their hermitian conjugates, that are the inverse to Eq.~(\ref{eq:5:5.25}).
 An elementary calculation now yields 
\begin{equation}
\langle a_k^{\dag} a_l^{\dag}\rangle = \langle a_k a_l\rangle^{\ast}
= \chi_k^{(j)}\psi_l^{(j)\ast}. \label{eq:5:5.46} 
\end{equation}
Thus for the vacuum Berry potential, we obtain the formula 
\begin{subequations}
\begin{equation}
A_i = \frac{1}{4}i\left[\chi_k^{(m)}\psi_l^{(m)\ast}\partial_i
Z_{kl} -\chi_k^{(m)\ast}\psi_l^{(m)}\partial_i Z_{kl}^{\ast}\right].
\label{eq:5:5.47} 
\end{equation}

Any other matrix element, $(A_i)_{\nu 0}$ of the Berry potential can be
calculated by similar techniques. In fact, the only non-vanishing elements
of this type are $(A_i)_{11,0}$, $(A_i)_{20,0}$, and $(A_i)_{02,0}$, where,
for example, $\nu=20$ means a state with two correlated bosons of type 1
and none of type 2. We quote formulas for these matrix elements that follow
from the same elementary techniques that furnished Eq.~(\ref{eq:5:5.47}).
We thus have 
\begin{eqnarray}
(A_i)_{11,0}&=&i[11\!:\!Q|\partial_i |0\!:\!Q] 
= -i[\partial_i\; 11\!:\!Q|0\!:\!Q]
= -i[0\!:\!Q|b_1\partial_i(b_2 )|0\!:\!Q]\} \nonumber \\
&=&-i 
\left[\sum_{j}\psi_j^{(1)\ast}\partial_i\chi_j^{(2)\ast}
-\chi_j^{(1)\ast}\partial_i\psi_j^{(2)\ast}\right].
\label{eq:5:5.48}
\end{eqnarray}
Similarly, remembering the normalization of the states, we find
\begin{equation}
(A_i)_{20,0}= \frac{1}{i\sqrt{2}}\left[\sum_{j}
\psi_j^{(1)\ast}\partial_i\chi_j^{(1)\ast}
-\chi_j^{(1)\ast}\partial_i\psi_j^{(1)\ast}\right].
\label{eq:5:5.490}
\end{equation}
\end{subequations}
Finally, the matrix element $(A_i)_{02,0}$ is obtained from (\ref{eq:5:5.490})
simply by replacing all superscripts 1 by 2.

In order to implement fully the BO approximation, it is necessary to
evaluate these matrix elements in terms of the explicit solutions of the
model displayed in Eqs.~(\ref{eq:5:5.37}).
It is also appropriate to remark here that in consequence of these
relations and Eq.~(\ref{eq:5:5.44}) we can write 
\begin{equation}
Z_{jk} = \exp(-i\phi)\tilde{Z}_{jk},\;\;\tilde{Z}_{jk}={\rm real} 
\label{eq:5:5.491} 
\end {equation}

The calculation of the Berry potentials as defined above is now 
straightforward. In terms of purely real quantities, we find, 
for example, 
\begin{equation}
A_i = (\partial_i \phi)[(\tilde{\chi}_1^{(1)}\psi_1^{(1)}+\tilde{\chi}_1^{(2)}
\psi_1^{(2)})\tilde{Z}_{11} + (\tilde{\chi}_1^{(1)}\psi_1^{(1)}
-\tilde{\chi}_1^{(2)}\psi_1^{(2)})\tilde{Z}_{12}]. \label{eq:5:5.492}
\end{equation} 
Using the real form of Eq.~(\ref{eq:5:5.44}), this
expression can be simplified to 
\begin{eqnarray}
\vec{A} &=& \vec{\partial}\phi[(\tilde{\chi}_1^{(1)})^2 +(\tilde{\chi}_1^{(2)})^2]
\nonumber\\
 &=& \frac{G_1^2}{2}\left[
\frac{1}{(\omega+\Omega^{(1)})(\omega+\Omega^{(1)}-2G_0Q)+(G_0^2-G_1^2)Q^2}
\right. \nonumber \\ && \left.
+\frac{1}{(\omega+\Omega^{(2)})(\omega+\Omega^{(2)}+2G_0Q)+(G_0^2-G_1^2)Q^2}
\right](-Q_2,Q_1)_i.
\label{eq:5:5.493} 
\end{eqnarray}
In contrast to the previous model the vector potential is not
singular at the origin, but take the finite value
\begin{equation}
\vec{A}(0) = \frac{G_1^2}{4 \omega(0)^2}(-Q_2,Q_1).
\end{equation}

Turning to the off-diagonal matrix elements, we find by similar calculations
\begin{subequations}
\begin{eqnarray}
(A_i)_{11,0}&=&0, \label{eq:5:5.494} \\
(A_i)_{20,0}&=& \sqrt{2}\exp(-i\phi)[(\partial_i\phi)\psi_1^{(1)}
\tilde{\chi}_1^{(1)} +i(\psi_1^{(1)}\partial_i\tilde{\chi}_1^{(1)}
- \tilde{\chi}_1^{(1)}\partial_i\psi_1^{(1)}]
\nonumber\\ & = &
-\sqrt{2}\exp(-i\phi)\left(\psi_1^{(1)}\right)^2 f^{(1)}
\left[\partial_i\phi+i(\partial_iQ) \partial_Q \ln f^{(1)} \right]
\nonumber\\ & = &
\sqrt{2}\exp(-i\phi)
\frac{G_1[(\Omega^{(1)}+\omega)-G_0Q]}
{2[(\omega+\Omega^{(1)})(\omega+\Omega^{(1)}-2G_0Q)+(G_0^2-G_1^2)Q^2]}
\nonumber\\ && \times
\left[\frac{1}{Q}(-Q_2,Q_1)+i \frac{1}{Q}(Q_1,Q_2) Q\partial_Q \ln f^{(1)} \right]
, \label{eq:5:5.495}\\ 
(A_i)_{02,0}&=& -\sqrt{2}\exp(-i\phi)
\frac{G_1[G_0Q+(\Omega^{(2)}+\omega)]}
{2[(\omega+\Omega^{(2)})(\omega+\Omega^{(2)}-2G_0Q)+(G_0^2-G_1^2)Q^2]}
\nonumber\\ && \times
\left[\frac{1}{Q}(-Q_2,Q_1)+i \frac{1}{Q}(Q_1,Q_2) Q\partial_Q \ln f^{(2)} \right].
 \label{eq:5:5.496}
\end{eqnarray}
\end{subequations}
These off-diagonal potentials are also regular as $Q$ goes to zero,
for reasonably well behaved $\omega(Q)$.

\begin{figure}
\centerline{\includegraphics[width=7cm]{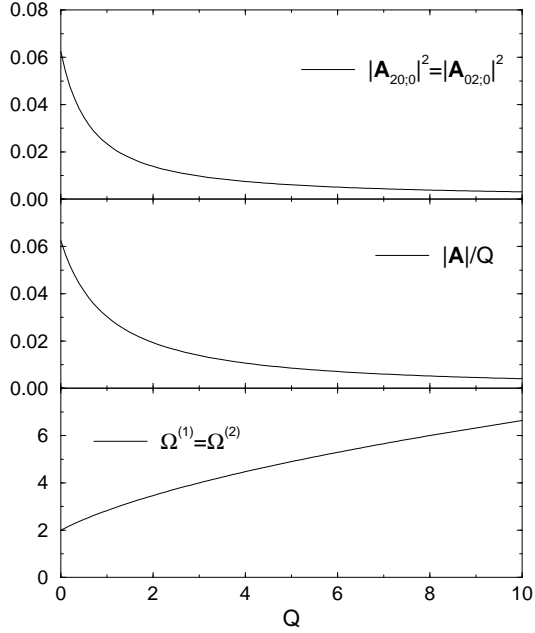}}
\caption{Parameters in the collective Hamiltonian for $G=1$, $\omega_0=2$. 
The lower panel shows the frequencies $\Omega^{(j)}$, the middle panel
shows the size of the diagonal Berry potential, and the
upper panel shows the square of the off-diagonal 
Berry-potentials.
\label{fig:5:G1om2}}
\end{figure}

We now study the numerical solution of the adiabatic Hamiltonian,
\begin{equation}
H_{\mathrm{C}} = \frac{1}{2}\left[(\vec{P}-\vec{A})^2 + |\vec{A}_{20,0}|^2 +
|\vec{A}_{02,0}|^2\right]
+\frac{1}{2}Q^2 + \frac{1}{2}\left[\Omega^{(1)}(Q)+\Omega^{(2)}(Q)\right],
\end{equation}
where we have once again made the choice $M=1,V(Q)=\frac{1}{2}Q^2$.
For simplicity we take $G_0=0$, and write $G_1=G$. 
For this special choice $\Omega^{(1)}=\Omega^{(2)}=\Omega$ and
$(A_i)_{02,0}=-(A_i)_{20,0}$.
A simple form for $\omega(Q)$, 
chosen such that $\Omega$ is positive definite,
is 
\begin{equation}
\omega(Q) = \omega_0+GQ.
\end{equation}
In Fig.~\ref{fig:5:G1om2}
we show some of the relevant quantities
in the collective Hamiltonian for the choice $G=1$, $\omega_0=2$. 

We diagonalize the collective Hamiltonian by first going to
circular coordinates, and obtain a radial equation using angular momentum
conservation, which holds obviously by construction for the exact
Hamiltonian but is also a property of the BOA.
We then map the $Q$ values
from the interval $[0,\infty]$ to the interval $[0,1]$. Finally
we make a finite difference approximation to the radial equation, and solve
the approximate equation by matrix diagonalization.

Solving the complete problem, without making the adiabatic approximation
is somewhat involved. Since the Hamiltonian is invariant under the 
interchange $a^\dagger_1 \leftrightarrow a^\dagger_2$, we introduce
new operators that are invariant under this parity transformation,
\begin{equation}
a^\dagger_\pm = \frac{1}{\sqrt 2} (a^\dagger_1 \pm a^\dagger_2).
\end{equation}
The interaction Hamiltonian takes the simple form
\begin{equation}
H_{{\rm int}} = -G_0 (a^\dagger_+a_+ -a^\dagger_-a_-) 
-\frac{G_1}{2} (Q_+[a_+^2-a_-^2] + Q_-[a_+^{\dagger2}-a_-^{\dagger2}]).
\label{eq:5:Hintnew}
\end{equation}
We now make the expansion
\begin{equation}
\Psi = \sum_{K,\kappa} \psi_{K,\kappa}(Q) 
\frac{\left(a^\dagger_+\right)^{2K-2\kappa}\left(a^\dagger_-\right)^{2\kappa}}
{\sqrt{(2K-2\kappa)!(2\kappa)!}}|0]
\end{equation}
(The even powers in this equation constitute the only combination
that includes the vacuum for the fast degrees of freedom.)

An approximate solution is obtained by limiting the sum over $K$, while 
summing over all allowed $\kappa$. At the same time we expand 
$\psi_{K,\kappa}(Q)$ in a finite number of
circular harmonic oscillator eigenfunctions
$\langle Q| nm \rangle$. 
We denote the $m$ value used in the expansion of
$\psi_{K,\kappa}(Q)$ by $m_{K,\kappa}$.
The interaction does not couple states of different $\kappa$ for fixed $K$.
We thus find that 
$m_{K,\kappa}=m_K=m_{K-1}-1$. Thus the value $m_0$ is 
a constant of the motion, and can be used to specify different
solutions. This quantity corresponds directly to the value of $m$ in the
collective Hamiltonian.
We have performed matrix diagonalizations for $\omega_0=2,10$ and $G=1$,
using harmonic oscillator states up to principal
quantum number 70, and $K$ values up to 20. 
(This corresponds to a 61,401 by 61,401 matrix.)
The resulting matrix,
which is very sparse, was diagonalized using a Lanczos algorithm.
The eigenvalues were checked for convergence by comparing
a calculation with smaller cut-offs on $n$ and $K$ against one
with larger cutoffs. In Tables \ref{tab:5:3} and \ref{tab:5:4}
we compare a few selected ground state energies of the complete
collective Hamiltonian with the exact solution. The splitting
between states with opposite values of $m$ is completely due
to the vector potential. For the case $\omega_0=2$, where the
adiabatic approximation is of questionable validity,
we see that the size of the splitting
is very close to the exact value. The difference is probably due 
to non-adiabatic effects that can not be completely neglected
for $\omega_0=2$. For the case $\omega_0=10$ the correspondence is
much closer, but the size of the 
Berry vector potential is much smaller as well.

\begin{table}
\caption{A comparison between the exact numerical ground state
energies and those for the collective Hamiltonian. $\omega_0=2,G=1$,
and the values of $m$ are listed in the table.\label{tab:5:3}}
\begin{center} 
\begin{tabular}{l|llll}
 & $m=-5$ & $m =5$ & $m=-1$ & $m=1$ \\
\hline
exact & 9.54715 & 9.73101 & 4.97309 & 5.03405 \\
collective & 9.56722 & 9.74656 & 4.98669 & 5.04375 
\end{tabular}
\end{center} 
\end{table}

\begin{table}
\caption{ A comparison between the exact numerical ground state
energies and those for the collective Hamiltonian. $\omega_0=10,G=1$,
and the values of $m$ are listed in the table.\label{tab:5:4}}
\begin{center} 
\begin{tabular}{l|ll}
 & $m=-5$ & $m =5$ \\
\hline
exact & 18.0745 & 18.0918 \\
collective & 18.0750 & 18.0922
\end{tabular}
\end{center}
\end{table}

\subsection{Born-Oppenheimer approximation and Berry phase: Fermi-Dirac
statistics\label{sec:5.3}}

\subsubsection{Introduction\label{sec:5.3.1}}

In this section, we consider the extension of the ideas presented
in Sec.~\ref{sec:5.1.3}
to the nuclear case. A short route that suggests itself is to follow
the scheme carried out in Sec.~\ref{sec:4.1} which is to transcribe previous results,
such as Eqs.~(\ref{eq:5:13.2.50}) - (\ref{eq:5:13.2.51}) to the nuclear case.
A major
objection to this course (among others) is that the small oscillations
Hamiltonian, $H_{\mathrm{NC}}$ of Eq.~(\ref{eq:5:13.2.51}) will not accurately describe
the first quantum corrections to the quantized classical Hamiltonian.
This was a principal result (quoted) in Sec.~\ref{sec:4.3.6}.
We therefore set for the goal of
this section the derivation {\it ab initio} of a quantum theory of
large amplitude collective, applicable to the nuclear case.
We shall again rely on the Born-Oppenheimer approximation (BOA),
but a major difference compared with
the previous considerations is that we shall
not introduce a localized basis for the intrinsic degrees of freedom.
The contents of this section represent a major revision of the ideas
presented in Refs.~\cite{28,29}.

\subsubsection{Definition of the effective Hamiltonian\label{sec:5.3.2}} 

We study the problem defined by the shell-model Hamiltonian 
\begin{equation} 
H = h_{\alpha\beta}a^{\dag}_\alpha a_\beta +\frac{1}{4}V_{\alpha\beta 
\gamma\delta}a^{\dag}_\alpha a^{\dag}_\beta a_\delta a_\gamma , 
\label{eq:5:13.2.1f}
\end{equation}
expressed in terms of the usual fermion creation (destruction) operators, 
$a^{\dag}_\alpha (a_\alpha)$ describing the shell-model orbit $\alpha$, 
and in terms of the hermitian one- and two-particle Hermitian matrices $h$ and 
$V$; the latter, as written, is also antisymmetric, separately, in the initial 
and in the final pair of indices. 

In order to carry out the program we have in view, we assume once more that
there exists a basis of states that is a product of localized states for
the collective coordinates and ``oscillator-like'' states for the intrinsic
coordinates, each complete in its own subspace
\begin{eqnarray}
|Q^i ,\nu\!:\!Q\rangle &=& |Q)|\nu\!:\!Q], \label{eq:5:13.2.1}\\
\sum_\nu |\nu\!:\!Q][\nu\!:\!Q|&=& 1_{\mathrm{NC}}, \label{eq:5:13.2.2} 
\end{eqnarray}
where $1_{\mathrm{NC}}$ is the unit operator in the non-collective space.
In terms of this basis set, which remains to be characterized in more 
detail, we assume the existence of a Hilbert space of collective states, 
$|n\rangle$, that can be represented in the general BO 
form
\begin{equation} 
|n\rangle=\sum_\nu \int dQ\,|Q)(Q|n\nu)|\nu\!:\!Q]. \label{eq:5:13.2.3}
\end{equation}
In contrast to the molecular case, or to specially chosen simple 
examples, it is generally not possible to specify ahead of time
the Hamiltonian of 
which the $|\nu\!:\!Q]$ are the state vectors.
Ultimately, as a consequence of the further development of the theory, 
we shall be able to obtain an approximate characterization 
of the space of the fast variables, adequate 
for most of our needs. In the usual BOA or adiabatic approximation, 
to which the considerations that follow 
pertain, we restrict the set $\nu$ in (\ref{eq:5:13.2.3}) to the ground-state
value $0$. Nevertheless, we shall see that the formalism forces the
excited intrinsic states back into the picture.
 
The next step is to derive an effective Hamiltonian for the 
collective subspace, defined within the BOA by the equation 
\begin{eqnarray} 
&&(n|H_{\rm eff}(Q,P)|n')\equiv\langle n|H|n'\rangle 
= \int dQdQ' (n|Q)[0\!:\!Q|(Q|H|Q')|0\!:\!Q'] (Q'|n'), \label{eq:5:13.2.5}
\end{eqnarray}
which has utilized the BOA in the form of Eq.~(\ref{eq:5:13.2.3}) with the
sum restricted to the single term $\nu=0$. 
We wish $H_{\rm eff}$ to depend only on the collective variables, $Q$, 
and their canonically conjugate momenta, $P$. The method is to make a
moment expansion of the matrix element $(Q|H|Q')$, which is still an
operator in the intrinsic space. We define
\begin{eqnarray} 
(Q|H|Q') &\equiv& K(\bar{Q},\tilde{Q}), \label{eq:5:13.2.40} \\
\bar{Q}&=& \frac{1}{2}(Q+Q'),~~~\tilde{Q}=Q-Q'. \label{eq:5:bartilde}
\end{eqnarray}
With the assumption that the matrix elements (\ref{eq:5:13.2.40}) are
peaked in the differences of collective coordinates,
we can carry out an expansion in terms of the delta function and
its derivatives with respect to such coordinates, 
\begin{eqnarray}
K(\bar{Q},\tilde{Q}) &=& K^{(0)}(\bar{Q})\delta(\tilde{Q})
+ K^{(1,i)}(\bar{Q})(-i\tilde{\partial}_i)\delta(\tilde{Q})
\nonumber\\&&
+\frac{1}{2} K^{(2,ij)}(\bar{Q})(-i\tilde{\partial}_i)(-i\tilde{\partial}_j)
\delta(\tilde{Q})+\cdots. \label{eq:5:13.2.41}
\end{eqnarray}
In this expression the various coefficient functions are examples of the
set of moments 
\begin{equation}
K^{(n,i_1\cdots i_n)}(\bar{Q}) =\int d\tilde{Q}(-i)^n \tilde{Q}^{i_1}\cdots 
\tilde{Q}^{i_n}K(\bar{Q},\tilde{Q}). \label{eq:5:moment} 
\end{equation} 
We expect the
moment expansion for the collective variables to generate a convergent 
series in powers of the reciprocal of the number of particles participating
in the collective motion. 

Inserting this expansion into Eq.~(\ref{eq:5:13.2.5}),
we easily succeed in identifying the operator
$H_{\rm eff}(Q,P)$. In recording its form we need first to define the following
 elements
\begin{eqnarray}
(ABC)_{00} &=&\sum_{\nu,\nu'} A_{0\nu}B_{\nu\nu'}C_{\nu' 0},
\label{eq:5:13.2.45}\\
(D_i)_{\nu\nu'} &=& P_i\delta_{\nu\nu'} - (A_i)_{\nu\nu'},\label{eq:5:13.2.46}\\
i\partial_i |\nu\!:\!Q] &=& \sum_{\nu'} |\nu'\!:\!Q](A_i)_{\nu'\nu},
\label{eq:5:13.2.47}
\end{eqnarray}
where (\ref{eq:5:13.2.47}) defines the Berry phase potentials. We thus
find that
\begin{eqnarray}
H_{\rm eff}&=& \bar{V}(Q) +\frac{1}{8}\{D_i,\{D_j,\bar{B}^{ij}\}\}_{00},
\label{eq:5:13.2.42}\\
\bar{V}(Q)&=& [0\!:\!Q|K^{(0)}(Q)|0\!:\!Q] =K^{(0)}_{00}, \label{eq:5:13.2.43} \\
\bar{B}^{ij}(Q)_{\nu\nu'} &=& [\nu\!:\!Q|K^{(2,ij)}(Q)|\nu'\!:\!Q].
\label{eq:5:13.2.44}
\end{eqnarray}
In reaching this result, we have assumed that the first moments vanish
by time reversal invariance.

\subsubsection{Relation of the effective Hamiltonian to matrix elements of
the density operator\label{sec:5.3.3}}

The aim of this section is to express the moments that occur in the effective
Hamiltonian (\ref{eq:5:13.2.42}) in terms of elements
that can be determined dynamically. 
Toward this end we apply the Kerman-Klein factorization method \cite{78}.
In this method, the evaluation of matrix elements of the Hamiltonian
of the type
$\langle n|H|n'\rangle$ is predicated on the following factorization of the 
two-body density matrix (summation convention)
\begin{eqnarray}
\langle n|a_\alpha^{\dag}a_\beta^{\dag}a_\delta a_\gamma|n'\rangle &\cong&
\frac{1}{2}[\langle n|a_\beta^{\dag}a_\delta |n''\rangle\langle n''|
a_\alpha^{\dag}a_\gamma|n'\rangle \nonumber \\
&&-(\beta\leftrightarrow \alpha)-(\delta\leftrightarrow \gamma)
+ (\beta\leftrightarrow \alpha,\delta\leftrightarrow\gamma)],
\label{eq:5:13.A.1}
\end{eqnarray} 
which is assumed to be valid when pairing correlations are neglected. The 
simplest qualitative arguments in favor of this approximate factorization
are: (i) it satisfies the Pauli principle; (ii) by suitable choice of the 
states $|n\rangle$, it leads to equations that preserve the symmetries 
broken by standard Hartree-Fock theory, namely, translation and rotation
invariance; (iii) it reduces to Hartree-Fock theory in the appropriate limit. 
It should be remarked, however, that there is a certain danger in taking the
factorization too literally, since there are certainly important neglected
two-particle correlations. Furthermore, the space of states to which it
can be applied must be
restricted, since including too large a space must lead eventually
to over-counting. At the present stage of development of the applications,
this does not represent an imminent danger.
The validity of the ensuing theory, as we shall observe,
only requires the accuracy of
two special averages of this factorization to be valid, the one carried out
with the matrix elements of the two-body potential that determines
energy matrix elements and the
equations of motion and the one that leads to Pauli principle restrictions
on the single particle density matrix.

In terms of the generalized density matrix elements,
\begin{eqnarray} 
\rho\langle\alpha n'|\beta n\rangle&=& \langle n|a_\beta^{\dag}
a_\alpha |n\rangle, \nonumber\\ 
&=& \int dQdQ' (n|Q)[0\!:\!Q|(Q|a_\beta^{\dag}a_\alpha|Q')|0\!:\!Q']
(Q'|n'), \label{eq:5:density2}
\end{eqnarray}
the application of (\ref{eq:5:13.A.1}) allows us to write
\begin{eqnarray}
\langle n'|H|n\rangle &=& h_{\alpha\beta}\,\rho\langle\beta,n|\alpha,n'\rangle
\nonumber \\
&&+ \frac{1}{2}V_{\alpha\beta\gamma\delta}\,\rho\langle\delta,n|\beta,n''
\rangle\,\rho\langle\gamma,n''|\alpha,n'\rangle. \label{eq:5:KKF}
\end{eqnarray}
By introducing a moment expansion analogous to (\ref{eq:5:13.2.41})
\begin{eqnarray}
(Q|a_\beta^{\dag}a_\alpha|Q') &=& [\rho^{(0)}_{\alpha\beta}(\bar{Q})
+\rho^{(1i)}_{\alpha\beta}(-i\tilde{\partial}_i) 
+\frac{1}{2}\rho^{(2ij)}_{\alpha\beta}(-i\tilde{\partial}_i)
(-i\tilde{\partial}_j)]
\delta(\tilde{Q}), \label{eq:5:13.3.1} \\
\rho^{(ni_1 ... i_n)}_{\alpha\beta}(\bar{Q}) &=&
\int d\tilde{Q} (-i)^n \tilde{Q}^{i_1}...\tilde{Q}^{i_n}(Q|a_\beta^{\dag}
a_\alpha|Q'), \label{eq:5:13.3.2}
\end{eqnarray}
a procedure imitating that which leads to (\ref{eq:5:13.2.42}),
now yields the equation
\begin{eqnarray}
\rho\langle\alpha n'|\beta n\rangle &=& \int dQ (n|Q)[
(\rho^{(0)}_{\alpha\beta})_{00} +\frac{1}{2}\{D_i,
(\rho^{(1i)}_{\alpha\beta}\}_{00} \nonumber \\
&& +\frac{1}{8}\{D_i,\{D_j,\rho^{(2ij)}_{\alpha\beta}\}\}_{00}](Q|n').
\label{eq:5:13.3.3}
\end{eqnarray}

We are now in possession of the tools necessary to calculate 
the moments that define the effective Hamiltonian in terms of the moments
of the density matrix. Toward this end, we insert Eq.~(\ref{eq:5:13.3.3})
into Eq.~(\ref{eq:5:KKF}), seeking to transform the result into a moment
expansion form, in order to identify the required elements. This leads
to difficulties associated with the fact that we have retained matrix
elements of the Berry potential involving excited intrinsic states.
This is inconsistent with our restricted BOA for the ground
band. Therefore, in continuing this discussion, we choose to
retain only matrix elements
in which all intrinsic states are unexcited. (We shall obtain the general
result below.) This means that we approximate
the matrix element of the product $(ABC)_{00}$ by the product of the
ground-state matrix elements $A_{00}B_{00}C_{00}$. By suitable integration
by parts and rearrangement, providing in essence a derivation of the
leading terms for the convolution theorem for the moment expansion, we can
obtain the desired result.
For cataloging purposes, we shall say that an $n{\em th}$ moment of the 
density matrix is of order $n$. In this spirit we record here only the 
contributions of order zero, one, and two to the corresponding moments
of the Hamiltonian matrix. We thus find
\begin{subequations}
\begin{eqnarray}
K^{(0)}(Q)_{00} &=& h_{\alpha\beta}\rho_{\beta\alpha}^{(0)}(Q)_{00}
+\frac{1}{2}V_{\alpha\beta\gamma\delta} 
\rho^{(0)}_{\gamma\alpha}(Q)_{00}\rho^{(0)}_{\delta\beta}(Q)_{00},
\label{eq:5:13.3.4} \\
K^{(1,i)}(Q)_{00} &=& h_{\alpha\beta}\rho_{\beta\alpha}^{(1,i)}(Q)_{00} 
\nonumber \\ &&+\frac{1}{2}V_{\alpha\beta\gamma\delta} 
[\rho_{\gamma\alpha}^{(0)}(Q)_{00}\rho_{\delta\beta}^{(1,i)}(Q)_{00} 
 + 
\rho_{\gamma\alpha}^{(1,i)}(Q)_{00}\rho_{\delta\beta}^{(0)}(Q)_{00}],
\label{eq:5:13.3.5}\\
K^{(2,ij)}(Q)_{00} &=& h_{\alpha\beta}\rho_{\beta\alpha} 
^{(2,ij}(Q)_{00}\nonumber \\
&& + \frac{1}{2}V_{\alpha\beta\gamma\delta} 
 \left[\rho_{\alpha\gamma}^{(1,i)}(Q)_{00} 
\rho_{\delta\beta}^{(1,j)}(Q)_{00} +\frac{1}{2} \rho_{\gamma\alpha}^{(0)}(Q)_{00} 
\rho_{\delta\beta}^{(2,ij)}(Q)_{00} 
\right. \nonumber\\&&\qquad \left.
 +\frac{1}{2}\rho_{\gamma\alpha}^{(2,ij)}(Q)_{00} 
\rho_{\delta\beta}^{(0)}(Q)_{00}\right] . \label{eq:5:13.3.6}
\end{eqnarray}
\end{subequations}
In addition to the terms recorded, there are higher order contributions
to each moment that arise naturally in the calculation. It is easily seen,
however, that the leading corrections vanish and thus the first terms that
contribute to each of the recorded expressions are two
orders smaller than those that have been displayed.

\begin{aside}
The appropriate generalization of the discussion given
above is obtained by requiring that there should be a set of 
collective bands, one for each intrinsic state included. Thus, we write
\begin{eqnarray}
|n\lambda\rangle &=& \sum_\nu\int dQ|Q)|\nu\!:\!Q](Q\nu|n\lambda)
\label{eq:5:13.3.7}
\end{eqnarray}
It follows that
\begin{eqnarray}
\langle n\lambda|H|n'\lambda'\rangle &=& \sum_{\nu\nu'}\int dQ
(n\lambda|Q\nu)H_{{\rm eff}}(Q,P)_{\nu\nu'}(Q\nu'|n'\lambda'),
\label{eq:5:13.3.8} \\
H_{{\rm eff}}(Q,P)_{\nu\nu'} &=& K^{(0)}(Q)_{\nu\nu'}
+\frac{1}{2}\{D_i, K^{1i}(Q)\}_{\nu\nu'} 
 +\frac{1}{8}\{D_i,\{D_j,K^{2ij}(Q)\}\}_{\nu\nu'}. \label{eq:5:13.3.9}
\end{eqnarray}
{\it Pro forma}, we have included the first moment term. 
Equation (\ref{eq:5:13.3.9}) should be compared with the limiting case
(\ref{eq:5:13.2.42})-(\ref{eq:5:13.2.44}). From this comparison, it becomes clear
how the formulas of this section will generalize, as soon as we specify that 
the generalization of Eq.~(\ref{eq:5:13.A.1}) is
\begin{eqnarray}
\langle n\lambda|a_\alpha^{\dag}a_\beta^{\dag}a_\delta a_\gamma|n'\lambda'
\rangle &\cong& \frac{1}{2}[\langle n\lambda|a_\beta^{\dag}a_\delta|n''\lambda''
\rangle\langle n''\lambda''|a_\alpha^{\dag}a_\gamma|n'\lambda'\rangle
\nonumber \\ && -(\beta\leftrightarrow\alpha) -(\delta\leftrightarrow\gamma)
+(\beta\leftrightarrow\alpha,\delta\leftrightarrow\gamma)]. 
\label{eq:5:13.3.10}
\end{eqnarray} 
\end{aside}

\subsubsection{Equations of motion; Pauli principle restrictions\label{sec:5.3.4}} 

For the remainder of this discussion, we limit our considerations to
the ground band. 
The collective Hamiltonian has been shown to be determined by the low-order
moments of the Hamiltonian matrix. The latter have been shown to be
determined by the low-order moments of the generalized density matrix.
In this section we exhibit the equations of motion from which the
latter may be computed. We may surmise that these must be related to
the decoupling conditions for large amplitude collective motion first
derived in Sec.~\ref{sec:2}. These conditions were not transcribed to the nuclear
many-body problem in Sec.~\ref{sec:4.1}, but rather only their consequences in the
form of the GVA and the LHA. Therefore this expectation will be investigated
and verified below in Sec.~\ref{sec:5.4}.

Our starting point here are the equations of motion for the
generalized density matrix as given in Ref.~\cite{78}. The technique is to
calculate the commutator $[a_\beta^{\dag}a_\alpha,H]$ in two ways and then
equate the results. Taking matrix elements between the states
$\langle n|$ and $|n'\rangle$, on the one hand we use the straightforward
evaluation by approximate sum over states, for example,
\begin{equation}
\langle n|a_\beta^{\dag}a_\alpha H|n'\rangle =
\langle n|a_\beta^{\dag}a_\alpha|n''\rangle\langle n''|H|n'\rangle.
\label{eq:5:13.4.1}
\end{equation}
Following the procedure that has been explained, using the BO representation
of the states and the moment expansions, we reduce the last expression to
the matrix element of an effective operator in the collective space.
For the second evaluation, we carry out the commutator explicitly and then
use the factorization (\ref{eq:5:13.A.1}). We find
\begin{eqnarray}
2\langle n|[a_\beta^{\dag}a_\alpha, H]|n'\rangle &=&
\left [
{\cal H}\langle\alpha n|\gamma n''\rangle \rho\langle\gamma n''|\beta n'\rangle 
+
\rho\langle\alpha n|\gamma n''\rangle {\cal H}\langle\gamma n''|\beta n'\rangle 
\right]\nonumber\\
&&- [(\alpha\gamma\leftrightarrow\gamma\beta)], \label{eq:5:13.4.2} \\
{\cal H}\langle \alpha n|\beta n'\rangle &=& h_{\alpha\beta}\delta_{nn'}
+V_{\alpha\gamma\rho\delta}\rho\langle\delta n|\gamma n'\rangle.
\label{eq:5:13.4.3}
\end{eqnarray}
This expression is also reduced to the matrix element
of an effective operator.

Useful consequences of these considerations are obtained by equating terms
of the same order in the moments. In this connection, the derivative
$\partial_i A^{(n)}(Q)$ is of the same order as the moment itself, and the
commutator $[\rho^{(n)},{\cal H}^{(n')}]$ is of order $n+n'+1$. We record
the first three orders with a notation that not only suppresses the
single-particle indices, but {\it every} quantity in the following
expressions is actually a $00$ matrix element in the sense defined, for
example, in Eq.~(\ref{eq:5:13.2.43}). We thus find
\begin{subequations}
\begin{eqnarray}
i\rho^{(1,i)}_{\alpha\beta}(Q)\partial_i \bar{V}(Q,q)&=& [\rho^{(0)},
{\cal H}^{(0)}_{\alpha\beta}], \label{eq:5:13.110} \\
-i[(\partial_j \rho^{(0)})\bar{B}^{ij} -\rho^{(2,ij)}\partial_j \bar{V}] &=& 
[\rho^{(0)},{\cal H}^{(1,i)}]+[\rho^{(1,i)},{\cal H}^{(0)}], 
\label{eq:5:13.111} \\
 i[ \rho^{(1,k)}\partial_k \bar{B}^{ij} -(\partial_k 
\rho^{(1,i)})\bar{B}^{jk} -(\partial_k \rho^{(1,j)}) 
\bar{B}^{ik}] & 
=& [\rho^{(0)},{\cal H}^{(2,ij)}] +[\rho^{(2,ij)},{\cal H}^{(0)}]\nonumber \\
&&+[\rho^{(1,i)},{\cal H}^{(1,j)}] +[\rho^{(1,j)},{\cal H}^{(1,i)}]. 
\label{eq:5:13.112}
\end{eqnarray}
\end{subequations}

The zeroth moment equation (\ref{eq:5:13.110}) has the form of a constrained
Hartree-Fock equation. We 
have not yet established, however, that solutions are to be sought in the 
space of Slater determinants. We shall establish this by looking at the
Pauli principle restrictions. As described in \cite{78}, the Pauli
principle restriction associated with the decomposition (\ref{eq:5:13.A.1}) is
\begin{equation} 
\rho\langle\alpha n|\beta n'\rangle =
\frac{1}{2}\rho\langle\alpha n|
\gamma n''\rangle\rho\langle\gamma n''|\beta n'\rangle
+\frac{1}{2}\rho\langle\gamma n|
\beta n''\rangle\rho\langle\alpha n''|\gamma n'\rangle. \label{eq:5:13.113}
\end{equation}
When the zeroth, first, and second moments of this equation are computed in 
lowest order approximation, these turn out to be indistinguishable from the 
well-known relations that follow from the equation $\rho^2=\rho$, evaluated
by an expansion, $\rho=\rho^{(0)} +\rho^{(1)} +\rho^{(2)} +\cdots$,
except that in
the relations below the superscripts refer to a moment of a distribution 
rather than to the order of smallness; in fact 
we have already made this identification. (They also refer to the
$00$ matrix element with respect to the intrinsic state.) The results are
\begin{subequations}
\begin{eqnarray} 
(\rho^{(0)})^2 &=& \rho^{(0)}, \label{eq:5:13.114} \\ 
\sigma^{(0)}\rho^{(1,i)}\sigma^{(0)}&=&\rho^{(0)}\rho^{(1,i)} 
\rho^{(0)}=0, \label{eq:5:13.115} \\ 
\rho^{(0)}\rho^{(2,ij)}\rho^{(0)}&=&-\rho^{(0)}(\rho^{(1,i)}\rho^{(1,j)} 
+\rho^{(1,j)}\rho^{(1,i)})\rho^{(0)}, \label{eq:5:13.116} \\ 
\sigma^{(0)}\rho^{(2,ij)}\sigma^{(0)}&=& 
\sigma^{(0)}(\rho^{(1,i)}\rho^{(1,j)} 
+\rho^{(1,j)}\rho^{(1,i)})\sigma^{(0)}, \label{eq:5:13.117} \\ 
\sigma^{(0)} &=& 1 - \rho^{(0)}. \label{eq:5:idem} 
\end{eqnarray} 
\end{subequations}
We note that the Pauli principle puts no restrictions on the $(ph)$
and $(hp)$ elements of $\rho^{(2,ij)}$. If we assume that these values
can be chosen to be zero, then it becomes easy to generalize the
pattern established in lowest order by the equations above. The odd
moments have only non-vanishing $(ph)$ and $(hp)$ elements (and are to
be determined dynamically), whereas the even moments have no such
elements and their $(pp)$ and $(hh)$ elements are determined by the
non-vanishing elements of lower order. This pattern turns out to be
correct for the representation of the density matrix that we use in
practice.
 
\subsection{Equivalence of equations of motion to decoupling conditions\label{sec:5.4}}

The purpose of this section is to show that
the first three moments of the equations of motion for the density matrix,
as embodied in Eqs.~(\ref{eq:5:13.110})-(\ref{eq:5:13.112}) are equivalent 
to the so-called decoupling conditions of large amplitude adiabatic collective
motion. These conditions, together with the canonicity condition,
revisited below,
provide the basis for the algorithms that have been applied in practice. 
The importance of this proof is that the algorithms in question determine
the ingredients of the collective Hamiltonian, called $H_{\rm eff}$ 
in this paper, namely the potential energy and the mass tensor. Since
the algorithms are formulated on the basis of a purely classical theory
of collective motion, this shows that by following the reasoning of 
this paper, we have derived a quantum theory of collective motion without
{\em ad hoc} requantization of the kinetic energy. In our formalism,
the symmetric or Weyl form of the kinetic energy is prescribed.

\begin{aside}
Below we shall need the canonicity condition in the form
 \begin{eqnarray}
 \frac{\partial Q^i}{\partial\rho_{hp}} &=&-i\frac{\partial\rho_{ph}} 
 {\partial P_i} \nonumber \\
&=& -i\rho_{ph}^{(1,i)} =-i\rho_{hp}^{(1,i)\ast}. \label{eq:5:13f}
\end{eqnarray}
That we have correctly identified the first moment of the density matrix
becomes evident if one calculates the Wigner transform of the density matrix. 
The moments are then the coefficients of an expansion
in powers of the momentum. 
\end{aside}

The form of the decoupling conditions that we shall study correspond to 
the case that the system has no constants of the motion in addition to 
the energy. Repeating, for convenience, the conditions in the form
\begin{subequations}
\begin{eqnarray} 
{ V}_{,\alpha}&=& \bar{V}_{,i}f^i_{,\alpha}, \label{eq:5:13.5.1} \\ 
{ B}^{\alpha\beta}f^i_{,\beta}&=& \bar{B}^{ij}g^\alpha_{,j}, 
\label{eq:5:13.5.2} \\ 
\bar{B}^{ij}_{,\alpha}&=&\bar{B}^{ij}_{,k}f^k_{,\alpha}.\label{eq:5:13.5.3}
\end{eqnarray}
\end{subequations}
In order to carry out a demonstration of the equivalence of
Eqs.~(\ref{eq:5:13.110})-(\ref{eq:5:13.112}) to
Eqs.~(\ref{eq:5:13.5.1})-(\ref{eq:5:13.5.3}),
it is convenient to consider the latter in the form appropriate for
complex canonical coordinates. 

We turn then to the derivation promised, first considering 
Eq.~(\ref{eq:5:13.110}). In view of the structure of this equation, 
only the $ph$ or $hp$ matrix elements are 
non-vanishing. We have, for example,
\begin{subequations}
\begin{eqnarray}
{\cal H}^{(0)}_{ph} &=&-i\rho^{(1,i)}_{ph}\bar{V}_{,i}. \label{eq:5:13.5.4}\\
{\cal H}^{(0)}_{ph} &=& 
(\partial V/\partial\rho_{hp}^{(0)}), \label{eq:5:13.5.5}
\end{eqnarray}
\end{subequations}
We can identify
(\ref{eq:5:13.5.4}) with (\ref{eq:5:13.5.1}) provided we take note of
(\ref{eq:5:13f}).
 
Consider next Eq.~(\ref{eq:5:13.111}). This equation has non-vanishing
$pp'$ and $hh'$ elements that will be considered first. Recalling the 
fact that $\partial_i\rho^{(0)}$ has only $ph$ and $hp$ non-vanishing 
elements and using Eqs.~(\ref{eq:5:13.116}) and (\ref{eq:5:13.117}) relating the 
$hh'$ and $pp'$ matrix elements of the second moment of the density 
matrix to the non-vanishing elements of the first moment, it is a 
straightforward exercise to see that the $pp'$ and $hh'$ elements of 
(\ref{eq:5:13.111}) are a consequence of the preceding equation,
(\ref{eq:5:13.110}). To study the $ph$ and $hp$ parts of (\ref{eq:5:13.111}) 
we need to know the quantities $\rho^{(2,ij)}_{ph}$ and 
$\rho^{(2,ij)}_{hp}$. We shall drop these quantities as a consequence 
of the argument given previously: There are no 
kinematical constraints on these quantities requiring them to be 
non-vanishing, and it also appears to be consistent with the 
dynamics to the order that we are working to do so. 
In any event, omitting these terms, we are left, for example, 
with the equation
\begin{eqnarray}
-B^{php'h'}(i\rho^{(1,j)}_{p'h'})&=&(\partial_j\rho^{(0)}_{ph})\bar{B}^{ij}, 
\label{eq:5:13.5.6} \\ 
B^{php'h'}&=&\frac{1}{2}({\cal H}^{(0)}_{pp'}+{\cal H}^{(0)}_{p'p})
        \delta_{hh'} \nonumber \\ 
&& -\frac{1}{2}({\cal H}^{(0)}_{h'h} +{\cal H}_{hh'}^{(0)}) 
\delta_{pp'} 
 +\frac{1}{2}(V_{ph'hp'} + V_{hp'ph'}
 -V_{pp'hh'}- V_{hh'pp'}) . \label{eq:5:13.5.7}
\end{eqnarray}
This result can be identified with (\ref{eq:5:13.5.2}). 
 
We turn finally to the analysis of the structure of Eq.~(\ref{eq:5:13.112}) 
The details are somewhat more tedious, and can be found in Appendix B of
Ref.~\cite{28}. There we show that the $ph$ elements
can be identified with Eq.~(\ref{eq:5:13.5.3}). 
Altogether we have shown that we can utilize the classical theory
developed previously and apply this theory to the nuclear case 
by means of the ``dictionary'' developed in Sec.~\ref{sec:4.1}. 
This dictionary is justified by the considerations of this section.

To summarize the theory that has been developed, we have placed the
quantization of the classical theory on a more solid footing and have,
in addition, included two sets of quantum corrections. In Sec.~\ref{sec:4.3}
we have shown how to include quantum corrections to the collective
potential energy arising from the oscillations of the non-collective
coordinates, and also given a nuclear application. In this section,
we have indicated how to include the effects of the Berry potential
arising within the BOA from the projection of the full Hamiltonian onto
the collective subspace. Though we have described several applications
of the non-nuclear theory developed in Sec.~\ref{sec:4.1.3}, we do not, to date,
have any applications of the new results of this section.

\newpage
 \section{Large amplitude collective motion at finite excitation energy}\label{sec:6}

 \subsection{Formal theory\label{sec:6.1}}

 \subsubsection{Introduction\label{sec:6.1.1}}

 The previous sections of this review were based on the assumptions,
 first, that the degrees of freedom of a many-particle system could be
 decomposed into two subsets, collective (slow, relevant) and
 non-collective (fast, irrelevant), and second, that we were at zero
 temperature. This last condition implies that the excitation energy,
 if any, remains concentrated in the collective degrees of freedom.  In
 this section we shall show how the previous theory can be generalized
 to a situation where the total amount of energy in the system is so
 high that the exchange of energy between the two subsets of degrees of
 freedom is inevitable.  The study of finite many-body systems at
 finite temperature or excitation energy is a complex problem with a
 long history. We shall not attempt to review it here, since much is
 not relevant for the present discussion.  Early work may be traced
 from Ref.~\cite{GR3}.  In broad terms, this work dealt with general
 quantum or semi-classical formulations of transport theory from which
 one sought to extract an extended one-body approximation that,
 minimally, included dissipation.  More recently, powerful algorithms
 designed to treat one and two-body correlations on an equal footing
 have been formulated and applied, using either a density matrix, i.e.,
 one-time Green's function \cite{C1,C2,C3} or a real-time Green's function
 approach \cite{RT1,RT2,C4}.  This approach, however, is not yet capable of
 dealing with problems of large amplitude collective motion.

 The aim of this section is to show how to extend the purely
 spectroscopic approach to collective motion, that has so far dominated
 this review, to the case that where exchange of energy between the
 collective and non-collective degrees of freedom can be described in
 the BOA.  The exchange itself is only treated approximately.  The
 account that follows is based on Ref.~\cite{31}.
 To the extent that the formulation is fully self-consistent and deals
 with large amplitude motion, it goes beyond the existing literature.
 In the respect that collisions are treated in two extreme scenarios
 only, it is less ambitious than some of the previous work.  The
 development that follows also contains a number of concepts distinct
 from anything encountered in the earlier sections of this review or in
 the literature, in particular the concept of mixed states introduced
 to extend the BOA to finite excitation energy and the way that the
 classical decoupling theory extends to finite excitations.  We shall
 again deal with the standard nuclear shell model Hamiltonian as we
 have done throughout this review.

 \subsubsection{Concept of thermal states\label{sec:6.1.2}}

 We begin with a discussion of the basis of states for the many-body
 system that will be utilized in the following developments, leading to
 the concept of a mixed state.  In the notation of Sec.~\ref{sec:5}, we
 assume that we can introduce a localized basis,
 \begin{equation}
   |Q,q\rangle = |Q)|q],   \label{eq:6:15.2}
 \end{equation}
 where $Q$ refers to the collective coordinates and $q$ to the
 non-collective coordinates.  As in Sec.~\ref{sec:5.3}, we shall adopt
 a generalized BO picture of the eigenstates of the Hamiltonian,
 designating them as $|n,\nu\rangle$.  For low excitation energies, we
 may assume that $n$ represents quantum numbers of a collective band
 and that $\nu$ is an ordering number for these bands that is
 associated with the state of excitation of the fast (non-collective)
 system.  For higher excitation energies, we shall continue to use the
 same designation, but its physical meaning is perhaps less clear,
 except when a simple spectrum (such as a set of harmonic oscillators)
 may be associated with the non-collective spectrum.  As before, we
 represent the states $|n,\nu\rangle$ in the form
 \begin{subequations}
 \begin{eqnarray}
   |n,\nu\rangle
   &=&
   \sum_{\nu'}\int dQ|Q,\nu'\rangle\langle Q,\nu'|n,\nu\rangle,
 \label{eq:6:15.3}  \\
   |Q,\nu\rangle
   &=&
   |Q)|\nu\!:\!Q]\equiv |Q)|\nu],
 \label{eq:6:15.4}  \\
   |\nu\!:\!Q]
   &=&
   \int dq\,|q][q|\nu\!:\!Q].
 \label{eq:6:15.5}
 \end{eqnarray}
 We have
 \begin{equation}
   \langle Q,\nu|Q',\nu'\rangle =\delta(Q-Q')\,\delta_{\nu\nu'}.
 \label{eq:6:15.6}
 \end{equation}
 \end{subequations}

 According to Eq.~(\ref{eq:6:15.4}), at $T=0$ we are dealing with a
 pure state with the full $N$-body density
 \begin{equation}
    \hat{D} =  |Q)|\nu]\,[\nu|(Q|.
 \label{eq:6:15.6b}
 \end{equation}
 At finite excitation energy, this is replaced by an incoherent average 
 in the intrinsic space yielding a mixed state
 \begin{equation}
    \hat{D}
    =
    |Q)\underbrace{\sum_{\nu\nu'} |\nu]d_{\nu\nu'}[\nu'|}_{\hat{d}}\,(Q| ,
 \label{eq:6:fullD}
 \end{equation}
 where the density in the intrinsic subspace has been abbreviated as
 $\hat{d}$. This is some arbitrary density at the moment. Below it will
 be specialized to a density valid for the characterization of the 
 BOA. Note that the development below depends for its
 validity only on the form of (\ref{eq:6:fullD}).
 A further specialization is to a
 density $\hat{d}(Q,T)$, describing a system to which a temperature
 can be assigned at all times, even if the system is not in overall
 equilibrium.  We refer to this case as local or constrained equilibrium.
 This will be one of the special cases treated in this development.

 We now introduce a construct that can be thought of as providing some
 quantum-mechanical underpinning for our ultimately classical  
 considerations.  We aim at a mixed description of coherent
 collective dynamics combined with an intrinsic state.
 We thus break the bilinear full density (\ref{eq:6:fullD}) back again
 into co- and contravariant linear pieces yielding a
 ``mixed state'', in analogy to the thermal states first introduced by
 Umezawa \cite{no71a},
 \begin{equation}
     |Q,d\rangle
     =
     |Q) \hat{d}^{1/2},
 \label{eq:6:halfD}
 \end{equation}
 where $\hat{d}^{1/2}$ is chosen as the Hermitian  square-root of the
 intrinsic density $\hat{d}$.
 We can also write this state in the form
 \begin{equation}
     |Q,d\rangle
     =
     |Q) \sum_{\nu\nu'} |\nu]a_{\nu\nu'}[\nu'|,
 \label{eq:6:dgoesa}
 \end{equation}
 and see that the coefficients $a_{\nu\nu'}$ therein are just the
 elements of $\hat{d}^{1/2}$. They fulfill the conditions
 \begin{equation}
    \sum_{\tilde{\nu}} a_{\nu\tilde{\nu}} a^*_{\nu'\tilde{\nu}}
    =
    d_{\nu\nu'}.
 \label{eq:6:agoesd}
 \end{equation}
 The state (\ref{eq:6:halfD}) has the
 property, essential for our needs, that its dual product reproduces the
 full density (\ref{eq:6:fullD}), i.e.,
 \begin{equation}
    \hat{D}
    =
    |Q,d\rangle \langle Q,d|
    =
    |Q)\hat{d}\,(Q|,
 \label{eq:6:repro}
 \end{equation}
 and with it all expectation values, products, etc.

 We explain in general outline
 the way in which the concept of thermal state will be exploited below.
 The problem that will arise is the evaluation of the classical limit of
 a matrix element $\langle Q,d|\hat{A}|Q'd'\rangle$.  With respect to
 the collective coordinate $Q$, we follow the well-worn path of introducing
 the Wigner transform.  Suppressing for the instant the intrinsic structure,
 we therefore compute
 \begin{eqnarray}
 A({Q},P) &=& \int d\tilde{Q}\exp(-iP\tilde{Q})
       \langle Q_1|\hat{A}|Q_2\rangle,  \label{eq:6:WT1} \\
 \tilde{Q} &=& Q_1-Q_2, \;\;
 {Q}  =\frac{1}{2}(Q_1+Q_2).
 \end{eqnarray}
 Provided that the ingredient matrix elements of $\hat{A}$ and $\hat{B}$
 are strongly peaked in $\tilde{Q}$ and slowly varying in ${Q}$,
 we then have for the classical limit of a product,
 \begin{eqnarray}
 C({Q},P) &=& \int d\tilde{Q}dQ_3\exp(-iP\tilde{Q})
 \langle Q_1|\hat{A}|Q_3\rangle\langle Q_3|\hat{B}|Q_2\rangle
  \cong A({Q},P)B({Q},P),  \label{eq:6:WT2}
 \end{eqnarray}
 i.e., the Wigner transform of a matrix product is approximately the
 product of the individual Wigner transforms.

 There remains the question of how we take the classical limit with
 respect to the variables $d$ that define the intrinsic behavior.  In
 the zero temperature (fully coherent) limit, this state is
 characterized by a set of non-collective coordinates $q$, usually
 treated in the small vibration limit.  In the classical limit, these
 become the conjugate pairs $(q,p)$.  The theory to be developed
 suggests that also at finite temperature one can extract a coherent
 set of non-collective variables.  There remains, however, an
 additional structure, to which we assign a label $f$, to be identified
 later, from the theoretical development as a set of occupation
 numbers.  Thus we write the thermal states as
 \begin{equation}
 |Q,d\rangle \equiv |Q,q,f\rangle.  \label{eq:6:WT3}
 \end{equation}
 These are still mixed states in the sense that we cannot form linear
 combinations of them.  It is furthermore assumed that for transition
 matrix elements between different thermal states, the value of $f$ is
 fixed, whereas $q$ shares the same behavior as $Q$.  Finally this
 means that for the full physics the quantity (\ref{eq:6:WT1}) is
 replaced by the quantity $A({Q},P,{q},p,f)$.  The way in which these
 constructs play their role will be seen in the following discussion.

 \subsubsection{Basic dynamical equations\label{sec:6.1.3}}

 What follows is an refined version of the Klein-Kerman approach
 described in Sec.~\ref{sec:5.3}.  We study the dynamics of the system
 starting with the equation of motion for the ``particle-hole''
 operator
 \begin{equation}
   \hat{\rho}_{\alpha\beta}
   =
   a_\beta^{\dag}a_\alpha,
 \label{eq:6:15.9}
 \end{equation}
 namely
 \begin{eqnarray}
   i\frac{d}{dt}\hat{\rho}_{\alpha\beta}
  &=&
   [\hat{\rho}_{\alpha\beta},\hat{H}] \nonumber \\
   &=&
   h_{\alpha\gamma}\hat{\rho}_{\gamma\beta}
   -\hat{\rho}_{\alpha\gamma}h_{\gamma\beta}
   +\frac{1}{2}V_{\alpha\gamma\delta\epsilon}a^{\dag}_\beta a^{\dag}_\gamma
   a_\epsilon a_\delta -\frac{1}{2}V_{\delta\epsilon\beta\gamma}a^{\dag}_\delta
   a^{\dag}_\epsilon a_\gamma a_\alpha.   \label{eq:6:15.10}
 \end{eqnarray}
 At the next stage we take the matrix element of this equation
 between two different
 thermal states (with the same value of $f$, see above).
 To evaluate matrix elements of two body operators, we introduce a
 generalized factorization (letting $Q$ stand for $Q,q,f$),
 \begin{eqnarray}
   \langle Q|a^{\dag}_\beta a^{\dag}_\gamma a_\epsilon a_\delta|Q'\rangle
   &=&
   \frac{1}{2}\{\langle Q|a^{\dag}_\gamma a_\epsilon|Q''\rangle
   \langle Q''|a^{\dag}_\beta a_\delta|Q'\rangle
  -(\epsilon\leftrightarrow \delta)
     -(\beta\leftrightarrow\gamma) \nonumber \\
      &&\quad +(\epsilon\leftrightarrow\delta,
     \beta\leftrightarrow\gamma)\}
    +c^{(2)}(\delta\epsilon Q'|\gamma\beta Q).    \label{eq:6:15.18}
 \end{eqnarray}
 This equation should first of all be viewed as a definition of
 $c^{(2)}$, the correlated part of the two-body density matrix.
 In all past applications, we have simply dropped this
 term, so that the resulting equation becomes an approximation,  the
 generalized density matrix or ``Kerman-Klein'' approximation.  For the  
 class of problems under study,
 we shall eventually have to go beyond these previous treatments,
 though in this presentation, we shall make simplifying assumptions that
 will allow us to postpone the problem of including $c^{(2)}$.

 Taking the Wigner transform of Eq.~(\ref{eq:6:15.18}) and applying
 the convolution theorem for a product in the approximate form (\ref{eq:6:WT2}),
 we obtain the equations
 that are fundamental to our further studies,
 \begin{equation}
   i\frac{d}{dt}\rho_{\alpha\beta}
   =
   [{\cal H},\rho]_{\alpha\beta}+{\cal I}_{\alpha\beta}.
 \label{eq:6:15.19}
 \end{equation}
 Here ${\cal H}$ is the mean field Hamiltonian
 \begin{subequations}
 \begin{eqnarray}
   {\cal H}_{\alpha\beta}(Q,P,q,p,f)
   &=&
   h_{\alpha\beta} +{\cal V}_{\alpha\beta}(Q,P,q,p,f),
 \label{eq:6:15.20}  \\
   {\cal V}_{\alpha\beta}(Q,P,q,p,f)
   &=&
   V_{\alpha\gamma\beta\delta}\rho_{\delta\gamma}(Q,P,q,p,f).
 \label{eq:6:15.21}
 \end{eqnarray}
 \end{subequations}
 The last term in (\ref{eq:6:15.19}), often designated simply as the
 collision term, represents, in fact, all the physics consequent upon
 the inclusion of the two-particle correlation function $c^{(2)}$,
 and thus is not included in
 the mean field approximation.  In this sense Eq.~(\ref{eq:6:15.19})
 is still exact.

 We have already alluded to the theoretical and practical advances that
 have been made in the study of the problem of the collision term
 \cite{C1,C2,C3,RT1,RT2,C4}.  Our purposes are best served by reference
 to the treatment of Ayik \cite{Ayik}.  He shows that in a
 weak-coupling approximation the diagonal part of the collision term
 (in the basis of natural orbitals discussed below) is the well-known
 collision term in the Boltzmann-Uhlenbeck (BU) equation, whereas the
 off-diagonal part provides the explicit two-body friction term in the
 equations of motion.  (It provides as well a change of the
 self-consistent field in the general non-Markovian situation.)

 In the development that follows, we have chosen to neglect the off-diagonal
 collision term.  This means that the remaining theory can only describe
 one-body friction, though we shall also later introduce, somewhat artificially,
 a limiting case of the effect of collisions. 

 \subsubsection{Density matrix in the basis of natural orbitals\label{sec:6.1.4}}

 Equation (\ref{eq:6:15.19}) is the basic dynamical equation
 from which we want to extract  the description of the collective motion
 in terms of the density matrix, $\rho(Q,P,q,p,f)$, that defines the
 collective manifold, where we shall shortly identify $f$ as a collection
 of occupation numbers.  We divide this
 procedure into several distinct steps.
 In the first step, carried out in this section, we introduce the description
 of the density matrix in terms of natural orbitals and thus identify the 
 elements that are singled out for further study.  The equations satisfied by
 these elements are determined by the application of Eq.~(\ref{eq:6:15.19}).
 We study what can be learned from this equation without specifying in detail
 the classical dynamics of the collective coordinates.
 In the representation
 of natural orbitals, the density matrix takes the form
 \begin{equation}
   \rho = \sum_a n_a |\varphi_a)(\varphi_a|,
 \label{eq:6:15.22}
 \end{equation}
 where the $|\varphi_a)$ are a complete set of orthonormal functions and the
 $n_a$ are correspondingly the occupation numbers for those orbits. As a
 consequence, we shall show in this section that the equations of motion
 can be decomposed into two subsets.  The first, which describes the
 time rate of change of the single-particle basis, will be seen to have the
 form of Hamilton's classical equations of motion, generalizing the
 previous considerations at zero temperature.   The second set, often
 called master equations, describes the rate of change of the occupation
 numbers brought about by collisions.

 To obtain equations for the elements contained in the form (\ref{eq:6:15.22}),
 namely, the single-particle
 wave functions and the occupation numbers, we study this form in
 conjunction with the equation of motion (\ref{eq:6:15.19}).  For instance,
 we would like to find
 the equation satisfied by the single-particle functions $|\varphi_a)$.
 Toward this end, it is convenient to consider
 an infinitesimal change in $\rho$. To conserve the norm of the single-particle
 functions, we have
 \begin{equation}
   \delta |\varphi_a)
   =
   \sum_{b\neq a}|\varphi_b)\delta r_{ba},
   \;\;\;\;
   \delta r_{ba}^{\ast}
   =
   - \delta r_{ab}.
 \label{eq:6:15.23}
 \end{equation}
 As a consequence, we can write
 \begin{eqnarray}
   \delta \rho
   &=&
   \sum_{b>a}|\varphi_b)\delta r_{ba}(n_a -n_b)(\varphi_a|  
   +\sum_{b>a}|\varphi_a)\delta r_{ba}^{\ast}(n_a -n_b)(\varphi_b|
   +\sum_a |\varphi_a)\delta n_a(\varphi_a|,
 \label{eq:6:15.24}
 \end{eqnarray}
 an expression that clearly separates off-diagonal and diagonal contributions.

 As already specified above, we study the off-diagonal pieces
 in an approximation which
 suppresses the off-diagonal elements of the collision term.  This
 approximation underlies our subsequent treatment of the off-diagonal
 elements of the equations of motion and is an essential simplification.
 {}From Eqs.~(\ref{eq:6:15.19}) and (\ref{eq:6:15.24}) we can therefore write
 (together with the complex conjugate relation)
 \begin{eqnarray}
  i\frac{d}{dt}r_{ba} &=& {\cal H}_{ba},    
   \;\;
   b> a,
 \label{eq:6:15.25}
 \end{eqnarray}
 where, with $W$ equal to the instantaneous Hartree-Fock energy,
 \begin{eqnarray}
   {\cal H}_{ba}
   &=&
   \frac{\delta W}{\delta \rho_{ab}}
   =
   \frac{\delta W}{\delta r_{ab}}\frac{1}{(n_b -n_a )},
 \label{eq:6:15.26} \\
   W
   &=&
   \trace h\rho +\frac{1}{2}\trace\trace \rho V\rho .
 \label{eq:6:WHF}
 \end{eqnarray}
 If we understand that the eigenvalues $n_a$ have been arranged in descending
 order and that henceforth we follow the convention $b>a$ (and therefore
 $n_a >n_b$), it follows that Eq.~(\ref{eq:6:15.25}) can be rewritten
 \begin{eqnarray}
   i\frac{ds_{ba}}{dt}
   &=&
   \frac{\delta W}{\delta s_{ba}^{\ast}},  \;\;
   \delta s_{ba} =
   \sqrt{n_a -n_b}\,\delta r_{ba}.    \label{eq:6:15.27}
 \end{eqnarray}
 These equations and their complex conjugates demonstrate that the
 off-diagonal elements of the equations of motion are of the form of
 Hamilton's canonical equations of motion for the complex 
 canonical coordinates $s_{ba}$ and $is^{\ast}_{ba}$.  

 Though this result is to some extent similar to what we have found
 previously at zero temperature, there are two essential
 differences.  The first is that the number of degrees of freedom of the 
 equivalent dynamical system is much larger than at zero temperature,
 comprising particle-particle and hole-hole pairs in addition to
 particle-hole pairs.  The second is that, as just shown, the Hartree-Fock
 energy serves as Hamiltonian only at fixed occupation number.
 It is now apparent that it is the occupation numbers that constitute
 the elements of the set $f$, hitherto unspecified, on which the density
 matrix elements depend.

 It remains for us to specify the dynamics of the
 occupation numbers.  According to Eqs.~(\ref{eq:6:15.19}) and (\ref{eq:6:15.22}),
 we have the equation
 \begin{equation}
 i\frac{dn_a}{dt} = {\cal I}_a(\vec{n},\vec{s}), \label{eq:6:occ1}
 \end{equation}
 wherein the right hand side, the ``collision term'' is specified as a
 function of the sets $\vec{n}=\{n_a\}$, $\vec{s}=\{s_{ba}\}$.  Together
 with (\ref{eq:6:15.27}) we have arrived at the following formulation:  Mean
 field theory with collisions is equivalent to two sets of equations for
 the elements of the one-particle density matrix in the basis of natural
 orbitals. One set, (\ref{eq:6:15.27}), describes the rates of change of the
 orbitals and is of Hamiltonian form.  The second, (\ref{eq:6:occ1}), is for the
 rates of change of the occupation numbers.  Together these independent
 elements define an initial value problem which describes the relaxation
 of a system of fermions initially perturbed away from equilibrium.

 A question that one may legitimately ask at this point is whether we would
 be completely stymied if we were to reinstate some approximation to the
 discarded two-body dissipation, because if this were the case the ultimate
 value of the current work would be seriously compromised.  We argue here
 that, in principle, there is no impediment to adding a friction term to
 Eq.~(\ref{eq:6:15.25}).  We can still identify canonical coordinates
 associated with the remaining parts of the equation, and later carry out
 the canonical transformation that will play such an essential role in
 separating collective from non-collective degrees of freedom.  The new terms
 will be carried along and transformed and ultimately add  (two-body) friction
 terms to the final equations, additional to those that we shall
 derive from the one-body friction concept.

 In Sec.~\ref{sec:6.1.6}, we shall study transformations from the set $\vec{s}$
 (or $\vec{r})$ to the preferred canonical set $(Q,P,q,p)$.  Imagine
 for the moment that this has been done.
 For fixed occupation numbers, we thus can consider $r=r(Q,P,q,p)$.
 Consequently we can write (\ref{eq:6:15.25}) in a more explicit form by using
 \begin{eqnarray}
 \frac{dr_{ba}}{dt} &=& [\dot{Q}\partial_Q +\dot{P}\partial_P
	+\dot{q}\partial_q +\dot{p}\partial_p]r_{ba}  
	\equiv \dot{z}\partial_z r_{ba}.   \label{eq:6:Par1}
 \end{eqnarray}
 This explicit form allows us to display the single particle equations
 that determine the functions $\varphi$.  The most general form consistent 
 with (\ref{eq:6:15.25}) and (\ref{eq:6:Par1}), from which the former may be derived, is
 \begin{equation}
    [\epsilon_a-\mu  +i\dot{z}\partial_z]|\varphi_a)
    = {\cal H}|\varphi_a).
 \label{eq:6:15.29}
 \end{equation}
 This has the form of a constrained Hartree-Fock equation with
 eigenvalue $\epsilon_a$, and $\mu$ is the chemical potential implying
 conservation of particle number.  Equation (\ref{eq:6:15.29}) can be
 derived from the variational principle
 \begin{eqnarray}
 &&  \delta[{\cal W} -\sum_a \epsilon_a n_a (\varphi_a|\varphi_a)]  =   0 ,
 \label{eq:6:15.36}   \\
 &&  {\cal W}  \equiv
    W  -\sum_a (\varphi_a|n_a [-\mu +i\dot{z}\partial_z ]|\varphi_a),
 \label{eq:6:15.33}
 \end{eqnarray}
 where ${\cal W}$ is the constrained mean field energy, $W$ is the
 mean-field energy defined in (\ref{eq:6:WHF}), and the variations 
 are carried out with respect to the  single-particle functions. 

 For the special case of instantaneous local equilibrium 
 of the single-particle
 degrees of freedom, the previous considerations may be supplemented  
 by the requirement that we maximize the mean-field entropy
 \begin{equation}
   S
   =
   -\sum_a n_a {\ln}(n_a)-\sum_a (1-n_a){\ln}(1-n_a),
 \label{eq:6:15.32}
 \end{equation}
 with respect to the choice of the occupation numbers
 $n_a$, subject to a fixed value for the constrained mean-field energy
 ${\cal W}$.
 This yields after a standard manipulation the expected result
 \begin{eqnarray}
   n_a &=& \{1+\exp[\beta(\epsilon_a-\mu)]\}^{-1}, \label{eq:6:15.35} \\
   \epsilon_a &=& (\varphi_a|{\cal H}|\varphi_a)
  = h_a +\sum_b V_{abab}n_b.   \label{eq:6:en2}
 \end{eqnarray}

 We consider finally the question of conservation of the mean-field energy.
 Using the equations of motion (\ref{eq:6:15.27}), we find easily
 \begin{eqnarray}
 \frac{dW}{dt} &=& \sum_a \epsilon_a \frac{dn_a}{dt}, \label{eq:6:en1}
 \end{eqnarray}
 Obviously $W$ is conserved if we ignore collisions ($dn_a/dt=0$).
 This is the limit that is the natural consequence of the assumptions
 that we have made.  On the other hand, it has been shown that the
 right hand side of (\ref{eq:6:en1}) vanishes (approximately) when one
 substitutes the usual form \cite{Ayik} of the master equation for
 $\dot{n}_a$.  Thus the total mean-field energy continues to be
 conserved approximately even in the presence of collisions, which are
 responsible, nevertheless, for the exchange of energy between the
 collective and non-collective degrees of freedom.  An alternative
 point of view that we shall investigate in some detail in the model
 study to be described in Sec.~\ref{sec:6.2}, is that we can simply set
 the right hand side of (\ref{eq:6:en2}) to zero, and view this as a
 condition that we impose on the motion.

 We shall utilize this point of view for the
 strong-collision limit, which we define as the one in which, in a
 relaxation process, the system passes only through states of local
 equilibrium.  As we shall point out in the discussion to be given
 in Sec.~\ref{sec:6.1.6},
 in this case the dynamic-thermal collective manifold is specified by values
 of $Q$ and $\beta$, and the combinations
 \begin{equation}
 {\cal E}_a(Q,\beta) \equiv \epsilon_a -\mu   \label{eq:6:15.400}
 \end{equation}
 will be determined as functions of $Q$ and $\beta$.
 Under these conditions, as we now compute, the conservation-of-energy
 condition will determine a relationship between $\dot{\beta}$ and $\dot{Q}$
 that will serve as one of the driving equations for the description of
 the relaxation process.  Thus, from the vanishing of the right hand side
 of (\ref{eq:6:en1}), by substituting (\ref{eq:6:15.35}), at the same time taking
 (\ref{eq:6:15.400}) into account, we obtain
 \begin{equation}
 \dot{\beta} = -\frac{\sum_a \beta{\cal E}_a \frac{\partial {\cal E}_a}
 {\partial Q}\frac{\exp(\beta{\cal E}_a)}{[1+\exp(\beta{\cal E}_a)]^2}}
 {\sum_a[{\cal E}_a^2 +\beta{\cal E}_a\frac{\partial{\cal E}_a}
 {\partial\beta}]
 \frac{\exp(\beta{\cal E}_a)}{[1+\exp(\beta{\cal E}_a)]^2}} \dot{Q}.
 \label{eq:6:15.401}                    \end{equation}
 In the strong collision limit, as we have defined it,  this equation
 replaces the entire panoply of rate equations for the occupation numbers.

 \subsubsection{Classical equations of motion with dissipation\label{sec:6.1.5}}

 A logical next step in the development might be the derivation of the
 canonical transformation to collective and intrinsic coordinates that
 provides optimum decoupling of the collective pairs $(Q,P)$ from the
 non-collective pairs $(q,p)$.  We postpone these consideration to the
 next section.  In this section, we shall assume that this
 step has been carried out.  We then use the resulting equations of motion
 to ``eliminate'' the non-collective degrees of freedom from the equations
 of motion leading to equations of motion for the collective coordinates
 that contain explicit dissipative terms.

 We thus turn to a derivation of the equations for a dissipative collective
 dynamics from ``first principles''.  
 This is also a subject with a long history.  For a recent
 review with extensive bibliography, see \cite{Abe}.  We shall not 
 attempt to do any justice to this topic, but simply present the material
 in the form that we require it.

 We suppose the full many-particle system, 
 with all degrees of freedom included,
 to be described classically by the Hamiltonian
 \begin{eqnarray}
 H &=& V(Q,q,p) + \frac{1}{2}B^{ij}(Q,q,p)P_i P_j    +A^i (Q,q,p)P_i.
 \label{eq:6:clham}     
 \end{eqnarray}
 So far,
 we have expanded only 
 in powers of the collective momenta.  The choice of variables in 
 Eq.~(\ref{eq:6:clham}) can be considered to result from a canonical 
 transformation at fixed occupation numbers 
 from the variables $s_{ba}$, $is_{ba}^\ast$
 that have been identified previously as canonical.  Thus Eq.~(\ref{eq:6:clham})
 can be viewed (to the second order in $P$) as formally equivalent
 to the original mean-field Hamiltonian
 $W$ for a fixed set of occupation numbers.  Further details concerning the
 conditions that determine the mapping
 \begin{equation}
 s_\alpha = s_\alpha(Q,P,q,p)   \label{eq:6:trans1}
 \end{equation}
 are discussed in the next section.

 In order to be able to present our arguments as explicitly as
 possible, we specialize the form (\ref{eq:6:clham}) to the small amplitude
 approximation in the intrinsic space.  In this approximation it becomes
 \begin{eqnarray}
 H &=& V(Q) +\frac{1}{2}B^{ij}(Q)P_i P_j  
  +\frac{1}{2}V_{,ab}(Q)q^a q^b +\frac{1}{2}B^{ab}(Q)p_a p_b \nonumber \\
 && +V_{,a}(Q)q^a + B^{ai}(Q)p_a P_i.   \label{eq:6:clham1}
 \end{eqnarray}
 Here $V_{,a}(Q)$ and $V_{,ab}(Q)$ 
 are first and second partial derivatives of $V$
 evaluated at $q=0$.  The problem of determining the potential and 
 mass coefficients that occur in this expression is precisely the problem of 
 determining the canonical transformation from the original form of 
 the mean field Hamiltonian.

 Limiting further discussion of the equations of motion to terms of the
 first order in $q$ and $p$, we thus have
 \begin{subequations}
 \begin{eqnarray}
 \dot{Q}^i &=& B^{ij}P_j +B^{ia}p_a,  \label{eq:6:cleom1} \\
 \dot{P}_i &=& -V_{,i} -V_{,ia}q^a, \label{eq:6:cleom2} \\
 \dot{q}^a &=& B^{ai}P_i +B^{ab}p_b,  \label{eq:6:cleom3} \\
 \dot{p}_a &=& -V_{,a} -V_{,ab}q^b.   \label{eq:6:cleom4}
 \end{eqnarray}
 \end{subequations}
 Without further loss of generality, we introduce normal coordinates in the 
 intrinsic space. This allows us to replace the quantities $B^{ab}$
 by unity and the quantities $V_{,ab}$ by $\omega_a^2$ and suppose that
 all remaining force and mass coefficients refer to the new coordinates.
 The distinction between covariant and contravariant indices now becomes
 irrelevant in the intrinsic space.  Henceforth all such indices will be 
 written as subscripts.

 At this point, we could 
 study the initial value or relaxation problem by adjoining
 the master equations (\ref{eq:6:occ1}) to (\ref{eq:6:cleom1})-(\ref{eq:6:cleom4}).
 For a system with a finite number of degrees of freedom, this might well
 be a practical and straightforward calculation.
 An application of interest would be to consider a set of initial
 conditions in which all the energy was concentrated in the collective
 degrees of freedom.  We would then ask how in the course of time
 this energy is transferred to the other degrees of freedom.  For a small
 number of coordinates overall, we would expect energy to re-concentrate
 from time to time in the collective coordinates.  As the number of
 non-collective coordinates increases, we expect this recurrence time to
 increase.  Beyond some point it becomes more sensible to talk about friction
 and a relaxation time. (For an analytically solvable model illustrating
 these concepts, see \cite{Fano}.)

 For the remainder of this discussion, we shall, however,
 consider the limit of
 very long recurrence time, where the concept of friction enters the
 description.  This case is treated by eliminating
 the intrinsic variables from the equations of motion
 for the collective variables.

 The arguments to be developed will be simplified considerably
 if we work only to the lowest non-trivial 
 order in $B^{ai}$, $V_{,a}$, and $V_{,ai}$. 
 This assumption is consistent with the procedure by which we shall define the
 canonical transformation from $s_\alpha$, $is_\alpha^{\ast}$ to $Q,P,q,p$,
 which assumes that for good decoupling to occur, these quantities
 must be small enough that hitherto they have been neglected.
 {}From Eqs.~(\ref{eq:6:cleom3}) and (\ref{eq:6:cleom4}), we thereby deduce the equation
 \begin{eqnarray}
 \ddot{q}_a &=& -\omega_a^2 q_a +X_a(Q), \label{eq:6:cleom5} \\
 X_a(Q) &=& -V_{,a} -B_a^i V_{,i},   \label{eq:6:defX}
 \end{eqnarray}
 with solution
 \begin{eqnarray}
 q_a(t) &=& q_a^{(h)}(t) +\int_0^t\, dt'\sin\omega_a(t-t')\frac{X_a(t')}
 {\omega_a},  \label{eq:6:solI} \\
 q_a^{(h)} &=& q_a(0)\cos\omega_a t +\frac{p_a(0)}{\omega_a}\sin\omega_a t.
 \end{eqnarray}
 Integrating by parts, we have
 \begin{eqnarray}
 q_a(t) &=& q_a^{(h)}(t) +\frac{X_a(t)}{\omega_a^2} 
 -\frac{X_a(0)}{\omega_a^2}\cos \omega_a t   \nonumber \\
 && -\int_0^t\, dt' \cos\omega_a(t-t')\frac{\partial_i X_a(t')}
 {\omega_a^2}\dot{Q}^i(t').   \label{eq:6:solHq}
 \end{eqnarray}
 For the intrinsic momenta, we can derive the corresponding expression
 \begin{equation}
 p_a = \dot{q}_a^{(h)}-B^{ai}P_i +\int_0^t\, dt'X_a(t')\cos\omega_a(t-t').
 \label{eq:6:solHp}                        \end{equation}
 The reader should take note that the quantities $X_a(t)$ depend on $t$,
 in general through their dependence on both the collective coordinates
 $Q^i(t)$ and the occupation numbers $n_a(t)$.  In the manipulations leading,
 for example, to Eq.~(\ref{eq:6:solHq}), we have taken the occupation numbers
 to be time independent, corresponding to the no-collision scenario.  The
 more complex situation where we lift this restriction will first be
 considered in connection with the model studied in Sec.~\ref{sec:6.2}.

 There is, naturally, no sign of irreversibility in these equations.
 We now consider the assumptions that will lead to a simple (Markov)
 description of dissipation.  For this purpose we must form the sums
 $V_{,ia}q^a$ and $B^{ia}p_a$ that appear in
 Eqs.~(\ref{eq:6:cleom2}) and (\ref{eq:6:cleom1}),
 respectively.  At this point we find it convenient to replace
 these first-order equations
 by second-order equations.  Working to only first order in $\dot{Q}^i$, we
 find with the help of Eqs.~(\ref{eq:6:cleom2}) and (\ref{eq:6:cleom4})
 \begin{subequations}
 \begin{eqnarray}
 \ddot{Q}^i &=& B^{ij}\dot{P}_j +B^{ia} \dot{p}_a
	 +\frac{\partial B^{ia}}{\partial Q_k}\dot{Q}^k p_a  \nonumber \\
 &=& F^i +  F^{ia} q_a -G^{ia}_k\dot{Q}^k p_a, \label{eq:6:so1} \\
 F_i &=& -B^{ij}V_{,j} - B^{ia} V_{,a},  \label{eq:6:so2} \\
 F^{ia} &=& -B^{ij}V_{,ja} -B^{ia}\omega_a^2, \label{eq:6:so3} \\
 G^{ia}_k &=& - \frac{\partial B^{ia}}{\partial Q^k}.  \label{eq:6:so4}
 \end{eqnarray}
 \end{subequations}

 We infer from Eq.~(\ref{eq:6:so1}) that $Q^i$ contains both the low frequencies of
 the collective motion and the high frequencies of the non-collective
 motion.  For a macroscopic description of the collective motion, it is
 appropriate to coarse-grain in time, in order to average over the effects
 of the high frequencies.  For this purpose, we assume that
 we can choose a time $\tau_O$, which relative to a time $\tau_{\mathrm{C}}$,
 characteristic
 of the collective motion, and a time $\tau_{\mathrm{NC}}$ characteristic of the
 intrinsic motion, satisfies the inequality
 \begin{equation}
 \tau_{\mathrm{NC}}<< \tau_0 << \tau_{\mathrm{C}}.  \label{eq:6:torder}
 \end{equation}
 Under these conditions, if we average Eq.~(\ref{eq:6:so1}) over a time interval
 $\tau_O$, the terms depending on the frequencies of the collective motion alone
 are essentially unaffected.

 The existence of times satisfying Eq.~(\ref{eq:6:torder}) is subject to serious
 question, particularly because of the influence of avoided level crossings
 as the system evolves in shape space, i.e., as it traverses the collective
 manifold.  But this problem also arises for the treatment of collective motion
 at zero temperature.  The answer lies not in the single-particle spectrum
 but in the spectrum of the local harmonic equation derived in Sec.~\ref{sec:6.1.6}.
 There is no basis for introducing the concept of collective motion for a
 system unless one or at most a few frequencies are low-lying and separated
 by a gap from the remaining frequencies.  This gap, in the nuclear case,
 is unlikely to be more than an order of magnitude under optimum conditions,
 not leaving much room, it appears, for $\tau_0$.  We believe that it is
 incorrect, however, to associate the reciprocal of this gap with $\tau_{\mathrm{NC}}$.
 To obtain a conventional friction tensor, it is necessary to assume a very
 broad distribution of frequencies characterizing the interaction between
 the collective and the non-collective coordinates, peaked at a frequency
 which is more properly identified with $(1/\tau_{\mathrm{NC}})$.  The following
 derivations require that this assumption on time scales be valid.

 To understand what happens to the high
 frequencies, we assume that $Q^i(t)$ can be written in the form
 \begin{eqnarray}
 Q^i(t) &=&  Q^i_0(t) + \delta Q^i(t).   \label{eq:6:hilo}
 \end{eqnarray}
 This decomposition is defined by the requirement that the coarse-grain
 value of $\delta Q^i$ vanish, which we write as
 \begin{equation}
 \langle \delta Q^i \rangle = 0.     \label{eq:6:tav1}
 \end{equation}
 On the other hand to the order of accuracy to which we shall solve
 Eq.~(\ref{eq:6:so1}),
 i.e., to second order in the coupling between the collective and 
 non-collective spaces we also need the value of 
 \begin{equation}
 \langle \delta Q^i(t) \delta Q^j(t)\rangle \neq 0.  \label{eq:6:neq}
 \end{equation}

 The arguments necessary to evaluate this quantity, described in
 Ref.~\cite{31}, will not be reproduced here.  These arguments, as well as
 the further arguments that are given below, transform Eq.~(\ref{eq:6:so1}),
 after coarse-graining, into the form
 \begin{eqnarray}
 \ddot{Q}^i &=& F^i +\delta F^i + F^i_{\mathrm{fluc}} -{\cal F}^i_j \dot{Q}^j,
 \label{eq:6:eomf}   
 \end{eqnarray}
 In this equation, the term $F^i_{\mathrm{fluc}}$ is the one that has its
 origin in the correlation (\ref{eq:6:neq}).  Its value is determined by the
 distribution of initial values of the intrinsic variables.  For the
 interesting case that these variables are initially unexcited this term
 vanishes.  The second term,
 \begin{equation}
 \delta F_i = \sum_a\frac{F^{ia}X_a}{\omega_a^2}, \label{eq:6:def101}
 \end{equation}
 representing an additional conservative force arising from the coupling
 of the collective coordinates to the low frequency part of the intrinsic
 motion, has its origin in the second term of (\ref{eq:6:solHq}).  (Working to
 first order in the collective velocity, there is no corresponding
 contribution from the second term of (\ref{eq:6:solHp}).)  Finally the friction
 term arises from a standard white-noise argument applied to the integrals
 in (\ref{eq:6:solHq}) and (\ref{eq:6:solHp}).

 \begin{aside}
 We review the argument that leads to the friction term of
 (\ref{eq:6:eomf}) by studying the term
 \begin{eqnarray}
 &&-\sum_a F^{ia}\int_0^t dt'\cos\omega_a(t-t')\frac{\partial_j X_a(t')}
 {\omega_a^2}\dot{Q}^j(t') \label{eq:6:avfa}  \\
 && \rightarrow -{\cal F}_{j1}^i\dot{Q}^j.    \label{eq:6:evalfa}
 \end{eqnarray}
 To extract such a frictional term without memory, we proceed as follows.
 We assume
 that we can replace the sum over $a$ by an integral over $\omega$,
 \begin{equation}
 \sum_a F^{ia}(t)\frac{\partial_j X_a(t')}{\omega_a^2} \rightarrow
 \int d\omega\Phi^i_j(\omega,t,t'),   \label{eq:6:phir}
 \end{equation}
 where $\Phi^i_j$, though necessarily an integrable distribution,
 is a very broad, slowly varying function of $\omega$.  For the purposes
 of evaluating the integral over $\omega$ (which we now do first),
 \begin{equation}
 I^i_j(t,t') \equiv \int_0^\infty
 d\omega \cos\omega(t-t') \Phi^i_j(\omega,t,t'),    \label{eq:6:int}
 \end{equation}
 we further assume that we can treat $\Phi^i_j$ as a function,
 $A^i_j(t,t')$, of the times alone (white noise assumption).
 To this approximation we have
 \begin{equation}
 I^i_j = \pi A^i_j(Q(t))\delta(t-t').
 \end{equation}
 {}From these considerations, we then conclude that 
 \begin{equation}
 {\cal F}^i_{j1}(Q) = \frac{1}{2}\pi A^i_j(Q),   \label{eq:6:calf1}
 \end{equation}
 where we have used the formula
 \begin{equation}
 \int_0^t dt'f(t')\delta(t'-t) =\frac{1}{2} f(t).
 \end{equation}

 We find a corresponding contribution from
 \begin{eqnarray}
  && -\sum_a G^{ia}_j \dot{Q}^j 
 \int dt' \cos\omega_a(t-t') X_a(t')   \nonumber  \\
 &&\rightarrow -{\cal F}^i_{j2}\dot{Q}^j. \label{eq:6:evalg2}
 \end{eqnarray}
 Here ${\cal F}^i_{j2}$ can be defined in analogy to ${\cal F}^i_{j1}$,
 Eq.~(\ref{eq:6:calf1}). 
 \end{aside}

 If we write Eq.~(\ref{eq:6:eomf}) in the form ($B_{ij}$ is the mass matrix
 inverse to $B^{ij}$)
 \begin{eqnarray}
 B_{ij}\ddot{Q}^j &=& -\partial_i {\cal V} -{\cal F}_{ij} \dot{Q}^j,
 \label{eq:6:eomff} \\
 {\cal F}_{ij} &=& B_{ik}{\cal F}^k_j  ,
 \end{eqnarray}
 thus defining the potential energy, ${\cal V}$, of the conservative forces,
 we can associate the expression
 \begin{equation}
 E_M =\frac{1}{2}B_{ij} \dot{Q}^i\dot{Q}^j +{\cal V}(Q)   \label{eq:6:meche}
 \end{equation}
 with the mechanical energy of collective motion.  The dissipative terms
 proportional to ${\cal F}$ lead to a loss of mechanical energy for which we
 obtain the standard energy-flow equation
 \begin{equation}
 \dot{E}_M = -{\cal F}_{ij} \dot{Q}^i\dot{Q}^j.    \label{eq:6:eflow}
 \end{equation}

 The quantities needed for the computation of the results developed in this
 section are all obtainable, in principle, from the theory explained below in
 Sec.~\ref{sec:6.1.6}.
 Given $Q^i(0)$ and $\dot{Q}^i(0)$, Eq.~(\ref{eq:6:eomff}) allows us to discuss the
 relaxation of the collective mechanical degrees of freedom.
 Here we must
 remember that the macroscopic parameters in this equation also depend
 on the occupation numbers, that we have assumed to be independent of time,
 as required by the mean-field approximation.
 The other case considered, in which the
 system relaxes through a sequence of states of local equilibrium
 will be studied in detail for
 the model described in Sec.~\ref{sec:6.2}.  Further discussion of these scenarios
 will also be found below, in relation to the problem of constructing the
 appropriate collective manifold associated with each of these cases.

 \subsubsection{Decomposition of mean-field Hamiltonian into collective and
 non-collective parts\label{sec:6.1.6}}

 The considerations of the previous subsection were based on transformation
 of the mean-field Hamiltonian into the form (\ref{eq:6:clham1}) together with
 the associated equations of motion (\ref{eq:6:cleom1})-(\ref{eq:6:cleom4}).  In this 
 section we describe how this transformation can be effected.  The 
 procedure is an extension of the calculations carried out 
 at zero temperature, as we shall describe.
 As has already been emphasized, the transformation will be based
 on the identity
 \begin{equation}
 H(Q,P,q,p,n_a) =W(s_\alpha(Q,P,q,p),s_\alpha^{\ast},n_a). \label{eq:6:equiv}
 \end{equation}

 In principle, we could adopt any of the methods developed in Sec.~\ref{sec:2}.
 In practice the considerations of this section will be based on the LHA
 without curvature, described in Sec.~\ref{sec:2.3.4}. 
 The procedure is to re-express the equations of motion (\ref{eq:6:15.27}) 
 by writing for the left hand side (as we have done before)
 \begin{equation}
 \dot{s}_\alpha =[\dot{Q}^i\partial_{Q^i}+\dot{P}_i\partial_{P_i}
 +\dot{q}^a\partial_{q^a} +\dot{p}_a\partial_{p_a}]s_\alpha
 \label{eq:6:rewrite1}      
 \end{equation}
 and then substituting in this expression the equations
 of motion (\ref{eq:6:cleom1})-(\ref{eq:6:cleom4}).
 On the right hand side of Eq.~(\ref{eq:6:15.27}), we expand in powers of
 $P^k q^l p^m$, with
 $k+l+m \leq 1$ and then equate corresponding powers on both sides.
 The resulting set of equations contains the density matrix non-linearly
 and its first derivatives linearly.  For the determination
 of the off-diagonal
 density matrix and of its first derivatives with respect to $Q$ and $P$
 at a given point, the equations derived thus far do not suffice.  The
 necessary additional equations are provided by differentiating
 the equation of motion with respect to $Q^i$  ($n_a$ fixed)
 and afterwards setting
 $P=Q=p=0$.  In order to obtain closure from this step, we must ignore
 second derivatives of $s_\alpha$.  The resulting set of equations then
 contains only zeroth and first derivatives of the density matrix.
 By extending these considerations, 
 a procedure can be formulated for including second and higher derivatives.

 Continuing the technical development, we introduce the definitions
 \begin{subequations}
 \begin{eqnarray}
 \partial_i &=& \partial_{Q^i},  \;\;
 \partial_a = \partial_{q^a}, \\
 \partial^i &=& \partial_{P_i},  
 \partial^a = \partial_{p_a}, \\
 \partial_0 &=& -V_{,i}\partial^i -V_{,a}\partial^a, \label{eq:6:part0} \\
 \partial^{1i} &=& B^{ij}\partial_j +B^{ia}\partial_a, \;\;
 \partial^{1a} = B^{aj}\partial_j + B^{ab}\partial_b,\label{eq:6:part2} \\
 \partial_{2i} &=& -V_{,ij}\partial^j -V_{,ib}\partial^b, \;\;
 \partial_{2a} = -V_{,aj}\partial^j -V_{,ab}\partial^b. \label{eq:6:part4}
 \end{eqnarray}
 \end{subequations}
 In terms of these definitions, the procedure we have described yields
 the equations
 \begin{subequations}
 \begin{eqnarray}
 \partial_0 s_\alpha &=& -i\frac{\partial W}{\partial s_\alpha^{\ast}}
 \equiv -iS_\alpha,   \label{eq:6:chf}   \\
 \partial^{1\mu}s_\alpha &=& -i[M_{\alpha\beta}\partial^\mu 
 s_\beta +L_{\alpha\beta}\partial^\mu s_\beta^{\ast}],  \label{eq:6:rpa1} \\
 \partial_{2\mu}s_\alpha &=& -i[M_{\alpha\beta}\partial_\mu
 s_\beta +L_{\alpha\beta}\partial_\mu s_\beta^{\ast}],  \label{eq:6:rpa2} \\
 M_{\alpha\beta} &=& \frac{\partial^2 W}{\partial s_\alpha^{\ast}
 \partial s_\beta},  \;\;
 L_{\alpha\beta} = \frac{\partial^2 W}{\partial s_\alpha^{\ast}
 \partial s_\beta^{\ast}}, \label{eq:6:def11} 
 \end{eqnarray}
 \end{subequations}
 where it is understood that all quantities
 are functions only of $Q$ and of the occupation numbers,
 and in which we have combined the sets $i$ and
 $a$ into a single index set $\mu$.
 To the above we adjoin the complex conjugate set.

 \begin{aside}
 The explicit expressions for the quantities $S$, $L$,
 and $M$ that appear in Eqs.~(\ref{eq:6:chf}-\ref{eq:6:def11})
 can be obtained starting from the mean-field Hamiltonian 
 using formulas given in Sec.~\ref{sec:6.1.4}.  Notice that the
 index $\alpha$ actually refers to a pair $(b,a),\, b>a$.  Recalling,
 in particular, 
 the definitions (\ref{eq:6:def11}), we find
 \begin{eqnarray}
 S_{ba} &=& \sqrt{n_a -n_b}{\cal H}_{ba},
 \label{eq:6:eval1} \\
 M_{badc} &=& M_{dcba}^{\ast} \nonumber \\
 &=& \frac{1}{\sqrt{(n_a -n_b)(n_c -n_d)}} 
 [{\cal H}_{bd}\delta_{ca}(n_c-n_b)
 -{\cal H}_{ca}\delta_{db}(n_a-n_d)] \nonumber \\
 && +V_{bcad} \sqrt{(n_a -n_b)(n_c -n_d)}, \label{eq:6:eval2} \\
 L_{badc} &=& L_{dcab}  \nonumber \\
 &=& \frac{1}{2\sqrt{(n_a-n_b)(n_c-n_d)}}[{\cal H}_{da}\delta_{cb}
 -{\cal H}_{bc}\delta_{da}] 
 [(n_a-n_b)-(n_c-n_d)]  \nonumber \\
 && +V_{dbac}\sqrt{(n_a-n_b)(n_c-n_d)}.
 \label{eq:6:eval3}
 \end{eqnarray}
 It is easily seen that in the zero temperature limit, when the occupation
 numbers correspond to the choice $(b,a)\rightarrow(ph),\, n_p=0,\, n_h=1$, 
 that the matrices reduce to pieces of the well-known RPA matrix, $M$
 to the Tamm-Dancoff (shell-model) matrix and $L$ to a two-particle, 
 two-hole matrix element of $V$, associated with the inclusion of 
 ground-state correlations. 
 \end{aside}

 We now ask what, if anything, is actually determined by
 the above set of equations.
 Without further restriction they are identities (to linear order in
 all variables but $Q$) satisfied by {\it any} canonical transformation from
 the original to the new set of canonical variables.  For example, any 
 transformation satisfying the Lagrange bracket conditions
 \begin{subequations}
 \begin{eqnarray}
 \partial_\mu s_\alpha\partial^\nu s_\alpha^{\ast}
 -\partial^\nu s_\alpha\partial_\mu s_\alpha^{\ast} &=& -i\delta^\mu_\nu , 
 \label{eq:6:lb1} \\
 \partial_\mu s_\alpha\partial_\nu s_\alpha^{\ast}
 -\partial_\nu s_\alpha\partial_\mu s_\alpha^{\ast} &=& 0, \label{eq:6:lb2} \\ 
 \partial^\mu s_\alpha\partial^\nu s_\alpha^{\ast}
 -\partial^\nu s_\alpha\partial^\mu s_\alpha^{\ast} &=& 0 , \label{eq:6:lb3} 
 \end{eqnarray}
 \end{subequations}
 will automatically guarantee the satisfaction of Eqs.~(\ref{eq:6:chf})-(\ref{eq:6:rpa2}).  

 What we actually want is a canonical transformation that minimizes
 the coupling between the collective and the non-collective spaces.  At
 zero temperature, with the neglect of curvature
 this dictated the imposition 
 of the decoupling conditions
 \begin{eqnarray}
 V_{,a}(Q) &=& \partial_a W =[\partial_a s_\alpha\partial_{s_\alpha}
	       +\partial_a s_\alpha^{\ast}\partial_{s_\alpha^{\ast}}]W =0,
 \label{eq:6:200} \\
 B^{ai} &=& \partial^a \partial^i W =0.   \label{eq:6:201}
 \end{eqnarray}
 The same conditions provide an extension of the zero temperature
 theory to finite excitation energy.  There is an additional practical
 complication that arises from the fact that the collective manifold
 depends not only on the collective coordinates $Q$ but on the assigned
 values of the occupation numbers.  This means that even if we limit
 ourselves to one collective coordinate, the collective manifold is
 many-dimensional at finite excitation energy, and thus its computation
 confronts us with a problem whose difficulty is of daunting proportions.
 It is for this reason that we have chosen to
 study two limiting cases of possible physical interest, each corresponding
 to a one-dimensional subspace of the space of occupation numbers.  One
 limit is that of collision-less motion, corresponding to fixed occupation
 numbers (in the local coordinate system) and one-body friction.
 For this case, the formal problem is the
 same as at zero excitation energy, since we deal with a case of fixed
 occupation numbers as we explore changes of the collective variables.
 The set of occupation numbers that interests us is that determined
 at equilibrium at a fixed
 temperature.  These remain fixed as we vary the collective coordinates,
 and therefore the resulting manifold is constructed from a series of
 collective paths with fixed entropy.
 The other limit is that of two-body collisions establishing
 local thermodynamic
 equilibrium on a time scale small compared to that associated with the
 collective motion.
 Thus, as opposed to the collision-less regime, we want the
 solutions of the LHA only at points where the occupation numbers satisfy
 Eq.~(\ref{eq:6:15.35}), with local values of single-particle energies and
 chemical potential, adjusted as part of the calculation.

 The submanifolds of occupation numbers for which we construct solutions
 can be thus be characterized as follows.
 In the collision-less case we have, zero
 referring to equilibrium,
 \begin{equation}
 n_a = n_a(Q_0,T_0).    \label{eq:6:ex1}
 \end{equation}
 Since $Q_0=Q_0(T_0)$, this is a one-dimensional subspace with a given
 entropy.  In the strong collision limit
 \begin{equation}
 n_a = n_a(Q,T),    \label{eq:6:ex2}
 \end{equation}
 is the Fermi distribution (\ref{eq:6:15.35}).  The derivation of the latter
 by maximization of the local entropy guarantees that for each $Q$, we have
 replaced the set $n_a$ by the single variable, $T$, defining the local
 equilibrium.  In principle (though not in practice) nothing prevents us
 from calculating the entire manifold of canonical transformations and then
 identifying the two special subspaces afterwards.

 Before continuing with the technical details,
 the most important additional point to keep in mind 
 is that after we complete the decoupling algorithm to be described below
 the solution found will not, in general, satisfy the decoupling conditions
 (\ref{eq:6:200}) and (\ref{eq:6:201})
 exactly (except for the singular case of exactly decoupled motion when
 there is no dissipation).  As explained below,
 the transformation found will, however,
 allow the evaluation of the quantities that occur in these conditions.
 Their non-vanishing values are, as we have seen in the previous section,
 essential to the development of dissipative behavior.
 Thus all the quantities in the Hamiltonian (\ref{eq:6:clham1}) can be evaluated,
 The formulas required for this purpose are obtained by inserting
 the solutions for the
 density matrix and its various first derivatives into an expansion of the
 right hand side of (\ref{eq:6:equiv}).  The simplest example of such a calculation
 is the potential energy
 \begin{equation}
 V(Q) = W(s_\alpha(Q,0,0,0), s_\alpha^{\ast},n_a).  \label{eq:6:par1}
 \end{equation}
 Next in complexity is the first derivative
 \begin{equation}
 V_{,a}(Q) =\partial_a W = \frac{\partial W}{\partial s_\alpha}\partial_a
 s_\alpha  \;\; +{\rm c.c.},     \label{eq:6:par2}
 \end{equation}
 where the right hand side is evaluated at $P=q=p=0$.  The remaining
 parameters that we need are all second derivatives,
 \begin{eqnarray}
 B^{\mu\nu} &=& \partial^\mu\partial^\nu W  \nonumber \\
  &\cong& \frac{\partial^2 W}{\partial s_\alpha \partial s_\beta}
  \partial^\mu s_\alpha \partial^\nu s_\beta
  +\frac{\partial^2 W}{\partial s_\alpha \partial s_\beta^{\ast}}
  \partial^\mu s_\alpha \partial^\nu s_\beta^{\ast} +{\rm c.c.}, \label{eq:6:par3}
 \end{eqnarray}
 where we have (consistently) neglected second derivatives of the density
 matrix. In order to evaluate Eq.~(\ref{eq:6:par2}) and all the elements of
 Eq.~(\ref{eq:6:par3}), it is evident that we need all the solutions of the 
 LHA equations (\ref{eq:6:rpa1}) and (\ref{eq:6:rpa2}),
 and not only the collective ones that play a special role
 in the self-consistent calculation.

 We review briefly the properties
 and solutions of Eqs.~(\ref{eq:6:chf})-(\ref{eq:6:rpa2}).  When
 the decoupling conditions are enforced, these equations simplify to the
 forms
 \begin{eqnarray}
 -V_{,i} \partial^i s_\alpha &=& -iS_\alpha ,  \label{eq:6:chf1} \\
 B^{ij}\partial_j s_\alpha &=& -i[M_{\alpha\beta}\partial^i 
 s_\beta +L_{\alpha\beta}\partial^i s_\beta^{\ast}], \label{eq:6:rpa11} \\
 -V_{,ij}\partial^j s_\alpha &=& -i[M_{\alpha\beta}\partial_i 
 s_\beta +L_{\alpha\beta}\partial_i s_\beta^{\ast}]. \label{eq:6:rpa21}
 \end{eqnarray}
 They can be simplified further
 by the consistent assumptions that $M$ and $L$ are real symmetric
 matrices and that the partial derivatives are either real or imaginary,
 \begin{eqnarray}
 \partial_i s_\alpha^{\ast} &=& \partial_i s_\alpha, \label{eq:6:real} \\
 \partial^i s_\alpha^{\ast} &=& -\partial^i s_\alpha. \label{eq:6:imag} 
 \end{eqnarray}
 This allows us to eliminate the partials of $s_\alpha^{\ast}$. The formalism
 now consists of the constrained Hartree-Fock equation (\ref{eq:6:chf1}) and two
 equivalent eigenvalue equations obtained by combining (\ref{eq:6:rpa11})
 and (\ref{eq:6:rpa21}), of which one is 
 \begin{equation}
 -(VB)_i^j \partial_j s_\alpha = [(L-M)(L+M)]_{\alpha\beta}
 \partial_i s_\beta,  \label{eq:6:eigen}
 \end{equation}
 and the other, for $\partial^i s_\alpha$, is the transpose of (\ref{eq:6:eigen}).
 This implies that $VB$ and $BV$ have the same diagonal form (if there 
 are no degeneracies, as we assume).  These eigenvalue equations plus the
 corresponding Lagrange bracket conditions are equivalent to the eigenvalue
 problem of the thermal RPA \cite{R,ST1,ST2,RT3}, but in our case the
 theory applies outside of thermal equilibrium.

 As emphasized previously, the solution of the system obtained
 utilizes an algorithm of the
 same general structure as required at zero temperature.  The solutions
 of the eigenvalue problem automatically satisfy the homogeneous
 Lagrange bracket
 equations, whereas the inhomogeneous brackets provide normalization
 conditions.  The solution procedure is usually started at a point of
 thermal equilibrium where the Hartree-Fock equations decouple from
 local harmonic equations.  At a general point, the
 algorithm requires an iteration between the constrained Hartree-Fock 
 equations and the eigenvalue equations.  This is the essence of the LHA.
 As remarked earlier, we are confining our attention to two cases.  If,
 starting from equilibrium, we keep the occupation numbers constant
 as we change $Q$,
 the analogy with the calculation at zero temperature is rather complete.
 If, on the hand we keep the temperature fixed, at every point there is
 the additional task of finding self-consistent values of the occupation
 numbers.  This type of calculation was also required for thermal
 Hartree-Fock.

 We summarize the argument developed in this section.  Given
 the mean-field Hamiltonian at finite excitation energy,
 we have described a method for introducing an optimal
 decomposition of the total space into collective and non-collective
 coordinates.  In general these spaces will not be exactly decoupled,
 but as a result of the actual calculations, we can evaluate the
 coupling terms.
 In other words, we can find explicit forms of Eqs.~(\ref{eq:6:cleom1})
 - (\ref{eq:6:cleom4}) and from these project out the collective subspace
 that leads to a dissipative dynamics, the procedure described in detail
 in Sec.~\ref{sec:6.1.5}.

 The practical
 end of our labors is a set of classical equations of motion for the 
 collective coordinates that includes necessary dissipative parameters.
 When combined with the equations that specify the rate of change of the
 occupation numbers, they
 describe the relaxation of a perturbed collective subsystem and the
 leakage of energy to the non-collective degrees of freedom.
 In Sec.~\ref{sec:6.2} below, we shall apply these ideas to a simple model.

 \subsection{Application to a model\label{sec:6.2}}

 \subsubsection{The model and its mean-field limit\label{sec:6.2.1}}

 We illustrate some of the concepts introduced  in the previous formal
 development by applying them to a Hamiltonian with the symmetry of
 SU(2)$\times$SU(2) (a doubled Lipkin model),
 in which the latter, exemplifying the slow, collective degrees of freedom,
 interacts with a set of harmonic oscillators representing the
 fast, non-collective variables.
 We remind the reader that what is termed the Lipkin model 
 \cite {L1,39} utilizes
 two single-particle levels of equal degeneracy distinguished only by
 an energy separation.  The destruction operators for the fermions in
 the upper level are called $a_{p+}$, those in the lower level $a_{p-}$,
 respectively.  Each level has degeneracy $2N$ .  We set
 $j=N-(1/2)$ and $-j\leq p\leq j$. 
 The only operators allowed in the Hamiltonian are the coherent sums
 \begin{eqnarray}
 J_+ =J_-^\dagger &=& \sum_{p=-j}^{j}a_{p+}^\dagger a_{p-}
 =J_1 +iJ_2 , \label{eq:6:jayplus} \\
 J_3 &=& \frac{1}{2}\sum_p(a_{p+}^\dagger a_{p+} - a_{p-}^\dagger a_{p-})
 \nonumber \\
 &=& \frac{1}{2}(N_+ - N_-) ,  \label{eq:6:jayzero}
 \end{eqnarray}
 which obey the commutation relations of scaled angular momentum operators,
 \begin{eqnarray}
 [J_i,J_j] &=& i\delta_{ijk}J_k, \label{eq:6:jaycom}
 \end{eqnarray}
 where $\delta_{ijk}$ is the alternating symbol in three dimensions.

 We study a problem in which two Lipkin systems, with
 operators labeled $J_i$ and $K_i$, respectively, are coupled to an array
 of harmonic oscillators.  This model can be considered a generalization
 of the model of Caldeira and Leggett \cite{CL}. To illustrate the limit of no collisions,
 it would have sufficed to consider only a single SU(2) algebra, but to
 also study the case of collisions that instantaneously equilibrate a
 constrained system locally, as we shall see, we need at least
 two such systems.  We choose the Hamiltonian
 \begin{eqnarray}
 H &=& h_1 J_3 +  h_2 K_3  -V_1 J_1^2 - V_2 K_1^2 \nonumber \\
 && -UJ_1 K_1 - q_J J_1 -q_K K_1 + \sum_i\frac{1}{2}(p_i ^2 +\omega_i q_i^2),
  \label{eq:6:ham100}        \\
 q_J &=& \sum_i c_i q_i, \;\; q_K = \sum_i d_i q_i.  \label{eq:6:def100}
 \end{eqnarray}
 Here $h_1,h_2$ are single-particle energies, and $V_1,V_2,U,c_i,d_i$
 are interaction strengths.

 Notice that for the Hamiltonian (\ref{eq:6:ham100}) there is separate number
 conservation for each  SU(2) subspace.  The associated quantities, 
 $N_J$ and $N_K$ are the sums
 \begin{eqnarray}
 N_J &=& N_{J+} +N_{J-},   \label{eq:6:en1_a} \\
 N_K &=& N_{K+} + N_{K-}.  \label{eq:6:en2_a}
 \end{eqnarray}

 The equations of motion that follow from the Hamiltonian (\ref{eq:6:ham100})
 are
 \begin{subequations}
 \begin{eqnarray}
 \dot{J}_1 &=& -h_1 J_2, \label{eq:6:eom11} \\
 \dot{J}_2 &=& h_1 J_1 +V_1\{J_1,J_3\}  +UJ_3 K_1 +q_J J_3, \label{eq:6:eom21} \\
 \dot{J}_3 &=& -V_1\{J_1,J_2\} - UJ_2 K_1, \label{eq:6:eom31} \\
 \dot{K}_1 &=& -h_2 K_2, \label{eq:6:eom41} \\
 \dot{K}_2 &=& h_2 K_1 +V_2\{K_1,K_3\}  +UJ_1 K_3 +q_K K_3, \label{eq:6:eom51} \\
 \dot{K}_3 &=& -V_2\{K_1,K_2\} - UJ_1 K_2, \label{eq:6:eom61} 
 \end{eqnarray}
 \end{subequations}
 We next average these equations over a suitably chosen mixed state by
 decomposing any operator $X$ as a sum of its average and of a
 fluctuating part whose average vanishes,
 \begin{eqnarray}
 X &=& \langle X \rangle +\delta X,  \label{eq:6:av1}  \\
 \langle XY\rangle &=& \langle X\rangle\langle Y\rangle
			+\langle \delta X \delta Y \rangle. \label{eq:6:av2}
 \end{eqnarray}
 If we neglect two operator correlations such as those that have the form
 of the second term of Eq.~(\ref{eq:6:av2}), we arrive at the mean-field
 approximation to our equations.  If we suppress angular brackets,
 and replace anticommutators by twice the product in either order, these
 equations have the same form as Eqs.~(\ref{eq:6:eom11})-(\ref{eq:6:eom61}).  We
 shall employ the symbol $W$ for the mean-field Hamiltonian that gives
 rise to these equations of motion in the Poisson bracket formulation of
 classical mechanics.  In the formal development, we devoted
 much attention
 to the nature of the mixed state used in the averaging and shall
 not repeat these considerations here.

 In the remainder of the text, we should deal overtly only with this
 classical limit, the collision-less case.
 Nevertheless, it is also possible,
 with a suitable additional assumption, to incorporate the limiting case
 of strong collisions, characterized by the assumption that the system is
 always in a state of constrained equilibrium, characterized by a local
 temperature.

 To continue study of the mean-field limit, we introduce the
 representation in which the single-particle density matrix is diagonal,
 i.e., the representation of natural orbitals.
 This requirement takes on an especially simple form for the model under
 study, where only special linear combinations of the density matrix elements
 occur, namely subsets having the symmetry of SU(2).  Since the ``angular
 momentum'' associated with each subspace is separately conserved, we
 can rotate the axes in each of these spaces independently to make the
 associated conserved vector point in the three direction.  For example, we
 introduce a rotation of the vector $\vec{J}$, according to the equation
 \begin{equation}
 J_i = A_{ij}R_j,   \label{eq:6:rot}
 \end{equation}
 where the $A_{ij}$ are the matrix elements of the orthogonal array (the
 reason for this particular choice will be seen below)
 \begin{equation}
 \vec{A} = \left( \begin{array}{ccc} 
	    \cos\theta & -\sin\theta\sin\psi & \sin\theta\cos\psi \\
	   0  & \cos\psi & \sin\psi \\
	    -\sin\theta & -\cos\theta\sin\psi  &\cos\theta\cos\psi
	\end{array} \right).  \label{eq:6:mat1}
 \end{equation}
 The angles $\theta$ and $\psi$ are to be chosen so that 
 in the new coordinates only $R_3$ is non-vanishing.
 With corresponding
 definitions for the vector $\vec{K}$, with angles $\alpha,\beta$
 we may therefore write
 \begin{subequations}
 \begin{eqnarray}
 J_1 &=& \sin\theta\cos\psi R_3,  \label{eq:6:400} \\
 J_2 &=& \sin\psi R_3,  \label{eq:6:500} \\
 J_3 &=& \cos\theta\cos\psi R_3, \label{eq:6:600} \\ 
 K_1 &=& \sin\alpha\cos\beta S_3, \label{eq:6:700} \\
 K_2 &=& \sin\beta S_3, \label{eq:6:800}\\
 K_3 &=& \cos\alpha\cos\beta S_3. \label{eq:6:900}
 \end{eqnarray}
 \end{subequations}

 Before substituting these values into the mean-field Hamiltonian, we
 notice that the transformation Eqs.~(\ref{eq:6:400})-(\ref{eq:6:900}), permits us to
 separate each triple of variables into a canonical pair and a measure of
 occupation in the new representation. Thus by computation of the Poisson
 brackets for {\it fixed} $R_3$ and $S_3$, we verify that the variables
 \begin{subequations}
 \begin{eqnarray}
 && Q_1=\theta,\; P_1=R_3\sin\psi,  \label{eq:6:1100} \\
 && Q_2= \alpha, \; P_2= S_3\sin\beta   \label{eq:6:1200}
 \end{eqnarray}
 \end{subequations}
 are sets of classical canonical variables.  In the adiabatic approximation,
 in which we work only to second order in the collective momenta $P_1, P_2$,
 Eqs.~(\ref{eq:6:400})-(\ref{eq:6:900}) are replaced by the equations
 \begin{subequations}
 \begin{eqnarray}
 J_1 &=& \sin Q_1(R_3 - \frac{1}{2R_3}P_1^2), \label{eq:6:1300} \\
 J_2 &=& P_1, \label{eq:6:14} \\
 J_3 &=& \cos Q_1(R_3 -\frac{1}{2R_3}P_1^2), \label{eq:6:1500}
 \end{eqnarray}
 \end{subequations}
 with corresponding equations for $K_i$.

 To reach the starting point for our further deliberations, we substitute
 these equations into $W$ and expand the latter 
 as a series in $P^k q_i^l p_i^m$, where $k+l+m\leq 2$.  This means that
 we are limiting our further attention to the large amplitude,
 small-velocity domain for the collective motion and to a (local)
 simple-harmonic approximation for the non-collective variables.
 Continuing to use the same symbol for the approximate form,  we find
 \begin{subequations}
 \begin{eqnarray}
 W &=& \frac{1}{2}B_1(Q_1,Q_2)P_1^2  +\frac{1}{2}
 B_2(Q_1,Q_2)P_2^2 \nonumber \\
 && + V((_1,Q_2) +\sum_i \frac{1}{2}(\omega_i^2 q_i^2 + p_i^2) 
 -q_J R_3\sin Q_1 -q_K S_3\sin Q_2,  \label{eq:6:1600} \\
 V(Q_1,Q_2) &=& h_1\cos Q_1 R_3 -V_1 \sin^2 Q_1 R_3^2  
  +h_2\cos Q_2 S_3 - V_2\sin^2 Q_2 S_3^2 \nonumber \\
 && - U\sin Q_1\sin Q_2 R_3 S_3, \label{eq:6:1700}   \\
 B_1 (Q_1,Q_2) &=& -\frac{h_1}{R_3}\cos Q_1 +2V_1\sin^2 Q_1 
 +U\sin Q_1\sin Q_2 \frac{S_3}{R_3},  \label{eq:6:1800} \\
 B_2 (Q_1,Q_2) &=& -\frac{h_2}{S_3}\cos Q_2 +2V_2\sin^2 Q_2 
 +U\sin Q_1\sin Q_2 \frac{R_3}{S_3}.   \label{eq:6:1900a}
 \end{eqnarray}
 \end{subequations}
 {}From the Hamiltonian (\ref{eq:6:1600}), we deduce the equations of motion
 (to first order in $P_a$)
 \begin{subequations}
 \begin{eqnarray}
 \dot{Q}_1 &=& B_1 P_1, \label{eq:6:1900} \\
 \dot{P}_1 &=& -\partial_{Q_1}V +q_J R_3\cos Q_1, \label{eq:6:2000}  \\
 \dot{Q}_2 &=& B_2 P_2, \label{eq:6:21} \\
 \dot{P}_2 &=& -\partial_{Q_2}V +q_K S_3\cos Q_2, \label{eq:6:2200} \\
 \dot{q}_i &=& p_i, \label{eq:6:2300} \\
 \dot{p}_i &=& -\omega_i^2 q_i +c_i R_3\sin Q_1 +d_i S_3\sin Q_2, \label{eq:6:2400}
 \end{eqnarray}
 \end{subequations}
 and $R_3$ and $S_3$ are time independent in the strict mean-field limit.

 As the last topic of this section, we are now in a position to define
 precisely the two limiting cases for our treatment of ``collisions''.
 We calculate the time rate of change of the mean-field energy, allowing
 a possibly non-vanishing value for the rates of $R_3$ and $S_3$. Using the
 equations of motion (\ref{eq:6:1900})-(\ref{eq:6:2300}), we find
 \begin{subequations}
 \begin{eqnarray}
 \dot{W} &=& 2\epsilon_{1+}\dot{R}_3 +2\epsilon_{2+}\dot{S}_3, \label{eq:6:2500} \\
 2\epsilon_{1+} &=& \frac{\partial W}{\partial R_3} = h_1\cos Q_1
 -2V_1\sin^2 Q_1 R_3  \nonumber \\
 &&- U\sin Q_1\sin Q_2 S_3 -q_J\sin Q_1,  \label{eq:6:2600} \\
 2\epsilon_{2+} &=& \frac{\partial W}{\partial S_3} = h_2\cos Q_2
 -2V_2\sin^2 Q_2 S_3  \nonumber \\
 &&- U\sin Q_1\sin Q_2 R_3 -q_K\sin Q_2.  \label{eq:6:2700} 
 \end{eqnarray}
 \end{subequations}
 Of course, if we insist strictly on the mean field limit, then $\dot{R}_3$
 and $\dot{S}_3$ vanish. (We could have added these requirements to the list
 of equations of motion that follow from $W$.)   Then the right hand side of
 Eq.~(\ref{eq:6:2500}) vanishes and $W$ is conserved, as we expect.

 In the first part of this section, we have also pointed out another
 means by which the mean-field energy can be conserved, namely by
 having the sum of the terms on the right hand side of
 Eq.~(\ref{eq:6:2500}) vanish, rather than the individual
 summands. This means that we must postulate master equations for $R_3$
 and $S_3$ that satisfy the constraint imposed.  This can be done by
 assuming that collisions are so effective that the system comes to
 local equilibrium in a time short even compared with a characteristic
 period of the collective motion.  This implies that the system can
 only pass through configurations characterized by a local temperature
 $T$ (or reciprocal temperature $\beta$). Such points are characterized
 by local occupations ($a=1,2$)
 \begin{equation}
 N_{a\pm} =N_a \frac{1}{1+\exp\beta(\epsilon_{a\pm} -\mu_a)}. \label{eq:6:2800}
 \end{equation}
 Here $\epsilon_{a+}=-\epsilon_{a-}$ are the single particle energies
 defined in Eqs.~(\ref{eq:6:2600}) and (\ref{eq:6:2700}), and $\mu_a$ are chemical
 potentials to be determined by number conservation.  Utilizing the latter
 and the definitions of $R_3$ and $S_3$, the condition to be exploited in the
 strong collision case becomes
 \begin{eqnarray}
 0 &=& 2\epsilon_{1+}\dot{R}_3 +2\epsilon_{2+}\dot{S}_3 
 = \sum_a 2\epsilon_{a+}\dot{N}_{a+}.  \label{eq:6:2900}
 \end{eqnarray}

 There are two problems that must be resolved before Eq.~(\ref{eq:6:2900}) can be
 applied to the study of non-equilibrium motion.
  The first is how to associate values of $T$ with values of $Q_a$.
 The answer here emerges from the procedure for the construction
 of the collective manifold described in detail at the start of
 Sec.~\ref{sec:6.2.4}.  The essential additional step in this process is the
 introduction of a heat bath in order to establish equilibrium values
 of the collective coordinates.  Second comes the realization
 that we are only interested in evaluating Eq.~(\ref{eq:6:2900}) on the
 collective manifold, where, as determined in the next section,
 $q_i$ is specified as a function of $Q_a$. This implies that
 the time derivative of
 Eq.~(\ref{eq:6:2800}) is non-vanishing in consequence only of its dependence 
 on $Q_a$ and $T$.  Thus from the consequent application of
 Eqs.~(\ref{eq:6:2800}) and (\ref{eq:6:2900}), we find the condition
 \begin{eqnarray}
 \dot{\beta} &=& -\frac{\dot{T}}{T^2}  
 =-\frac{\sum_{ab}(\epsilon_{a+}\partial_{Q_b}N_{a+})\dot{Q}_b}
       {\sum_a(\epsilon_{a+}\partial_\beta N_{a+})}.  \label{eq:6:17.300}
 \end{eqnarray}
 This equation will be adjoined to the dissipative equations of
 motion to be derived in Sec.~\ref{sec:6.2.4} to form a basis for a study of the
 non-equilibrium behavior of the combined system.

 \subsubsection{Definition of the collective manifold\label{sec:6.2.2}}

 We have assumed until now that the $Q_a$ describe the collective degrees of
 freedom and the $q_i$ the non-collective ones. To justify this assumption
 we examine the equations of motion (\ref{eq:6:2300}) and (\ref{eq:6:2400}).
 If the $q_i$
 are a suitable choice of non-collective variables, then the right hand sides
 of these equations must vanish on the collective manifold.
 This yields as the equations that define this manifold
 \begin{eqnarray}
 p_i &=& 0,  \label{eq:6:3100} \\
 \omega_i^2 q_i &=& c_i R_3\sin Q_1 +d_i S_3\sin Q_2.  \label{eq:6:3200}
 \end{eqnarray}
 If we substitute these equations into the mean-field Hamiltonian (\ref{eq:6:1600}),
 we obtain the classical Hamiltonian of the collective manifold.  The
 quantization of this expression at zero temperature provides a quantum
 basis for the study of the collective spectrum of the associated system.

 On the other hand, if we substitute these values into the equations of
 motion (\ref{eq:6:1900})-(\ref{eq:6:2200}), and work as always to first order in the
 momenta, we obtain the classical equations of
 motion on the collective manifold ($\partial_a =(\partial/\partial_{Q_a})$)
 \begin{subequations}
 \begin{eqnarray}
 \dot{Q}_1 &=& B_1 P_1 ,   \label{eq:6:3300} \\
 \dot{P}_1 &=& -\partial_1 V +A_{cc}R_3^2\sin Q_1\cos Q_1  
 +A_{cd}R_3 S_3\cos Q_1\sin Q_2, \label{eq:6:3400} \\
 \dot{Q}_2 &=& B_2 P_2, \label{eq:6:3500} \\
 \dot{P}_2 &=& -\partial_2 V +A_{cd}R_3 S_3\sin Q_1\cos Q_2 
 +A_{dd}S_3^2\sin Q_2 \cos Q_2,   \label{eq:6:3600}   \\
 A_{cc} &=& \sum_i\frac{c_i^2}{\omega_i^2},  \label{eq:6:3700} \\
 A_{cd} &=& \sum_i\frac{c_i d_i}{\omega_i^2},  \label{eq:6:3800} \\
 A_{dd} &=& \sum_i\frac{d_i^2}{\omega_i^2}.  \label{eq:6:3900}
 \end{eqnarray}
 \end{subequations}
 In the next section, we shall study the coupling to the non-collective modes.
 When we ``eliminate'' the latter, we expect, under suitably prescribed
 conditions, that the above equations will be modified mainly by the addition
 of dissipative terms.

 We shall thus end up with a combined classical-quantum
 description, in that the collective variables are treated classically,
 but the occupation numbers remain as variables obeying Fermi-Dirac
 statistics.  The expanded adiabatic collective manifold is defined by
 values of $Q_a,R_3,S_3$ and is thus a four-dimensional manifold.
 The two extreme scenarios that we have described for the treatment of
 collisions select different three-dimensional spaces (in addition to
 avoiding study of dynamical master equations for the time-rate-of-change
 of the occupation numbers).

 \subsubsection{Dissipative equations of motion\label{sec:6.2.3}}

 We study next the equations of
 motion including full coupling to the non-collective modes, with the aim of
 eliminating the latter in favor of a description of the
 collective modes including dissipation.
 We describe how dissipation in the collective motion can arise
 irrespective of the treatment of the occupation numbers (with characteristic
 differences between the two scenarios considered.)  This is an explicit
 form of the procedure described in some detail in the preceding subsection,
 and has been discussed in Ref.~\cite{31_2}.
 The solution of
 Eqs.~(\ref{eq:6:2300}) and (\ref{eq:6:2400}) is 
 \begin{eqnarray}
 q_i(t) &=& q^{(h)}(t) +\int dt^{\prime}\sin\omega_i(t-t^{\prime})
 \frac{1}{\omega_i}  \nonumber\\
 &&\qquad\qquad \times \left[c_i R_3(t')\sin Q_1(t')
 + d_i S_3(t')\sin Q_2(t')\right],  \label{eq:6:3900a} \\
 q^{(h)}(t) &=& q_i(0)\cos\omega_i t +\frac{p_i(0)}{m_i\omega_i}
 \sin\omega_i t, \label{eq:6:4000} 
 \end{eqnarray}
 where $q^{(h)}(t)$ is the solution of the homogeneous equation.         
 Integrating by parts , we transform (\ref{eq:6:3900a})
 to the form 
 \begin{eqnarray}
 q_i(t) &=& q_i^{(h)}(t) -\frac{1}{\omega_i^2}
 \left(c_i\sin Q_1 R_3 +d_i\sin Q_2 S_3\right)  \nonumber \\
 &&+\frac{1}{\omega_i^2}\left(c_i\sin Q_1(0)R_3(0) 
 +d_i\sin Q_2(0)S_3\right)\cos\omega_i t       \nonumber \\
 &&+\frac{1}{\omega_i^2}\int_{0}^{t} dt^{\prime}\cos\omega_i(t-t^{\prime})
 \nonumber \\
 && \qquad\qquad\times\frac{d}{dt^{\prime}}\left[c_i\sin Q_1(t')R_3(t') 
 +d_i\sin Q_2(t')S_3(t')\right]  \label{eq:6:4000a}
 \end{eqnarray}

 We substitute these equations into the equations of motion (\ref{eq:6:2000})
 and (\ref{eq:6:2200}). Furthermore, neglecting terms of second and higher order
 in $P_a$, we convert to second order equations
 \begin{eqnarray}
 \ddot{Q}_1 = -B_1(\partial_1 V +q_J R_3\cos Q_1) , \label{eq:6:4100} \\
 \ddot{Q}_2 = -B_2(\partial_2 V +q_K S_3\cos Q_2) . \label{eq:6:4200}
 \end{eqnarray}
 The introduction of second-order equations in this section has as its purpose
 to make contact with the elementary description of friction within the
 the framework of Newton's laws.
 For purposes of numerical calculation, first order equations are, of course,
 more convenient.  These will be quoted when needed.

 It is essential to appreciate at this point that Eqs.~(\ref{eq:6:4100})
 and (\ref{eq:6:4200}) contain both low frequencies characteristic of the collective
 modes and high frequencies characteristic of the non-collective motion.
 Our aim at this point is to average out the high-frequency behavior.
 This procedure was described in Sec.~\ref{sec:6.1}.  We noted that
 the high frequency behavior of the collective variables was driven by the
 initial value of the non-collective coordinates and momenta, and that our
 task was simplified by assuming that these are initially unexcited.
 With this omission, we are left, upon elimination of the non-collective
 coordinates, with the pair of equations of motion (recalling
 the definitions (\ref{eq:6:3700})-(\ref{eq:6:3900}))
 \begin{subequations}
 \begin{eqnarray}
 \ddot{Q}_1 &=& B_1\left(-\partial_1 V +A_{cc}R_3^2\sin Q_1\cos Q_1 
 +A_{cd}R_3 S_3\cos Q_1\sin Q_2\right) \nonumber \\
 &&-B_1 R_3\cos Q_1\sum_i\int_0^t dt'\cos\omega_i(t-t') 
 \nonumber\\&& \qquad\qquad \times 
 \left[\frac{c_i^2}{\omega_i^2}\frac{d}{dt'}(\sin Q_1(t')R_3(t'))
 +\frac{c_i d_i}{\omega_i^2}\frac{d}{dt'}(\sin Q_2(t')S_3(t')\right],
 \label{eq:6:4700} \\
 \ddot{Q}_2 &=& B_2\left(-\partial_2 V +A_{cd}R_3 S_3\sin Q_1\cos Q_2 
 +A_{dd}S_3^2\cos Q_2\sin Q_2\right) \nonumber \\
 &&-B_2 S_3\cos Q_2\sum_i \int_0^t dt'\cos\omega_i(t-t') 
 \nonumber \\&& \times  \qquad\qquad 
 \left[\frac{c_i d_i}{\omega_i^2}\frac{d}{dt'}
 (\sin Q_1(t')R_3(t'))  
 +\frac{d_i^2}{\omega_i^2}\frac{d}{dt'}(\sin Q_2(t')S_3(t')\right].  \label{eq:6:4800}
 \end{eqnarray}
 \end{subequations}
 Having swept away any possible dependence on the initial conditions of
 the non-collective coordinates, we observe as well that if we were to
 discard the last term of Eq.~(\ref{eq:6:4700}) and of (\ref{eq:6:4800}), the latter
 would reduce
 to the equivalent of Eqs.~(\ref{eq:6:3300})-(\ref{eq:6:3600}), namely the
 equations of motion
 for the collective coordinates on the collective manifold.  The new terms,
 which we now study, will be the source of dissipation.

 We consider here only the
 limit where the array of non-collective oscillators provides a source of
 white noise.  This treatment will give rise to conventional friction terms.
 To achieve this end, we replace the sum on $i$
 by an integral over $\omega$ and the various pivotal summands
 $(c_i^2/\omega_i^2)$, $(c_id_i/\omega_i^2)$, ($d_i^2/\omega_i^2)$ 
 by broad, slowly varying, distributions $\Phi_{11}(\omega)$,
 $\Phi_{12}(\omega)$, and $\Phi_{22}(\omega)$, respectively.  Here we use
 the explicit form
 \begin{equation}
 \frac{c_i^2}{\omega_i^2}\rightarrow \Phi_{11}(\omega)
 =\frac{A_{11}}{\omega^2 +\omega_0^2},    \label{eq:6:4900}
 \end{equation}
 and with the specification that the width, $\omega_0$,
 is the same for all the
 distributions, we also introduce constants $A_{12}$ and $A_{22}$.
 We thus encounter integrals of the form (subscripts understood)
 \begin{eqnarray}
 \int d\omega \Phi(\omega) &=& (\pi A)/(2\omega_0) \equiv C,  \label{eq:6:5000} \\
 \lefteqn{\int_0^t dt'\,\int d\omega\Phi(\omega)\cos\omega(t-t')\dot{X}(t')}
 \nonumber\\
 &=&D\left[\dot{X}(t) 
 -\exp(-\omega_0 t)\dot{X}(0) 
 -\int dt'\exp[-\omega_0(t-t')]\ddot{X}(t')\right],   \label{eq:6:5100} \\
  D &=& C/\omega_0.    \label{eq:6:5200}
 \end{eqnarray}

 If we suppose, consistently, that $(\Omega/\omega_0)<<1$, where $\Omega$
 is a frequency characteristic of the collective motion, then we may
 approximate Eq.~(\ref{eq:6:5100}) by its first term.  The second term
 is a transient
 that decays in a time short compared to $(1/\Omega)$ and the last term is
 down by a factor of $(\Omega/\omega_0)$.  These results allow us to
 replace Eqs.~(\ref{eq:6:4700}) and (\ref{eq:6:4800}) by the dissipative equations
 \begin{eqnarray}
 \ddot{Q}_1 &=& B_1\left(-\partial_1 V +C_{11}R_3^2\sin Q_1\cos Q_1 
  +C_{12}R_3 S_3\cos Q_1\sin Q_2\right) 
 \nonumber \\&&
 -B_1 R_3\cos Q_1 
 \left[D_{11}\frac{d}{dt}
 (\sin Q_1 R_3) +D_{12}\frac{d}{dt}(\sin Q_2 S_3)\right],  \label{eq:6:5300}  \\
 \ddot{Q}_2 &=& B_2\left(-\partial_2 V +C_{12}R_3 S_3\sin Q_1\cos Q_2 
  +C_{22}S_3^2\cos Q_2\sin Q_2\right) 
 \nonumber\\&&
 -B_2 S_3\cos Q_2 
 \left[D_{12}\frac{d}{dt}
 (\sin Q_1 R_3) +D_{22}\frac{d}{dt}(\sin Q_2 S_3)\right]. \label{eq:6:5400}
 \end{eqnarray}

 These equations may be written in the form
 \begin{eqnarray}
 M_a\ddot{Q}_a &=& -\partial_{Q_a}V_{\mathrm{C}} -\sum_b {\cal F}_{ab}\dot{Q}_b
 -{\cal F}_{aR}\dot{R}_3 -{\cal F}_{aS}\dot{S}_3,   \label{eq:6:5900}
 \end{eqnarray}
 where
 \begin{subequations}
 \begin{eqnarray}
 M_a &=& (1/B_a),   \label{eq:6:5500} \\
 V_{\mathrm{C}} &=& V -\frac{1}{2}C_{11}R_3^2 \sin^2 Q_1 -C_{12}R_3 S_3 \sin Q_1\sin Q_2
 -\frac{1}{2}C_{22}S_3^2\sin^2 Q_2,  \label{eq:6:5600} \\
 {\cal F}_{11} &=& R_3^2 D_{11}\cos^2 Q_1,  \label{eq:6:57} \\
 {\cal F}_{12} &=& {\cal F}_{21} = R_3 S_3 D_{12}\cos, Q_1\cos Q_2,
 \label{eq:6:5800} \\
 {\cal F}_{22} &=& S_3^2 D_{22}\cos^2 Q_2, \label{eq:6:5800a} \\
 ({\cal F}_{1R},{\cal F}_{1S}) &=& R_3 \cos Q_1(D_{11}  \sin Q_1, 
  D_{12}  \sin Q_2), \\
 ({\cal F}_{2R},{\cal F}_{2S}) &=& S_3 \cos Q_2 (D_{12}  \sin Q_1, 
  D_{22} \sin Q_2).
 \end{eqnarray}
 \end{subequations}

 These equations suggest the definition of the mechanical energy of collective
 motion, $W_{\mathrm{C}}$,
 \begin{equation}
 W_{\mathrm{C}} = \frac{1}{2}\sum_a M_a\dot{Q}_a^2 + V_{\mathrm{C}}, \label{eq:6:17.600}
 \end{equation}
 that in consequence of the equations of motion (\ref{eq:6:5900}) satisfies an
  energy-flow equation
 \begin{eqnarray}
 \frac{dW_{\mathrm{C}}}{dt} &=& -\sum_{a,b} {\cal F}_{ab}\dot{Q}_a\dot{Q}_b 
  -\sum_a {\cal F}_{aR}\dot{Q}_a\dot{R}_3   
 -\sum_a {\cal F}_{aS}\dot{Q}_a\dot{S}_3.      \label{eq:6:flow}
 \end{eqnarray}
 When the occupation numbers are constant, this reduces to a familiar energy
 flow equation in the presence of friction.

 \subsubsection{Numerical procedures\label{sec:6.2.4}}

 There are two separate parts of the numerical calculation.
 First we must characterize the collective space by computing all of the
 macroscopic functions, such as the reciprocal mass functions,
 $B_a$, in terms of the collective 
 coordinates $Q_a$ and one other variable.  For the no collision scenario,
 which we shall call scenario 1, this variable is the set of entropy values
 associated with equilibrium configurations.  For the strong collision
 limit, which we shall call scenario 2, this is the set of
 equilibrium temperature values (i.e., all temperatures).
 Second, we must solve the classical dissipative equations of motion for the
 two scenarios, when the system is initially in a non-equilibrium
 configuration.

 We describe the procedure that defines the collective space.
 The macroscopic functions
 are surfaces in the two-dimension space $Q_1, Q_2$ for each 
 temperature $T$ defined at the dynamical equilibrium point. Let us first of
 all define this point: it is obtained by iteration involving the following
 steps:
 \begin{enumerate}
 \item
 Starting from a set of trial values for $Q_a, R_3, S_3$ (for 
 example the ones corresponding to equilibrium at zero temperature for the
 uncoupled system, which are known), calculate the single particle energies 
 given by Eqs.~(\ref{eq:6:2600}) and (\ref{eq:6:2700}), or more explicitly
 \begin{eqnarray}
 \epsilon_{1+} &=& \frac{1}{2}  h_1\cos Q_1 
 -\frac{1}{2} [(2V_1+C_{11})\sin^2 Q_1 R_3  
 +(U+C_{12})\sin Q_1 \sin Q_2 S_3]  \label{eq:6:eps1} \\ 
 \epsilon_{2+} &=& \frac{1}{2}  h_2\cos Q_2 
 -\frac{1}{2}[ (2V_2+C_{22}) \sin^2 Q_2 S_3  
 +(U+C_{12})\sin Q_2 \sin Q_1 R_3 ].   \label{eq:6:eps2}
 \end{eqnarray}
 \item
 Calculate new values of $R_3, S_3$ from the equations
 \begin{subequations}
 \begin{eqnarray}
 R_3 &=& \frac{N_1}{2}\left[\frac{1}{1 + \exp(\frac{\epsilon_{1+}}{T})}
	 -\frac{1}{1 + \exp(-\frac{\epsilon_{1+}}{T})}\right], \label{eq:6:R3}\\
 S_3 &=& \frac{N_2}{2}\left[\frac{1}{1 + \exp(\frac{\epsilon_{2+}}{T})}
	 -\frac{1}{1 + \exp(-\frac{\epsilon_{2+}}{T})}\right].\label{eq:6:S3}
 \end{eqnarray}
 \end{subequations}
 (As previously remarked, in this model, the chemical potentials $\mu_a$ vanish.)
 \item
 Calculate new values of the coordinates from the equations
 \begin{subequations}
 \begin{eqnarray}
 \cos Q_1 &=& -\frac{h_1}{( 2V_1 + C_{11} ) R_3 + ( U + C_{12} ) S_3
 (\frac{\sin Q_2}{\sin Q_1})},\label{eq:6:Q1} 
 \\    
 \cos Q_2 &=& -\frac{h_2}{( 2V_2 + C_{22} ) S_3 + ( U + C_{12} ) R_3
 (\frac{\sin Q_1}{\sin Q_2})}.\label{eq:6:Q2}
 \end{eqnarray}
 \end{subequations}
 \item
 If these new values of $R_3, S_3, Q_1, Q_2$  coincide
 with the trial values, then we have the solution for the equilibrium
 point which we shall denote by the superscript zero; otherwise, return to
 step 1.
 \end{enumerate}

 Once the equilibrium point is obtained, we now define the surface
 $Q_1, Q_2$. Let us first move along $Q_1$. Starting from $Q_1^0, Q_2^0$, 
 increase $Q_1^0$ by $\Delta Q_1$ to a new point $Q_1=Q_1^0+\Delta Q_1, 
 Q_2=Q_2^0$. The solution for this point will be obtained by iteration
 involving the following steps, starting with the ingredients of the
 previous point.
 \begin{enumerate}
 \item
 In scenario 2, calculate new values of $R_3, S_3$ using 
 Eqs.~(\ref{eq:6:R3}) and (\ref{eq:6:S3}). Remove this step in scenario 1
 where these quantities are constant.
 \item
 Calculate new values of the single particle energies using
 Eqs.~(\ref{eq:6:eps1}) and (\ref{eq:6:eps2}).
 \item
 If these values do not agree with the inputs, return to step 1;
 otherwise continue to define the next $Q_1$ point. In this way, we
 get a path along $Q_1$ for a fixed value of $Q_2$.
 \item
 To generate the whole surface, come back to the equilibrium point 
 $Q_1^0, Q_2^0$ and in the same manner define a nearby point $Q_1^0,
 Q_2= Q_2^0 + \Delta Q_2$ from which, using the 3 steps above, we can define
 another path along $Q_1$ for the new value of $Q_2$. Thus, path by path, we 
 create a whole surface corresponding, in scenario 2, to a given value of
 the temperature and, in scenario 1, to the fixed values of $R_3, S_3$ of
 the equilibrium point. 
 \end{enumerate}

 The above procedure defines a number of macroscopic functions
 such as $B_a, R_3, S_3$ as well as the potential and total
 energies $V, W_{\mathrm{C}}$ which are functions of three variables at discrete 
 points of a net. These variables are $Q_a$ and the entropy $S$ for
 scenario 1, $Q_a,T$ for scenario 2.
 Values at intermediate points that are needed for the
 dynamical problem can be obtained by interpolation, using appropriate
 subroutines. The latter define at the same time the partial derivatives of
 these functions.

 We are now ready to discuss the initial value problem.
 Focusing on scenario 2, the problem to be solved is the following:
 Given at time $t=0$ a point
 in this 3-dimension space, say at $Q_a^i, T^i$ moving with initial velocities 
 defined by $P_a^i$, we want to study its motion, namely obtain $T, Q_a, P_a$
 at a later time $t$.  In order to use integration subroutines, we need the
 functions $\dot{T}, \dot{Q}_a, \dot{P}_a$. They are given explicitly by
 the following equations:
 \begin{enumerate}
 \item
 First of all, we have the two equations for $\dot{Q}_a$
 \begin{equation}
 \dot{Q}_a = B_a P_a,\, a=1,2. \label{eq:6:dotQ}
 \end{equation}
 \item
 The time variation of $T$ is given by Eq.~(\ref{eq:6:17.300}) namely
 \begin{equation}
 \dot{T} =
  \frac{\epsilon_1 \sum_a\frac{\partial{R}_3}{\partial{Q}_a} \dot{Q}_a
 +\epsilon_2 \sum_a\frac{\partial{S}_3}{\partial{Q}_a}\dot{Q}_a}
 {\epsilon_1 \frac{\partial{R}_3}{\partial{T}}+
 \epsilon_2 \frac{\partial{S}_3}{\partial{T}}}.\label{eq:6:Tdot}
 \end{equation}
 \item
 {}From the above equations, one has
 \begin{subequations}
 \begin{eqnarray}
 \dot{R}_3 &=& \frac{\partial{R}_3}{\partial{T}}\dot{T} +
 \sum_a \frac{\partial{R}_3}{\partial{Q}_a} \dot{Q}_a, \label{eq:6:dotR}  \\
 \dot{S}_3 &=& \frac{\partial{S}_3}{\partial{T}}\dot{T} +
 \sum_a \frac{\partial{S}_3}{\partial{Q}_a} \dot{Q}_a. \label{eq:6:dotS}
 \end{eqnarray}
 \end{subequations}
 \item
 Finally, the dynamical equations for $P_a$ are
 \begin{subequations}
 \begin{eqnarray}
 \dot{P}_1 &=& h_1 R_3\sin Q_1 + ( 2V_1+ C_{11} ) R^2_3 \sin Q_1\cos Q_1
 \nonumber\\
 && + ( U + C_{12} ) R_3 S_3 \cos Q_1 \sin Q_2 \nonumber\\   
 && - R_3 \cos Q_1  D_{11} (\cos Q_1 R_3 \dot{Q}_1 + \sin Q_1 \dot{R}_3)
 \nonumber\\    
 && -R_3 \cos Q_1   D_{12} (\cos Q_2 S_3 \dot{Q}_2 + \sin Q_2 \dot{S}_3), 
 \label{eq:6:dotP1} \\
 \dot{P}_2 &=& h_2 S_3\sin Q_2+ ( 2V_2+ C_{22} ) S^2_3 \sin Q_2\cos Q_2
 \nonumber\\    
 && + ( U + C_{12} ) R_3 S_3 \cos Q_2 \sin Q_1 \nonumber\\
 && - S_3 \cos Q_2  D_{22} (\cos Q_2 S_3 \dot{Q}_2 + \sin Q_2 \dot{S}_3)
 \nonumber\\    
 && - S_3 \cos Q_2  D_{12} (\cos Q_1 R_3 \dot{Q}_1 + \sin Q_1 \dot{R}_3). 
 \label{eq:6:dotP2}
 \end{eqnarray}
 \end{subequations}
 \end{enumerate}
 The above equations completely define the motion of the system 
 point. The equations for scenario 1 are simpler in that $R_3, S_3$
 are constant.

 \subsubsection{Numerical results and discussion\label{sec:6.2.5}}

     As a test, we have carried out the calculation using the following 
 parameters

 \begin{eqnarray}
 h_1 &=& 2 , \,\, h_2 = 1.5, \nonumber\\
 N_1 &=& N_2 = 30, \nonumber\\
 V_1 &=& V_2 = U = 0.10, \nonumber\\
 C_{11} &=& C_{22} = 0.30 \,, \, C_{12} = 0, \nonumber\\
 \omega_0 &=& 5.
 \end{eqnarray}
 We have deliberately chosen a model which is not symmetric between the two
 modes in which the coupling is defined by the parameter $U$ while the 
 dissipation is characterized by $C_{11}, C_{22}$ with $C_{12} = 0$. 
 (This parameter is somewhat redundant, in view of the existence of
 $U$ with which it appears as a sum in the formulas in the text.) 

\begin{figure}
 \centerline{\includegraphics[width=7cm]{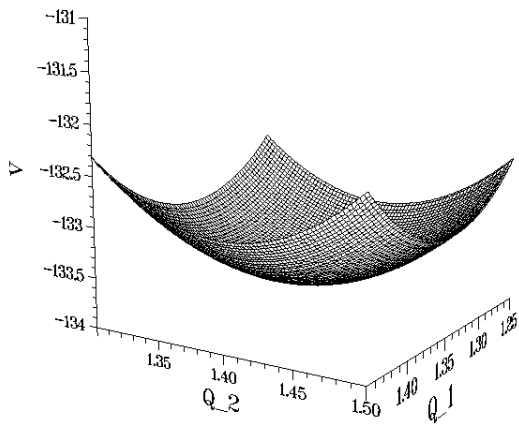}}
 \caption{The potential energy surface $V$ for scenario 1 corresponding to the
 temperature $T=1$ at equilibrium. The unit for the collective coordinates
 is $0.005$.
 \label{fig:6:1}}
 \end{figure}

\begin{figure}
 \centerline{\includegraphics[width=7cm]{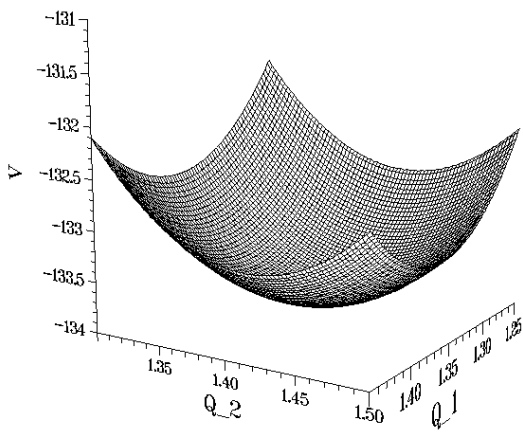}}
 \caption{ same as for Fig.~\ref{fig:6:1} for scenario 2.
 \label{fig:6:2}}
 \end{figure}

\begin{figure}
 \centerline{\includegraphics[width=7cm]{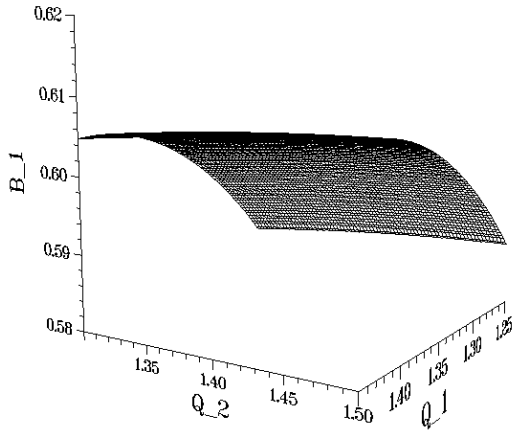}}
 \caption{The inverse mass $B_1$ for scenario 1 for $T=1$.
 \label{fig:6:3}}
 \end{figure}

\begin{figure}
 \centerline{\includegraphics[width=7cm]{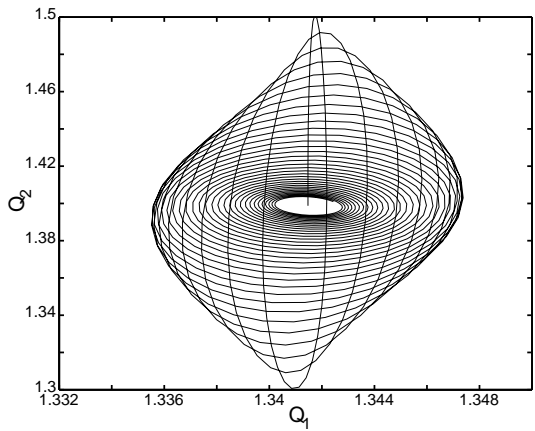}}
 \caption{The trajectory in coordinate space up to time $t=25$.
 \label{fig:6:4}}
 \end{figure}

\begin{figure}
 \centerline{\includegraphics[width=7cm]{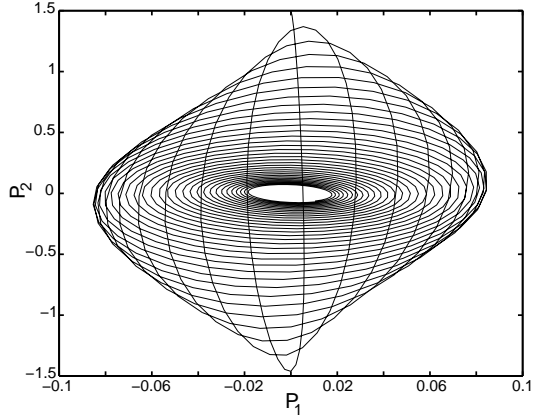}}
 \caption{Evolution of the momenta.
 \label{fig:6:5}}
 \end{figure}

\begin{figure}
 \centerline{\includegraphics[width=7cm]{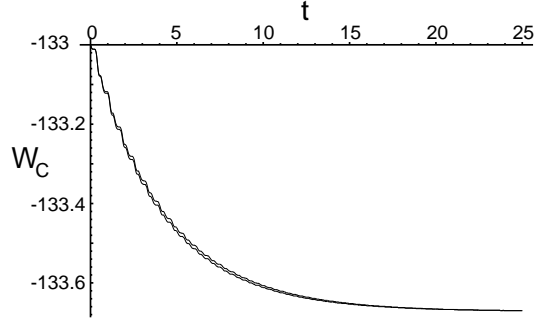}}
 \caption{Variation of the total collective energy $W_{\mathrm{C}}$ for the two scenarios.
 \label{fig:6:6}}
 \end{figure}

\begin{figure}
 \centerline{\includegraphics[width=7cm]{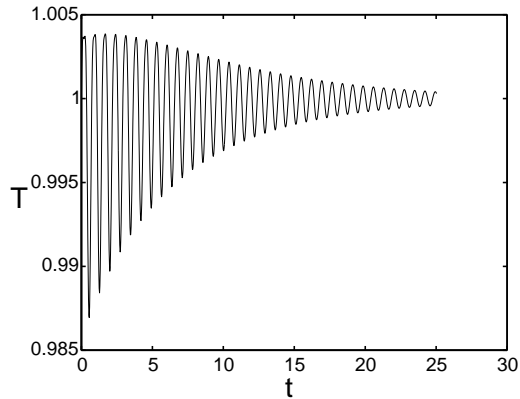}}
 \caption{Variation of the temperature $T$ with time. 
 \label{fig:6:7}}
 \end{figure}

     We assume that the motion starts from the equilibrium point
 corresponding to the chosen temperature, $T=1$, namely, with the
 given parameters,    
 $ Q_1^i = 1.3415, \, Q^i_2 = 1.3937 $, and that the initial values of momenta
 are $P_1^i=0$, $P_2^i=1.5$.   Some results of our calculation are shown in
 Figs.~\ref{fig:6:1}-\ref{fig:6:7}.

 Fig.~\ref{fig:6:1} and Fig.~\ref{fig:6:2} give the potential energy
 surfaces for $T=1$ for the two scenarios. Aside from the equilibrium
 point, there is no one-to-one correspondence between the coordinates
 $Q_1, Q_2$ for the two scenarios because one corresponds to a fixed
 $T$ while the other to a fixed set of values of $R_3, S_3$.  Notice
 that, for scenario 2, the equilibrium point, for non-zero temperature,
 is not the point of lowest value of the potential energy, but instead
 is the minimum of the free energy $ F = V - ST$.

 In Fig.~\ref{fig:6:3}, we display the inverse mass function $B_1$,
 which is seen to be a smooth, slowly varying function of the
 coordinates.  This provides some justification for the approximations
 we have made in deriving the final equations of motion.  See also the
 discussion in reference to Fig.~\ref{fig:6:6}.

 Figure \ref{fig:6:4} shows the trajectory of the system point. It starts from 
 the equilibrium point and after a long time will return
 at the end (we stop the calculation at $t=25$) to the same point which,
 as said above, is the one with lowest free energy. 

 Figure \ref{fig:6:5} shows how the momenta are transferred from one
 mode to the other. This motion will also stop after a long
 time. Figs.~\ref{fig:6:4} and \ref{fig:6:5} are obtained for scenario
 2.  Both figures demonstrate that we are dealing with underdamped
 motion, which was to be expected, since for our derivation to be
 justified the frictional forces must be weak compared to the
 conservative forces.  At this point it is necessary to explain why our
 equations predict that the system returns to its original starting
 temperature rather than to a higher one resulting from the dissipation
 of collective kinetic energy.  First we shall prove that it does so,
 and then we shall explain where this energy has gone.

 Consider first scenario 1, where we have not displayed the figures
 corresponding to Figs.~\ref{fig:6:4} and \ref{fig:6:5}, but where the
 end results are the same.  The adiabatic collective manifold is
 defined by the variables $Q_a$ and the entropy $S$, given as a
 function of the occupation numbers for our model by the formula,
 \begin{equation}
 S = - \sum_{a\pm} [n_{a\pm}\ln(n_{a\pm}) 
  +(1-n_{a\pm})\ln(1-n_{a\pm})].  \label{eq:6:ess}
 \end{equation}
 It is important to emphasize that though in general the $n_a$ can
 change, as long as $S$ remains constant, the mean-field approximation
 limits matters further to fixed sets of $n_{a\pm}$ determined by
 equilibrium configurations (in order to guarantee that $R_3$ and $S_3$
 remain constant).  Recall as well that the allowed values of $n_a$ are
 associated with equilibrium temperatures, $T_{{\rm eq}}$, and we might
 further put the same subscript on $S$ to emphasize the restriction on
 permitted values of $n_{a\pm}$.  We visualize the collective manifold
 by associating with each macroscopic function a surface in $Q_1$,
 $Q_2$ with each value of $S_{{\rm eq}}$ or, equivalently, $T_{{\rm
 eq}}$.  Figure \ref{fig:6:1} is such a representation for the
 mechanical energy, $V_{\mathrm{C}}$, which for our system may also be
 identified as the internal energy when
 expressed as a function of the $Q_a$ and $S$, whose minima determine
 the equilibrium values of $Q_a$.

 Now consider the relaxation phenomenon.  To be able to compare results for
 the two scenarios we have chosen identical initial conditions --  equilibrium
 values of variables defining the collective submanifold and an initial
 collective kinetic energy.  The system dissipates the kinetic energy
 as it also moves through non-equilibrium values of $V_{\mathrm{C}}$.  But in this
 mean field approximation it must at all times maintain the initially
 assigned value of the equilibrium entropy.  Therefore when it returns to
 equilibrium, it returns to the equilibrium configuration of the coordinates
 associated with the initial occupation numbers, namely the initial
 temperature.

 Where has the collective kinetic energy gone?  The answer is, in fact,
 trivial and could have been given without going through the thermodynamic
 argument of the previous paragraph.  We simply have to remember that
 according to Eqs.~(\ref{eq:6:R3}) and (\ref{eq:6:S3}), we
 are describing equilibrium by a canonical distribution, and therefore we have
 put the system into contact with a heat bath at temperature $T$.
 Consequently
 whenever we disturb the system from its equilibrium point simply by
 a collective kick (collective kinetic energy), this energy will eventually
 dissipate to the heat bath and the system return to its original
 equilibrium thermodynamic state.  But this implies that as long as we
 displace the system from equilibrium with the given choice of initial
 conditions, the final state will be the same independent of the choice of
 scenario, i.e., independent of the choice of master equations, and
 should hold for any choice of master equations that have the canonical
 distribution as their static solution.

 Matters are slightly more complicated if we start the system at a point
 with non-equilibrium values of the coordinates $Q_a$.  In the mean field
 scenario, we must,nevertheless, choose values of $R_3$ and $S_3$ that
 correspond to a common temperature, which is the final equilibrium
 temperature.  For scenario 2, we must specify a starting {\em local}
 temperature and let the dynamics tell us the final equilibrium temperature.

 Figure \ref{fig:6:6} gives the evolution of the total collective
 energies $W_{\mathrm{C}}$ for the two scenarios. The
 evolution is not quite the same but the near overlap of the two curves
 means that the total spaces (including the temperature) are almost
 identical. This may be understood by noting that the latter are
 defined starting from the same equilibrium points and furthermore that
 we are looking at the motion close enough to these points. The curves
 have been obtained using Eq.~(\ref{eq:6:17.600}).  We checked that the
 same values are obtained by direct integration of
 Eq.~(\ref{eq:6:flow}).  This reflects the fact that the inverse mass
 functions are weakly coordinate dependent. Figures
 \ref{fig:6:4},\ref{fig:6:5},\ref{fig:6:6} all show the dissipative
 character of the motion.

 Finally, Fig.~\ref{fig:6:7} exhibits the variation of the temperature
 $T$. It starts from $T=1$ and ends up at the same value when the
 motion stops.  The quasi-periodic structure of the curve reflects the
 underdamping.  One should not be surprised by the fact that local
 temperatures dip below its equilibrium value.  This can be understood
 as a consequence of Eq.~(\ref{eq:6:Tdot}).  Most simply, if we had
 reversed the sign of the initial momenta, the system would have moved
 immediately in the direction of smaller local temperature.

 In summary, we have made a number of compromises in the treatment of
 our model that suggest further studies to circumvent these
 limitations.  The simplest that comes to mind is to investigate the
 dynamics of the self-contained system without introducing the
 white-noise approximation.  For example, we can study the
 initial-value problem in which the initial energy resides completely
 in the collective subspace.  We then integrate the equations of motion
 to examine the flow of energy between that space and the
 non-collective space as the number of oscillators in the latter is
 increased.  We expect to find that for a sufficiently large
 non-collective space the collective energy all dissipates eventually,
 and we may even be able to extract effective frictional constants to
 describe average effects of this process.  For an initial study, it
 would suffice to take a simple Lipkin model in the mean-field
 approximation.

\newpage
\section{Survey of literature and summary}\label{sec:7}
 
\subsection{Survey of literature\label{sec:7.1}} 

In making the decision to write this review completely from the
unitary point of view of the authors, we have slighted a rich
development.  Much of this took place in the decade preceding our
initial efforts in the field. However, research of sometimes
considerable depth, if with narrower participation, has continued to
the present. The discussion which follows represents our attempt to
redress this imbalance. We shall attempt a survey of the literature,
that is relatively complete in its attributions, but relatively sparse
on specific details. From our narcissistic point of view, we can
divide the literature into three compartments, papers that have had an
impact on the field as a whole but very little influence on our own
efforts, papers not widely noticed that have been helpful to us, and
work that is both well-known and has been inspirational to us.  In the
course of the survey that follows, we shall try to acknowledge
indebtedness in so far as we are aware of it.

Though their approach has never been brought to full fruition in the
sense in which it was originally envisaged, the work of Baranger and
Veneroni \cite{no3} (the publication of their full report occurred a
number of years after their work was already widely known) must be
credited with stimulating much of the extensive activity and
development that took place in this field over a period of more than a
decade. Their special contribution was to introduce a new
decomposition of the density matrix, $\rho(t)$, associated with
time-dependent Hartree-Fock theory into time-even and time-odd parts
that could be identified as canonical coordinates and canonical
momenta, respectively, in the adiabatic limit. Specifically, they
introduced the decomposition
\begin{equation}
\rho(t) = \exp[i\chi(t)]\rho_0(t)\exp[-i\chi(t)]. \label{eq:7:16.1}
\end{equation}
Here $\rho_0(t)$ is a time-even Slater determinant appropriate for the
description of the unconstrained equilibrium shape of a system with an
even number of nucleons, or of the conditional equilibrium shape
achieved when shape constraints are imposed. In the adiabatic limit,
the exponential factors can be expanded, and it is then shown that the
Hartree-Fock equations have the form of Hamilton's classical equations
of motion with the matrix elements of $\rho_0$ serving as coordinates
and those of $\chi$ as canonical momenta. However the authors do not
reach the concept that the introduction of collective coordinates can
be associated with a canonical transformation, nor do they formulate a
useful algorithm for the construction of the collective subspace. They
do suggest a time-dependent calculation designed to identify
collective behavior.  In a useful paper \cite{42}, to
which we shall return, Goeke, Reinhard, and Rowe describe this
procedure, but find that it does not work when applied to a simple
(non-nuclear) model. We quote without discussion other papers that
have studied some aspects of adiabatic TDHF theory using the
decomposition (\ref{eq:7:16.1}) \cite{no18,no19,no20,no21}.

This decomposition was also
studied further in Ref.~\cite{no4}, in which it is pointed out that
the decomposition (\ref{eq:7:16.1}) is a useful starting point for the derivation
of the hydrodynamic limit of TDHF. Some of the authors of the latter 
report made an extensive effort to apply their ideas. The outcome of
this effort \cite{no45,no46,no47} 
was a method of computing what we have called the 
covariant collective mass without having to invert the given 
contravariant mass tensor, sometimes a formidable task in 
the nuclear physics context. 
This method is known as the double cranking
method, and was suggested earlier, e.g., by Villars \cite{no48}, though
not previously implemented. We outline the essential steps of this
method. We first remind the reader that to first order in the collective
momenta, the decoupling conditions have the form 
\begin{eqnarray}
{\cal H}_{ph} &=& -i\rho_{ph}^{1i}\partial_i V
 =\frac{\partial Q^i}{\partial\rho_{hp}} =f^i_{ph}\lambda_i,
 \label{eq:7:16.2} \\
B^{php'h'}f^i_{ph} &=& B^{ij}\partial_j\rho_{ph}, \label{eq:7:16.3} \\
B^{ij} &=& f^i_{ph}B^{php'h'}f^j_{p'h'}. \label{eq:7:16.4}
\end{eqnarray}

\begin{aside}
In passing, we describe a global method for (possibly) solving these equations that is
distinct from any that we have proposed previously. First solve 
the cranked Hartree-Fock equation (\ref{eq:7:16.2}) with prescribed cranking
operators $f^i_{ph}$. This yields a density matrix $\rho(Q^i)$, from
which derivatives with respect to $Q^i$ may be computed. From (\ref{eq:7:16.4})
a trial value of $B^{ij}$ may be obtained. Thus a trial value of the 
right hand side of (\ref{eq:7:16.3}) is known. By inverting the matrix
$B_{php'h'}$ we thus obtain improved values for the cranking operators,
\begin{eqnarray}
f^i_{ph} &=& B_{php'h'}B^{ij}\partial_j\rho_{p'h'}. \label{eq:7:16.5}
\end{eqnarray}
We can now cycle back and forth between (\ref{eq:7:16.2}) and (\ref{eq:7:16.3})
until convergence is hopefully achieved. It should be noted, however,
that in a test calculation in Ref.~\cite{42}, using the landscape model,
convergence could not be achieved with this method. 
\end{aside}

Now suppose we want to do a calculation without full self-consistency,
in particular one with prescribed cranking operators, chosen on the
basis of physical plausibility, as has been the case for most of the 
history of the subject. The question that arises is how we should
compute the collective mass tensor. The formula (\ref{eq:7:16.4}), using the 
prescribed cranking operators is not sufficiently accurate since it
does not, in general, contain the (self-consistent) cranking limit.
The alternative formula for the covariant mass tensor
\begin{eqnarray}
B_{ij} &=& \partial_i\rho_{ph}B_{php'h'}\partial_j\rho_{p'h'}, \label{eq:7:16.6}
\end{eqnarray}
can be shown to be the self-consistent cranking result provided the 
derivatives $\partial_i\rho_{ph}$ are computed from the solution of
(\ref{eq:7:16.2}). But this evaluation also requires the calculation of the
inverse of $B^{php'h'}$, and it is precisely the difficulty of this
calculation that discourages the application of the self-consistent
procedure that involves (\ref{eq:7:16.5}) as the essential step.

The work under discussion avoids the difficulty of inversion (but renounces
the goal of self-consistency) by the following procedure. We utilize
the formulas
\begin{eqnarray}
\dot{\rho} &=& \frac{\partial\rho}{\partial Q^i}\dot{Q^i}
 +\frac{\partial\rho}{\partial P_i}\dot{P}_i \label{eq:7:16.7}\\
\frac{\partial Q^i}{\partial\rho_{hp}} &=& -i\frac{\partial\rho_{ph}}
{\partial P_i}, \label{eq:7:16.8}
\end{eqnarray}
the second of which will be recognized as a canonicity condition previously
utilized, and apply the equation of motion for $\dot{P}_i$. As a consequence
the TDHF equation,
in the representation in which $\rho$ is diagonal (working always
to first order in velocity or momentum), takes the form
\begin{equation}
{\cal H}_{ph}-\lambda_j f^j_{ph}-i\dot{Q^j}\partial_j\rho_{ph} = 0.
\label{eq:7:16.9} 
\end{equation}
This equation may be called a double cranking equation with cranking
parameters $\lambda_i$ and $\dot{Q^i}$. It can be solved as follows:
First set $\dot{Q^i}$ to zero, so that we have the usual single cranking.
{}From the solution of this problem, we calculate $\partial_i \rho_{ph}$,
which defines the cranking operator associated with the Lagrange
multipliers $\dot{Q^i}$. For each value of $Q$ (suppressing indices)
we then crank on the
velocity. We thus obtain $\rho(Q,\dot{Q})$. For small $\dot{Q}$,
this should have the form
\begin{eqnarray}
\rho(Q,\dot{Q}) &=& \rho(Q) + \frac{\partial\rho}{\partial\dot{Q^i}}
 \dot{Q^i}, \label{eq:7:16.10} \\
\frac{\partial\rho_{ph}}{\partial\dot{Q}^j} &=& \frac{\partial\rho_{ph}}
{\partial P_i}\frac{\partial P_i}{\partial\dot{Q}^j}
=if^i_{ph} B^{ij}. \label{eq:7:16.11}
\end{eqnarray}
{}From (\ref{eq:7:16.11}) we can calculate $B^{ij}$.
The results of this formulation have been applied to the study of 
isoscalar quadrupole modes in light nuclei.

In this method, the
collective coordinate is prescribed, but the associated canonical momentum
is determined from the calculation. Because full self consistency is not
attempted, the choice of collective pair is not optimal.

The utility of deriving conditions for characterizing large amplitude
adiabatic collective motion by solving the TDHF equations as a power series
in the collective momenta was first recognized by Villars \cite{67}. In
his work, he derived the TDHF version of the three decoupling conditions,
namely the force, mass, and curvature conditions. However, he was
ambivalent about the curvature condition, emphasizing correctly that
to second order in the momentum, extended point transformations would
have to be considered. Since he confined his attention to point
transformations, he felt that the inclusion of this condition was
inconsistent. The equations he adopted are therefore summarized in
(\ref{eq:7:16.2})-(\ref{eq:7:16.4}). The global method he suggested for solving
these equations is precisely that described above. As already stated
this method was
tested on a simple model in \cite{42} and shown to be unstable.
Stabilization is achieved by converting these equations into the LHA.
This conversion has been
described repeatedly in this work, and is one of the foundations of our
program. In a further paper \cite{74},
the theory was extended to include coupling between the collective and
the non-collective spaces.

The influence of the concepts introduced by Villars on our own work is
manifest. Together with the less well-known papers \cite{32,33} which
emphasize the clarity that can be achieved by first considering large
amplitude collective motion within a purely classical framework, this
work has provided the theoretical framework within which our development
is contained.

Among the works also based on an expansion of TDHF in powers of the
collective momenta, we shall discuss the theoretical formulation
of Goeke, Reinhard, and their collaborators \cite{nu1,no6,no11,42}.
These authors also adopt as their basic equations the zero and first
order conditions (\ref{eq:7:16.2}) and (\ref{eq:7:16.3}). By eliminating
$f^i_{ph}$ between these two equations, they obtain the condition
\begin{equation}
B^{php'h'}{\cal H}_{p'h'} = \lambda_i B^{ij}\partial_j\rho_{ph}.
\label{eq:7:16.12}
\end{equation}
For the case of a single collective coordinate, the right hand side of
(\ref{eq:7:16.12}) becomes a constant times $\partial_Q\rho_{ph}$. The constant
can be absorbed into the scale of $Q$, and thus we obtain a first-order
differential equation to determine a change in the density matrix when we
change $Q$ by a small amount. However, if our starting point is not on
a collective path, the one-dimensional manifold that we generate will not
be the optimized path we seek. To find a suitable starting point, the
authors formulate a validity criterion specifying that such a point
should approximately satisfy the condition of second order in the
momentum that emerges from the fundamental analysis. They also show how to
solve (\ref{eq:7:16.12}) in such a way that starting at an arbitrary point, one
can descend to a point on the collective path. They have applied this
method with reasonable success to a number of examples
of ion-ion scattering at low energies
\cite{nu2,no13,12a,12b,no51,no52}. The agreement with experiment is
improved if some quantum corrections, obtained by the method of generator
coordinates, are included. The application of this method to the theory
of large amplitude collective motion will be discussed below.

A second body of work to which we are indebted is that of Rowe and
collaborators \cite{no7,34,no38}. The first of these papers contains,
as far as we know, the first formulation of the LHA without curvature,
associated with a stable local method for obtaining a solution that we have,
in effect, utilized in our work. The second paper cited is a careful
mathematical analysis of the properties of valleys and fall lines (lines
in the direction of the gradient of the potential) and a proof of why the
curvature criterion of Goeke and Reinhard, mentioned above, is relevant
for the determination of the collective path. Our formulation of the
concept(s) of multi-dimensional valleys came after a study
of this work. The purpose of Ref.~\cite{42} was to illustrate
the various methods, then known, for constructing a collective
path. After 
definitions of the proposals of Rowe-Basserman (LHA), Villars,
Goeke-Reinhard (GK), and Baranger-Veneroni, these are all applied to the 
landscape model in two dimensions, from which one aims to decouple one
collective mode. As discussed above, only the LHA and the method of
GK ``work''. The former can be applied to decouple a collective
manifold of any dimensionality (in principle), but the latter is restricted
in its application to one collective coordinate.

The idea of an expansion in powers of the momentum was also used 
effectively later in the work of Mukherjee and Pal \cite{no9,no9a}, 
who actually showed that the three 
decoupling conditions implied that a 
decoupled path must be a valley. This work was very influential on our 
own thinking, though in strict mathematical terms, it is a paraphrase 
of the earlier result of Rowe \cite{34}, 
who remarked that geodesic fall 
lines are decoupled. In the language of this review, this means that 
a path that satisfies the mass and force conditions and is also 
geodesic is decoupled, since the last condition is equivalent to also 
satisfying the curvature condition. Later work by these authors
\cite{no9b,no9c,no9d,no9e,no9f} is both contentious and displays no
real conceptual advance compared to the earlier work. For example, these
authors never realize that a decoupled manifold is perfectly well-defined
by the LHA, except that it is not a valley unless the curvature is
included. Also they never reached the conception that the theory as
formulated applies perfectly well to a multi-dimensional collective manifold.

We turn next to the accomplishments of Marumori, Sakata, and their
associates \cite{68,no40,no41,no54,no55,no56,no57}.  One can find in
their work the analogues of many of the concepts described in this
review: exactly and approximately decoupled manifolds, goodness of
decoupling condition, and stability condition, though the terminology
is often different from the one we use. Except for their early work,
which emphasized a concept called the ``Invariance Principle of the
Schr\"odinger Equation'', in practice a special treatment of the
variational principle for TDHF, the development has been parallel to
ours, as we have always emphasized.  (Their silence in this regard
has, on the other hand, been deafening.)  The algorithm associated
with their work, the self-consistent coordinate method (SCCM), is
local harmonic, closely tied to the TDHF formalism, and designed to
obtain a collective Hamiltonian to describe anharmonic vibrations.
Applications by these authors have been limited to schematic models,
but, as we shall discuss below, realistic applications of SCCM have
been carried out by others. Also, one of the present authors was
involved in an independent development of some of the early ideas of
Marumori within a completely quantum framework \cite{no62}.
An interesting comparison between the approaches by Sakata and Marumori, 
Goeke and Reinhard, and that by  Rowe, can be found in Ref.~\cite{PassosCruz},
which should be compared to our own work in Ref.~\cite{22}.

In their later work, which diverges completely from ours
\cite{no57a,no58,no59,no60,no61,no63a,no63b,no63c,no63d,no63e,no63f,no63g},
they undertook the program of studying TDHF as a non-separable
Hamiltonian system with the aim of incorporating advances in the
understanding of chaos into the nuclear problem.
They identified the
occurrence of resonant conditions under which the perturbative 
construction of the series of anharmonic terms 
failed to converge. Subsequently they 
modified their theory to take into account the transfer of energy 
between the collective and these 
special non-collective degrees of freedom. 
Finally, they have developed a version of transport theory to deal with 
this class of problems. The connection of their work to the more restricted
considerations presented in Sec.~\ref{sec:6} is not apparent.

Important theoretical contributions and some elementary applications have 
also been carried out by Yamamura and Kuriyama. In one series of papers 
\cite{no64,no65,no66,no66a,no67,no68}, they have also exploited the canonical
structure of TDHF and the essential role of canonical transformations
to introduce collective and intrinsic variables. Their work is,
however, still tied to the SCCM of Marumori, Sakata, {\it et al} and
to the anharmonic limit; no algorithm applicable to large amplitude
collective motion is given. In subsequent work \cite{no69,no70,no71},
they have carried through a deeper mathematical study of the same set
of ideas. All this work plus other uses of TDHF are reviewed in
Ref.~\cite{no71}, where one also finds a few applications. Following
this work, the authors, in collaboration with J.\ da Providencia,
undertook to investigate, using the methods of thermo field dynamics
\cite{no71a}, the application of TDHF to systems at finite excitation
energy. The development may be traced from
Refs.~\cite{no71b,no71c,no71d,no71e}, which represents the part of the
research most closely related to LACM.

The TDHF method in an expansion about equilibrium (and ultimately the SCCM)
have also been applied to the study of the wobbling motion of a triaxial
rotor by Kaneko \cite{Kaneko1,Kaneko2,Kaneko3}.
A related approach is discussed by Marshalek \cite{marsh96}.

We emphasize the parallelism of the basic theoretical structure in the
works of Yamamura and Kuriyama, of Marumori and Sakata, and of our
own efforts. Because of the way we have developed these ideas, both
theoretically and practically, the superficial reader may, however, encounter
difficulty in recognizing the overlap of ideas.

As already emphasized, there have been no applications of any of the
theories referred to in the paragraphs immediately preceding to
large-amplitude collective motion.  However, the SCCM, which is a
classical method for anharmonic vibrations that requires
requantization, has been applied very successfully to gamma vibrations
of deformed nuclei \cite{no72,no73,no74,Mat86,no76,no76a}. It has also
been applied to collective-noncollective quadrupole coupling
(anharmonic vibrations in the single-j shell model \cite{no77}, to the
description of two-phonon states in Ru and Se \cite{no77a}, and most
enticingly
\cite{FMM91}, in work
that cries for further study and extension, to a comparison of the
diabatic versus the adiabatic approximation. The model found in this
reference has been studied by our techniques in Ref.~\cite{30}, of which
no account has been included in this review. On the other hand in
Sec.~\ref{sec:4.4.1}, we have reviewed a study of a more general model incorporating
similar concepts. In the study of gamma
vibrations, particular attention was paid to the ambiguities of 
requantization, to the nature of the mode-mode coupling responsible for 
the anharmonicity in the gamma mode, 
and to the restoration of number conservation. Finally we
mention \cite{no78a} an application of SCCM to the spherical to
deformed phase transition in the Sm isotopes. Since this method cannot
deal with arbitrary large-amplitude distortions, the problem was
studied by expanding to fourth order in the quadrupole coordinates
and momenta.
Such a Hamiltonian is rich enough to describe a phase transition; whether
it is sufficiently accurate remains to be seen.
 
An elegant but formal study of the description of decoupled manifolds
within the framework of time-dependent Hartree-Bogoliubov theory and a
group structure dictated by the nuclear shell model has also been
carried out by Nishiyama and Komatsu \cite{no79,no80,no81,no81a,no82}.
Though they make no use of the canonical structure, their basic
approach can be understood by reference to the results found in
Sec.~\ref{sec:2.3} of this review, if we identify the local harmonic
equation developed there as {\em integrability} conditions for the
consistency of the decoupling conditions. We have emphasized how the
two sets of equations, consistency conditions and decoupling
conditions, must be solved in concert to provide integral surfaces
that are candidates for decoupled manifolds. In the work cited
\cite{no82}, only the power series option is suggested.
 
The concept of an approximately decoupled manifold occurs very naturally 
in the description of chemical reactions. In an adiabatic description, 
one can think of the reaction as involving the passage from a local 
minimum of the total potential energy of the reactants through a 
transition state that corresponds to a saddle point of the potential 
energy and then down to another local minimum corresponding to the 
final products. The reaction is thus described classically by a 
reaction path, corresponding to a one-dimensional approximately 
decoupled manifold \cite{no83}. An approximate Hamiltonian 
for the system, often sufficient for 
studying the given reaction, can be obtained by expanding the full 
Hamiltonian about the reaction path and retaining quadratic terms 
(the small oscillation approximation). The procedures to be found in the 
chemical literature differ only in the way the reaction path is 
determined. The literature on this subject is vast, but can be traced from
the papers to be cited. Most of the work
\cite{no83,no84,no85,40,no87,no88}
is based on a reaction path chosen as the fall line (direction of the
gradient of the potential) in both directions
from the saddle point. Thus there is a remarkable similarity to the 
method of Goeke and Reinhard described above. Quite recently several 
alternatives have been proposed. One method, the gradient extremal 
method \cite{no89,no90}, is tantamount to the LHA with the neglect of 
curvature corrections. A second proposal \cite{no91} is to use a 
suitably chosen straight line path of descent from the saddle point to 
each of the minima. This proposal is relevant to the situation that 
the fall line has large curvature, rendering it unlikely that the system 
will follow such a path. In this method, the kinetic energy can be 
brought to diagonal form. This means that the mass and curvature decoupling 
condition has been satisfied, at the expense, however, of a violation 
of the force condition.

Several field-theoretical applications of LACM techniques have also
been studied, in particular to tunneling
phenomena, and to the construction of collective coordinates for
solitons. The treatment of tunneling is discussed in a recent review
\cite{AHK+97::1}, where a valley approach is used, but in the action rather 
than in configuration space. The idea of using the LHA to generate 
the collective coordinates has been applied in Refs.~\cite{Walet1,Walet2}
to the Skyrme model. It was found to be a useful approach in the study of parts
of the collective manifold.

So far, all the work in nuclear physics quoted in this section has been
based on TDHF theory. Remarkably the first published derivation of the LHA,
that of Holzwarth and Yukawa \cite{60}, was based on the generator
coordinate method (GCM). In this method, we represent an eigenstate of
the many-body system in the form
\begin{eqnarray}
|n\rangle &=&\int dC F_n(C)|C\rangle, \label{eq:7:GC}
\end{eqnarray}
where $|C\rangle$ represents a continuous set of Slater determinants
parameterized by a set of generally complex coordinates $C$ (or real pairs
$Q,P$). Historically, the need for this doubling was discovered by the
application of this method to the problems of translational motion and to
the rotational motion of deformed nuclei \cite{no92}. For the translational
problem, for example, it was necessary to superpose not only
determinants centered at different points (center of mass coordinate), but
also to allow determinants with a continuous distribution of momenta of this
center of mass. Only then was it possible to extract from (\ref{eq:7:GC})
translational motion with the correct total mass. The general need for this
doubling was first fully explained in the works of Dobaczewski
\cite{no93,no94}.

In the traditional application of the GCM to the study of collective
motion, the manifold of states $|C\rangle$ is assumed known and the
state (\ref{eq:7:GC}) is considered as a trial function leading to the
well-known Hill-Wheeler \cite{HW53} integral equation for $F_n(C)$. To
be specific, consider the quadrupole degrees of freedom studied in the
Kumar-Baranger theory \cite{1a,1b,1c,1d,1e}. For this and for more
modern applications, the latter has been replaced by a two-step
procedure \cite{no96,no97}. In the first step, one constructs the
required manifold $|C\rangle$ as the solution of a constrained HF +
BCS problem with so-called time-even, i.e., velocity-independent,
constraints, such as the quadrupole moment and selected higher
moments.  This parallels the first step of the KB theory, except that
more realistic forces fitted to ground-state properties are
utilized. In the second step, one solves the Hill-Wheeler
equation. Compared to the procedures described in this review, and in
the bulk of the work noted in this section, it has the advantage that
no adiabatic approximation is necessary and thus application to
large-velocity phenomena is less questionable. On the other hand, the
description of the large-amplitude collective motion is not
self-consistent in our sense.  Furthermore, there remains a question
whether inertial properties of the collective degrees of freedom are
correctly described by this method, since we know from the simpler
examples of translation and rotation that in such cases the manifold
$|C\rangle$ needs to be constructed from a double cranking procedure
utilizing one velocity dependent constraint for each collective degree
of freedom. Nonetheless, an approach exists to construct a collective
potential and mass parameter from the GCM that has been widely applied,
see, e.g., \cite{BDFHM,Bobyk}.

Returning to the work of Holzwarth and Yukawa, these authors extended the
use of the state vector (\ref{eq:7:GC}) as a trial function by varying not
only the ``wave function'' $F_n(C)$, but also by varying the manifold
$|C\rangle$ so as to optimize its choice for each value of the set $C$.
Their argument, which we shall not reproduce, leads to the LHA without
curvature. Further study of the ideas introduced by HY has been carried
out by Goeke and Reinhard \cite{no98,no99,66}. Basically they have
shown that if one applies the generator
coordinate method to the space of Slater determinants that contains both
even and odd members, parameterized by pairs of
canonical coordinates, then by a suitable expansion in powers of the
momentum one essentially reaches the results obtainable from TDHF by
a momentum expansion. There is an additional bonus that zero point
fluctuations associated
with the collective coordinates can be computed; these were incorporated
in the applications to which we have previously alluded
\cite{no11,nu2,no13,12a,12b,no51,no52}.

For completeness, we remark on the occurrence of two sets of researches
on problems of large amplitude collective motion that utilize the path
integral method. One set involves H.\ Reinhardt and collaborators
(see the review in Ref.~\cite{no104a}),
and comprises two papers on the application of PIM to simple solvable
models \cite{no101,no102}, two formal papers deriving a theory of large
amplitude collective motion \cite{no103} and a theory of fission
\cite{no104}, and finally, one paper \cite{no105} showing how to extract
adiabatic TDHF by the path integral method. In the work of Levit, Negele,
{\it et al}, we encounter a formal development \cite{no106,no107}, which
then concentrates on the problem of fission \cite{no108}, culminating in
a method of solution that showed considerable promise \cite{no109,no110}.

In Sec.~\ref{sec:5} of this review, we have presented our version of what may
be termed the nuclear BO method. In this presentation, the basic ideas
have a clear association with the corresponding concepts utilized in
molecular physics. On the other hand our aim was to embed the results
previously derived from TDHF into a quantum framework. The only other
attempt to study large amplitude collective motion by a BO method was the
prior work of Villars \cite{no111}. The differences between this work and
ours appear to be greater than the similarities.
We both use basis vectors in Hilbert space
associated with localized collective and intrinsic operators. Villars
uses an expansion of the Hamiltonian in terms of collective momenta that
is an operator equivalent of the moment expansion. (We had, in fact noted
this relationship a long time ago \cite{no112}). On the other hand the
treatment of state vectors in \cite{no111} is faintly redolent of the
Hill-Wheeler representation. Villar's ideas have been applied with
success to an exactly solvable model \cite{no113} and to a fresh
theoretical study of rotations and vibrations \cite{no114a}.

We have also studied the effects of the BO approximation and the role
of the induced gauge fields, which are related to (avoided) level
crossings.  This area of physics has a long history, and is already
discussed in old work by Brink \cite{BrinkSchutte}. An interesting
modern perspective, in a framework similar to ours, can be found in
the work of Bulgac and Kusnezov
\cite{Bulgac1,BK1}.

\subsection{Summary of concepts and results\label{sec:7.2}} 
 
This review has been devoted first to the study of some 
properties of a classical Hamiltonian at most quadratic in the 
momenta with a non-singular coordinate-dependent 
mass tensor. After an elementary introduction, the theoretical structure
upon which most of the applications have so far been based was developed
in Sec.~\ref{sec:2}. The most general canonical
transformation that maintains this structure exactly is a point 
transformation; in so far as we restrict ourselves to such 
transformations, we are within the realm of Lagrangian dynamics.
On the other hand, if we view the Hamiltonian as
an approximate one resulting from an expansion of the exact Hamiltonian 
in powers of the momenta (adiabatic approximation), 
and think of a general canonical transformation 
similarly expanded, it is necessary to go one step beyond strictly point 
transformations to assure that final results do not depend on a 
particular initial choice of canonical coordinates. For want of a 
better term, we referred to such (approximate) canonical transformations 
as extended point transformations. 
 
Consider first strictly point transformations. Our aim was to discover 
if the given Hamiltonian admits motions confined to a submanifold of 
configuration space, called a decoupled manifold. 
The idea of such motions both generalizes the concept 
of separability and, for the case of a many-particle 
system, provides a natural definition of collective motion 
for systems where the original choice of coordinates corresponds to 
single-particle degrees of freedom. A parameterization of 
the manifold is provided by a subset (the collective subset) 
of a new set of coordinates 
introduced by a point canonical transformation, though it becomes 
immediately evident that any single choice of coordinates is not unique, the 
only invariant being the geometric structure itself. 
 
Our first strategic contribution was to insist on the separation of two 
sets of problems, one theoretical and the other practical. 
The first is that of stating precisely the conditions 
for a decoupled Lagrangian manifold to occur. There we found it 
necessary to distinguish between the case that none of the collective 
momenta is conserved and the case that one or more is. In the former 
instance, that engaged our full attention first, we found 
three conditions, 
of order zero, one, and two in the collective momentum, christened 
force, mass, and curvature condition, respectively. The mass and 
curvature conditions require that an exactly decoupled manifold be 
geodesic with respect to the mass tensor as metric. Stimulated in 
part by previous work, that we have generalized, 
we found {\em three} transformations of these conditions 
into exactly equivalent sets of conditions. In the first presented, 
the generalized valley formulation, the curvature condition was replaced 
by an infinite set of generalized force conditions. In the second 
transformation, we 
proved that the curvature condition can be replaced by an eigenvalue 
problem that generalizes the usual random phase approximation, 
reducing to it at critical points of the potential energy. This was 
called 
the Riemannian local harmonic formulation to distinguish it from the 
third formulation, the symplectic local harmonic formulation. In the 
latter it was shown that the force and mass conditions alone imply a 
different eigenvalue equation that replaces only the mass condition. 
The revisions of the formulations necessary when extended point 
transformations and/or additional constants of the motion must be 
considered were described rather more briefly than for the case 
considered above. 
 
Most Hamiltonians do not admit exactly decoupled manifolds. Yet the 
phenomenon of approximately decoupled collective motion is observed 
widely. From the point of view of the present work, these are motions 
that remain in some neighborhood of a submanifold either for a very 
long time or possibly forever. To uncover such manifolds, 
we described two different constructions or algorithms that 
provide submanifolds of preassigned dimensionality $K$ (in practice 
$K=1$ or 2). The nature of these manifolds must be such that they are 
exactly decoupled if the given Hamiltonian admits such motions. This was 
guaranteed by the way in which each algorithm is related to 
an associated formulation. Consider, 
for example, the one associated with the generalized valley formulation. 
Recall that the latter requires that 
an infinite number of generalized forces lie in the tangent plane to 
the decoupled manifold. In the algorithm, 
we replaced these \emph{consistency} 
conditions by a well-posed problem, namely, we defined a candidate 
surface of dimension $K$ to be the locus of points at which the 
``first" $K+1$ generalized forces determined a plane, generally 
distinct from the tangent plane. Various goodness-of-decoupling criteria 
suggested depend on the relative orientation of this plane with respect 
to the tangent plane, the two planes coinciding at each point if the 
decoupling is exact. 
 
The second algorithm was based on the local 
harmonic formulations, which require that the basis vectors of the 
tangent plane can chosen as eigenvectors of a second covariant 
derivative of the potential energy, the force itself being required 
as well to lie in this plane. The problem becomes well-posed 
when the plane attached to each point of a manifold, as determined 
from the solutions of the local harmonic equation, 
is not required to 
be tangent to the manifold. We showed, however, that for numerical 
purposes this algorithm could be formulated as a 
variant of the GVA. 
As illustrated in the numerical applications, different algorithms 
give different approximate results, with exceptions as noted. 
 
In addition to testing for goodness of decoupling, one must also test 
for local stability of motion. The latter is the more-or-less 
standard calculation, the same for both algorithms, 
of frequencies of small oscillation in directions 
normal to the manifold, to verify that they are real. In general such 
frequencies are position-independent only in the special case of 
separable motion; an imaginary frequency signals that $K$, the
dimensionality of the collective subspace, must be
increased.
 
Section III contains first applications of the general formalism to several
problems involving only a few total degrees of freedom. As a general
characterization, we may refer to these applications as the approximate
quantization of non-separable systems, in which we attempt, successfully
in all the examples considered, to decouple one or two collective
coordinates from a system with just a few additional degrees of freedom.
These applications include schematic models of nuclear fission and
chemical isomerization. It is clear that much more development and
application can be done along these lines.

However, in Sec.~\ref{sec:4} we turn to the problem that stimulated the theory
in the first place, the interest in LACM in nuclear physics. These
applications are based on TDHF where pairing can be neglected and on
TDHFB where it cannot. Our ability to apply the theory of Sec.~\ref{sec:2}
to these problems is based on the well-known property of these approximations
that they can be shown to have the classical Hamiltonian structure. The
formulas necessary for applications are first developed and then applied to
three sets of problems. We first successfully solve an exactly solvable
model of monopole vibrations and show how non-solvable extensions can be
studied. We then devote considerable attention to a study of the low lying
spectrum of $^{28}$Si. For reasons that have to do with the model chosen
-- the size of the configuration space and the form of the Hamiltonian --
the results of this investigation are only partially satisfactory.
Finally, in our third application, we lay the basis for a realistic
treatment of heavy nuclei designed to generalize the KB theory. In the light
of the results of this preliminary investigation and the remaining
contents of the review, this opens the most promising path for realistic
applications.

The theory developed in Sec.~\ref{sec:2} is a classical theory. The applications
to quantum mechanics are based on a requantization of a classical
collective Hamiltonian. In the course of our studies, we also remark on how
to include the next order of quantum corrections, which involve oscillations
of the non-collective coordinates. For the nuclear case, with its
Fermi-Dirac statistics, this involves a non-trivial new development whose
results we have quoted and applied. Both for the chemical isomerization
discussed in Sec.~\ref{sec:3} and for the case of Si, these corrections are of
considerable importance. In Sec.~\ref{sec:5}, we consider the basic theory anew,
from a quantum-mechanical viewpoint based on the study of the equations
of motion within the framework of a Born-Oppenheimer approximation.
In addition to regaining old results, we are able to incorporate the
effects of the Berry phase into our treatment. Several simplified models
that incorporate this new physics are worked out, but the inclusion
in realistic applications remains a task for the future.

Our final contribution, in Sec.~\ref{sec:6}, describes the generalization of the
theory of LACM to finite excitation energy so as to describe the exchange
of energy between collective and non-collective coordinates. Though
a formal structure has been elucidated and applied to a simple model,
it is clear that much more work is required before we can hope to apply
this theory to realistic cases.

We add a few words about future prospects.  The main reason for
optimism concerning the immediate future of our subject is that the ideas
explored in Sec.\ IV for heavy nuclei indicate that the class of problems
that stimulated the researches in this field in the first place, the low-energy spectroscopy of
deformed and transitional nuclei, is finally amenable
to an improved treatment compared to previous non-selfconsistent studies.

\section*{Acknowledgements}

We would like to acknowledge the important contribution of our
collaborators to the work reported here, especially Dr. Aurel Bulgac,
who made major contributions to the early stages of the work, and
Dr. Takashi Nakatsukasa, who was instrumental in the applications to
more realistic nuclear models.

This work was supported by a research grant (GR/L22331) from the
Engineering and Physical Sciences Research Council (EPSRC) of Great
Britain, and through a grant (PN 98.044) from ALLIANCE, the
Franco-British Joint Research programme.  The Laboratoire de Physique
Th\'eorique is a Unit\'e Mixte de Recherche du C.N.R.S., UMR 8627.

\appendix
\renewcommand{\theequation}{\Alph{section}.\arabic{equation}}
\setcounter{equation}{0}

\newpage
\section{Choice of canonical variables in the Hamiltonian formulation
of adiabatic time-dependent Hartree-Fock theory}\label{sec:A}

\subsection{Introduction\label{sec:A.1}}

In Sec.~\ref{sec:4.1}, we have provided a lexicon for the
transcription of the Hamiltonian theory of large amplitude collective
motion developed in Sec.~\ref{sec:2} into the language of TDHF. In
Sec.~\ref{sec:4.4}, we also required a corresponding relationship for
time-dependent Hartree-Fock-Bogoliubov theory (TDHFB). In this
appendix, we shall survey some of the ways of carrying out this
program. The identification can be made both independently of its
interest for LACM, using several different choices of canonical
variables \cite{43,43a,43b,44,44a}, as well as within the framework of
LACM \cite{no7,68,no3,no4,19}. In our own work, we have utilized
either the method developed in Refs.~\cite{43,44,44a}
or an even simpler method based on a observation first made in Ref.~\cite{19}.
Part of the material that follows is based on unpublished work of the
authors \cite{BKD1}.

Superficially, we can divide the choices into two sets. In one, as in
Ref.~\cite{43b}, the canonical variables can be associated directly with
time-dependent single-particle orbitals. In the others, the parameterization
is associated with the one-particle density matrix. We shall first address
briefly the sets that we do not use, and afterwards go into more detail on
those that play a role in our work.

\subsection{Kerman-Koonin variables\label{sec:A.2}} 
 
We want to recognize the TDHF equations as a form of Hamilton's equations.
It is convenient to start
from the variational formulation of the TDHF. The action $S$ has the form
\begin{eqnarray}
S &=& \int dt\, {\cal L}(t) 
= \int dt\,  \langle\Phi(t)| (i\partial_{t}-H)|\Phi(t)\rangle \nonumber \\ 
&=& \int dt \left\{
\sum_{h} [\langle\varphi_{h}(t)| i\partial_{t}|\varphi_{h}(t)\rangle
-W(\varphi,\varphi^{*})]
\right\} ,\label{eq:A:bu1}
\end{eqnarray}
where $\Phi(t)$ is a normalized Slater determinant, $\varphi_{h}(t)$ 
are the occupied 
single particle wave functions (s.p.w.f.), $H$ is the total Hamiltonian, 
and $W(\varphi,\varphi^{*})$ is the Hartree-Fock energy for the
determinant $\Phi(t)$ and thus a functional of the orbitals
$\varphi_{h}(t)$. Only the time dependence is explicitly 
shown. For applications to nuclear physics, we shall understand here
and in future sections that we are dealing with a non-relativistic
Hamiltonian of the standard form
\begin{eqnarray}
H &=& \sum_{a,b}h_{ab}a_a^{\dag}a_b
+\frac{1}{4}V_{abcd}a_a^{\dag}a_b^{\dag}
a_da_c, \label{eq:A:bu2}
\end{eqnarray}
where $a_\alpha,\,a_\alpha^{\dag}$ are destruction, creation operators for
nucleons in an arbitrary basis of single-particle states and $h$ and $V$
are Hermitian matrices, with $V_{\beta\alpha\gamma\delta}
=V_{\alpha\beta\delta\gamma}=-V_{\alpha\beta\gamma\delta}$. Also $h$ may be
the kinetic energy, or as is more likely for the applications that we shall
carry out, also include a potential term.

The TDHF equations are obtained from the variational principle
$\delta S=0$. By varying with respect to $\varphi_h^{\ast}$, for example,
we obtain the TDHF equation for the orbitals $\varphi_h$,
\begin{eqnarray}
i\dot{\varphi}_{h}(t) &=& {\cal H}(t)\varphi_{h}(t). \label{eq:A:bu3}
\end{eqnarray}
In the combination of coordinate and spin-isospin labels, summarized
as ``$x$'', ${\cal H}$ is the integral operator
\begin{subequations}
\begin{eqnarray}
{\cal H}(x,x';t) &=& \delta W/\delta\rho(x',x;t), \label{eq:A:bu4} \\
\rho(x,x';t) &=& \sum_h \varphi_h(x,t)\varphi_h^{\ast}(x',t), \label{eq:A:bu5}\\
W &=& \trace h\rho +\frac{1}{2}\trace V\rho\rho. \label{eq:A:bu6}
\end{eqnarray}
\end{subequations}
It follows that Eq.~(\ref{eq:A:bu3}) takes the form
\begin{eqnarray}
i\dot{\varphi}_h(x,t) &=& (\delta W)/\delta\varphi_h^{\ast}(x,t),
\label{eq:A:bu7}
\end{eqnarray}
which demonstrates that $\varphi_h, \varphi_h^{\ast}$ are complex
canonical coordinates. We decompose these into real coordinates and momenta
leading to Hamilton's equations in real form and to an altered Lagrangian,
\begin{subequations}
\begin{eqnarray}
\varphi(x,t) &=& \frac{1}{\sqrt{2}}\left(\xi_h(x,t)+i\pi_h(x,t)\right), \label{eq:A:bu8} \\
\dot{\xi}_h(x,t) &=& \delta W/\delta\pi_h(x,t), \label{eq:A:bu9} \\
\dot{\pi}_h(x,t) &=& -\delta W/\delta\xi_h(x,t), \label{eq:A:bu10} \\
{\cal L} &=& \sum_k\int dx\,\pi_h(x)\dot{\xi}_h(x)-W(\xi,\pi)
+ \text{ total time derivative}. \label{eq:A:bu11}
\end{eqnarray}
\end{subequations}
Equations (\ref{eq:A:bu9}) and (\ref{eq:A:bu10}) follow either by transformation
from (\ref{eq:A:bu7}) and its complex conjugate or by applying the variational
principle to (\ref{eq:A:bu11}).
 
{}From (\ref{eq:A:bu5}) we compute
\begin{eqnarray}
\rho(x,y,t) && =\sum_{h}\frac{1}{2}\left[\xi_{h}(x,t)\xi_{h}(y,t)+ 
\pi_{h}(x,t)\pi_{h}(y,t)\right] \nonumber \\ 
&& +\frac{i}{2}\sum_{h}\left[\pi_{h}(x,t)\xi_{h}(y,t)- 
\xi_{h}(x,t)\pi_{h}(y,t)\right] . \label{eq:A:bu12}
\end{eqnarray}
Therefore the time even part of $\rho$ is real and the time odd part is 
imaginary.

We shall be interested in the small-momentum limit of (\ref{eq:A:bu12}). We thus
write
\begin{eqnarray}
\rho(x,y;t) &=& \rho_{0}(x,y;t) +\rho_{1}(x,y;t) + \cdots \nonumber \\
&=& \sum_h (1/2)\xi_h(x,t)\xi_h(y,t) \nonumber \\
&& +(i/2)\sum_{h}[\pi_{h}(x,t)\xi_{h}(y,t)-
\xi_{h}(x,t)\pi_{h}(y,t)]+\cdots, \label{eq:A:bu13} \\
\rho_{0}^{2}(x,y) &=& \rho_{0}(x,y)=\frac{1}{2}\sum_{h}\xi_{h}(x)\xi_{h}(y).
\label{eq:A:bu14}
\end{eqnarray}
In lowest order, then, $(1/\sqrt{2})\xi_a(x.t),\,a=(h,p),$ are a set of
orthonormal functions, and from the equality $\rho^2=\rho$ applied to the
expansion (\ref{eq:A:bu13}) we learn that to first order the expansion of
$\pi_h(x,t)$ in terms of the complete set $\xi_a$ contains only unoccupied
orbitals,
\begin{eqnarray}
\pi_h(x,t) &=& \sum_p \xi_p(x,t)\pi_{ph}. \label{eq:A:bu15}
\end{eqnarray}
We shall not pursue the physics of these variables any further, since they
are not used in this work.

\subsection{Polar representation\label{sec:A.3}}
 
Another interesting set of canonical variables is obtained from 
a polar decomposition of the s.p.w.f.
\begin{equation} 
\varphi_{h}(x)=\tilde{\rho}^{1/2}_{h}(x)\exp[i\tilde{\pi}_{h}(x)],
\label{eq:A:bu16}
\end{equation}
with both $\tilde{\rho}^{1/2}_{h}(x)$ and $\tilde{\pi}_{h}(x)$ real.

\begin{aside}
There is an important ambiguity in defining these
quantities  when the s.p.w.f.\ $\varphi_{h}(x)$ 
has nodes. If one chooses
$\tilde{\rho}_{h}(x)=|\varphi_{h}(x)|^{2}$, 
the function $\tilde{\pi}_{h}(x)$ 
will be discontinuous at the nodes. However, if one {\it defines} 
the phase $\tilde{\pi}_{h}(x)$ as a 
continuous function,  $\tilde{\rho}^{1/2}_{h}(x)$ 
 must  change sign at the
nodes in order to compensate. Then, $\tilde{\pi}_{h}(x)$ 
 will represent the ``real'' phase of the 
s.p.w.f.\ and will not feature any jumps at the nodes.
\end{aside}

One way of establishing that $\tilde{\rho}_h$ and $\tilde{\pi}_h$ are
canonical variables is by showing that the Lagrange density (\ref{eq:A:bu11}),
when expressed in terms of the new variables, takes the form
\begin{equation}
{\cal L}= 
\sum_{h}\int dx\, \tilde{\pi}_{h}(x) \dot{\tilde{\rho}}_{}(x)- W 
(\tilde{\rho},\tilde{\pi}) +\text{total time derivative}. \label{eq:A:bu17}
\end{equation}
{}From this expression it is clear that 
$\tilde{\pi}_{k}(x)$ and $\tilde{\rho}_{k}(x)$ 
play the role of canonically 
conjugate momenta and coordinates for the TDHF problem. 
Another way is to show that the equations of transformation
\begin{eqnarray}
\xi_h &=& \sqrt{2}\tilde{\rho}_h^{{1}/{2}}\cos\tilde{\pi}, \label{eq:A:bu18}\\
\pi_h &=& \sqrt{2}\tilde{\rho}_h^{{1}/{2}}\sin\tilde{\pi}, \label{eq:A:bu19}
\end{eqnarray}
satisfy the Poisson bracket relations. In the small-momentum limit, these
equations become
\begin{eqnarray}
\xi_h &=& \sqrt{2}\tilde{\rho}^{{1}/{2}}, \label{eq:A:bu20} \\
\pi_h &=& \sqrt{2}\tilde{\rho}^{{1}/{2}}\tilde{\pi}_h. \label{eq:A:bu21}
\end{eqnarray}
By comparison with Eqs.~(\ref{eq:2:2.2}) and (\ref{eq:2:2.5}), we see
that these are the equations of a point transformation.

The polar representation under discussion can be gainfully employed
as a basis for a nuclear fluid dynamical description of collective motion
\cite{Hol1}, but entering this domain will take us too far afield of our
main purposes. The special case that the phase function is independent
of orbit, $\tilde{\pi}_n(x,t) =\chi(x,t)$ is a special case of the choice of
variables studied next, and has been shown to provide an elegant basis
for the study of irrotational flow \cite{GVVB}.

\subsection{Canonical variables of Baranger and Veneroni\label{sec:A.4}} 
 
The approach of Baranger and Veneroni (BV) \cite{no3,no4} 
to ATDHF is based entirely on the one body density matrix, 
$\rho (x,y;t)$. They 
proved that for any Slater determinant $\rho$ has the unique decomposition
\begin{equation}
\rho=\exp(i\chi)\rho_{0}\exp(-i\chi), \label{eq:A:bu22}
\end{equation}
where $\rho_{0}$ is a time even operator, and $\chi$ is a time-odd
operator. Notice that the equation 
$\rho^{2}_{0}=\rho_{0}$ imposes no conditions on $\chi$. In fact, different
choices for $\chi$ correspond to different physical situations. For example,
we have already noted above that choosing $\chi$ as a local operator
(point function) is the appropriate choice for the description of
irrotational flow. For the study of large amplitude collective motion in the
adiabatic limit, the appropriate choice is that $\chi$ have only $ph$
matrix elements in the representation in which $\rho_0$ is diagonal, as
expressed by the equation
\begin{equation} 
\rho_{0}\chi+\chi\rho_{0}=\chi. \label{eq:A:bu23}
\end{equation}

In the adiabatic limit we consider $\chi$ to be a first-order
quantity.  In this limit, we now demonstrate that $\rho_0(x,y;t)$ and
$\chi(x,y;t)$ are canonically conjugate pairs. (As implied by our
discussion above, this cannot be true generally.) Here we identify
$\rho_0$ with the quantity defined in (\ref{eq:A:bu14}). Comparing the
first-order term of (\ref{eq:A:bu13}) with the first order expansion
of (\ref{eq:A:bu22}), we conclude that
\begin{equation}
\pi_h(x,t) =\int dy\, \chi(x,y;t)\xi_h(y,t) \label{eq:A:bu24}
\end{equation}
This allows us to carry out a transformation of the essential kinetic term
of the variational principle, as follows
\begin{eqnarray}
\sum_h \int dx\,\pi_h(x,t)\dot{\xi}_h(x,t) &=& \sum_h \int dxdy\, \chi(x,y;t)
\xi_h(y,t)\dot{\xi}_h(x,t) \nonumber\\
&=& \int dxdy\,\chi(x,y;t)\dot{\rho}_0(y,x), \label{eq:A:bu25}
\end{eqnarray}
which follows if we remember that $\chi$ is a real symmetric matrix.
The last form establishes that $\chi$ and $\rho_0$ are canonically
conjugate in the adiabatic limit.

\subsection{Marshalek-Weneser and Blaizot-Orland variables\label{sec:A.5}} 
 
{}From the theorem of Thouless we know that any Slater determinant,
$|z\rangle$, that is not orthogonal to a given one, $|0\rangle$, can be
written in the (unnormalized) form 
\begin{eqnarray} 
| z\rangle &=& \exp [\sum_{ph}z_{ph}a_{p}^{\dag}a_{h}]| 0\rangle,
\label{eq:A:bu26} \\
a_{p}| 0\rangle &=& a_{h}^{\dag}| 0\rangle=0, \label{eq:A:bu27}
\end{eqnarray}
with $z_{ph}$ a set of complex numbers that will serve as variational
parameters to derive a version of TDHF. However, as we shall see below,
these are not canonical coordinates. Clearly, $h$ refers to the orbitals
occupied in $|0\rangle$ and $p$ to the complementary set.

{}From the definition of the density matrix element,
\begin{equation}
\rho_{ab} =\langle z|a_b^{\dag}a_a|z\rangle/\langle z|z\rangle, \label{eq:A:bu27a}
\end{equation}
we can derive the formulas (written in terms of a rectangular matrix $z$
composed of the complex numbers $z_{ph}$)
\begin{subequations}
\begin{eqnarray}
\rho_{ph} &=& [(1+zz^{\dag})^{-1}z]_{ph}=\rho_{hp}^{\ast}, \label{eq:A:bu28} \\
\rho_{hh'} &=& [(1+z^{\dag}z)^{-1}]_{hh'}, \label{eq:A:bu29} \\
\rho_{pp'} &=& [z(1+z^{\dag}z)^{-1}z^{\dag}]_{pp'}. \label{eq:A:bu30}
\end{eqnarray}
\end{subequations}

\begin{aside}
 We indicate a simple method for deriving these formulas.
By utilizing the definition (\ref{eq:A:bu26}), the properties (\ref{eq:A:bu27}),
and the operator commutation relations, one can readily derive the
equations
\begin{subequations}
\begin{eqnarray}
\rho_{hh'} &=& \delta_{hh'} -\sum_p\rho_{hp}z_{ph}, \label{eq:A:bu31} \\
\rho_{hp} &=& \sum_{h'}\rho_{hh'}z^{\dag}_{h'p}. \label{eq:A:bu32}
\end{eqnarray}
\end{subequations}
Together these equations imply (\ref{eq:A:bu28}) and (\ref{eq:A:bu29}). Starting
from the definition of $\rho_{pp'}$, we can derive a corresponding pair
that yield (\ref{eq:A:bu28}) and (\ref{eq:A:bu30}). 
\end{aside}

Turning to the variational principle, we remember that we should utilize
a normalized trial function,
\begin{equation}
|\Phi(z)\rangle = |z\rangle/\langle
 z|z\rangle^{{1}/{2}}. \label{eq:A:bu32a}
\end{equation}
A relatively simple calculation then yields the result
\begin{equation}
\int dt\, \langle\Phi(z)|i\partial_t|\Phi(z)\rangle =
\int dt\, \frac{i}{2}\trace (\rho \dot{z} -\dot{z}^{\dag}\rho), \label{eq:A:bu33}
\end{equation}
where $\trace AB =\sum_{hp}A_{hp}B_{ph}$ and the needed matrix elements of
$\rho$ are provided by (\ref{eq:A:bu28})-(\ref{eq:A:bu30}). 
It can be verified that the equations of motion for $z$ and
$z^{\dag}$ that follow from the variational principle
are not of Hamiltonian form. (Equation (\ref{eq:A:bu33}) does not
agree with Eq.~(9.118) of Ref.~\cite{44a}. The result given there follows
if one neglects the time dependence of the normalization factor in
(\ref{eq:A:bu32a}).)

A transformation that replaces the matrix $z$ by a matrix $\beta$ that
defines canonical pairs is, in fact, well-known \cite{43,44,44a}.
The transformation
\begin{equation}
z = \beta(1-\beta^{\dag}\beta)^{-{1/2}} \label{eq:A:bu34}
\end{equation}
replaces (\ref{eq:A:bu28})-(\ref{eq:A:bu30}) by the formulas
\begin{subequations}
\begin{eqnarray}
\rho_{ph} &=& [\beta(1-\beta^{\dag}\beta)^{{1/2}}]_{ph}
=\rho_{hp}^{\dag}, \label{eq:A:bu35} \\
\rho_{hh'} &=& [1-\beta^{\dag}\beta]_{hh'}, \label{eq:A:bu36} \\
\rho_{pp'} &=& [\beta\beta^{\dag}]_{pp'}. \label{eq:A:bu37}
\end{eqnarray}
\end{subequations}
Actually Eqs.~(\ref{eq:A:bu35})-(\ref{eq:A:bu37}), as a mapping of density matrix
elements onto complex canonical coordinates was first derived as the
classical limit of a ``boson mapping'' from the algebra of all pairs
$a^{\dag}a$ to particle bosons, as reviewed in \cite{39}.

Defining $|\Psi(\beta)\rangle=|\Phi(z)\rangle$, we can transform
(\ref{eq:A:bu33}) into the form
\begin{equation}
\int dt\, \langle\Psi(\beta)|i\partial_t|\Psi(\beta)\rangle =
\int dt\, \frac{i}{2} \trace (\beta^{\dag}\dot{\beta}
-\beta\dot{\beta}^{\dag}). \label{eq:A:bu38}
\end{equation}
This result clearly establishes $\beta$ and $\beta^{\ast}$ as complex
canonical coordinates. To apply this choice to LACM, we introduce real
canonical coordinates according to the standard equation
\begin{equation}
\beta_{ph} = \frac{1}{\sqrt{2}}(\xi_{ph} + i\pi_{ph}). \label{eq:A:bu39}
\end{equation}
We then expand the mean field energy to second order in the variables
$\pi$ in order to obtain a classical Hamiltonian quadratic in the momenta
as a starting point for the theory developed in Sec.~\ref{sec:2}.

The theory developed here can be extended to TDHFB. This extension is
carried out in Sec.~\ref{sec:4.4.2}, where it is first needed, and will
not be discussed here. We shall instead pass on to our last choice of
canonical variables, in some ways simplest of all.

\subsection{Canonical coordinates associated with instantaneous natural
orbitals\label{sec:A.6}}

Consider the TDHF equation
\begin{equation}
i\dot{\rho}_{ab} = [{\cal H},\rho]_{ab}. \label{eq:A:bu40}
\end{equation}
We choose to study this equation in the representation in which $\rho$ is 
instantaneously diagonal,
\begin{equation}
\rho_{ab} = \delta_{ab}\delta_{ah}, \label{eq:A:bu41}
\end{equation}
and consider the particle-hole matrix elements of (\ref{eq:A:bu40}) in this 
representation, utilizing (\ref{eq:A:bu41}). We thus find
\begin{equation}
i\dot{\rho}_{ph} = {\cal H}_{ph} = \delta H/\delta\rho_{hp}, \label{eq:A:bu42}
\end{equation}
that together with the complex conjugate equation identify $\rho_{ph}$
and $\rho_{hp}=\rho_{ph}^{\ast}$ as complex canonical coordinates.

For purposes of application, these variables may be used in a fashion that is
quite analogous to what we have described for the variables introduced
in the previous subsection. We set (though the left hand side is formally
zero)
\begin{equation}
\rho_{ph} = (1/\sqrt{2})(\xi_{ph} + i\pi_{ph}), \label{eq:A:bu43}
\end{equation}
and expand the mean field energy in powers of $\pi$ to second order in 
$\pi$ in order to establish a starting point for the theory developed in
Sec.~\ref{sec:2}. We state without proof that to this order the variables
under discussion are indistinguishable from those 
introduced in the previous subsection,
though this identity will break down in higher order. For this reason, the
further development of the theory and applications in Sec.~\ref{sec:4}
utilized the present choice, even though
the references on which they are based used the alternative choice.

We conclude this appendix with an account of how the considerations just
mentioned can be generalized to the TDHFB.
Toward this end
we introduce a set of quasi-particle (qp) operators $b_k^{\dag}$
related to the particle operators by a general HFB transformation
\begin{subequations}
\begin{eqnarray}
a_a^{\dag} &=& U_{ak}^{\ast}b_k^{\dag} +V_{ak}b_k, \label{eq:A:hfb1} \\
b_k^{\dag} &=& U_{ak}a_a^{\dag} + V_{ak}a_a, \label{eq:A:hfb2} \\
U^{\dag}U +V^{\dag}V &=& 1, \;\; UU^{\dag} +V^{\ast}V^T=1, \label{eq:A:com1} \\
U^T V+V^T U &=& 0, \;\; UV^{\dag} +V^{\ast}U^T =0, \label{eq:A:com2}
\end{eqnarray}
\end{subequations}
and define the qp vacuum state $|\Psi\rangle$ by the conditions
\begin{equation}
b_k|\Psi\rangle = 0. \label{eq:A:vac}
\end{equation}

To reach our goal as expeditiously as possible, we study the expectation
value in the state $|\Psi\rangle$ of the equation of motion for the
product $b_i b_j$. With the definition
\begin{equation}
K_{ji} =\langle\Psi|b_i b_j|\Psi\rangle, \label{eq:A:def1}
\end{equation}
we evaluate the right hand side of the equation
\begin{equation}
i\partial_t K_{ji} = \langle\Psi|[b_i b_j, H]|\Psi\rangle \label{eq:A:eom1}
\end{equation}
by reexpressing $H$ in normal form with respect to the qp operators and
imposing the condition (\ref{eq:A:vac}) and its hermitian conjugate. To
conserve particle number on the average, it is understood that we have
displaced the single-particle matrix elements according to
\begin{equation}
\epsilon_{ab} \rightarrow \epsilon_{ab} - \mu\delta_{ab}, \label{eq:A:mu}
\end{equation}
with $\mu$ the chemical potential.

The decomposition of the Hamiltonian (\ref{eq:A:bu2}) takes the form
\begin{eqnarray}
H &=& H^0 +H_{ij}^{11}b_i^{\dag}b_j +\frac{1}{2}[H_{ij}^{20}b_i^{\dag}
b_j^{\dag} +{\rm h.c.}] \nonumber \\
&& + \frac{1}{4}H_{ijkl}^{22}b_i^{\dag}b_j^{\dag}b_l b_k
+[H_{ijkl}^{31}b_i^{\dag}b_j^{\dag}b_k^{\dag}b_l + {\rm h.c.}] \nonumber \\
&& +[H_{ijkl}^{40}b_i^{\dag}b_j^{\dag}b_k^{\dag}b_l^{\dag} + {\rm h.c.}]
\label{eq:A:aa.1}
\end{eqnarray}
With the definitions
\begin{subequations}
\begin{eqnarray}
h &=& \epsilon + \Gamma, \label{eq:A:def2} \\
\Gamma_{ab} &=& V_{acbd}\rho_{dc} , \label{eq:A:def3} \\
\Delta_{ab} &=& \frac{1}{2}V_{abcd}\kappa_{cd}, \label{eq:A:def4} \\
\rho_{ab} &=& \langle\Psi|a_b^{\dag}a_a|\Psi\rangle =(V^{\ast}V^T)_{ab},
\label{eq:A:def5} \\
\kappa_{ba} &=& \langle\Psi|a_a a_b|\Psi\rangle =(V^{\ast}U^T)_{ba}
=-\kappa_{ab}, \label{eq:A:def6}
\end{eqnarray}
\end{subequations}
we obtain the following expressions for the coefficients in (\ref{eq:A:aa.1}),
\begin{subequations}
\begin{eqnarray}
H^0 &=& Tr [\epsilon\rho +\frac{1}{2}\Gamma\rho-\frac{1}{2}\Delta
\kappa^{\ast}], \label{eq:A:aa.2} \\
H^{11} &=& U^{\dag}hU - V^{\dag}h^T V +U^{\dag}\Delta V
-V^{\dag}\Delta^{\ast}U, \label{eq:A:aa.3} \\
H^{20} &=& 2U^{\dag}hV^{\ast} +U^{\dag}\Delta U^{\ast}
-V^{\dag}\Delta^{\ast}V^{\ast}, \label{eq:A:aa.4} \\
H^{40}_{ijkl} &=&\frac{1}{4}V_{abcd}U_{ai}^{\ast}U_{bj}^{\ast}
V_{dk}^{\ast}V_{cl}^{\ast}, \label{eq:A:aa.5} \\
H^{31}_{ijkl} &=& \frac{1}{2}V_{abcd}[U_{ai}^{\ast}V_{bl}V_{ck}^{\ast}
V_{dj}^{\ast} + U_{ai}^{\ast}U_{bj}^{\ast}U_{cl}V_{dk}^{\ast}],
\label{eq:A:aa.6}\\
H^{22}_{ijkl} &=& V_{abcd}[U_{ai}^{\ast}U_{bj}^{\ast}U_{ck}U_{dl}\nonumber\\
&& +V_{al}V_{bk}V_{cj}^{\ast}V_{di}^{\ast}
 +4V_{al}U_{bi}^{\ast}V_{dj}^{\ast}U_{ck}]. \label{eq:A:aa.7}
\end{eqnarray}
\end{subequations}
The evaluation of (\ref{eq:A:eom1}) thus yields the result
\begin{equation}
i\partial_t K_{ji} = H_{ji}^{20}. \label{eq:A:eom2}
\end{equation}

The final task of this argument is to show that (\ref{eq:A:eom2}) can be
recognized as
a form of Hamilton's equations of motion. Toward this end, we study the
variation of the mean-field energy $H^0$. According to Eq.~(\ref{eq:A:aa.2}),
we need expressions for $\delta\rho$ and $\delta\kappa$. To evaluate
these quantities, which are averages of particle operators, we transform
to qp operators and recognize one further result. We notice that not
only is the quantity $\langle b_j^{\dag}b_i\rangle=0$, where the average is
understood for the qp vacuum, but also
\begin{equation}
i\partial_t \langle b_j^{\dag}b_i\rangle = 0
\end{equation}
in consequence of the equations of motion and the definition of the
vacuum. Thus in our approximation the quantity in question may be taken
to be a vanishing constant. It follows from their definitions
(\ref{eq:A:def5}) and (\ref{eq:A:def6}) that
\begin{subequations}
\begin{eqnarray}
\delta\rho_{ba} &=& U_{ak}^{\ast}V_{bl}^{\ast}\delta K_{kl}^{\ast}
+V_{ak}U_{bl}\delta K_{lk}, \label{eq:A:delta1} \\
\delta\kappa_{ba} &=& V_{ak}^{\ast}V_{bl}^{\ast}\delta K_{kl}^{\ast}
+U_{ak}U_{bl}\delta K_{lk}. \label{eq:A:delta2}
\end{eqnarray}
\end{subequations}

By means of these relations we find that
\begin{equation}
2\delta H^0 = H_{kl}^{20}\delta K_{kl}^{\ast}
 +(H_{kl}^{20})^{\ast} \delta K_{kl}. \label{eq:A:delta}
\end{equation}
With the definition
\begin{equation}
{\cal K}_{kl} = (1/\sqrt{2})K_{kl},
\end{equation}
We have thus derived the equations of motion
\begin{equation}
i\partial_t {\cal K}_{kl} =(\delta H^0/\delta {\cal K}_{kl}^{\ast}),
\label{eq:A:eom3}
\end{equation}
establishing that ${\cal K}_{kl}$ and ${\cal K}_{kl}^{\ast}$ are pairs
of complex canonical variables, with $H^0$ playing the role of Hamiltonian.

\end{document}